\documentclass [b5paper,10pt,onecolumn,twoside,final] {book}

\usepackage [dvips] {graphicx}
\usepackage [footnotesize] {caption}
\usepackage [ps2pdf,bookmarks=true,bookmarksnumbered=true] {hyperref}

\newcommand {\MM} [1] {\ensuremath{#1}}

\newcommand {\tsp} [1] {\ensuremath{\mskip #1\thinmuskip}}

\newcommand {\IT} [1] {\MM{\ifmmode{#1}\else{\textit{#1}}\fi}}
\newcommand {\RM}  [1] {\MM{\ifmmode{\mathrm{#1}}\else{#1}\fi}}

\newcommand {\DIV} [2] {\MM{#1/#2}}
\newcommand {\SUB} [2] {\MM{#1\ensuremath{_{#2}}}}
\newcommand {\SUP} [2] {\MM{#1\ensuremath{^{#2}}}}
\newcommand {\SQRT}[1] {\MM{\ifmmode{\sqrt{#1}}\else{\ensuremath{\surd}#1}\fi}}
\newcommand {\ABS} [1] {\MM{\ensuremath{|}#1\ensuremath{|}}}
\newcommand {\FRAC}[2] {\MM{\ifmmode{\frac{#1}{#2}}\else{\SUP{\SUB{}{#2}}{\SUB{#1}{---}}}\fi}}

\newcommand {\DIVa}[2] {\MM{#1\ifmmode{\!}\else{}\fi/#2}}
\newcommand {\DIVb}[2] {\MM{#1/\ifmmode{\!}\else{}\fi#2}}
\newcommand {\DIVc}[2] {\MM{#1\ifmmode{\!}\else{}\fi/\ifmmode{\!}\else{}\fi#2}}
\newcommand {\DIVd}[2] {\MM{#1/\tsp{-0.5}#2}}
\newcommand {\DIVe}[2] {\MM{#1\ifmmode{\!}\else{}\fi/\tsp{-0.3}#2}}

\newcommand {\etameson} {\ensuremath{\eta}}
\newcommand {\mass} {\IT{m}}
\newcommand {\Mass} {\IT{M}}
\newcommand {\momentum} {\IT{p}}
\newcommand {\energy} {\IT{E}}
\newcommand {\temperature} {\IT{T}}
\newcommand {\pressure} {\IT{P}}
\newcommand {\Number} {\IT{N}}
\newcommand {\nucleon} {\IT{N}}
\newcommand {\edens} {\ensuremath{\varepsilon}}
\newcommand {\ndof} {\IT{g}}
\newcommand {\ndofq} {\SUB{\ndof}{\RM{q}}}
\newcommand {\ndofg} {\SUB{\ndof}{\RM{g}}}
\newcommand {\nflav} {\SUB{\IT{n}}{\RM{f}}}
\newcommand {\transverse} {\IT{T}}
\newcommand {\longitudinal} {\IT{L}}
\newcommand {\forward} {\IT{F}}
\newcommand {\pT} {\SUB{\momentum}{\transverse}}
\newcommand {\mT} {\SUB{\mass}{\transverse}}
\newcommand {\pL} {\SUB{\momentum}{\longitudinal}}
\newcommand {\eT} {\SUB{\energy}{\transverse}}
\newcommand {\eF} {\SUB{\energy}{\forward}}
\newcommand {\Tc} {\SUB{\temperature}{\RM{c}}}
\newcommand {\cspeed} {\IT{c}}

\newcommand {\femto} {\RM{f}}
\newcommand {\pico} {\RM{p}}
\newcommand {\nano} {\RM{n}}
\newcommand {\micro} {\ensuremath{\mu}}
\newcommand {\milli} {\RM{m}}
\newcommand {\centi} {\RM{c}}
\newcommand {\kilo} {\RM{k}}
\newcommand {\Mega} {\RM{M}}
\newcommand {\Giga} {\RM{G}}

\newcommand {\second} {\RM{s}}
\newcommand {\muA} {\micro\RM{A}}
\newcommand {\mus} {\micro\second}
\newcommand {\ns} {\nano\second}
\newcommand {\MOhm} {\Mega\ensuremath{\Omega}}
\newcommand {\degree} {\SUP{}{\circ}}
\newcommand {\Kelvin} {\degree\RM{K}}
\newcommand {\meter} {\RM{m}}
\newcommand {\fm} {\femto\meter}
\newcommand {\mum}{\micro\meter}
\newcommand {\mm} {\milli\meter}
\newcommand {\cm} {\centi\meter}
\newcommand {\km} {\kilo\meter}

\newcommand {\cmsquare} {\SUP{\cm}{2}}
\newcommand {\Volt} {\RM{V}}
\newcommand {\kV} {\kilo\Volt}
\newcommand {\MV} {\Mega\Volt}
\newcommand {\Vcm} {\DIVa{\Volt}{\cm}}
\newcommand {\barn} {\RM{b}}
\newcommand {\mb} {\milli\barn}

\newcommand {\pb} {\pico\barn}

\newcommand {\pbinv} {\SUP{\pb}{-1}}

\newcommand {\eV} {\tsp{0.1}\RM{e}\tsp{-0.3}\Volt}
\newcommand {\GeV} {\Giga\eV}
\newcommand {\GeVc} {\DIVe{\GeV}{\cspeed}}
\newcommand {\GeVcinv} {\SUP{(\GeVc)}{-1}}
\newcommand {\GeVcc} {\DIVe{\GeV}{\SUP{\cspeed}{2}}}
\newcommand {\GeVfmcube} {\DIVa{\GeV}{\SUP{\fm}{3}}}
\newcommand {\MeV} {\Mega\eV}
\newcommand {\MeVc} {\DIVe{\MeV}{\cspeed}}
\newcommand {\MeVcc} {\DIVe{\MeV}{\SUP{\cspeed}{2}}}
\newcommand {\megabyte} {\Mega\RM{B}}
\newcommand {\Tesla} {\RM{T}}
\newcommand {\cmmus} {\DIVd{\cm}{\mus}}
\newcommand {\rad} {\RM{rad}}
\newcommand {\mrad} {\milli\rad}
\newcommand {\Hz} {\RM{Hz}}
\newcommand {\MHz} {\Mega\Hz}
\newcommand {\fmc} {\DIVd{\fm}{\cspeed}}

\newcommand {\SP} {\ifmmode{\:}\else{\ }\fi}
\newcommand {\unitns} [1] {\MM{#1}}
\newcommand {\unit} [1] {\MM{\SP#1}}

\newcommand {\xcoord} {\IT{x}}
\newcommand {\ycoord} {\IT{y}}
\newcommand {\zcoord} {\IT{z}}
\newcommand {\der} {\RM{d}}
\newcommand {\dEdx} {\DIV{\der\energy}{\der\xcoord}}
\newcommand {\zvertex} {\SUB{\zcoord}{\RM{vertex}}}

\newcommand {\Znum} {\RM{Z}}
\newcommand {\Anum} {\RM{A}}
\newcommand {\Nch} {\SUB{\Number}{\RM{ch}}}
\newcommand {\Npart} {\SUB{\Number}{\RM{part}}}
\newcommand {\Ncoll} {\SUB{\Number}{\RM{coll}}}
\newcommand {\Ncollmean} {\ensuremath{\langle}\Ncoll\ensuremath{\rangle}}
\newcommand {\impact} {\IT{b}}
\newcommand {\rapidity} {\IT{y}}
\newcommand {\pseudorapidity} {\ensuremath{\eta}}
\newcommand {\tauzero} {\SUB{\ensuremath{\tau}}{0}}
\newcommand {\phiangle} {\ensuremath{\phi}}
\newcommand {\etacoord} {\ensuremath{\eta}}
\newcommand {\etacoordabs} {\ABS{\etacoord}}
\newcommand {\zvertexabs} {\ABS{\zvertex}}
\newcommand {\gama} {\ensuremath{\gamma}}
\newcommand {\EPS} {\ensuremath{\varepsilon}}
\newcommand {\Mgg} {\SUB{\Mass}{\gama\gama}}
\newcommand {\LESS} {\ensuremath{<}}
\newcommand {\LESSEQ} {\ensuremath{\leq}}
\newcommand {\GREATER} {\ensuremath{>}}
\newcommand {\GREATEREQ} {\ensuremath{\geq}}
\newcommand {\PLMN} {\ensuremath{\pm}}
\newcommand {\DELTA} {\ensuremath{\Delta}}
\newcommand {\TIMES} {\ensuremath{\times}}
\newcommand {\PI} {\ensuremath{\pi}}
\newcommand {\TO} {\ensuremath{\to}}
\newcommand {\SIGMA} {\ensuremath{\sigma}}

\newcommand {\piminus} {\SUP{\PI}{-}}
\newcommand {\piplusminus} {\SUP{\PI}{\PLMN}}
\newcommand {\pizero} {\SUP{\PI}{0}}
\newcommand {\svar} {\IT{s}}
\newcommand {\tvar} {\IT{t}}
\newcommand {\sbar} {\ensuremath{\bar{\svar}}}
\newcommand {\tbar} {\ensuremath{\bar{\tvar}}}
\newcommand {\sNN} {\SQRT{\SUB{\svar}{\nucleon\nucleon}}}
\newcommand {\momentumbold}{\textbf{\em p}}

\newcommand {\electron} {\IT{e}}
\newcommand {\proton} {\IT{p}}
\newcommand {\neutron} {\IT{n}}
\newcommand {\antiproton} {\ensuremath{\bar\proton}}
\newcommand {\antineutron} {\ensuremath{\bar\neutron}}
\newcommand {\deuteron} {\IT{d}}
\newcommand {\copper} {\RM{Cu}}
\newcommand {\copperion} {\SUP{\copper}{29+}}
\newcommand {\gold} {\RM{Au}}
\newcommand {\goldion} {\SUP{\gold}{79+}}

\newcommand {\collision} [2] {\MM{#1\ifmmode{ }\else{\ }\fi+\ifmmode{ }\else{\ }\fi#2}}

\newcommand {\protonproton} {\collision{\proton}{\proton}}
\newcommand {\antiprotonproton} {\collision{\antiproton}{\proton}}
\newcommand {\deuterongold} {\collision{\deuteron}{\gold}}
\newcommand {\goldgold} {\collision{\gold}{\gold}}
\newcommand {\coppercopper} {\collision{\copper}{\copper}}
\newcommand {\nucleonnucleon} {\collision{\nucleon}{\nucleon}}

\newcommand {\yr} [1] {#1}

\newcommand {\SMD} {\RM{SMD}}
\newcommand {\SMDe} {\SMD-\etacoord}
\newcommand {\SMDp} {\SMD-\phiangle}

\newcommand {\ADC} {\RM{ADC}}
\newcommand {\ADCtower} {\SUB{\ADC}{\RM{tower}}}
\newcommand {\ADCHighTower} {\SUP{\SUB{\ADC}{\RM{trigger}}}{\RM{HighTower}}}
\newcommand {\ADCPatchSum}  {\SUP{\SUB{\ADC}{\RM{trigger}}}{\RM{PatchSum}}}
\newcommand {\MUX} {\RM{HT6}}
\newcommand {\LUT} {\RM{LUT}}
\newcommand {\PED} {\RM{PED}}
\newcommand {\LUTPED} {\SUB{\PED}{\LUT}}
\newcommand {\PedestalShift} {\RM{Pedestal\tsp{0.5}Shift}}

\newcommand {\rBeamBg} {\IT{r}}

\newcommand {\FTPCRefMult} {\SUB{\Number}{\RM{FTPC}}}

\newcommand {\levelZero} {level-0}
\newcommand {\levelTwo} {level-2}

\newcommand {\Rcp} {\SUB{\IT{R}}{\IT{CP}}}
\newcommand {\RAB} {\SUB{\IT{R}}{\IT{AB}}}
\newcommand {\RAA} {\SUB{\IT{R}}{\IT{AA}}}
\newcommand {\RdA} {\SUB{\IT{R}}{\IT{dA}}}
\newcommand {\TAB} {\SUB{\IT{T}}{\IT{AB}}}
\newcommand {\TABmean} {\ensuremath{\langle}\TAB\ensuremath{\rangle}}

\newcommand {\ee} [1] {\SUP{10}{#1}}
\newcommand {\e} [1] {\MM{\TIMES\ee{#1}}}

\newcommand {\Eseed}{\SUB{\energy}{\RM{seed}}}
\newcommand {\Eadd} {\SUB{\energy}{\RM{add}}}
\newcommand {\Emin} {\SUB{\energy}{\RM{min}}}
\newcommand {\Nmax} {\SUB{\Number}{\RM{max}}}

\newcommand {\etatopi} {\DIV{\etameson}{\pizero}}
\newcommand {\epluseminus} {\MM{\SUP{\electron}{+}\SUP{\electron}{-}}}

\newcommand {\HighTowerOne} {HighTower-\tsp{-0.4}1}
\newcommand {\HighTowerTwo} {HighTower-\tsp{-0.3}2}

\hypersetup {
pdfauthor = {Oleksandr Grebenyuk},
pdftitle = {Neutral meson production in d+Au and p+p collisions at sqrt(s_NN) = 200 GeV in STAR},
pdfsubject = {Neutral meson production in high energy proton-proton and deuteron-gold collisions measured by the STAR experiment
at the Relativistic Heavy Ion Collider (RHIC)},
pdfkeywords = {relativistic heavy ion collisions,high transverse momentum,neutral pion production,nuclear modification factor},
pdfcreator = {LaTeX with hyperref package},
pdfproducer = {dvips + ps2pdf}
}

\begin {document}

\pagenumbering {alph}
\setcounter {page} {0}

\pagestyle {empty}

\begin {center}

{
\bf \boldmath \Large
\mbox {Neutral meson production in \deuterongold\ and \protonproton}
\mbox {collisions at \MM{\sNN{} = 200\unit{\GeV}} in STAR}
\unboldmath
}

\vfill

{\bf\large\mbox{Oleksandr Grebenyuk\rule{4.5pt}{0pt}}}

\mbox {\rule{0pt}{10.6pt}\ }

\end {center}

\pagebreak

\mbox{ }

\vfill

{
}

\newpage

\begin {center}

%
%
%
%
%
%
%
%
%
%
%

{
\bf \boldmath \Large
\mbox {Neutral meson production in \deuterongold\ and \protonproton}
\mbox {collisions at \MM{\sNN{} = 200\unit{\GeV}} in STAR}
\unboldmath
}

\vspace {48 pt}

{
\bf \boldmath \large
\mbox {Productie van neutrale mesonen in \deuterongold\ en \protonproton}
\mbox {botsingen bij \MM{\sNN{} = 200\unit{\GeV}} in STAR}
\unboldmath
}

\mbox {(met een samenvatting in het Nederlands)}

\vfill

{
\bf \large
\mbox {Proefschrift}
}

{
\sc
ter verkrijging van de graad van doctor

aan de Universiteit Utrecht

op gezag van de rector magnificus, prof.~dr.~J.C.~Stoof,

ingevolge het besluit van het college voor promoties

in het openbaar te verdedigen op

donderdag 29 november 2007 des ochtends te 10.30 uur

door

}

{
\bf \large
\mbox {Oleksandr Grebenyuk}
}

\mbox {geboren op 27 november 1980 te Ufa, Rusland}

\end {center}

\pagebreak

\mbox {Promotor: Prof. dr. Th. Peitzmann}

\mbox {Co\tsp{0.3}-promotor: Dr. M.\tsp{1.0}A.\tsp{1.0}J. Botje}

\vfill

{
\footnotesize \noindent
Dit werk maakt deel uit van het onderzoekprogramma van de Stichting voor Fundamenteel Onderzoek der Materie (FOM), financieel gesteund door de Nederlandse Organisatie voor Wetenschappelijk Onderzoek (NWO).
}

\addtolength {\oddsidemargin} {1cm}
\addtolength {\evensidemargin} {-1cm}

\pagestyle {plain}
\pagenumbering {roman}
\setcounter {page} {0}

\addtolength {\textheight} {20pt}
\tableofcontents
\addtolength {\textheight} {-20pt}

\cleardoublepage
\pagestyle {headings}
\pagenumbering {arabic}
\setcounter {page} {1}

\chapter {Heavy Ion Physics}

\section {Introduction} \label {se:introduction}

Colliding heavy ions in particle accelerators offers a unique
opportunity to study the strong interaction of matter in the regime of
extremely high densities and temperatures. It is believed that in such
collisions temperatures and densities are reached that prevailed in the
universe the first few microseconds after the Big Bang.

In the Standard Model of particle physics, the strong interactions
between the fundamental quark constituents of matter are described by
a field theory called Quantum Chromo Dynamics (QCD)~\cite {ref_qcd}\@.
In this theory, the quarks carry a strong charge, called \textit{color}, and the
strong force is mediated between the colored quarks by the exchange of
gluons, which are the quanta of the strong field. A very important
feature of QCD is that the gluons also carry color charge, so that they
do not only act as mediators but also themselves couple to the strong
force. It turns out that, as a consequence, the potential increases
with increasing distance between the color charges.  This is in sharp
contrast with the field theory of Quantum Electro Dynamics
(QED)~\cite {ref_qed}, where the force between electrically charged
particles is mediated by the electrically neutral photon. 
Here the potential vanishes for large distances.

The behavior of the strong coupling with varying distance,
which is related to the behavior of the potential as discussed above,
has profound phenomenological consequences. 
First, the coupling between the colored
quarks becomes weak at short distances, a property called \textit{asymptotic freedom}\@.
Such short distances are probed in hard scattering processes,
where the momentum exchange between the participating quarks is large.
Since the strong coupling is weak in the hard regime, the interaction
cross sections can be calculated in a framework called
\textit{perturbative QCD} (pQCD)\@. Because they are calculable, hard scattering processes 
form a unique probe of the constituents of matter,
while being, at the same time, a testing ground for the validity of
QCD\@. In this way, QCD has been firmly established as the correct
theory of the strong interaction in the last four decades by
performing a large variety of experiments on deep inelastic scattering
of electrons and muons on protons and neutrons and by the study of
electron-positron and proton-(anti)proton collisions at large centre
of mass energies in storage rings.

The strong coupling increases with the distance between the quarks, and
the interaction becomes, in fact, so strong, that in ordinary matter
the quarks are permanently confined to colorless hadrons. In this
regime of large distances \linebreak
or, equivalently, small momentum transfers,
pQCD breaks down, so that it \linebreak
cannot be used to calculate soft
scattering cross sections from first principles. \linebreak
However, recently much progress has been made in the understanding of the \linebreak
non-perturbative domain by so\tsp{0.3}-called \textit{lattice QCD} calculations, where
the QCD field equations are numerically solved on a discrete
space\tsp{0.3}-time lattice~\cite {ref_latticeqcd}\@.

\begin {figure} [tb]
\begin {center}
\includegraphics [width=0.8\textwidth] {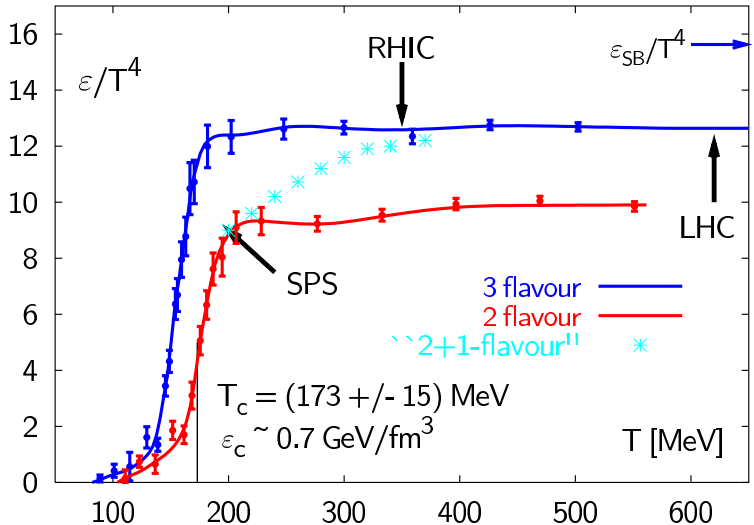}
\end {center}
\caption {The energy density \edens\ scaled by \MM{\SUP{\temperature}{4}}
  calculated from lattice QCD with \MM{(2,3)} degenerate quark flavors, as
  well as with two light and one heavy (strange) quark. The arrow on
  the right\tsp{0.2}-hand side shows the Stefan-\tsp{-0.2}Boltzmann limit for an ideal
  quark-gluon gas. The critical temperature \Tc\ and the
  temperatures which can presumably be reached by RHIC and LHC are also
  indicated.  Figure taken from~\cite {ref_karsch}\@.}
\label {fig:epsilon}
\end {figure}

One of the remarkable results of lattice QCD is the prediction that
hadronic matter at sufficiently high temperatures and densities will
undergo a phase transition to a state of quasi-free quarks and gluons. 
This deconfined dense state of matter is called a \textit{Quark Gluon Plasma} (QGP)\@. 
In Figure~\ref {fig:epsilon}
are shown lattice QCD calculations of the energy density~(\edens) 
divided by the fourth power of the temperature~(\temperature)~\cite {ref_karsch}\@. 
This dimensionless quantity is proportional to the effective number of degrees of freedom available in the medium. 
Below the critical temperature (\Tc) the medium consists mainly of confined hadrons, while above \Tc\ the
quarks and gluons become deconfined, causing a rapid increase in the number of degrees of freedom. 
Figure~\ref {fig:epsilon} shows that the phase transition occurs when nuclear matter is heated to a temperature
\Tc\ of about 175\unit{\MeV}, corresponding to an energy density of 0.7\unit{\GeVfmcube}\@.

In the limit of an ideal Stefan-\tsp{-0.3}Boltzmann gas, the equation of state
(EoS) of a QGP is given by
\begin {equation} \label {eq:sbeos} 
\SUB{\pressure}{\RM{SB}} = 
\FRAC{\SUB{\edens}{\RM{SB}}}{3} \qquad \RM{and} \qquad \SUB{\edens}{\RM{SB}} = 
\FRAC{\SUP{\PI}{2}}{30}\SP\ndof\SP\SUP{T}{4},
\end {equation}
where \SUB{\pressure}{\RM{SB}} is the pressure, \SUB{\edens}{\RM{SB}} the energy density, \ndof\ the
effective number of partonic degrees of freedom, and \temperature\ is the temperature~\cite {ref_qgp_equil}\@. 
The effective number of partonic degrees of freedom is given by
\begin {equation} \label {eq:defg}
   \ndof{} = \FRAC{7}{8}\SP\ndofq{} + \ndofg,
\end {equation}
where \ndofq\ and \ndofg\ are the degeneracies of, respectively, the quark and gluon states. 
Each quark flavor has a quark/antiquark state, 
two spin states, and three color states, 
whereas each gluon has two spin states and eight color states. 
The total degeneracy is, therefore, given by
\begin {equation} \label {eq:gtotal}
\ndof{} = \FRAC{7}{8} \TIMES{} \nflav{} \TIMES{} 2 \TIMES{} 2 \TIMES{} 3 + 2 \TIMES{} 8 = \FRAC{21}{2} \nflav{} + 16,
\end {equation}
which yields the value \MM{\ndof{} = 37 (95/2)} for an \MM{\nflav{} = 2 (3)} flavor QGP\@. 
This is an order of magnitude larger than for a hadron gas, where \MM{\ndof \ensuremath{\approx} 3}\@. 

The horizontal arrow in Figure~\ref {fig:epsilon} indicates the
Stephan-\tsp{-0.3}Boltzmann limit for a QGP with \MM{\nflav{} = 3} light flavors. 
The lattice QCD calculation shows that \DIV{\edens}{\SUP{\temperature}{4}} above \Tc\ remains
far below this limit, indicating that a QGP, according to these
calculations, does not behave as an ideal gas of quarks and gluons.

The possible existence of a QGP was conjectured before the advent of
lattice QCD calculations, and already in the \yr{1980}s experiments started
to look for signatures of this plasma in heavy ion collisions. This
initiated the rapidly developing field of heavy ion physics, and led to
a large series of experiments performed at the AGS in Brookhaven,
the ISR and the SPS at CERN, and, since the year \yr{2000}, at the
Relativistic Heavy Ion Collider (RHIC) at the Brookhaven National
Laboratory (BNL) in the USA\@.

\section {Heavy ion collisions} \label {se:hicollisions}

To describe a particle collision, we denote by
\SUB{\momentum}{\IT{A}} the 4\tsp{0.3}-momentum of particle \IT{A} moving along the beam (\zcoord\ axis),
and by \SUB{\momentum}{\IT{B}} the 4\tsp{0.3}-momentum of particle \IT{B} moving in the opposite direction. 
The Lorentz\tsp{0.2}-invariant measure of the square of the center-of-mass energy available in the collision is
\begin {equation}
\svar{} = \SUP{(\SUB{\momentum}{\IT{A}} + \SUB{\momentum}{\IT{B}})}{2}.
\end {equation}

The Lorentz\tsp{0.2}-invariant inclusive cross section of the scattering process 
$$
\IT{AB}\TO\IT{CX}
$$
is defined by
\begin {equation} \label {eq:xsecdef}
\energy{} \FRAC{\SUP{\der}{3}\SIGMA(\IT{AB}\TO\IT{CX})}{\der\SUP{\momentumbold}{3}} = 
\FRAC{\SUP{\der}{3}\SIGMA}{\pT\tsp{0.5}\der\pT\tsp{0.5}\der\rapidity\tsp{0.5}\der\phiangle},
\end {equation}
where \IT{C} is the final state particle being measured and \IT{X} denotes all other particles produced in the collison \cite {ref_hadron_inter}\@.
Because of azimuthal symmetry, it is convenient to separate longitudinal and transverse momentum components.
In Eq.~(\ref {eq:xsecdef}), \linebreak
\energy\ and \momentumbold\ are the energy and 3\tsp{0.3}-momentum,
\pT\ is the transverse component of the momentum, 
\phiangle\ is the azimuthal angle, and \rapidity\ is the rapidity of particle \IT{C} in the center-of-mass frame.
The rapidity is a measure of the longitudinal momentum component (\pL) and is defined by
\begin {equation}
\rapidity{} = \FRAC{1}{2} \ensuremath{\ln} \ensuremath{\left(} \FRAC{\energy + \pL}{\energy - \pL} \ensuremath{\right)}.
\end {equation}
The rapidity variable has the advantage of being additive under Lorentz boosts along the \zcoord\ axis.
Another commonly used variable is the pseudorapidity (\etacoord) defined by
\begin {equation}
\etacoord{} = - \ensuremath{\ln} \ensuremath{\tan} (\DIV{\ensuremath{\theta}}{2}),
\end {equation}
which is simply a measure of the polar angle ($\theta$) and does not depend on the particle mass.
This is, therefore, a convenient variable, since it can be calculated without knowing the particle identity.
In the limit \MM{\energy{} = \SQRT{\SUP{\momentum}{2} - \SUP{\mass}{2}} \ensuremath{\approx} \momentum{} \ensuremath{\gg} \mass} of very energetic particles, 
the pseudorapidity \etacoord\ approaches the rapidity \rapidity, because particle masses can then be neglected.

\begin {figure} [h!]
\begin {center}
\includegraphics [width=0.55\textwidth] {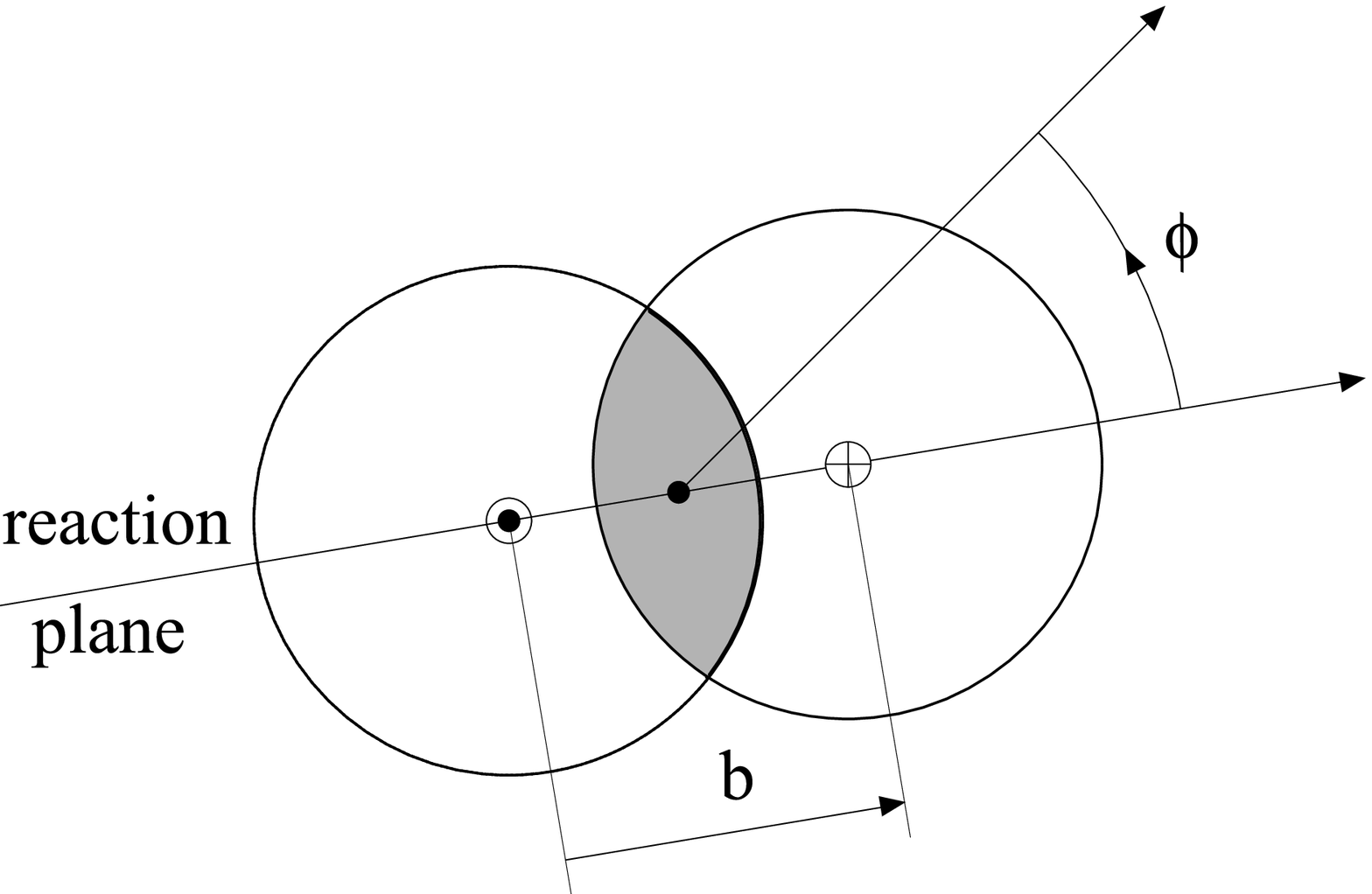}
\end {center}
\caption {Transverse view of two colliding nuclei defining the reaction plane, the impact \protect\linebreak
\mbox{parameter}~\impact,~and~the~\mbox{azimuthal}~\mbox{angle}~\phiangle~of~a~\mbox{produced}~\mbox{particle}~with~\mbox{respect}~to~the~\mbox{reaction}~plane.}
\label {fig:bdef}
\end {figure}

Because atomic nuclei are spatially extended objects, a characteristic of nucleus\tsp{0.3}-nucleus collisions is the impact parameter (\impact),
which is the transverse distance between the centers of the two colliding nuclei, as shown in Figure~\ref {fig:bdef}\@.

Other measures of the collision centrality are the number of participants (\Npart) and the number of binary collisions (\Ncoll)\@.
The number of binary collisions \Ncoll\ is defined as the number of individual inelastic nucleon-nucleon collisions 
that happened during the nucleus\tsp{0.3}-nucleus collision.
The number of participants \Npart\ is defined as the number of nucleons 
that suffered at least one inelastic collision with another nucleon.
The relation between the impact parameter \impact\ and the number of collisions \Ncoll\ or \Npart\ is 
calculable in the framework of the Glauber model~\cite {ref_glauber_original}\@.

Experimentally, the centrality of a heavy ion collision is estimated from a measurement of one or more quantities 
that vary monotonically with the \linebreak
impact parameter.
Such quantities are the charged particle multiplicity (\Nch),
the transverse energy (\eT) of all charged particles emitted near midrapidity, 
or the forward energy (\eF) measured close to the beam line.
The relation between the observables and the impact parameter is established by 
Monte Carlo event generators that model nuclear collisions at relativistic energies~\cite {ref_hijing}\@.

The range of impact parameters can be represented as a fraction of the total geometric cross section.
It is customary to define centrality classes as adjacent intervals in \impact\ that contain a certain percentile of the
differential cross section \DIV{\der\SIGMA}{\der\impact}\@.
For instance, a \MM{0}--\MM{5\unitns{\%}} centrality class contains events with five percent of the smallest impact parameters,
such that it corresponds to five percent of the total geometric cross section.

\section {Heavy ion physics at RHIC}

RHIC is a multipurpose colliding beam facility~\cite {ref_rhic_design_overview,ref_rhic_experiments_overview_nim}, 
capable of accelerating protons, deuterons, and heavy ions over a broad energy range.
At present, RHIC has delivered colliding beams of protons, deuterons, copper, 
and gold ions with beam energies of up to 100\unit{\GeV} per nucleon~\cite {ref_rhic_five_years_of_tracking,ref_rhic_run_overview_url}\@.

An estimate of the energy density in the created medium is obtained using the Bjorken formula~\cite {ref_bjorken_edens} 
\begin {equation} \label {eq:bjorken}
\SUB{\edens}{\RM{Bj}} = \FRAC {\der\eT} {\der\rapidity} \FRAC {1} {\cspeed\tauzero\tsp{0.5}\PI\SUP{\IT{R}}{2}},
\end {equation}
where \tauzero\ is the formation time and \IT{R} is the initial radius of the expanding \linebreak
system. Using the value \MM{\DIV{\der\eT}{\der\etacoord} = 503 \ensuremath{\pm} 2\unit{\GeV}} measured in 
central \goldgold\ 
collisions~\cite {ref_phobos_eT_AuAu130}
and taking \MM{\IT{R} = 1.2\SUP{A}{1/3}\unit{\fm}},
together with reasonable guess for the parameter value
\MM{\tauzero{} = 1\unit{\fmc}},
an initial energy density of about \MM{5\unit{\GeVfmcube}} is calculated.
This is well above the critical energy density of about 1\unit{\GeVfmcube}\ predicted by lattice QCD 
for a phase transition to the quark-gluon plasma, as shown in Figure~\ref {fig:epsilon}\@.
A major part of the physics program at RHIC is, therefore, to 
measure particle production in high energy nuclear collisions with the goal to study the properties
of the state of matter (presumably a QGP) produced in such collisions.

Particles emitted with large transverse momentum are important probes 
of the medium produced in the collision, 
because they most likely originate from high energetic partons that propagate
through and couple to the created medium 
and thus carry information about its properties.
A convenient way to observe medium-induced modification of particle production is to compare 
a nucleus\tsp{0.3}-nucleus collision (\collision{A}{B}) with an incoherent superposition of 
the corresponding number of individual nucleon-nucleon collisions (\nucleonnucleon)\@.
This is done via the \textit{nuclear modification factor} (\RAB), defined as the ratio of the particle yield in 
nucleus\tsp{0.3}-nucleus collisions and the yield in nucleon-nucleon collisions scaled with the number of binary collisions \Ncoll,
\begin {equation} \label {eq:rabdef}
\RAB{} = 
\FRAC 
    {\DIV{\SUP{\der}{2}\SUB{\Number}{\IT{AB}}}{\der\pT\tsp{0.5}\der\rapidity}} 
    {\TABmean{}\ \DIV{\SUP{\der}{2}\SUP{\SIGMA}{\protonproton}\tsp{-0.5}}{\der\pT\tsp{0.5}\der\rapidity}}.
\end {equation}
Here \TABmean\ is the nuclear overlap function that is related to the number 
of inelastic nucleon-nucleon collisions in one \collision{A}{B}\ collision through
\begin {equation} \label {eq:tabdef}
\TABmean{} \TIMES{} \SUB{\SUP{\SIGMA}{\nucleon\nucleon}}{\RM{inel}} = \Ncollmean.
\end {equation}
In the absense of medium effects, the nuclear modification factor is unity, while \MM{\RAB{} \ensuremath{<} 1} 
indicates a suppression of particle production in heavy ion collisions, 
compared to an expectation based on an incoherent sum of nucleon-nucleon collisions.

\begin {figure}[t]
\begin {center}
\includegraphics [width=0.8\textwidth] {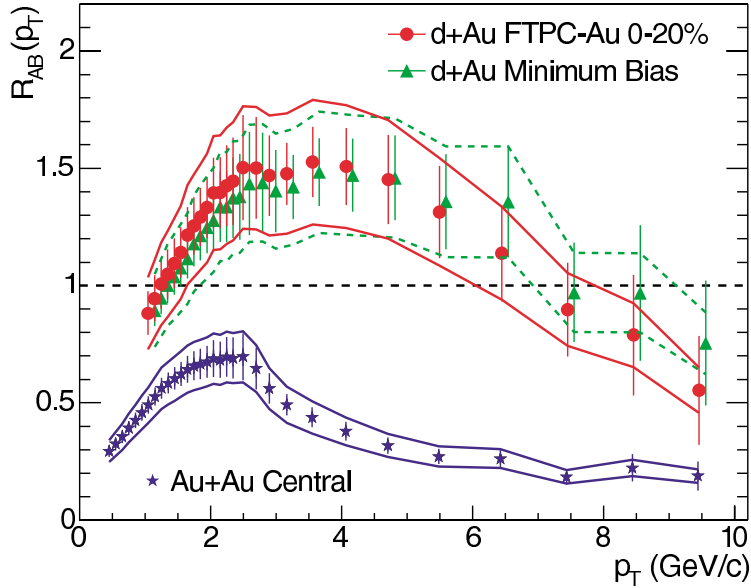}
\end {center}
\caption {Ratio \RdA\ of charged hadron production, as a function of \pT, measured by the STAR Collaboration in \deuterongold, and 
\RAA\ measured in central \goldgold\ collisions at \protect\linebreak
\MM{\sNN{} = 200\unit{\GeV}}\@.
Figure taken from~\protect\cite {ref_star_dAu_evidence}\@.}
\label {fig:star_RAB}
\end {figure}

In Figure~\ref {fig:star_RAB}
we show the ratio \RAA\ of charged hadron production, as a function of \pT, measured by the STAR Collaboration in central \goldgold\ collisions 
at \MM{\sNN{} = 200\unit{\GeV}}~\cite {ref_star_dAu_evidence} 
(the quantity \sNN\ is the center-of-mass energy of an individual nucleon-nucleon collision)\@.
It is evident that charged particle production in \goldgold\ collisions is significantly suppressed, compared to that 
in \protonproton\ collisions at the same center-of-mass energy, in particular at large \MM{\pT{} \ensuremath{\approx} 8\unit{\GeVc}}, 
where \RAA\ reaches a value of about 0.2\@.

Also shown in Figure~\ref {fig:star_RAB} is the nuclear modification factor measured in minimum bias (no centrality selection) 
and central \deuterongold\ collisions.
This measurement is important to distinguish between initial and final state effects.
Since we can safely assume that in \deuterongold\ collisions no hot and dense medium is created, 
the presence of a suppression would indicate 
initial state effects, such as nuclear modification of the parton densities in the gold nucleus.
It is seen from Figure~\ref {fig:star_RAB} that such suppression is absent in \deuterongold\ collisions, 
indicating that the suppression observed in \goldgold\ collisions is a final state effect 
caused by the dense medium created in such collisions.

A significant enhancement \MM{\RdA{} \GREATER{} 1} seen in \deuterongold\ collisions in the region \linebreak
\MM{2 \LESS{} \pT{} \LESS{} 7\unit{\GeVc}} in Figure~\ref {fig:star_RAB} 
can be explained by the so\tsp{0.3}-called Cronin \linebreak
effect~\cite {ref_cronin}\@.
This effect is likely caused by multiple scattering of the projectile parton inside the target nucleus, 
which acts as an additional transverse momentum kick of the parton, overpopulating the \MM{\pT{} \ensuremath{>} 2\unit{\GeVc}} region.
Since there is an indication in Figure~\ref {fig:star_RAB} of a possible suppression in \deuterongold\ collisions at \MM{\pT{} \ensuremath{>} 8\unit{\GeVc}}, 
it is interesting to measure the \RdA\ factor at even higher \pT\@. 
This thesis presents such a measurement.

\begin {figure} [t!]
\begin {center}
\includegraphics [width=\textwidth] {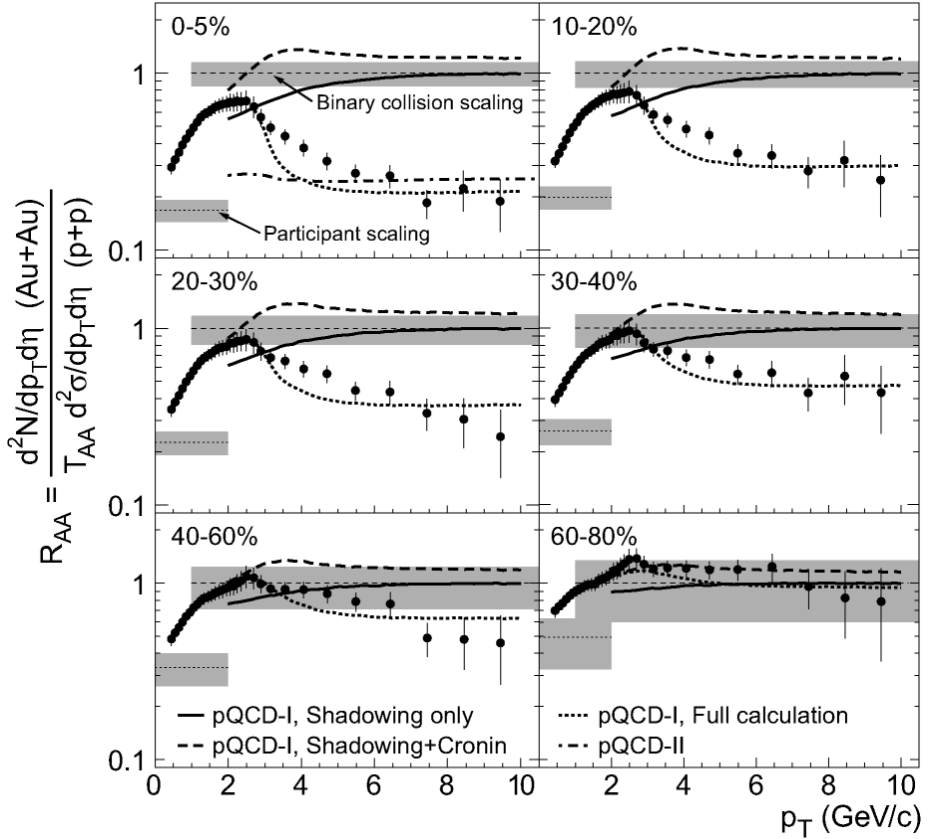}
\end {center}
\caption {Ratio \RAA\ of charged hadron production, as a function of collision centrality, 
measured by the STAR Collaboration in \goldgold\ collisions at \MM{\sNN{} = 200\unit{\GeV}}\@.
Figure taken from~\protect\cite {ref_star_AuAu_suppression}\@.}
\label {fig:star_AuAu_RAA_centr}
\end {figure}

\begin {figure}[t!]
\begin {center}
\includegraphics [width=\textwidth] {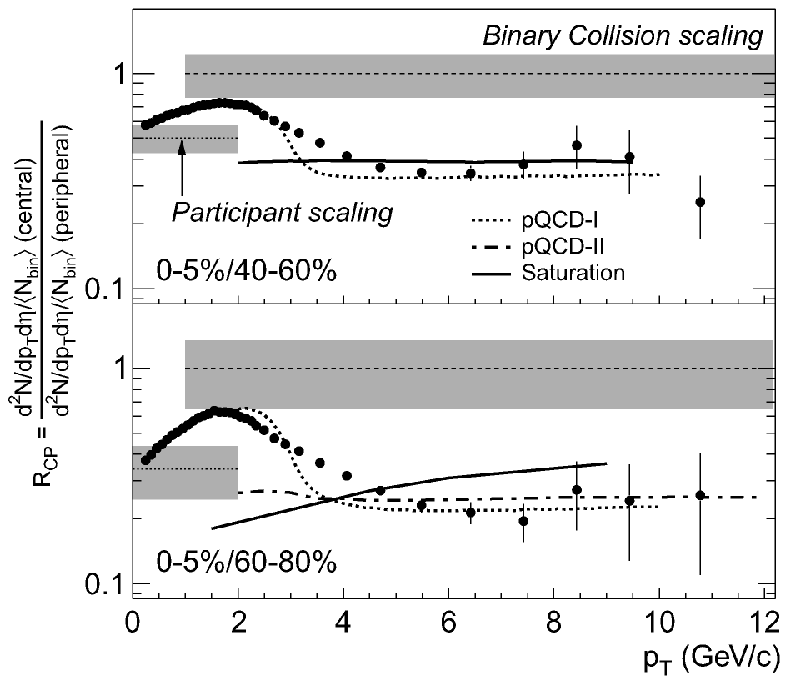}
\end {center}
\caption {Ratio \Rcp\ of the charged hadron production 
measured by the STAR Collaboration in \goldgold\ collisions at \MM{\sNN{} = 200\unit{\GeV}}\@.
Figure taken from~\protect\cite {ref_star_AuAu_suppression}\@.}
\label {fig:star_AuAu_RCP}
\end {figure}

For peripheral collisions, the number of participant nucleons is small and the creation of a dense medium is not expected.
This is illustrated in Figure~\ref {fig:star_AuAu_RAA_centr}, 
which shows the centrality dependence of \RAA\ for charged hadrons as measured by STAR in \goldgold\ collisions.
Indeed, the large suppression observed in central \linebreak collisions gradually vanishes with decreasing centrality.
This suggests that, \linebreak
instead of \protonproton\ interactions, peripheral collisions can be used as a reference.
This is done through the ratio of particle production in central (\IT{C}) and peripheral (\IT{P}) events:
\begin {equation} \label {eq:rcpdef}
\Rcp{} = 
\FRAC 
    {\SUB{\Ncollmean}{\IT{P}}} 
    {\SUB{\Ncollmean}{\IT{C}}} 
\FRAC 
    {\DIV{\SUP{\der}{2}\SUB{\Number}{\IT{C}}}{\der\pT\tsp{0.5}\der\rapidity}} 
    {\DIV{\SUP{\der}{2}\SUB{\Number}{\IT{P}}}{\der\pT\tsp{0.5}\der\rapidity}}.
\end {equation}
The advantage of this measure is that no \protonproton\ reference data are needed.
The disadvantage is that a stronger model dependence is introduced, because the uncertainties in \Ncollmean\ are
much larger for peripheral collisions.
In Figure~\ref {fig:star_AuAu_RCP}
is shown \Rcp\ for charged hadrons measured in \goldgold\ collisions by STAR~\cite {ref_star_AuAu_suppression}\@.

To provide a useful reference, it is important to measure particle production in nucleus\tsp{0.3}-nucleus interactions, as well 
as in the \protonproton\ collisions, under the same experimental conditions.
For instance, prior to having the first \protonproton\ collisions delivered by RHIC, both STAR and PHENIX collaborations have 
published the measurements of \RAA~\cite {ref_phenix_AuAu130,ref_star_AuAu130}
based on \protonproton\ and \antiprotonproton\ reference spectra \linebreak
obtained from a large body of world data, extrapolated to RHIC energies. \linebreak
These extrapolations yielded significant systematic uncertainties and more \linebreak
precise measurements of \RAA~\cite {ref_star_AuAu_suppression,ref_phenix_pi0_dAu}
only became available when \protonproton \linebreak
reference data were taken at RHIC in \yr{2001}--\yr{2002}\@.
\pagebreak

\begin {figure}[t!]
\begin {center}
\includegraphics [width=\textwidth] {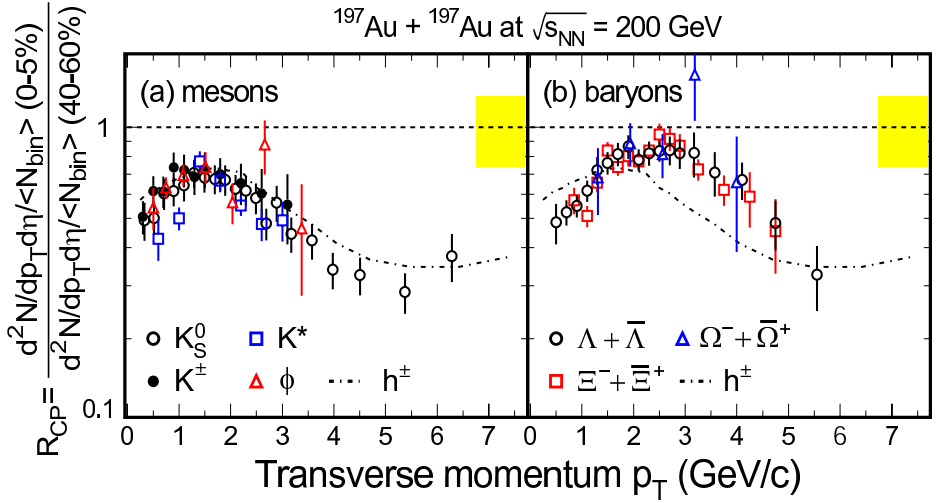}
\end {center}
\caption {Ratio \Rcp\ of identified hadron production
measured by the STAR Collaboration \protect\linebreak for mesons (a) and baryons (b) 
in \goldgold\ collisions at \MM{\sNN{} = 200\unit{\GeV}}\@.
Figure taken from~\protect\cite {ref_star_whitepaper}\@.}
\label {fig:star_RCP_meson_baryon}
\end {figure}

A detailed study of the intermediate- and high-\pT\ production of various hadron species shows that there is a systematic 
difference between meson and baryon production in \goldgold\ collisions, 
as illustrated in Figure~\ref {fig:star_RCP_meson_baryon}~\cite {ref_star_whitepaper}\@.
The \Rcp\ ratio for identified hadrons is shown separately for mesons (a) and baryons (b), 
and the clear difference between them suggests 
that the particle production in this \pT\ range depends not on the mass of the hadron 
but rather on the number of valence quarks contained within it.
This can be explained naturally in the quark recombination model for hadron formation, rather than fragmentation.
We do not discuss this model here and refer to 
\cite {ref_quark_recomb_model_1,ref_quark_recomb_model_2,ref_quark_recomb_model_3,ref_quark_recomb_model_4,ref_quark_recomb_model_5,ref_quark_recomb_model_6} 
for details.
The measurement of \Rcp\ for neutral pions and eta mesons would also be interesting in context of this observation.

This thesis presents a baseline measurement with the STAR detector of \linebreak
neutral pion and eta meson production in \protonproton\ and
\deuterongold\ collisions~at~a~\mbox{center-of-} \linebreak
mass energy of \MM{\sNN{} = 200\unit{\GeV}}\@.
The neutral pion spectrum complements that \linebreak
of the charged pions measured in STAR 
in the range \MM{0.35 \LESS{} \pT{} \LESS{} 10\unit{\GeVc}}~\cite {ref_star_idhadrons}
and extends up to \MM{\pT{} = 17\unit{\GeVc}}\@. 
Preliminary results of this analysis have been published in~\cite {ref_mischke_pi0,ref_grebenyuk_pi0}\@.
Also presented in this thesis are the first measurements by STAR of \etameson\ meson production.
\clearpage

\section {Proton-proton collisions}

In QCD, the hadronic interactions are described in terms of the interactions of their constituent partons.
The inclusive cross section of the reaction 
$$
\IT{AB}\TO\IT{CX}
$$
is calculated as the weighted sum of differential cross sections 
of all possible parton scatterings that can contribute~\cite {ref_hadron_inter}:
\begin {equation} \label {eq:sigma}
\SUB{\energy}{\IT{C}} 
\FRAC
    {\SUP{\der}{3}\SIGMA(\IT{AB}\tsp{-1.0}\TO\tsp{-1.0}\IT{CX})}
    {\der\SUP{\SUB{\momentumbold}{\IT{C}}}{3}}
= 
\tsp{-0.5}
\SUB{\ensuremath{\sum}}{\IT{abcd}} 
\SUP{\SUB{\ensuremath{\int}}{0}}{1}\tsp{-2.0}\der\SUB{\xcoord}{\IT{a}} 
\SUP{\SUB{\ensuremath{\int}}{0}}{1}\tsp{-2.0}\der\SUB{\xcoord}{\IT{b}}
\tsp{1.0}
\SUP{\SUB{\IT{f}}{\IT{A}}}{\IT{a}}(\SUB{\xcoord}{\IT{a}}) 
\SUP{\SUB{\IT{f}}{\IT{B}}}{\IT{b}}(\SUB{\xcoord}{\IT{b}})
\tsp{1.0}
\FRAC
    {1}
    {\PI\SUB{\zcoord}{\IT{c}}}
\FRAC
    {\der\SIGMA(\IT{ab}\tsp{-1.0}\TO\tsp{-1.0}\IT{cd})}
    {\der\tbar}
\tsp{1.0}
\SUP{\SUB{\IT{D}}{\IT{c}}}{\IT{C}}(\SUB{\zcoord}{\IT{c}}).
\end {equation}
Here \MM{\SUP{\SUB{\IT{f}}{\IT{A}}}{\IT{a}}(\SUB{\xcoord}{\IT{a}})} is the parton density function (PDF) that gives
the probability that hadron \IT{A} contains a parton \IT{a} which carries the fraction 
\MM{\SUB{\xcoord}{\IT{a}} = \DIV{\SUB{\IT{q}}{\IT{a}}}{\SUB{\momentum}{\IT{A}}}} of its momentum.
A similar definition applies to the density \MM{\SUP{\SUB{\IT{f}}{\IT{B}}}{\IT{b}}(\SUB{\xcoord}{\IT{b}})}\@.
The cross section \DIV{\der\SIGMA}{\der\tbar} of the hard partonic scattering 
$$
\IT{ab}\TO\IT{cd}
$$
is calculated in pQCD\@.
The invariant kinematic variables for the partonic sub\tsp{0.4}-process are
$$
\sbar{} = \SUP{(\SUB{\IT{q}}{\IT{a}} + \SUB{\IT{q}}{\IT{b}})}{2}
$$
$$
\ 
\tbar{} = \SUP{(\SUB{\IT{q}}{\IT{a}} - \SUB{\IT{q}}{\IT{c}})}{2}
,
$$
where \SQRT{\sbar} is the partonic center-of-mass energy 
and \SQRT{-\tbar\,} is the momentum transfer from \IT{a} to \IT{c}\@.
The fragmentation function \MM{\SUP{\SUB{\IT{D}}{\IT{c}}}{\IT{C}}(\IT{z})} in Eq.~(\ref {eq:sigma}) describes the probability that 
a given parton \IT{c} produces a final state hadron \IT{C} carrying a momentum fraction 
\MM{\SUB{\zcoord}{\IT{c}} = \DIV{\SUB{\momentum}{\IT{C}}}{\SUB{\IT{q}}{\IT{c}}}}\@.

It follows from the above that the cross section calculations rely on two \linebreak
inputs --- parton densities \SUP{\SUB{\IT{f}}{\IT{A}}}{\IT{a}} 
and fragmentation functions \SUP{\SUB{\IT{D}}{\IT{c}}}{\IT{C}}\@.
These functions \linebreak
are non-perturbative, so that they cannot be calculated in QCD from first principles.
However, they represent a properties of individual hadrons independent \linebreak
of the process in which they participate.
Parton densities and fragmentation \linebreak
functions can, therefore, be obtained from an analysis of a large variety of
scattering data.

A widely used set of parton densities is obtained by the CTEQ Collaboration
from a global QCD analysis of a large body of experimental data~\cite {ref_cteq6}\@.
The global fit, together with a detailed treatment of published experimental uncertainties, 
resulted in an excellent agreement with all available data.
An alternative popular parametrization is MRST~\cite {ref_mrs}\@.

The fragmentation functions \MM{\SUP{\SUB{\IT{D}}{\IT{c}}}{\IT{C}}(\IT{z})} 
can be obtained directly from the process 
\MM{\epluseminus\TO(\gama,\IT{Z})\TO\IT{CX}}, 
in which the initial state has no hadrons.
Such annihilation processes have been measured at many \epluseminus\ 
colliders over a wide range of center-of-mass energies.
The most recent parametrizations of fragmentation functions are 
KKP~\cite {ref_kkp_ff}, BKK~\cite {ref_bkk_ff}, BFGW~\cite {ref_bfgw_ff}, and Kretzer~\cite {ref_kretzer_ff}\@.

The cross sections for the individual partonic sub\tsp{0.4}-processes 
are calculated in pQCD with no additional input,
except for the strong coupling constant \SUB{\ensuremath{\alpha}}{\IT{S}}\@.
These calculations are usually performed at next-to\tsp{0.4}-leading order (NLO), or even at next-to\tsp{0.4}-next-to\tsp{0.4}-leading order (NNLO)\@.

\begin {figure} [tb]
\centerline {\hbox {
\includegraphics [width=0.8\textwidth] {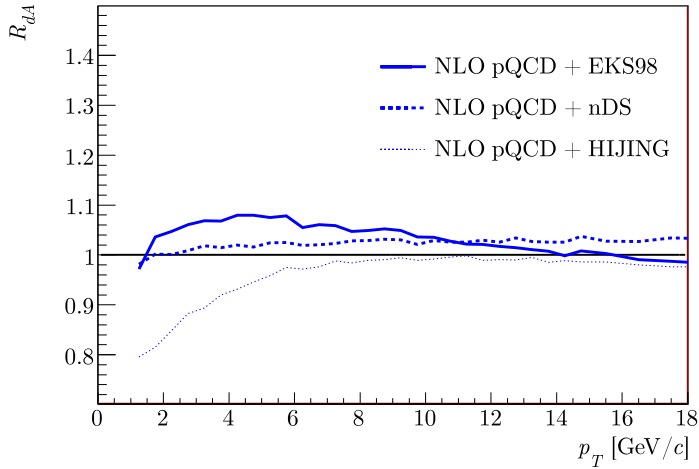}
}}
\caption {Nuclear shadowing effects on the \RdA\ ratio, calculated with 
EKS98~\cite {ref_EKS_shadowing}, nDS~\cite {ref_nDS_shadowing}, and HIJING~\cite {ref_HIJING_shadowing} 
shadowing parametrizations.}
\label {fig_shadowing}
\end {figure}

An important initial state effect in the heavy ion collisions is the modification of parton distribution functions inside nuclei.
It is well known, that the quark structure functions at low fractional momentum are depleted in a nucleus relative to a free nucleon.
This depletion is commonly referred to as nuclear shadowing.
In Figure~\ref {fig_shadowing}
we show the shadowing effects in \deuterongold\ collisions on the \RdA\ ratio~\cite {ref_xinnianwang_private}, 
calculated with various parametrizations --- EKS98~\cite {ref_EKS_shadowing}, nDS~\cite {ref_nDS_shadowing}, and HIJING~\cite {ref_HIJING_shadowing}\@.
It is also a motivation for the present analysis to observe the nuclear shadowing and differentiate between models, 
although the required experimental precision may be prohibitively high.
\enlargethispage{\baselineskip}

\chapter {The experiment}

\section {RHIC accelerator complex}

The STAR experiment is located at the Brookhaven National Laboratory (BNL) on Long Island, USA\@. 
An important part of the physics program of the \linebreak
Laboratory is carried out at the Relativistic Heavy Ion Collider (RHIC)\@. 
This is a multipurpose colliding beam facility~\cite {ref_rhic_design_overview,ref_rhic_experiments_overview_nim}, 
capable of accelerating protons, deuterons, and heavy ions over a broad energy range 
from the injection energy per nucleon of \MM{10\unit{\GeV}} up to the top energy 
of \MM{100\unit{\GeV}} for heavy ions and \MM{250\unit{\GeV}} for protons.

\begin {figure} [p]
\centerline {\hbox {
\includegraphics [width=0.85\textwidth] {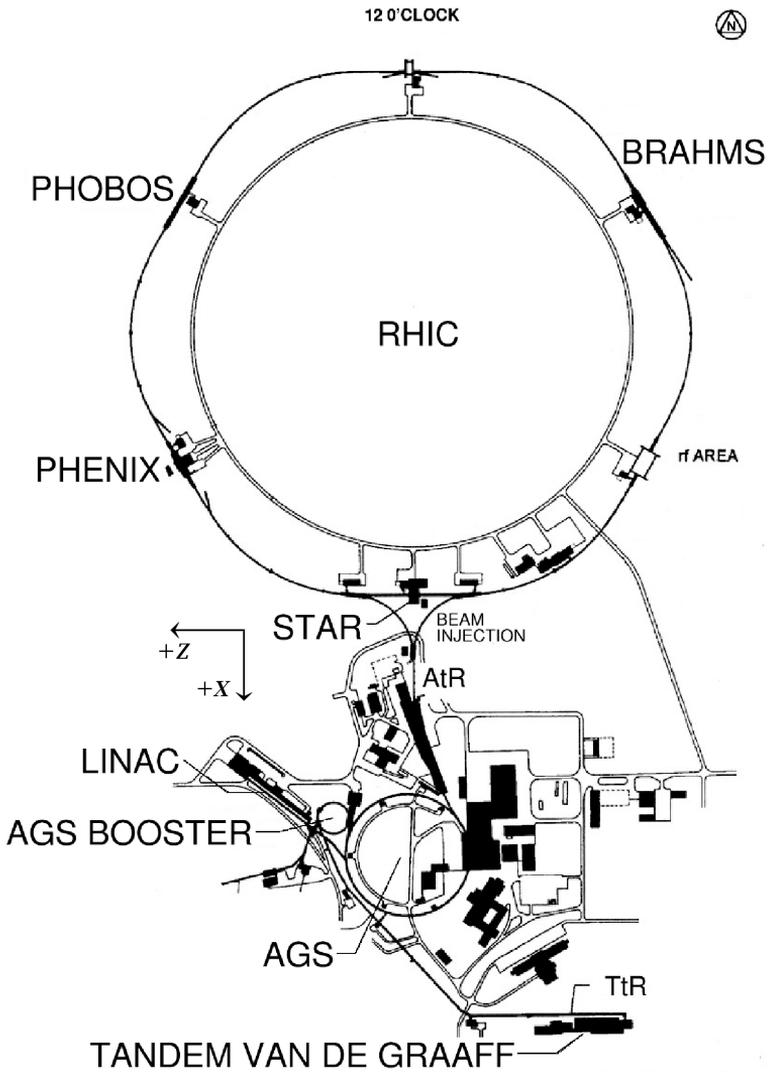}
}}
\caption {Layout of the RHIC accelerator complex. 
Shown are the locations of the STAR, PHENIX, PHOBOS, and BRAHMS experiments around the RHIC ring.
Also indicated in the figure is the STAR coordinate system with the positive
\zcoord\ axis pointing in the West direction.
Figure taken from~\protect\cite {ref_rhic_design_overview}\@.
}
\label {fig_rhic_complex}
\end {figure}

The layout of the accelerator complex is shown in Figure~\ref {fig_rhic_complex}\@.
Heavy ions are accelerated in the Tandem Van de Graaff accelerator, the Booster, the Alternating
Gradient Synchrotron (AGS), and in the RHIC accelerator itself.
The Linac serves to accelerate protons, which are then injected into the Booster.
Below we will give a short description of each component of the accelerator complex.

\paragraph {Tandem Van de Graaff generator}

Gold ions with unit negative charge are generated in the Pulsed
Sputter Ion Source which delivers \MM{250\unit{\muA}} pulses of \MM{600\unit{\mus}} duration each. 
The ions are then accelerated in the Tandem Van de Graaff generator from the ground to \MM{+14\unit{\MV}} potential.
They pass a set of stripping foils where they acquire a unit positive charge, and are subsequently \linebreak
accelerated again to the ground potential. 
The \MM{1\unit{\MeV}} ions leaving the Tandem are stripped further to a charge of \MM{+32}\@. 
There are two identical Tandems available to provide two different ion species simultaneously 
(presently deuterium and copper in addition to gold)\@.

\paragraph {LINAC}

The LINAC serves to accelerate protons to an energy of \MM{200\unit{\MeV}}, which are injected directly into the Booster.

\paragraph {Booster synchrotron}

The \MM{600\unit{\mus}} long Tandem pulse is injected into the Booster,
after which the particles are captured into six bunches and accelerated to an energy of \MM{95\unit{\MeV}}\@. 
Gold ions, when they are extracted from the Booster, are stripped to the charge \MM{+77}, 
leaving only two tightly bound \IT{K}\tsp{-0.7}-shell electrons 
to be stripped at a later stage in the acceleration chain.

\paragraph {AGS}

From the Booster, \MM{24} bunches are injected into the AGS 
and rearranged into four final bunches 
containing \SUP{10}{9} ions each. Those bunches are accelerated to an energy of about \MM{10\unit{\GeV}}\@.
When transferred to the RHIC accelerator, the ions are fully stripped to the charge \MM{+39} in case of copper and to \MM{+79} in case of gold.

\paragraph {RHIC accelerator}

The final stage of acceleration takes place in the RHIC synchrotron, where beams are circulating in two rings in opposite directions. 
The rings have a circumference of \MM{3.83\unit{\km}} and are equipped with independent bending and focusing magnets and 
RF cavities. This provides the capability of operating the accelerator with two beams of unequal species. 
The bending magnets are superconductive and cooled by liquid helium. 
The complete cooling of the rings from room temperature to the operating temperature of \MM{4.6\unitns{\Kelvin}} takes about ten days.

Up to \MM{120} bunches can be injected in each ring and accelerated to an energy between \MM{30} and \MM{100\unit{\GeV}}\@. 
After acceleration, the bunches are transferred to the storage RF system, 
which maintains the bunch length at \MM{1.52\unit{\meter}} or \MM{5\unit{\ns}}\@. The lifetime of a stored beam is about \MM{10} hours, 
whereafter the beam is dumped and a new fill begins.
A chosen pattern of empty buckets provides a sample of unpaired bunches crossing each interaction region for beam-\tsp{-0.3}background studies.

Beams are made to cross at six points along the ring, 
four of which are used by the experiments STAR~\cite {ref_star_overview_nim}, PHENIX~\cite {ref_phenix_overview_nim}, 
PHOBOS~\cite {ref_phobos_overview_nim}, and BRAHMS~\cite {ref_brahms_overview_nim}\@.
Of the remaining two crossing points, one is occupied by the RF system, while the other is not used at present.
\pagebreak

\paragraph {RHIC performance}

To date, RHIC has delivered a variety of colliding beams of protons, deuterons, copper (\copperion), 
and gold ions (\goldion)~\cite {ref_rhic_five_years_of_tracking,ref_rhic_run_overview_url}\@. \\
In Table~\ref {tab:1} 
\begin {table} [t!]
\begin {center}
\caption {\normalsize RHIC runs in the years \yr{2000}\tsp{0.3}--\yr{2007}\@.}
\label {tab:1}
\begin {tabular*} {\textwidth} {lccr@{.}lr@{$\:\times\:$}lc}
Run   & Year                 & Particle        & \multicolumn {2} {c} {Beam energy}           & \multicolumn {2} {c} {Integrated}       & Average beam \\
      &                      & species         & \multicolumn {2} {c} {per nucleon}         & \multicolumn {2} {c} {luminosity}       & polarization \\
      &                      &                 & \multicolumn {2} {c} {[\unitns{\GeV}]} & \multicolumn {2} {c} {[\unitns{\pbinv}]}  & [\unitns{\%}] \\
\hline
Run-\tsp{-0.4}1\rule[10pt]{0pt}{0.6pt} & \yr{2000} & \goldgold & 27 & 9                     & \multicolumn {2} {c} {$<$~$10^{-9}$}         &              \\
      &                      & \goldgold       & 65 & 2 & 20 & $10^{-6}$       &              \\
\hline
Run-\tsp{-0.3}2\rule[10pt]{0pt}{0.6pt} & \yr{2001}--\yr{2002} & \goldgold & 100 & 0 & 258 & $10^{-6}$      &              \\
      &                      & \goldgold       & 9 & 8 & 0.4 & $10^{-6}$      &              \\
      &                      & \protonproton   & 100 & 0                          & \multicolumn {2} {c} {1.4\tsp{0.5}}            & 14         \\
\hline
Run-3\rule[10pt]{0pt}{0.6pt} & \yr{2002}--\yr{2003} & \enspace\thinspace\thinspace\deuterongold & 100 & 0 & 73 & $10^{-3}$       &              \\
      &                      & \protonproton   & 100 & 0                          & \multicolumn {2} {c} {5.5\tsp{0.5}}            & 34         \\
\hline
Run-4\rule[10pt]{0pt}{0.6pt} & \yr{2003}--\yr{2004} & \goldgold & 100 & 0 & 3740 & $10^{-6}$     &              \\
      &                      & \goldgold       & 31 & 2 & 67 & $10^{-6}$       &              \\
      &                      & \protonproton   & 100 & 0                          & \multicolumn {2} {c} {7.1\tsp{0.5}}            & 46         \\
\hline
Run-5\rule[10pt]{0pt}{0.6pt} & \yr{2004}--\yr{2005} & \coppercopper & 100 & 0 & 42.1 & $10^{-3}$     &              \\
      &                      & \coppercopper   & 31 & 2 & 1.5 & $10^{-3}$      &              \\
      &                      & \coppercopper   & 11 & 2 & 0.02 & $10^{-3}$             &              \\
      &                      & \protonproton   & 100 & 0                          & \multicolumn {2} {c} {\tsp{0.5}29.5\mbox{\ \ }}           & 46         \\
      &                      & \protonproton   & \mbox{\ \,}204 & 9               & \multicolumn {2} {c} {0.1\tsp{0.5}}            & 30         \\
\hline
Run-6\rule[10pt]{0pt}{0.6pt} & \yr{2006}            & \protonproton   & 100 & 0   & \multicolumn {2} {c} {\tsp{0.5}93.3\mbox{\ \ }}           & 58         \\
      &                      & \protonproton   & 31 & 2                           & \multicolumn {2} {c} {\tsp{0.5}\mbox{\ }1.05}           & 50         \\
\hline
Run-\tsp{-0.4}7\rule[10pt]{0pt}{0.6pt} & \yr{2006}--\yr{2007} & \goldgold       & 100 & 0                           & 7250 & $10^{-6}$     &              \\
\end {tabular*}
\end {center}
\end {table}
we list the RHIC runs from the beginning of operations in the year~\yr{2000}\ up to the year~\yr{2007}\@.
The \protonproton\ runs provide reference data for the heavy ion physics program, as well as data to measure the proton spin structure at RHIC\@.
For the latter purpose, the proton beams are polarized, reaching degrees of up to \MM{60\unitns{\%}} in~\yr{2006}\@.
The data used in this thesis were taken in the \deuterongold\ run in~\yr{2002}/\yr{03} and \protonproton\ in~\yr{2005}, 
both at center-of-mass energies of \MM{200\unit{\GeV}}\@.
\clearpage
\addtolength {\textfloatsep} {-10pt}
\section {STAR detector}

\begin {figure} [tb]
\centerline {\hbox {
\includegraphics [width=0.95\textwidth] {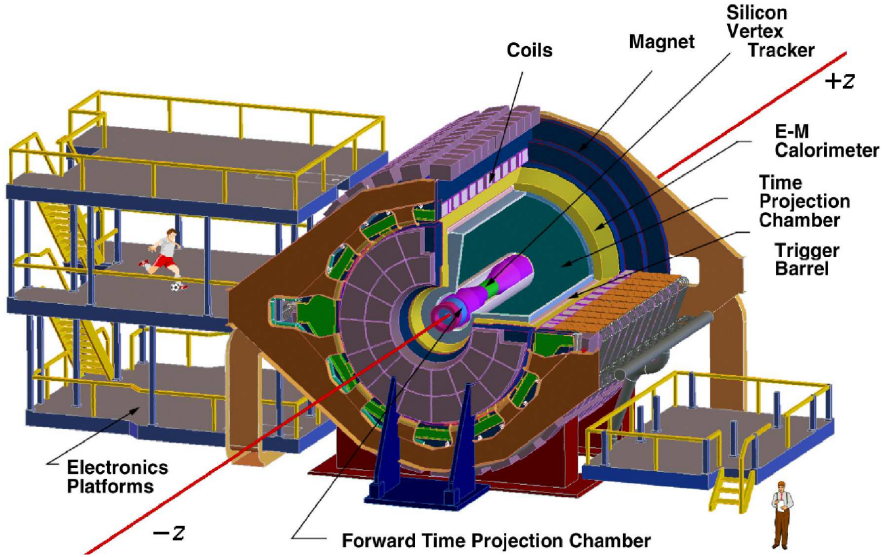}
}}
\caption {Perspective view of the STAR detector. Figure taken from~\protect\cite {ref_star_overview_nim}\@.}
\label {fig_star_perspective_view}
\end {figure}

The STAR detector (Solenoidal Tracker At RHIC)~\cite {ref_star_overview_nim} was designed primarily for measurements of hadron production 
in heavy ion and proton-proton collisions over a large solid angle.
For this purpose, large acceptance high granularity tracking detectors are placed inside a large volume magnetic field.
A perspective view of the detector is shown in Figure~\ref {fig_star_perspective_view}, 
\begin {figure} [tb]
\centerline {\hbox {
\includegraphics [width=0.78\textwidth,height=0.47\textwidth] {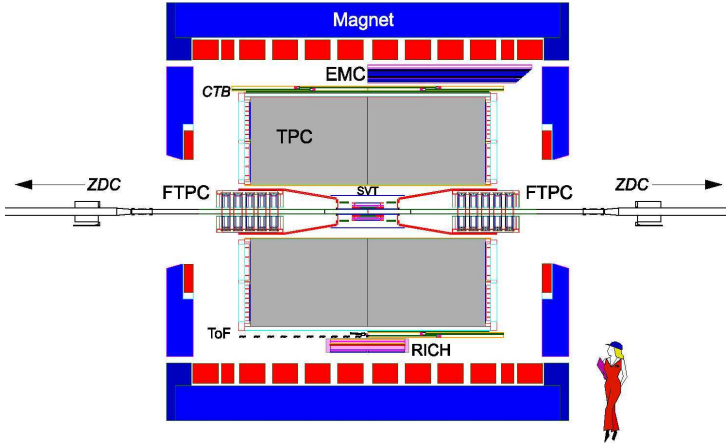}
}}
\caption {Cutaway side view of the STAR detector, as configured in \yr{2001}\@. 
Figure taken from~\protect\cite {ref_star_overview_nim}\@.}
\label {fig_star_2001}
\end {figure}
and a cutaway side~view~in~Figure~\ref {fig_star_2001}\@.

The barrel tracking detectors in STAR are a Silicon Vertex Tracker, surrounding the beam pipe, (SVT, not used in this analysis) 
and a large volume Time Projection Chamber (TPC), with an inner radius of \MM{0.5\unit{\meter}}, 
an outer radius of \MM{2\unit{\meter}}, and a length of \MM{4.2\unit{\meter}}\@.
The TPC covers a pseudorapidity range of \MM{\etacoordabs{} \LESS{} 1.8} and 
is designed to reconstruct the very high multiplicity events produced in heavy ion collisions.
These multiplicities can reach up to \MM{1000} charged tracks per unit rapidity
in a central \goldgold\ collision at the largest beam energies.
High granularity tracking in the forward and backward regions is achieved by two Forward TPCs (FTPC), 
each covering a range of \MM{2.5 \LESS{} \etacoordabs{} \LESS{} 4} in pseudorapidity.

\enlargethispage{\baselineskip}

For trigger purposes, the TPC is surrounded by a layer of scintillator tiles (Central Trigger Barrel, CTB, not used in this analysis)\@	.

To trigger on the energy deposited by high transverse momentum photons, electrons, and electromagnetically decaying hadrons, 
a Barrel Electromagnetic Calorimeter (BEMC)~\cite {ref_emc_nim} 
was incrementally added to the STAR setup from the year \yr{2001}\ to \yr{2005}\@. 

The calorimeter surrounds and covers the full acceptance of the TPC and CTB\@.
An Endcap Electromagnetic Calorimeter (EEMC)~\cite {ref_eemc_nim} was installed in \yr{2002}--\yr{2003}\ 
to cover the pseudorapidity range \MM{1 \LESS{} \etacoord{} \LESS{} 2}\@.
In the data taking \linebreak
period covered by this thesis, only the West half of the BEMC was fully operational 
(\MM{0 \LESS{} \etacoord{} \LESS{} 1})\@.

The STAR barrel detectors are placed inside a room temperature solenoidal magnet with maximum field of \MM{0.5\unit{\Tesla}}\@. 
The inner dimensions of the magnet are \MM{5.8\unit{\meter}} in length and \MM{5.27\unit{\meter}} in diameter.

To provide a minimum bias trigger and to measure centralities in heavy ion collisions, 
two sampling calorimeters (ZDC) are placed in the RHIC tunnel at \MM{18\unit{\meter}} from the interaction point.
Tiled arrays of scintillator counters (Beam-Beam Counter, BBC) are mounted around the beam pipe 
at a distance of \MM{3.7\unit{\meter}} from the interaction point, 
to provide a minimum bias trigger in \protonproton\ collisions.
The detector subsystems relevant for the present analysis are briefly described in the following sections.
We refer to Chapter~\ref {ref_chapter_BEMC} for a detailed description of the BEMC, which plays a central role in the analysis.

\enlargethispage{2\baselineskip}

Throughout this thesis we will use a Cartesian coordinate system defined as follows: 
\zcoord\ pointing along the beam in the West direction (see Figure~\ref {fig_rhic_complex}), \linebreak
\ycoord\ pointing upward, right\tsp{0.2}-handed.

\addtolength {\textfloatsep} {10pt}

\subsection {Time Projection Chamber}

The Time Projection Chamber~\cite {ref_star_tpc} is the central tracking device in STAR\@. 
It allows one to track charged particles, measure their momenta, and identify the 
particle species by measuring the ionization energy loss \dEdx\@.

\begin {figure} [tb]
\centerline {\hbox {
\includegraphics [width=0.8\textwidth] {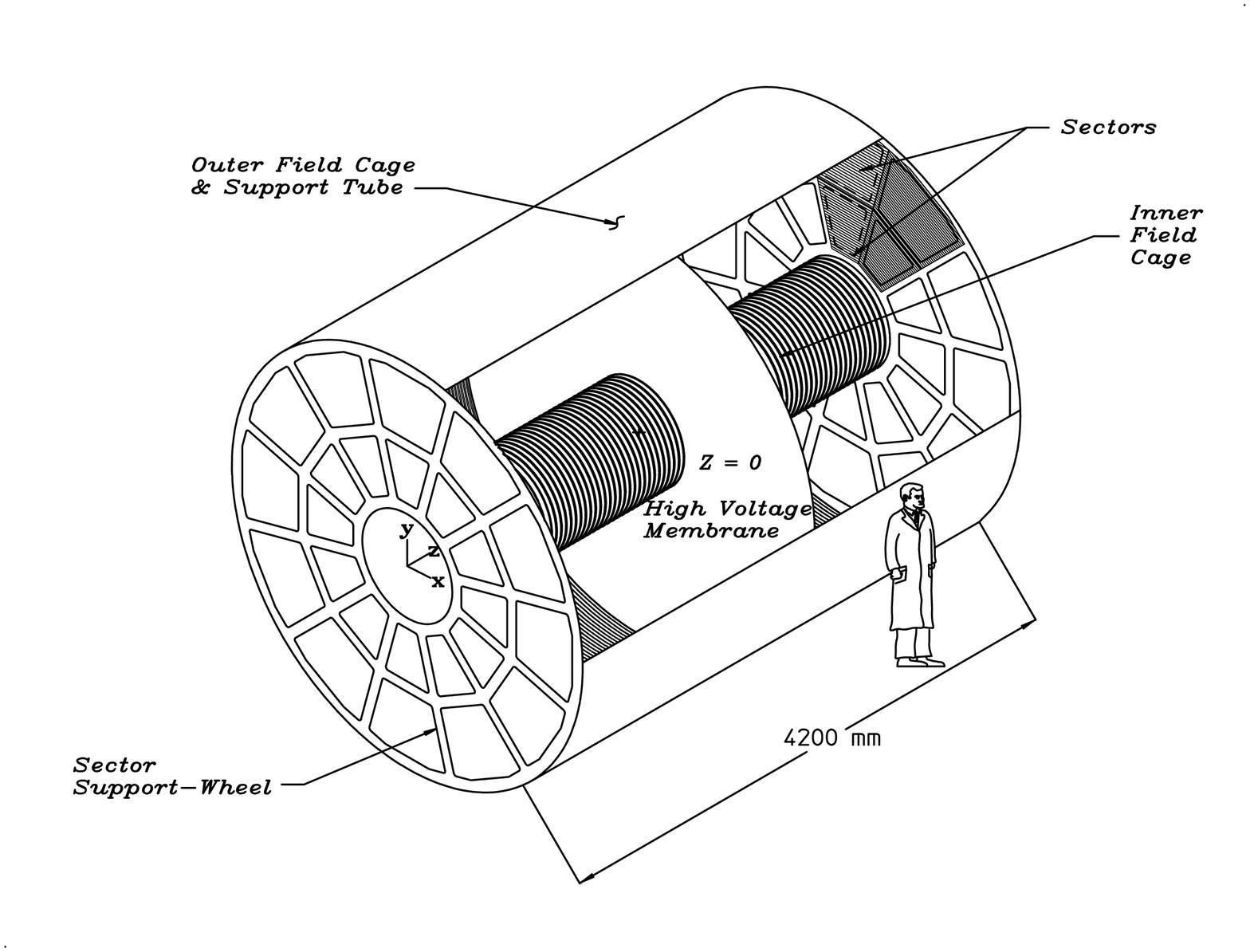}
}}
\caption {Schematic perspective view of the STAR TPC\@. Figure taken from~\protect\cite {ref_star_tpc}\@.}
\label {fig_star_tpc_general}
\end {figure}

A schematic layout of the TPC is shown in Figure~\ref {fig_star_tpc_general}\@.
The TPC barrel measures \MM{4.2\unit{\meter}} in length and has an inner radius of \MM{0.5\unit{\meter}} 
and an outer radius of \MM{2\unit{\meter}}\@. 
The TPC acceptance covers \MM{\PLMN1.8} units in pseudorapidity and full azimuth. 
Particles are identified over a momentum range from \MM{100\unit{\MeVc}} to \MM{1\unit{\GeVc}}, and 
their momentum is measured in the range from \MM{100\unit{\MeVc}} to \MM{30\unit{\GeVc}}\@.

The TPC is a gas filled cylindrical volume with a well defined uniform electric field gradient of about \MM{135\unit{\Vcm}}\@.
The secondary electrons released by ionizing particles along their path drift in the electric field towards the readout endcaps. 
The electric field is generated between a central membrane held at \MM{28\unit{\kV}} \linebreak
potential and the endcaps, which are held at ground potential. 
A uniform field gradient is maintained by concentric equi-potential field cage cylinders biased via \MM{2\unit{\MOhm}} resistors.
The drift volume is filled with a gas mixture of \MM{10\unitns{\%}} methane and \MM{90\unitns{\%}} 
argon, which is held slightly above atmospheric pressure. 
The drift \linebreak
velocity is \MM{5.45\unit{\cmmus}}, and the maximum drift time from the central membrane to endcap is \MM{38.5\unit{\mus}}\@.

The endcaps are instrumented with Multi\tsp{0.3}-\tsp{-0.3}Wire Proportional Chambers \linebreak
(MWPC) with pad readout. 
The transverse coordinates of a track are reconstructed from the hits in the MWPCs, while
the \zcoord\ coordinate is reconstructed from a measurement of the drift time.
The total drift time of \MM{5.45\unit{\mus}} is sampled by the readout electronics
in \MM{512} time buckets.

In each endcap, the MWPCs are arranged in \MM{12} sectors, each consisting of inner and outer sub\tsp{0.3}-sector.
The inner sub\tsp{0.3}-sectors are in the region of highest track density and are, therefore, 
optimized for better two\tsp{0.3}-track resolution, while the outer sub\tsp{0.3}-sectors are optimized 
for better performance in the measurement of \dEdx\@.

In the analysis presented in this thesis, the TPC is used as a charged particle veto in the identification of photons in the BEMC\@.
Samples of electrons reconstructed in the TPC serve to calibrate the energy response of the BEMC\@.

\subsection {Forward TPC modules}

\begin {figure} [t!]
\centerline {\hbox {
\includegraphics [width=0.7\textwidth,clip] {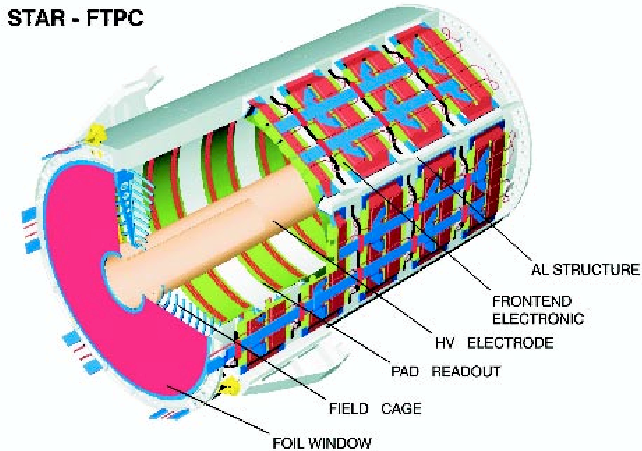}
}}
\caption {Perspective view of a STAR FTPC\@. Figure taken from~\protect\cite {ref_star_ftpc}\@.}
\label {fig_star_ftpc}
\end {figure}

Two Forward Time Projection Chambers~(FTPC)~\cite {ref_star_ftpc}~\mbox{extend}~the~STAR~\mbox{tracking} \linebreak
capability to the pseudorapidity range \MM{2.5 \LESS{} \etacoordabs{} \LESS{} 4}\@.
The layout of the FTPC is shown in Figure~\ref {fig_star_ftpc}\@.
Each FTPC is a cylindrical volume with a \linebreak
diameter of \MM{75\unit{\cm}} and a length of \MM{120\unit{\cm}}, with radial drift field and 
pad readout chambers mounted on the outer cylindrical surface.
Two such detectors are \linebreak
installed partially inside the main TPC on both sides of the interaction point.
The FTPC is capable of reconstructing all charged tracks (typically \MM{1000}) traversing the detector in a central \goldgold\ event.

In this thesis, the forward charged track multiplicity recorded in the FTPCs is used as a measure of the centrality in \deuterongold\ collisions.

\subsection {Zero Degree Calorimeter}

\begin {figure} [tb]
\centerline {\hbox {
\includegraphics [height=0.9\textwidth,angle=270] {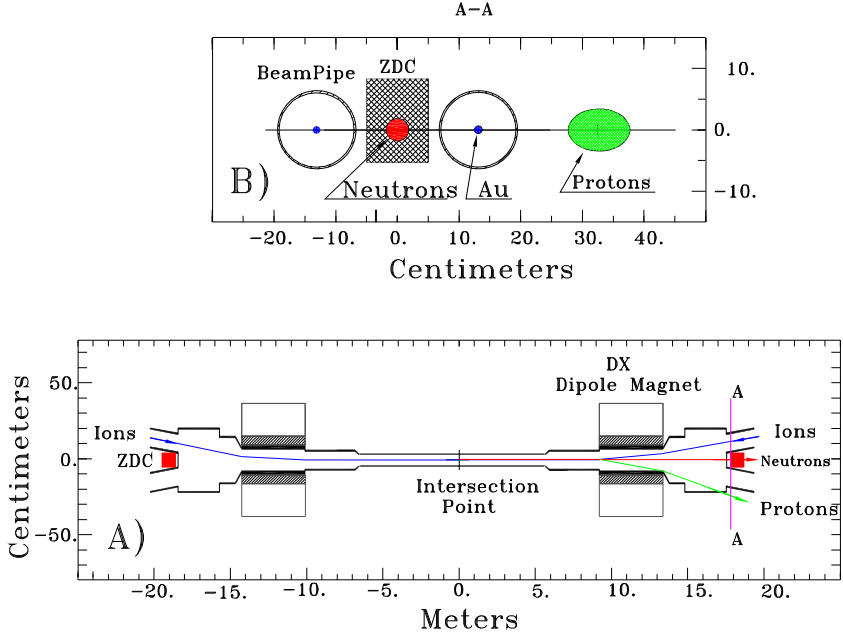}
}}
\caption {
Transverse view of the collision region, indicating the beam pipes, the Zero Degree Calorimeter, and the impact regions
of neutrons and charged fragments with \MM{\DIV{\Znum}{\Anum} = 1} (top)\@.
Top view, showing the position of the ZDC modules behind the DX dipole magnets in between the two RHIC beam pipes (bottom)\@. 
Figure taken from~\protect\cite {ref_rhic_zdc}\@.
}
\label {fig_rhic_zdc_view}
\end {figure}

In addition to the STAR barrel detectors, a sampling calorimeter is placed at a distance of \MM{18\unit{\meter}} from the interaction point 
in the RHIC tunnel on both sides of the experimental hall, as shown in Figure~\ref {fig_rhic_zdc_view}\@.
These Zero Degree Calorimeters (ZDC)~\cite {ref_rhic_beam_instrumentation,ref_rhic_zdc} are used 
to provide the minimum bias trigger and to measure centralities in heavy ion collisions.
Furthermore, identical ZDC detectors are installed at each of the four RHIC experiments, providing 
comparable collision rate measurements to monitor the RHIC luminosity.

The ZDC detector measures the total energy of the unbound neutrons emitted from the nuclear fragments after a collision.
The charged fragments of the collision are bent away by the RHIC dipole magnets DX\@. In the upper plot of Figure~\ref {fig_rhic_zdc_view} 
is shown a transverse view at the front face of the ZDC, indicating the position of the two beam pipes, 
the neutron spot inside the ZDC acceptance, and the spot of deflected fragments with \MM{\DIV{\Znum}{\Anum} = 1}\@.

\begin {figure} [tb]
\centerline {\hbox {
\includegraphics [width=0.5\textwidth] {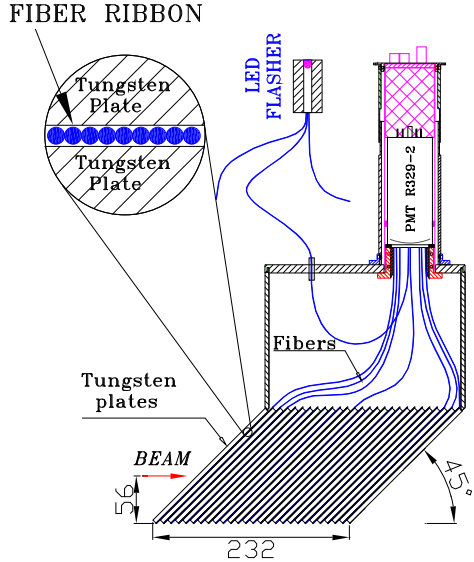}
}}
\caption {Mechanical design of the ZDC modules. Figure taken from~\protect\cite {ref_rhic_zdc}\@.}
\label {fig_rhic_zdc_design}
\end {figure}

The mechanical layout of the ZDC is shown in Figure~\ref {fig_rhic_zdc_design}\@. 
It consists of alternating layers of tungsten absorber and Cherenkov fibers with a total length of about \MM{0.7\unit{\meter}}\@. 
The transverse dimension of \MM{\xcoord{} \TIMES{} \ycoord{} = 10 \TIMES{} 13.6\unit{\cmsquare}} corresponds to 
an angular acceptance of about \MM{2.5\unit{\mrad}} around the forward direction.
\enlargethispage{\baselineskip}

In this thesis, we do not use the ZDC for centrality measurement, and refer to~\cite {ref_rhic_zdc} for details on such a measurement.
For the \deuterongold\ data used in the present analysis, the ZDC provided a minimum bias trigger 
by requiring the detection of at least one neutron in the \gold\ beam direction. 
The acceptance of this trigger corresponds to \MM{95 \PLMN{} 3\unit{\%}} of the 
total \deuterongold\ geometrical cross section, as determined from detailed simulations of the ZDC acceptance~\cite {ref_star_dAu_evidence}\@.

\subsection {Beam-Beam Counter}

To provide a minimum bias trigger in \protonproton\ collisions, Beam-Beam Counters (BBC)~\cite {ref_star_rel_lum_measur_bbc,ref_star_local_polar_bbc} 
are mounted around the beam pipe beyond both poletips of the STAR magnet at a distance of \MM{3.7\unit{\meter}} from 
the interaction point. 
The BBC also serves to reject beam-gas events at the trigger level 
and to measure the beam luminosity in \protonproton\ runs.

The detector
\begin {figure} [t!]
\centerline {\hbox {
\includegraphics [width=0.8\textwidth] {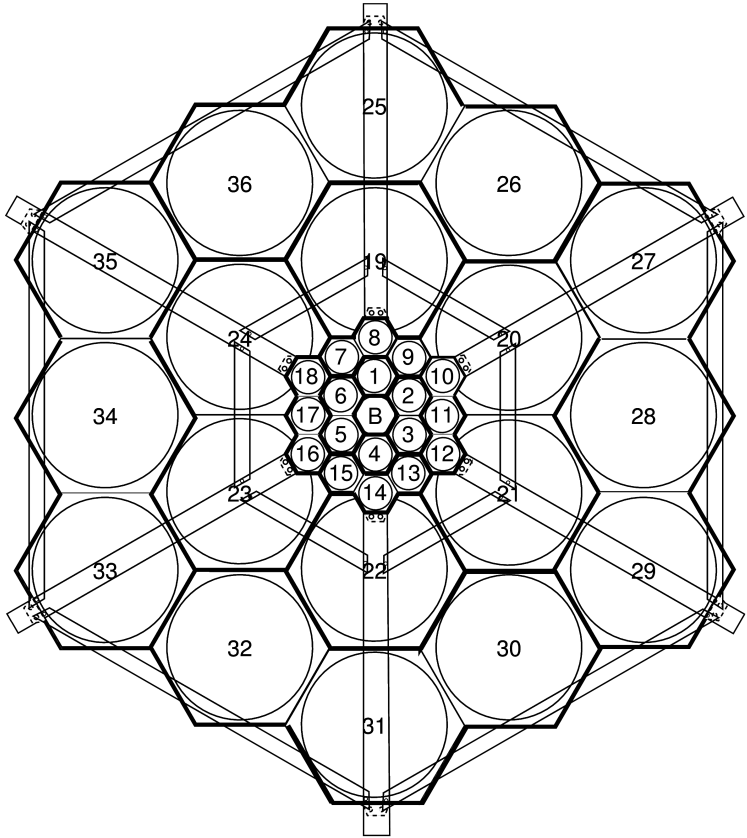}
}}
\caption {Schematic view of the BBC detector. Figure taken from~\protect\cite {ref_star_local_polar_bbc}\@.}
\label {fig_star_bbc_schematic}
\end {figure}
consists of two sets of hexagonal scintillator tiles, see Figure~\ref {fig_star_bbc_schematic}\@.
A ring with radius between \MM{9.6} and \MM{48\unit{\cm}} is formed by \MM{18} small tiles,
while \MM{18} large tiles on the outside
cover a radius between \MM{38} and \MM{193\unit{\cm}}\@. 
The small and large tile arrangements cover the pseudorapidities 
\MM{3.4 \LESS{} \etacoordabs \LESS{} 5.0} and \MM{2.1 \LESS{} \etacoordabs{} \LESS{} 3.6}, respectively.

In \protonproton\ runs, a minimum bias trigger is provided by a coincidence of 
signals in at least one of the \MM{18} small BBC tiles on each side of the interaction region.

The two BBC counters also record the time of flight, which provides a measurement of the \zcoord\ position of the interaction vertex 
to an accuracy of about \MM{40\unit{\cm}}\@.
Large values of the time of flight difference between the two BBC counters 
indicate the passage of beam halo, which is rejected at the trigger level.

A measurement of the counting rate in the BBCs allows for a determination of the absolute luminosity to an accuracy of about \MM{15\unitns{\%}}, 
the relative luminosities per run are determined to a precision of better than \SUP{10}{-3}~\cite {ref_star_rel_lum_measur_bbc,ref_star_local_polar_bbc}\@.

\chapter {STAR Electromagnetic Calorimeter}
\label {ref_chapter_BEMC}

\begin {figure} [tb]
\centerline {\hbox {
\includegraphics [width=1.05\textwidth] {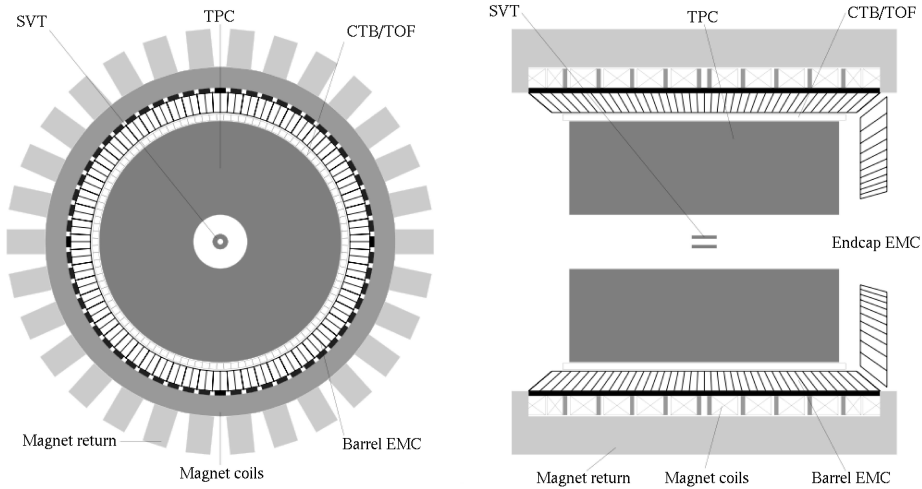}
}}
\caption {Cross-sectional and longitudinal view of the STAR detector, showing the layout of the BEMC\@.
Figure taken from~\protect\cite {ref_emc_nim}\@.}
\label {fig_bemc_view1}
\end {figure}

The Barrel Electromagnetic Calorimeter (BEMC)~\cite {ref_emc_nim} is a lead-scintillator \linebreak
sampling calorimeter, surrounding the STAR TPC as shown in Figure~\ref {fig_bemc_view1}\@.
The BEMC was installed in several stages during the period of \yr{2001}--\yr{2005}\@. 
Only the West half of the BEMC was fully operational during the \yr{2003}\ and \yr{2005}\ runs 
which provided the data presented in this thesis. 
The Endcap Calorimeter~\cite {ref_eemc_nim}, which is not used in the present analysis, 
was installed in the years \yr{2002}--\yr{2003}\@.

The BEMC is used to trigger on and to measure jets, leading hadrons, \linebreak
direct photons, and electrons from heavy quarks produced at large transverse momentum.
For this purpose, the BEMC provides large acceptance 
for photons, electrons, \pizero, and \etameson\ mesons in all colliding systems ranging from 
\protonproton\ up to \goldgold\@. 
In the next sections we will describe the BEMC in more detail.

\section {Mechanical layout}

The calorimeter is located inside the magnet coil and surrounds the TPC\@.
It covers a pseudorapidity range of \MM{\etacoordabs{} \LESS{} 1} and full azimuth, matching the TPC acceptance. 
The calorimeter is divided in two adjacent barrels, one positioned at the West half of the STAR detector 
(\MM{0 \LESS{} \etacoord{} \LESS{} 1}) and the
other one at the East half (\MM{-1 \LESS{} \etacoord{} \LESS{} 0})\@.
Each half-barrel has a length of \MM{293\unit{\cm}}, an inner radius of \MM{223\unit{\cm}}, and an outer radius of \MM{263\unit{\cm}}\@.

\begin {figure} [p]
\centerline {\hbox {
\includegraphics [width=\textwidth] {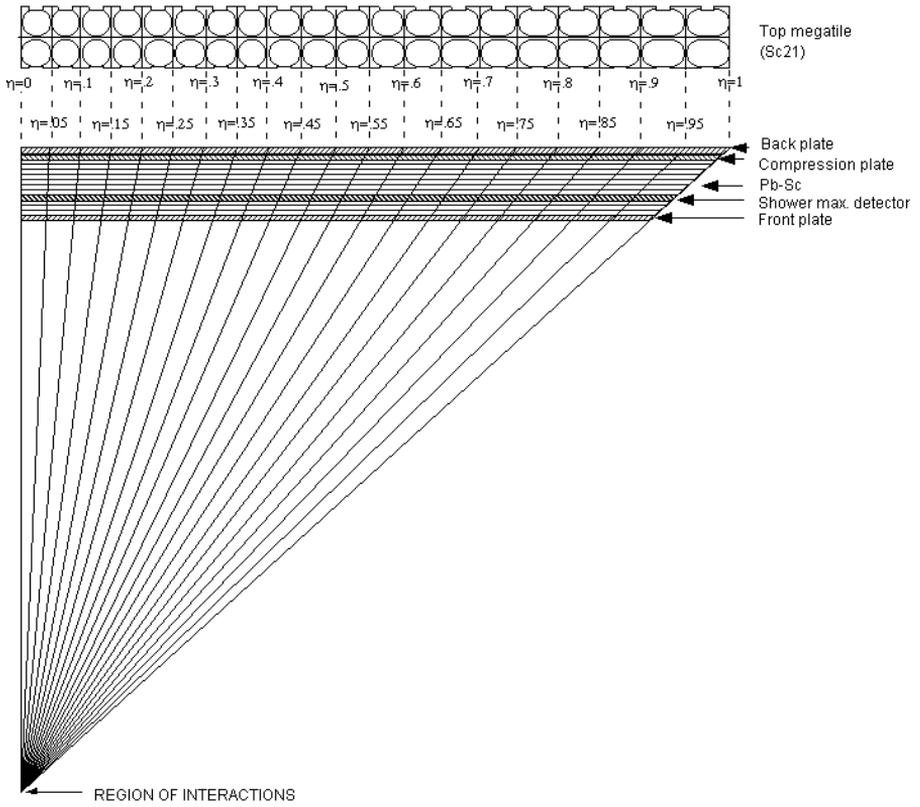}
}}
\caption {Side view of a calorimeter module and top view of a scintillator plate segmented into \MM{20 \TIMES{} 2} towers.
Figure taken from~\protect\cite {ref_emc_nim}\@.}
\label {fig_bemc_view2}
\end {figure}

\begin {figure} [p]
\centerline {\hbox {
\includegraphics [width=\textwidth] {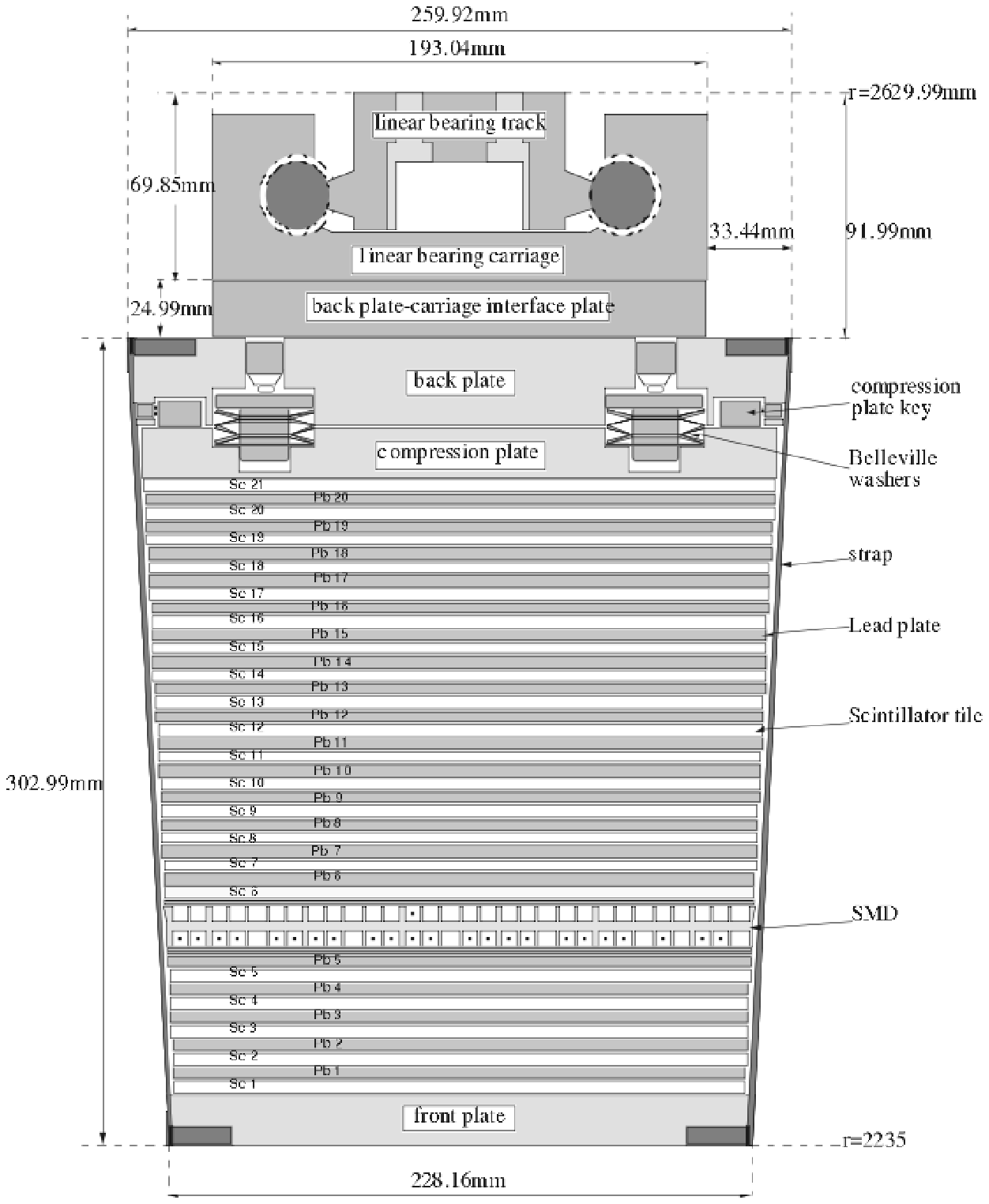}
}}
\caption {Transverse view of a calorimeter module, showing the inner layer of lead/scintillator stacks, 
the shower maximum detector (\SMD), 
the outer layer of stacks, and the carriage structure at the outer radius of the BEMC barrel.
Figure taken from~\protect\cite {ref_emc_nim}\@.}
\label {fig_bemc_view3}
\end {figure}

The half-barrel is azimuthally segmented into \MM{60} modules.
Each module is approximately \MM{26\unit{\cm}} wide and covers \MM{6} degrees (\MM{17\unit{\mrad}}) 
in azimuth and one unit in pseudorapidity.
The active depth is \MM{23.5\unit{\cm}}, to which is added \MM{6.6\unit{\cm}} of structural elements at the outer radius.
The longitudinal and transverse segmentation of a module is shown in Figure~\ref {fig_bemc_view2}, 
and the radial structure in Figure~\ref {fig_bemc_view3}\@.

The modules are segmented into \MM{40} projective towers of lead-scintillator stacks, \MM{2} in \phiangle\ and \MM{20} in \etacoord\@.
A tower covers \MM{0.05} in \MM{\DELTA\phiangle} and \MM{0.05} in \MM{\DELTA\etacoord}\@.
Each calorimeter half is thus segmented into a total of \MM{2400} towers.

Each tower consists of an inner stack of \MM{5} layers of lead and \MM{5} layers of scintillator,
and an outer stack of \MM{15} layers of lead and \MM{16} layers of scintillator. 
All these layers are \MM{5\unit{\mm}} thick, except the innermost two scintillator layers, which are \MM{6\unit{\mm}} thick.
A separate readout of these latter two layers provides the calorimeter preshower signal.
A Shower Maximum Detector (\SMD) is positioned between the inner and outer stacks, at a depth of appoximately \MM{5} radiation lengths.
The whole stack is held together by mechanical compression and friction between layers.

\section {Optical structure}

The plastic scintillator layers are machined as ``megatiles'', covering the full length and width of a module. 
These megatiles are segmented into \MM{40} optically isolated tiles, as shown in the top diagram of Figure~\ref {fig_bemc_view2}\@. 
The optical separation between the individual tiles is achieved by \MM{95\unitns{\%}} 
deep cuts in the scintillator filled with opaque epoxy. 
The optical crosstalk between adjacent tiles is reduced to a level of \MM{0.5\unitns{\%}} 
by painting a black line on the surface opposite to the isolation groove.
\enlargethispage{\baselineskip}

\begin {figure} [tb]
\centerline {\hbox {
\includegraphics [width=0.6\textwidth] {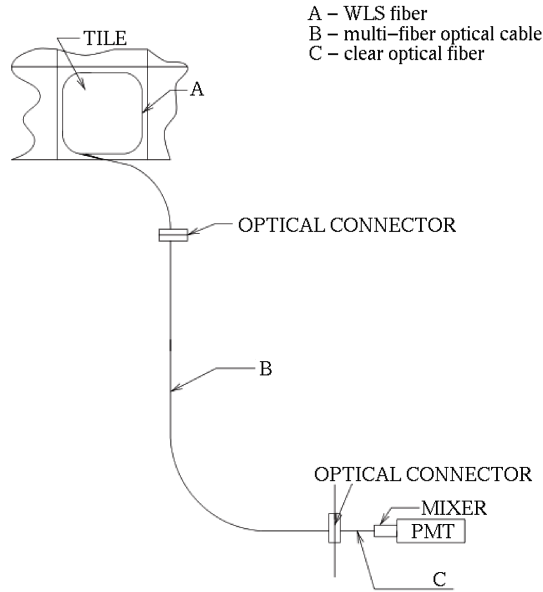}
}}
\caption {Optical readout scheme of a BEMC tower.
Figure taken from~\protect\cite {ref_emc_nim}\@.}
\label {fig_bemc_view4}
\end {figure}

The optical readout scheme is shown in Figure~\ref {fig_bemc_view4}\@.
The signal from each tile is collected by a wavelength shifting (WLS) fiber embedded in a \SIGMA-groove in the tile. 
The WLS fibers run along the outer surface of the stack and terminate in an optical connector mounted at the back-plate of the module. 
From the back-plate, \MM{2.1\unit{\meter}} long fibers run through the STAR magnet structure to the readout boxes mounted on the outer side of the magnet. 
In these boxes, the 21 fibers from the tiles of one tower are connected to a single photomultiplier tube (PMT)\@.
The PMTs are powered by Cockroft-$\!$Walton bases, which are remotely controlled over a serial communication line by the slow control software.

From layer by layer tests of the BEMC optical system,
together with an analysis of cosmic ray and test beam data, 
the nominal energy resolution of the calorimeter is estimated to be 
\MM{\DIV{\ensuremath{\delta}\energy}{\energy} = \DIV{15\unitns{\%}}{\SQRT{\energy[\unitns{\GeV}]}} \ensuremath{\oplus} 1.5\unitns{\%}}~\cite {ref_emc_beamtest}\@.

\section {Shower Maximum Detector}

The Shower Maximum Detector (\SMD) is a multi-\tsp{-0.1}wire proportional counter with strip readout.
It is located at a depth of approximately \MM{5.6} radiation lengths at \MM{\etacoord{} = 0} 
increasing to \MM{7.9} radiation lengths at \MM{\etacoord{} = 1}, 
including all material immediately in front of the calorimeter.

The purpose of the \SMD\ is to improve the spatial resolution of the calorimeter.
This is necessary because the transverse dimension of each tower (about \MM{10 \TIMES{} 10\unit{\cmsquare}})
is much larger than the lateral spread of an electromagnetic shower.
The improved resolution is essential to separate the two photon showers originating from the decay of high momentum \pizero\ and \etameson\ mesons.

\begin {figure} [p]
\centerline {\hbox {
\includegraphics [height=\textwidth,angle=270] {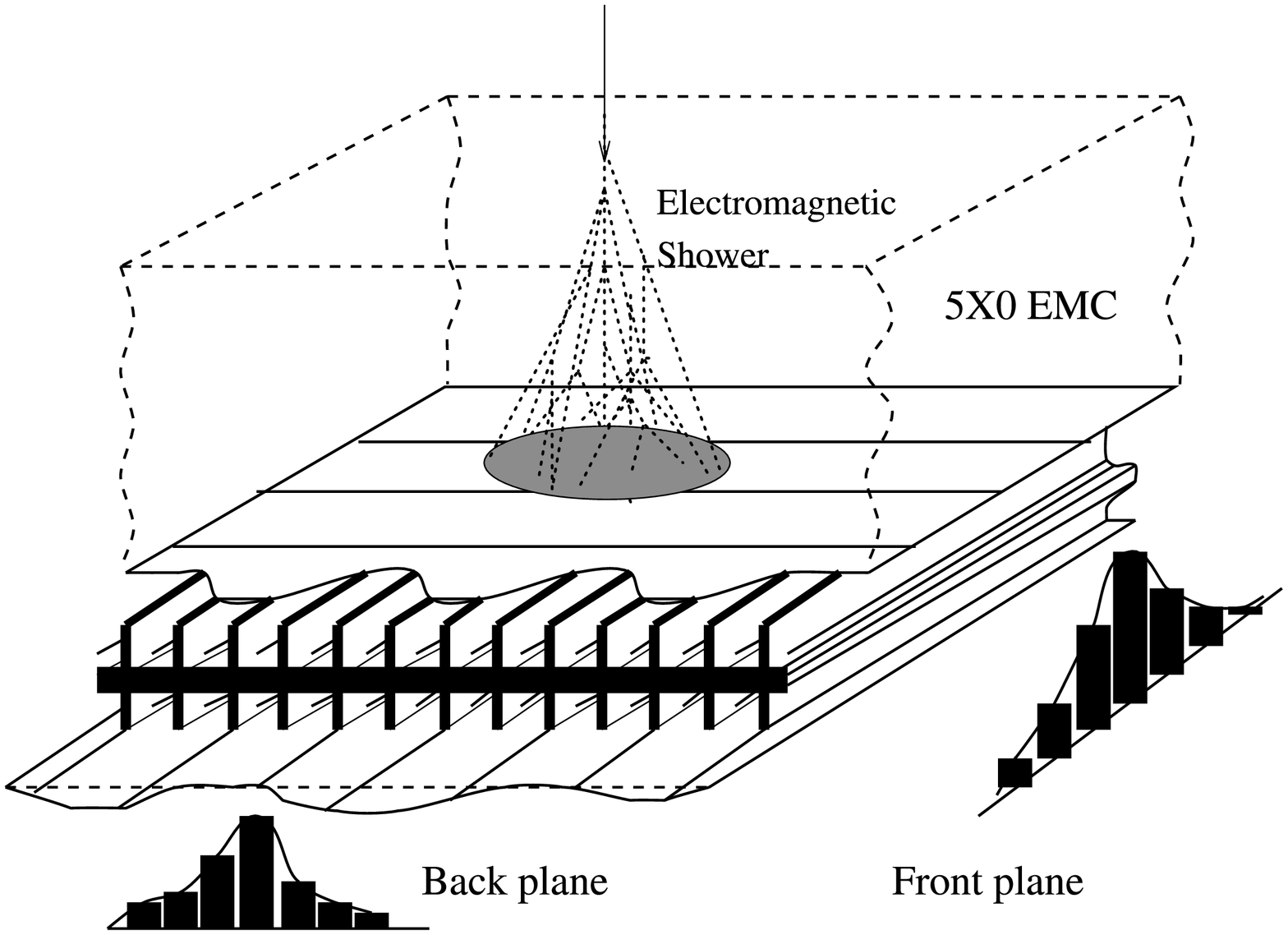}
}}
\centerline {\hbox {
\includegraphics [height=\textwidth,angle=270] {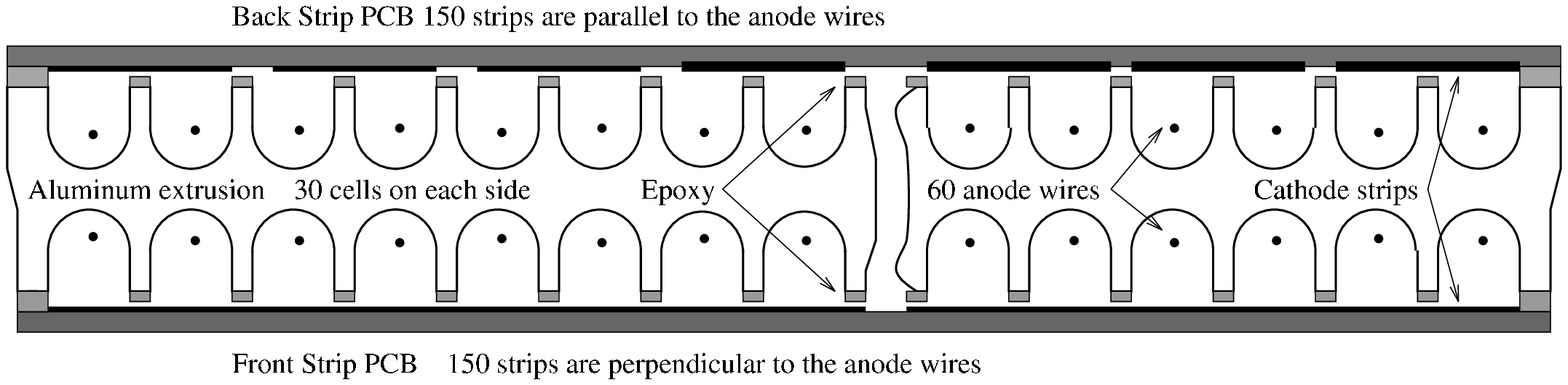}
}}
\caption {Schematic illustration of the \SMD, showing in the top figure 
a three dimensional view of the extruded aluminium profile, containing the anode wires, 
and the two readout pad planes, running parralel and perpendicular to the wires.
The profile of a BEMC shower, as recorded in the two \SMD\ pad planes, is shown by the histograms.
The bottom plot shows a transverse view of the aluminium extrusion, the anode wires, and the pad planes.
Figure~taken~from~\protect\cite {ref_emc_nim}\@.}
\label {fig_bemc_view5}
\end {figure}

The layout of the \SMD\ is shown in Figure~\ref {fig_bemc_view5}\@.
Independent cathode planes with strips along \etacoord\ and \phiangle\ directions allow the reconstruction of a two-dimensional image of a shower.
The coverage in \MM{\DELTA\etacoord{} \TIMES{} \DELTA\phiangle} is \MM{0.0064 \TIMES{} 0.1} 
for the \etacoord\ strips and \MM{0.1 \TIMES{} 0.0064} for the \phiangle\ strips.
There are a total of \MM{36000} strips in the full detector.

Beam test results at the AGS have shown that the \SMD\ has an approximately linear response versus energy.
The energy resolution in the \etacoord\ coordinate (front plane) is approximately 
\MM{\DIV{\ensuremath{\delta}\energy}{\energy} = \DIV{86\unitns{\%}}{\SQRT{\energy[\unitns{\GeV}]}} \ensuremath{\oplus} 12\unitns{\%}}, 
whereas that in the \phiangle\ coordinate (back plane) is worse by about \MM{3}--\MM{4\unitns{\%}}\@.
The position resolution is 
\MM{\SIGMA(\IT{r}\phiangle) = \DIV{5.6}{\SQRT{\energy[\unitns{\GeV}]}} \ensuremath{\oplus} 2.4\unit{\mm}} 
and 
\MM{\SIGMA(\IT{z}) = \DIV{5.8}{\SQRT{\energy[\unitns{\GeV}]}} \ensuremath{\oplus} 3.2\unit{\mm}}\@.

\section {Preshower Detector}

The first and second scintillating layers of each calorimeter module are used as a preshower detector (PSD)\@. 
To achieve a separate readout of these layers, two WLS fibers are embedded instead of one in the \SIGMA-groove of each tile.
This additional pair of fibers from the two layers illuminate a single pixel of a multi-anode PMT\@. 
A total of \MM{300} \MM{16}\tsp{0.3}-pixel multi-anode PMTs are used to provide the \MM{4800} tower preshower signals.

The preshower detector was fully instrumented and read out only in \yr{2006}, so that it could not be used in the present analysis.

\section {BEMC electronics}

The calorimeter is a ``fast'' detector in STAR, so that its ADCs can be read out on each RHIC bunch crossing.
The calorimeter data is also used in the STAR \levelZero\ trigger, in the form of the ``High Tower'' and ``Patch Sum'' trigger primitives.

The \levelZero\ HighTower trigger used in this analysis is a requirement that the energy deposited in any single calorimeter cell in the event 
exceeds a given threshold. This allows one to enhance the statistics at the high energy part of the spectrum.

The complete description of the BEMC electronics operation is given in Appendix~\ref {BEMC_electronics_operation}\@.

\chapter {Event reconstruction in STAR}
\label {ref_event_reconstruction}

\section {Data aquisition and trigger}
\label {section_triggers}

The STAR data aquisition system (DAQ)~\cite {ref_star_daq} receives the input from multiple detectors at various readout rates. 
The typical recorded event rate of \MM{100\unit{\Hz}} is limited by the drift time in the TPC (the slowest detector in STAR)\@.
The total event size can reach up to \MM{200\unit{\megabyte}} in \goldgold\ collisions. 
STAR takes data in runs of about half an hour duration, each having \MM{50}\tsp{0.3}--\MM{100\e{3}} events.

The STAR trigger~\cite {ref_star_trigger} is a pipelined system, capable to cope with the RHIC beam crossing frequency of \MM{10\unit{\MHz}}\@. 
The trigger processes information from fast detectors, such as the ZDC, BBC, CTB, or BEMC, 
and decides if the event should be read out and saved to tape.
Each event is categorized by multiple trigger criteria, and the events selected by various branches of the 
decision tree are written to tape, sharing the available \MM{100\unit{\Hz}} DAQ bandwidth.

The datasets used in the present analysis were taken in the \deuterongold\ run of \yr{2003}\ and the \protonproton\ run of \yr{2005}, 
see also Table~\ref {tab:1}\@.
The following trigger conditions had to be satisfied:

\label {syst_minbias_xsec}
\paragraph {Minimum bias (MinBias) trigger in \boldmath \deuterongold \unboldmath \ collisions} 
This condition required the presence of at least one neutron signal in the ZDC in the gold beam direction.
As given in~\cite {ref_star_dAu_evidence}, this trigger condition captured \MM{95 \PLMN{} 3\unit{\%}} 
of the total \deuterongold\ geometric 
cross section of \MM{\SUP{\SUB{\SIGMA}{\RM{hadr}}}{\deuterongold} = 2.21 \PLMN{} 0.09\unit{\barn}}\@.

\paragraph {MinBias trigger in \boldmath \protonproton \unboldmath \ collisions}
This condition required the coincidence of signals from two BBC tiles on the opposite sides of the interaction point.
Due to the dual-arm configuration, this trigger is sensitive to the non-singly diffractive (NSD) cross section, 
which is a sum of the non-diffractive and doubly diffractive cross section. 
The total inelastic cross section is a sum of the NSD and singly diffractive cross section.

A minimum bias cross section of 
\MM{\SUB{\SIGMA}{\RM{BBC}} = 26.1 \PLMN{} 0.2(\RM{stat}) \PLMN{} 1.8(\RM{syst})\unit{\mb}} 
was independently measured via Vernier scans in dedicated accelerator runs~\cite {ref_bbc_vernierscan}\@.
This trigger captured \MM{87 \PLMN{} 8\unit{\%}} of the \protonproton\ non-singly diffractive (NSD) cross section, as was determined 
from the detailed simulation of the BBC acceptance~\cite {ref_star_AuAu_suppression}\@.
Correcting the BBC cross section for the acceptance, we obtain a value for the NSD cross section of 
\MM{\SUP{\SUB{\SIGMA}{\RM{NSD}}}{\protonproton} = 30.0 \PLMN{} 3.5\unit{\mb}}\@.

\paragraph {HighTower trigger}
This condition required, in addition to the MinBias, an energy deposit above a predefined threshold in at least one calorimeter tower.
The purpose of this trigger is to enrich the sample with events that have a large transverse energy deposit.
Two different thresholds were applied, giving the \HighTowerOne\ and \HighTowerTwo\ datasets.
The values of these thresholds for the various runs are shown in Table~\ref {tab:ht_thresholds}\@.
\begin {table}
\caption {\normalsize HighTower trigger thresholds used in \deuterongold\ \yr{2003}\ and \protonproton\ \yr{2005}\ data runs.}
\label {tab:ht_thresholds}
\begin {center}
\begin {tabular} {ccc}
Dataset	                 & \HighTowerOne\ threshold & \HighTowerTwo\ threshold \\
	                 & [\unitns{\GeV}]         & [\unitns{\GeV}]         \\
\hline
\deuterongold\ \yr{2003} & 2.5 & 4.5 \\
\protonproton\ \yr{2005} & 2.6 & 3.5 \\
\end {tabular}
\end {center}
\end {table}

\section {STAR reconstruction chain}

The events recorded on tape are passed through the standard STAR reconstruction chain.
This reconstruction is performed routinely on the RHIC Computing Facility (RCF),
which is a large computing farm located at BNL\@.

The most important part of the data reconstruction at this stage is tracking 
in the TPC and FTPCs\@. Charged tracks are reconstructed in the main TPC 
using a Kalman filter~\cite {ref_star_tpc_kalman}, and in the FTPCs using a conformal mapping method~\cite {ref_tracking_conformal}\@.
The primary vertex is found by extrapolating and intersecting all reconstructed tracks. 
The vertex resolution in \zcoord\ is between \MM{0.9\unit{\mm}} and \MM{0.35\unit{\mm}} depending on the track multiplicity,
whereas in the transverse plane it is about \MM{0.5\unit{\mm}}\@.
Once the vertex has been found, all tracks that approach to it closer than \MM{3\unit{\cm}} are 
re\tsp{0.1}-fitted to include the vertex position as the origin.
Although the wire chambers are sensitive to almost \MM{100\unitns{\%}} of the drifting electrons, 
the overall tracking efficiency is only \MM{80}--\MM{90\unitns{\%}} due to fiducial cuts, track merging, bad pads, and dead channels. 
The momentum resolution of tracks worsens linearly with \pT\ from \MM{2\unitns{\%}} for \MM{300\unit{\MeVc}} pions 
to \MM{7\unitns{\%}} for \MM{4\unit{\GeVc}} pions.

Because the BEMC reconstruction is not yet performed in the standard STAR reconstruction,
the raw BEMC data are passed to the physics analysis.
The tower ADC data can be directly passed because they only take a small fraction of the event size.
The \SMD\ strip ADCs are zero-suppressed and then also passed to the analysis.
This scheme implies that removal of malfunctioning elements and a full calibration of the BEMC 
is performed as a part of the physics analysis.
This has the advantages that the reconstructed electron tracks in the TPC can be used to calibrate the energy response of the BEMC, 
and that the successive improvements in the BEMC calibration do not require re-generating the full dataset from the raw events on tape.

All data reconstruction and analysis in STAR is performed using the ROOT framework~\cite {ref_root_framework}\@.
The processing of a full dataset, such as \protonproton\ or \deuterongold, takes about three months.

\section {BEMC status tables}
\label {BEMC_status_tables_preparation}

A quality assurance (QA) procedure for the BEMC is routinely performed \linebreak
before the physics analysis, in order to remove malfunctioning detector components from the data 
and to correctly reproduce the time dependence of the detector acceptance in the Monte Carlo simulation.
This QA procedure results in timestamped status tables, which are used as an input to the physics analysis.
Below we describe the QA procedure performed for the BEMC towers, a similar procedure is applied to the \SMD\ strips.

\begin {figure} [tb]
\centerline {\hbox {
\includegraphics [width=\textwidth] {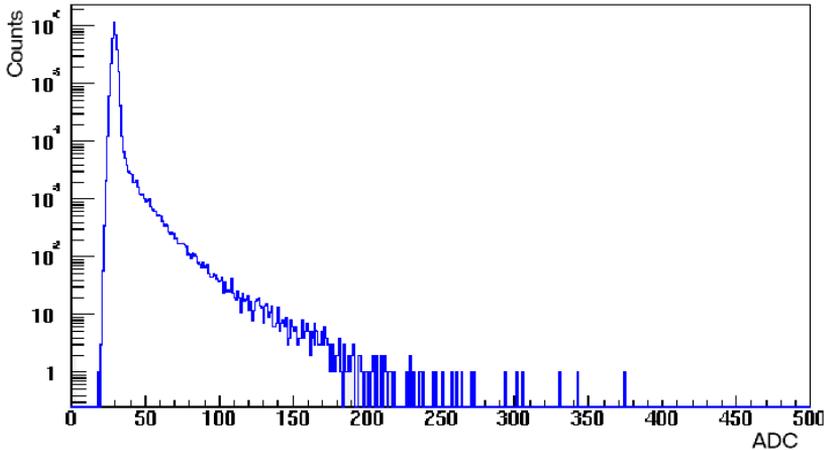}
}}
\caption {ADC spectrum of a BEMC tower, showing a pedestal located at \MM{30} ADC counts.}
\label {fig_adc_spectrum}
\end {figure}

For each run, the raw ADC spectra of all towers were accumulated 
and a number of criteria were applied to recognize common failure modes, 
such as the malfunctioning of entire readout boards and crates.
A typical ADC spectrum of a tower is displayed in Figure~\ref {fig_adc_spectrum}
and shows the signal distribution and the accumulation of ADC counts in absense of a signal (pedestal)\@.
The position of these pedestals provide the zero offset of the ADC measurement and are, \linebreak
together with the width, stored in time dependent tables for each tower.
Channels with anomalous pedestal positions and widths are flagged as bad.
The signal fraction was defined as the number of ADC counts that are more than six standard deviations above the pedestal.
Towers with a signal fraction smaller than \MM{0.001} are flagged as ``cold'' or ``dead'',
while those with a fraction above \MM{0.1} are marked as ``hot'' or ``noisy'' 
(the exact numbers are multiplicity dependent and are adjusted for each collision system)\@.
Saved are, as function of run number, the position of the pedestals, their widths, and flags indicating the status of each tower.
The average fraction of good towers was found to be about \MM{90\unitns{\%}} in the \yr{2003}\ \deuterongold\ run,
with run-to-run fluctuations of about \MM{2}--\MM{5\unitns{\%}}\@.
In the \yr{2005}\ \protonproton\ data the fraction of good towers was found to be about \MM{97\unitns{\%}}\@.

In the BEMC reconstruction performed in this analysis, the status tables were read in and used for pedestal subtraction of the ADC
signals and for removal of towers which were flagged as bad.

\section {BEMC energy calibration}
\label {energy_calibration}

The purpose of the energy calibration is to establish the relation between ADC counts and the energy scale in \unitns{\GeV}\@.
The calibration proceeds in two stages.
First, a relative calibration matches the gains of individual towers to achieve an overall uniform response of the detector.
A common scale between ADC counts and energy is then determined in a second absolute calibration step.
The relative tower-by-tower calibration is done using minimum ionizing particles (MIP), while
the absolute energy scale is determined from 
energy measurements of identified electrons in the TPC\@.

\subsection {MIP calibration}

A significant fraction (\MM{30}--\MM{40\unitns{\%}}) of high energy charged hadrons traversing the BEMC only deposit a small amount of energy in the towers, 
equivalent to a \MM{250}--\MM{350\unit{\MeV}} electron, largely due to ionization energy loss (minimum ionizing particles)\@.
The signal from these particles is usually well separated from the tower pedestals. 

\begin {figure} [b!]
\centerline {\hbox {
\includegraphics [width=0.7\textwidth] {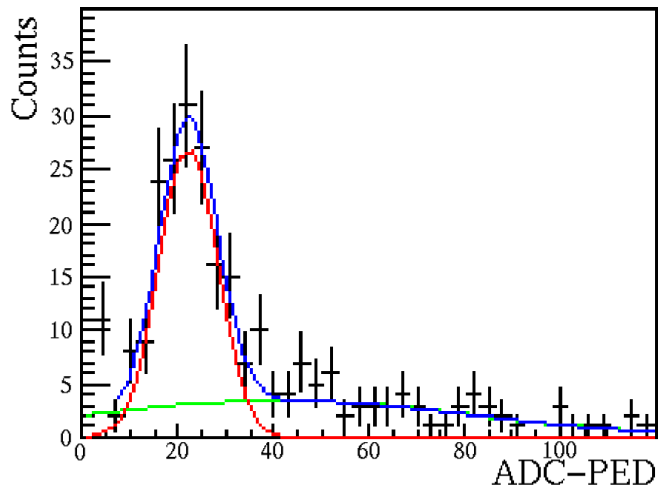}
}}
\caption {The BEMC tower response to MIP tracks. The curves indicate a fit to two Gaussians, one for the peak and one for the background.
Figure taken from~\protect\cite {ref_emc_calib2003}\@.}
\label {fig_mip}
\end {figure}

To identify MIP particles, TPC tracks of sufficiently large momentum above \MM{1.2\unit{\GeVc}} 
are extrapolated to the BEMC and the response spectra are accumulated, provided that
the track extrapolation is contained within one tower and 
that there are no other tracks found in a \MM{3 \TIMES{} 3} patch around this tower.
In Figure~\ref {fig_mip} 
is shown a tower ADC spectrum collected from the \deuterongold\ dataset, which clearly shows the position of the MIP peak
superimposed on a broad background~\cite {ref_emc_calib2003}\@.
The position of the fitted Gaussian is calculated for each tower and used to calculate the 
tower-by-tower gain corrections needed to equalize the detector.

The disadvantage of this method is that the calibration is performed at the low end of the scale, where the signal 
is more susceptible to noise and where the lack of lever arm does not allow to detect
possible non-linearities in the detector response.

\subsection {Electron calibration}

Because the electron momentum can be independently measured in the TPC, it is possible to calibrate 
the absolute energy scale of the calorimeter using the simple
relation for the ultra-relativistic electrons, \MM{\DIV{\energy}{\momentum} = 1}\@.

\begin {figure} [tb]
\centerline {\hbox {
\includegraphics [width=0.7\textwidth] {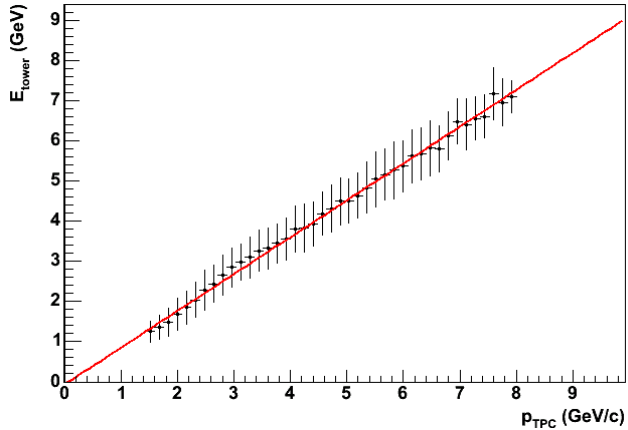}
}}
\caption {Electron energy measured in BEMC vs.~momentum measured in TPC\@. 
The first order polynomial fit determines the global calibration constant.
Figure taken from~\protect\cite {ref_emc_calib2003}\@.}
\label {fig_electron}
\end {figure}

Figure~\ref {fig_electron} 
shows the electron energy measured in the calorimeter versus its momentum measured in the TPC~\cite {ref_emc_calib2003}\@. 
The calorimeter response is quite linear up to \MM{8\unit{\GeV}}, and the 
global gain correction obtained from the linear fit is applied to all towers.

This method takes advantage of the well understood TPC detector for the precise measurement of the electron track momentum in a wide range. 
However, it requires high statistics to calibrate the high energy part of the spectrum, so that only one global 
calibration constant for the calorimeter is obtained at present.
The systematic year-to\tsp{0.4}-\tsp{-0.1}year uncertainty on the electron calibration was estimated to be \MM{5\unitns{\%}}~\cite {ref_spin_asymmetry}\@.

It has been found that the current calibration is less reliable at the edges of the calorimeter half-barrel, 
therefore, the tower signals from the two \etacoord-rings at each side are later removed from this analysis (see Section~\ref {cluster_cuts})\@.

This combination of the MIP-based equalization and electron-based absolute calibration is applied to the data after each running period, 
starting from \yr{2003}\ \deuterongold\ run.
The run dependent calibration constants are saved in the STAR database and automatically applied to the \ADC\ readout in the software.

\section {Event selection}
\label {event_selection}

The event selection starts with rejecting events where subdetectors needed for this analysis were not operational or malfunctioning.
In the following sub\tsp{0.3}-sections we will describe several additional selection criteria in detail.

\subsection {Beam background rejection}

\begin {figure} [tb]
\centerline {\hbox {
\includegraphics [width=0.5\textwidth,clip] {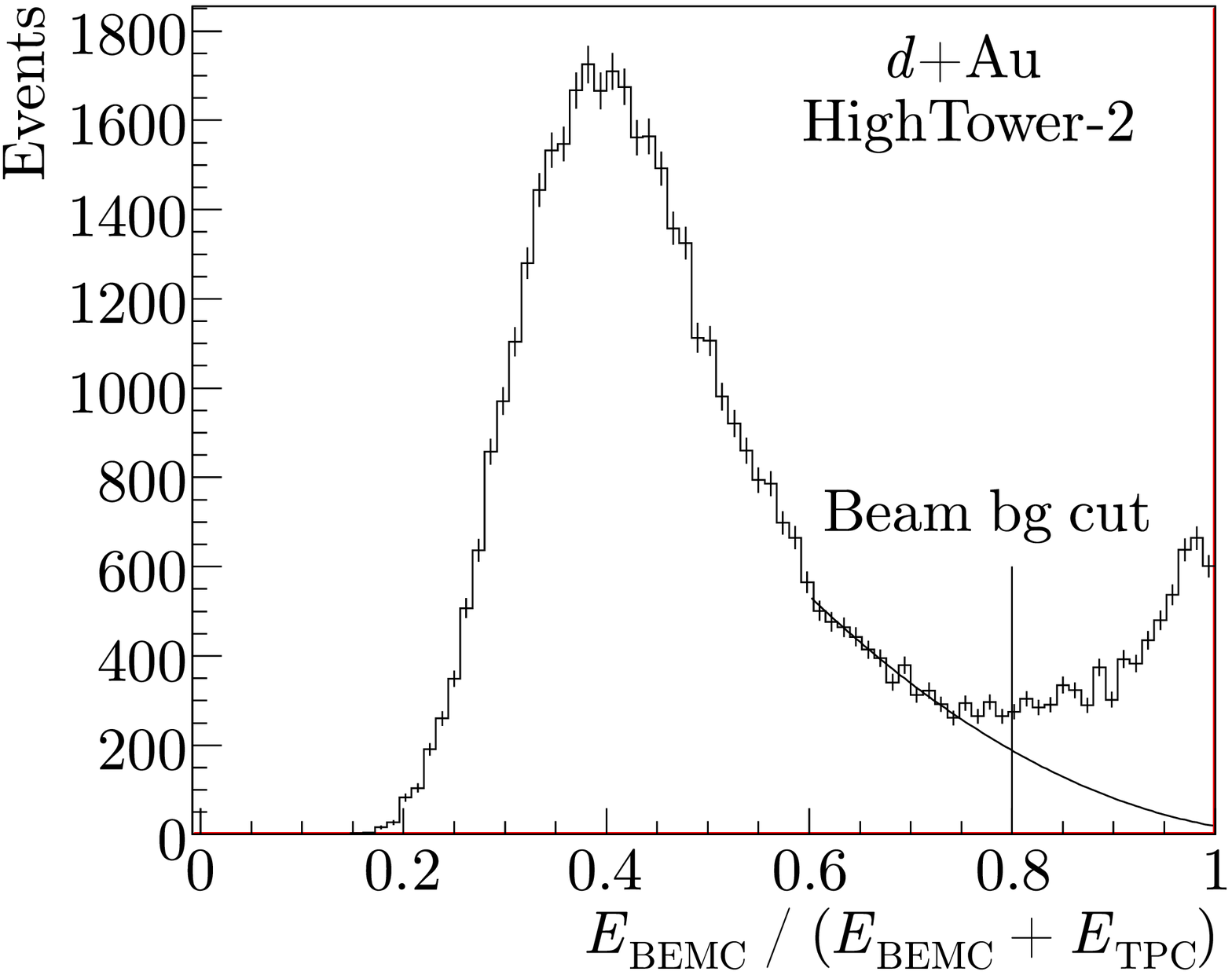}\includegraphics [width=0.5\textwidth,clip] {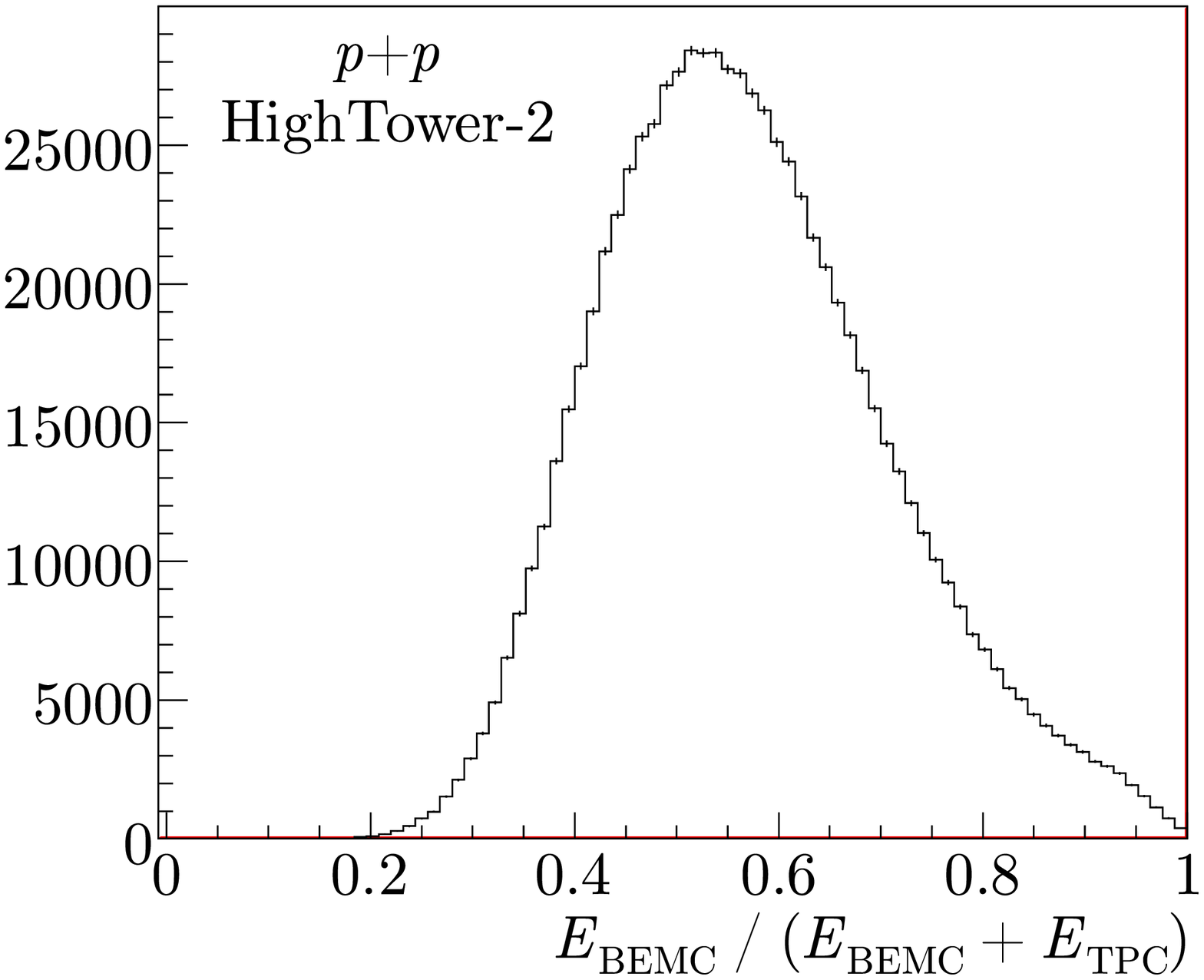}
}}
\caption {Distribution of \MM{\rBeamBg{} = \DIV{\SUB{\energy}{\RM{BEMC}}}{(\SUB{\energy}{\RM{BEMC}} + \SUB{\energy}{\RM{TPC}})}}, 
which shows beam background at \MM{\rBeamBg{} \GREATER{} 0.8} in \deuterongold\ events (left) 
and its absense in \protonproton\ events (right)\@.
The curve in the left hand plot indicates a polynomial fit used to estimate the false rejection rate in the \deuterongold~data.}
\label {fig_cuts_beambg_1}
\end {figure}

In \deuterongold\ events, interactions of gold beam particles with material 
approximately \MM{40\unit{\meter}} upstream from the interaction region
give rise to charged tracks that traverse the detector almost parallel to the beam direction. 
To identify events containing such background tracks, the ratio
$$
\rBeamBg{} = \FRAC {\SUB{\energy}{\RM{BEMC}}} {\SUB{\energy}{\RM{BEMC}} + \SUB{\energy}{\RM{TPC}}}
$$
is calculated, where \SUB{\energy}{\RM{BEMC}} is the total energy recorded in the BEMC 
and \SUB{\energy}{\RM{TPC}} is the energy of all charged tracks reconstructed in TPC\@.
In events containing background tracks, the ratio \rBeamBg\ tends to become large
because the background tracks give a large energy deposit in a calorimeter without being 
reconstructed in the TPC, since they do not point to the vertex.
This is shown in Figure~\ref {fig_cuts_beambg_1},
where the distribution of \rBeamBg\ is plotted for the \deuterongold\ and \protonproton\ datasets.
The peak near unity in the left\tsp{0.2}-hand plot indicates the presence of beam halo in \deuterongold\ collisions, 
and events with \MM{\rBeamBg{} \GREATER{} 0.8} were removed from the \deuterongold\ analysis.
This cut rejected \MM{3.4\unitns{\%}} of MinBias and \MM{13\unitns{\%}} of \HighTowerTwo\ triggered events.
From a polynomial fit to the \deuterongold\ distribution in the region \MM{\rBeamBg{} = 0.6}--\MM{0.8} 
(curve in Figure~\ref {fig_cuts_beambg_1}), 
the false rejection rate was estimated to be \MM{3.6\unitns{\%}} in the \deuterongold\ \HighTowerTwo\ data 
and less than \MM{1\unitns{\%}} in the other datasets.

The cut was not applied to the \protonproton\ data since here the beam background is almost absent,
as can be seen in the right\tsp{0.2}-hand plot of Figure~\ref {fig_cuts_beambg_1}\@.

During the summer in \yr{2006}, additional shielding walls were installed in STAR to reduce this beam background to a negligible level.

\subsection {Vertex reconstruction}

\begin {figure} [tb]
\centerline {\hbox {
\includegraphics [width=\textwidth] {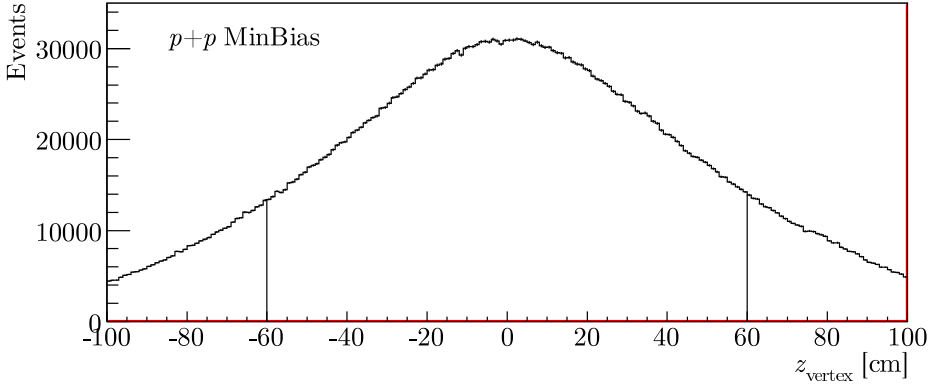}
}}
\caption {Vertex distribution in the \protonproton\ MinBias dataset.
Vertical lines indicate the cut used in the analysis.}
\label {fig_vertex_z_tpc}
\end {figure}

The event vertex is reconstructed to an accuracy of better than a millimiter in 
the \zcoord\ direction, from the tracks reconstructed in the TPC\@.
The distribution of the vertex \zcoord\ coordinate in the \protonproton\ MinBias data is shown in Figure~\ref {fig_vertex_z_tpc}\@.

Events with \MM{\zvertexabs{} \GREATER{} 60\unit{\cm}} were rejected in the analysis, 
as indicated by the vertical lines in Figure~\ref {fig_vertex_z_tpc}\@.
This cut is applied because the amount of material traversed by a particle increases dramatically at large values of \zvertexabs\@.
As a consequence, the TPC tracking efficiency drops for vertices located far from the center of the detector.

\begin {figure} [p]
\centerline {\hbox {
\includegraphics [width=0.9\textwidth] {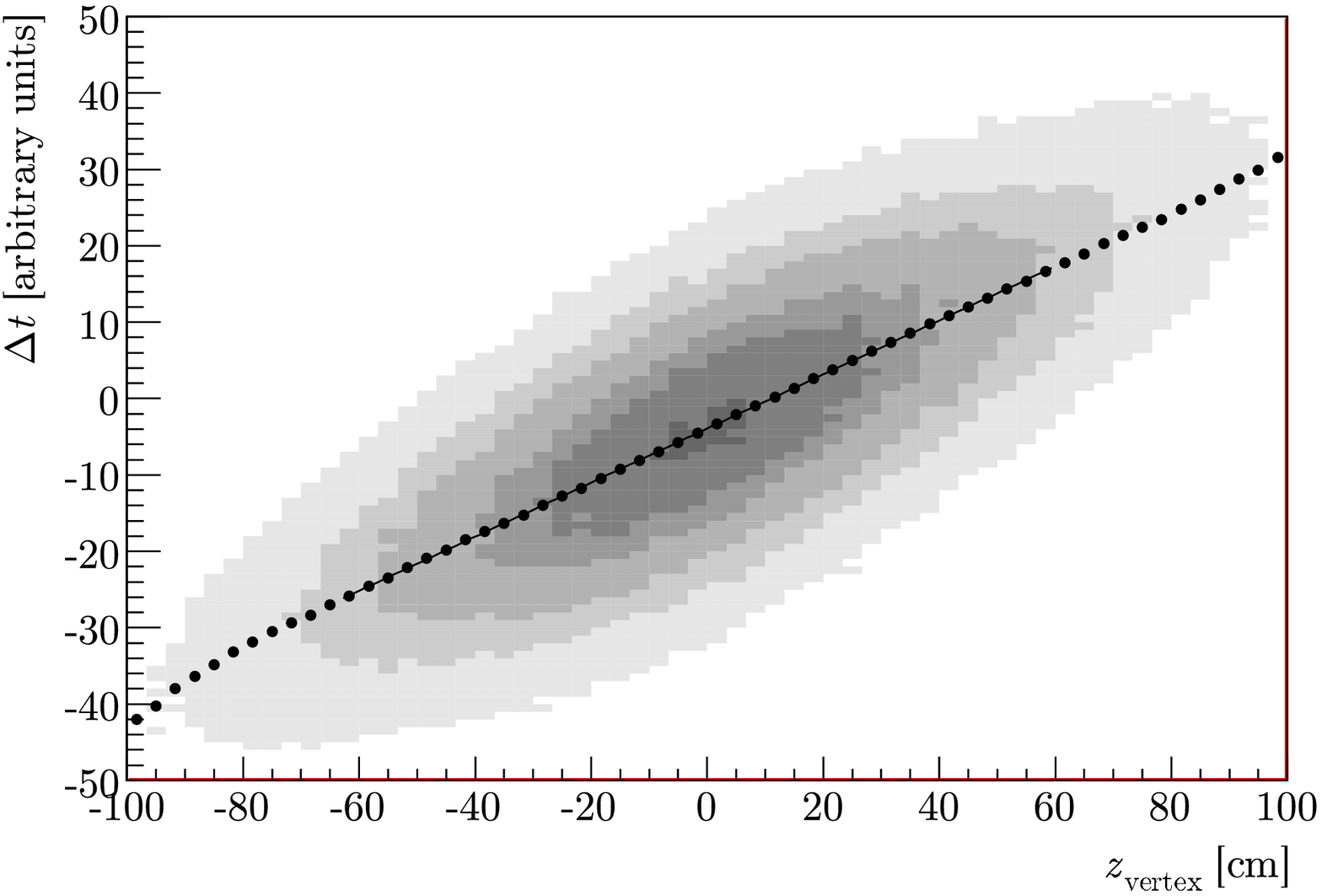}
}}
\centerline {\hbox {
\includegraphics [width=0.9\textwidth] {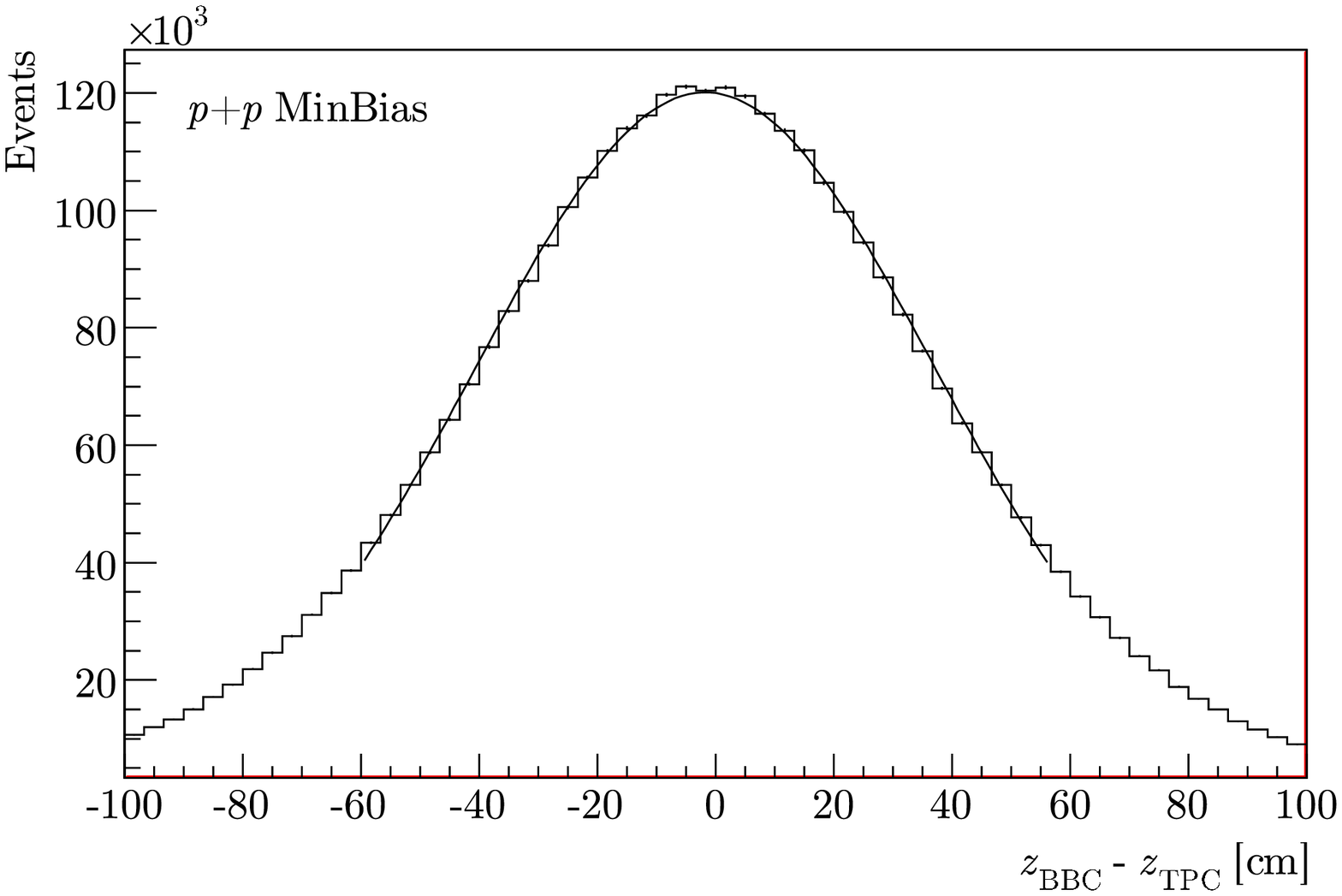}
}}
\caption {The correlation between the BBC time difference and \zvertex\ (top)\@.
Dots represent the positions of a fitted Gaussian in each vertical slice.
The straight line indicates a linear fit used to calibrate the BBC readings.
The distribution of \MM{\SUB{\zcoord}{\RM{BBC}} - \SUB{\zcoord}{\RM{TPC}}} and a Gaussian fit (bottom)\@.
}
\label {fig_vertex_bbc_corr}
\end {figure}

In the HighTower trigger data, the track multiplicity is almost always sufficient for a TPC vertex reconstruction,
but this is not so in the \protonproton\ and \deuterongold\ minimum bias data.
Since the \protonproton\ minimum bias trigger is based on coincidences in the BBC, we can use the timing information of the BBC
to reconstruct a vertex for every event, even when the TPC vertex reconstruction failed (about \MM{35\unitns{\%}} of the minimum bias events)\@.
The timing information from the BBC was calibrated against the \zcoord\ vertex coordinate reconstructed in the TPC, 
as illustrated in Figure~\ref {fig_vertex_bbc_corr} (top),
where we show the correlation between the BBC time difference \MM{\DELTA\IT{t}} and \zvertex\ in the TPC\@.
The straight line in the plot corresponds to a linear fit
$$
\zvertex{} = \IT{a}\DELTA\IT{t} + \IT{b}
$$
yielding \MM{\IT{a} = 2.824 \PLMN{} 0.003\unit{\cm}} per ADC count and 
\MM{\IT{b} = 11.00 \PLMN{} 0.02\unit{\cm}}\@.
In the bottom plot of Figure~\ref {fig_vertex_bbc_corr} 
we show the distribution of \MM{\SUB{\zcoord}{\RM{BBC}} - \SUB{\zcoord}{\RM{TPC}}}, together with a Gaussian fit.
From this fit we obtain the BBC vertex resolution of \MM{40\unit{\cm}}\@.

Whereas \protonproton\ events without a TPC vertex can be recovered by using the BBC timing information,
this cannot be done for \deuterongold\ events because the BBC is not in the trigger and timing information may be absent.
Since the \pizero\ reconstruction requires the presence of vertex, the \deuterongold\ events without a TPC vertex 
are removed from the analysis.
The vertex finding efficiency was determined from detailed Monte Carlo simulation of the full \deuterongold\ events 
and was found to be \MM{93 \PLMN{} 1\unit{\%}} in the \MM{\PLMN{} 60\unit{\cm}} window~\cite {ref_star_dAu_evidence}\@.
This result is used to correct the \deuterongold\ data for vertex inefficiencies, 
as will be explained in Section~\ref {cross_section_calculation}\@.
\clearpage

\subsection {HighTower trigger condition}
\label {ref_HT_filter}

The HighTower-triggered data are filtered using a software implementation of the HighTower trigger. 
In this filter, the highest tower ADC value found in the event is required to exceed 
the same \HighTowerOne\ (\HighTowerTwo) threshold as the one that was used during the run. 
This filter is needed to remove events that were falsely triggered due to the presence of noisy channels (hot towers)\@. 
Such channels are identified offline in a separate analysis and recorded in a database as described Section~\ref {BEMC_status_tables_preparation}\@.
This software filter also serves to make the trigger efficiency for Monte Carlo and real data as close as possible.

\section {Centrality selection in \textit{d}${} + {}$Au data}
\label {section_centrality_selection}

To measure the centrality in \deuterongold\ collisions, we use the correlation between 
the impact parameter of the collision and the charged track multiplicity in the forward direction.
This correlation was established from a Monte Carlo Glauber simulation ~\cite {ref_star_AuAu130,ref_glauber_2} using, 
as an input, the Woods-Saxon nuclear matter density for the gold ion~\cite {ref_hulthen_1} 
and the Hulth\'{e}n wave function of the deuteron~\cite {ref_hulthen_2}\@.
In this simulation, the inelastic cross section of an individual nucleon-nucleon collision 
was taken to be \MM{\SUP{\SUB{\SIGMA}{\RM{inel}}}{\nucleon\nucleon} = 42\unit{\mb}}\@.
The produced particles were then \linebreak
propagated through a full GEANT simulation of the STAR detector and the charged track
multiplicity was recorded, together with the number of nucleon-nucleon collisions simulated by the event generator.

\begin {figure} [tbp]
\centerline {\hbox {
\includegraphics [width=\textwidth] {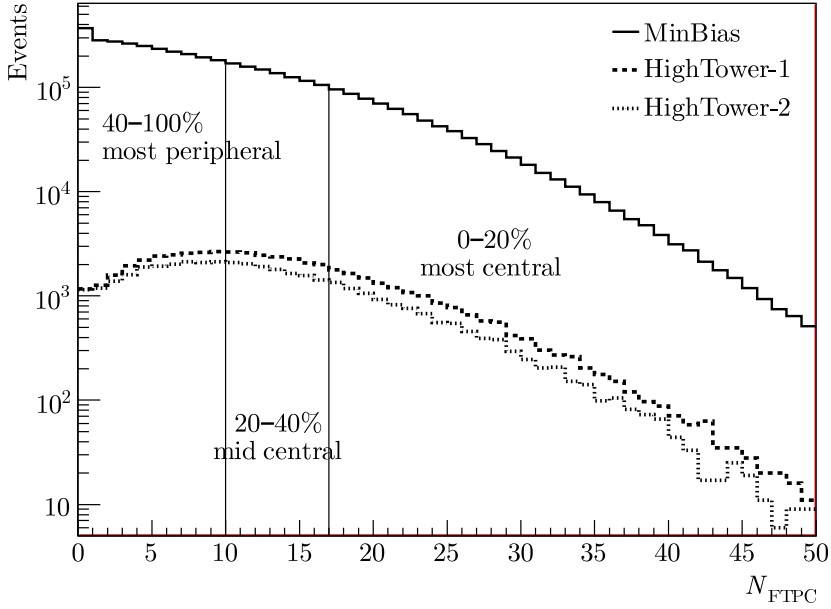}
}}
\caption {Centrality selection in \deuterongold\ data, based on the FTPC multiplicity.
Three centrality classes are defined, containing \MM{0}--\tsp{-0.3}\MM{20\unitns{\%}} most central, 
\MM{20}--\MM{40\unitns{\%}} mid central, and \protect\linebreak
\MM{40}--\tsp{-0.4}\MM{100\unitns{\%}} most peripheral events.}
\label {fig_ftpc_centrality}
\end {figure}

For the event-by-event centrality determination, we measured the multiplicity (\FTPCRefMult) of tracks
reconstructed in the FTPC\tsp{0.3}-East acceptance (in the \gold\ beam direction), 
following the centrality binning scheme used in other STAR publications~\cite {ref_star_dAu_evidence,ref_mmiller_thesis}\@.
The following quality cuts were applied to the reconstructed
tracks: $\!$(i) at least \MM{6} hits are required on the track;
(ii) \MM{\pT{} \LESS{} 3\unit{\GeVc}}, to guarantee
that the track is fully contained in the FTPC acceptance, and
(iii) distance of closest approach (DCA) to the vertex should be less than \MM{3\unit{\cm}}\@.
The multiplicity distributions obtained from the \deuterongold\ dataset are shown in Figure~\ref {fig_ftpc_centrality}
for the MinBias, \HighTowerOne, and \HighTowerTwo\ triggers.

Based on the measured multiplicity, the events were separated into three centrality classes: 
\MM{0}--\tsp{-0.3}\MM{20\unitns{\%}} most central, \MM{20}--\MM{40\unitns{\%}} mid central, and \MM{40}--\tsp{-0.4}\MM{100\unitns{\%}} most peripheral, 
as illustrated by the vertical lines in Figure~\ref {fig_ftpc_centrality}\@.

Table~\ref {tab_centrality_classes}
\begin {table}[tbp]
\begin {center}
\caption {\normalsize Centrality classes defined for the \deuterongold\ data and the corresponding \Ncollmean\ values~\protect\cite {ref_mmiller_thesis,ref_star_dAu_evidence}\@.
The errors given for \Ncollmean\ indicate the systematic uncertainty.}
\label {tab_centrality_classes}
\begin {tabular} {r@{\ }lcr@{\,$\pm$\,}l}
\multicolumn {2} {c} {Centrality class} & \FTPCRefMult\ range & \multicolumn {2} {c} {\Ncollmean} \\
\hline
\deuterongold & minimum bias                     & ---                 &  7.5 & 0.4 \\
\MM{0}--\tsp{-0.3}\MM{20\unitns{\%}} & most central        & $< 10$              & 15.0 & 1.1 \\
\MM{20}--\MM{40\unitns{\%}} & mid central        & 10--16              & 10.2 & 1.0 \\
\MM{40}--\tsp{-0.4}\MM{100\unitns{\%}} & most peripheral   & $\geq 17$           &  4.0 & 0.3 \\
\hline
\multicolumn {2} {c} {\protonproton\ \ }         & ---                 & \multicolumn {2} {c} { \mbox{\ \,\tsp{-0.25}}1 } \\
\end {tabular}
\end {center}
\end {table}
lists the \FTPCRefMult\ ranges that defined the centrality classes, and the corresponding 
mean number of binary collisions \Ncollmean\ in each class, obtained from the Glauber model.
\label {syst_glauber_model}
\enlargethispage{\baselineskip} In the table are also listed the systematic uncertainties 
on \Ncollmean, which are estimated by varying the Glauber model parameters.

\chapter {Neutral meson reconstruction}

The goal of this analysis is to measure \pizero\ and \etameson\ production in \deuterongold\ and \protonproton\ collisions.
The \pizero\ and \etameson\ are identified by their decay 
$$
\pizero\TO\gama\gama \quad \RM{and} \quad \etameson\TO\gama\gama
.
$$
These decay modes have branching ratios of \MM{0.988} and \MM{0.392}, respectively~\cite {ref_pdg}\@.
The BEMC is used to detect the decay photons, as will be described in the next sections.
The lifetime of the \pizero\ is \MM{\ensuremath{\tau} = 8.4\e{-17}\unit{\second}}, which corresponds 
to a decay length \MM{\cspeed\ensuremath{\tau} = 0.025\unit{\mum}}\@.
The lifetime of the \etameson\ is even shorter (\MM{7\e{-19}\unit{\second}})\@.
Therefore, we can assume that the decay photons originate from the primary vertex.
For each event, the invariant mass 
\begin {equation} \label {eq:inv_mass}
\Mgg{} = \SQRT{2\tsp{0.5} \SUB{\energy}{1} \SUB{\energy}{2} (1 - \ensuremath{\cos\psi})}
\end {equation}
is calculated for all pairs of photons detected in the BEMC\@.
Here \SUB{\energy}{1} and \SUB{\energy}{2} are the energies of the decay photons and \ensuremath{\psi} is 
the opening angle between them, as measured in the laboratory system.

The reconstructed masses are accumulated in invariant mass spectra,
where the \pizero\ and the \etameson\ show up as peaks around their nominal masses.
These peaks are superimposed on a broad distribution of combinatorial background,
which originates from photon pairs that are not produced by the decay of a single parent particle.

In Table~\ref {tab:4}
\begin {table}
\begin {center}
\caption {\normalsize Statistics used in the analysis after the event selection.}
\label {tab:4}
\begin {tabular} {r@{\ }lrrr}
\multicolumn {2} {c} {Dataset}               & \multicolumn {3} {c} {Number of events} \\
\multicolumn {2} {c} {}                      &       MinBias &  \HighTowerOne &  \HighTowerTwo \\
\hline
\multicolumn {2} {c} {\ \deuterongold}         & 164\SP608 &     53\SP154 &     40\SP974 \\
\MM{0}--\tsp{-0.3}\MM{20\unitns{\%}}   & most central    & 21\SP382 &     12\SP567 &      8\SP744 \\
\MM{40}--\tsp{-0.4}\MM{100\unitns{\%}} & most peripheral & 108\SP904 &     33\SP201 &      24\SP658 \\
\hline
\multicolumn {2} {c} {\protonproton\ \ }     & 4\SP433\SP817 &    920\SP567 &    872\SP811 \\
\end {tabular}
\end {center}
\end {table}
we list the number of events in all datasets used in the analysis after the event selection procedures
described in section~\ref {event_selection} were applied.

\section {BEMC clustering}

The first step in the invariant mass reconstruction is to find clusters of energy deposits in the calorimeter.
The purpose of the cluster finding algorithm is to group adjacent hits that are likely to have originated from a single incident photon.
The algorithm is applied to the BEMC tower and preshower signals, as well as to the signals from each of the two \SMD\ layers.

The clustering algorithm starts by accumulating a list of cluster seeds 
that contains all hits in a module with an energy deposit above a certain threshold (\Eseed)\@.
Starting from the most energetic seed in the list, an energy ordered list of module 
hits is searched for those adjacent to the present cluster.
When such a hit is found, then, provided that it is above a second threshold (\Eadd), 
it is added to the cluster and removed from the list.
The clustering stops when either a pre-defined maximum cluster size (\Nmax) is reached or no more adjacent hits are found. 
The clustering algorithm then proceeds to process the next most energetic seed.
At the end, clusters with total energy below the third threshold (\Emin) are discarded. 
Note, that, by construction, the clusters are confined within a module and cannot be shared by adjacent modules.
However, the likelihood of cluster sharing between modules is considered to be low 
since the modules are physically separated by about \MM{12\unit{\mm}} air gaps.
In Table~\ref {tab:default_thresholds} 
\begin {table} [tbp]
\begin {center}
\caption {\normalsize Cluster finder threshold values used in the analysis.}
\label {tab:default_thresholds}
\begin {tabular} {cr@{.}lr@{.}lr@{.}lc}
Detector  & \multicolumn{2}{c}{\Eseed} & \multicolumn{2}{c}{\Eadd} & \multicolumn{2}{c}{\Emin} & \Nmax \\
          & \multicolumn{2}{c}{[\unitns{\GeV}]} & \multicolumn{2}{c}{[\unitns{\GeV}]} & \multicolumn{2}{c}{[\unitns{\GeV}]} & \\
\hline
Towers    & \mbox{\ }0&35          & 0&035         & \mbox{\ }0&02          & 4     \\
Preshower & \mbox{\ }0&35          & 0&035         & \mbox{\ }0&02          & 4     \\
\SMDe     & \mbox{\ }0&2           & 0&0005        & \mbox{\ }0&1           & 5     \\
\SMDp     & \mbox{\ }0&2           & 0&0005        & \mbox{\ }0&1           & 5     \\
\end {tabular}
\end {center}
\end {table}
we list the threshold values used in the clustering algorithm for all four detectors.

\begin {figure} [tbp]
\centerline {\hbox {
\includegraphics [width=0.7\textwidth] {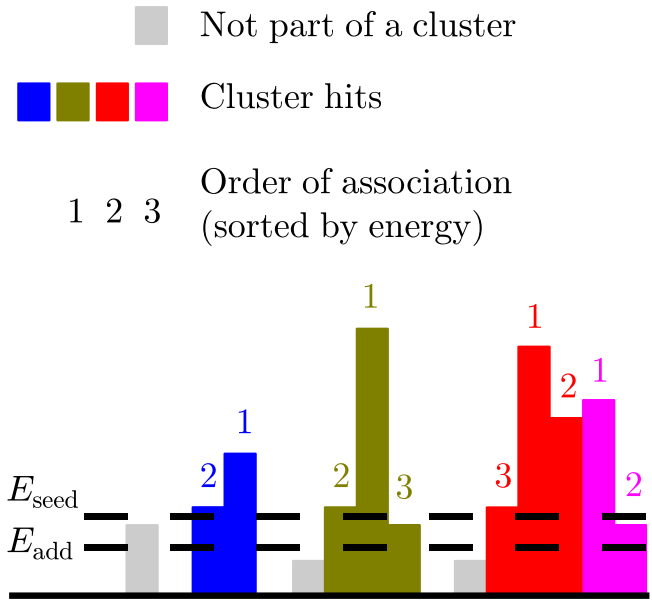}
}}
\caption {Schematic view of hit topologies in the BEMC and the assignment of hits to clusters by the algorithm described in the text.}
\label {fig_cluster_finder}
\end {figure}

In Figure~\ref {fig_cluster_finder} 
we show the assignments made by the algorithm on several possible one\tsp{0.3}-dimensional cluster topologies.
Note, that the rightmost hit pattern in this figure shows a double\tsp{0.3}-peak structure, which is splitted into two adjacent clusters by the algorithm.
However, statistical fluctuations in single photon signals may also be the cause of a double\tsp{0.3}-peak structure. 
In such a case, the cluster splitting by the algorithm becomes a source of background, as will be discussed in Section~\ref {section_low_mass}\@.

The readout of the \SMD\ \etacoord\ and \phiangle\ planes is one\tsp{0.3}-dimensional, so that there is no ambiguity in what is considered to be an adjacent hit.
The calorimeter tower readout is two-dimensional, and two hits are considered to be adjacent when they share a side and not when they share only a corner.

The cluster position in the \etacoord\ and \phiangle\ coordinates is calculated as the
energy weighted mean position of the participating hits.
In this calculation, the
geometrical center of the detector element is taken as the hit position.

\begin {figure} [p]
\centerline {\hbox {
\includegraphics [width=\textwidth] {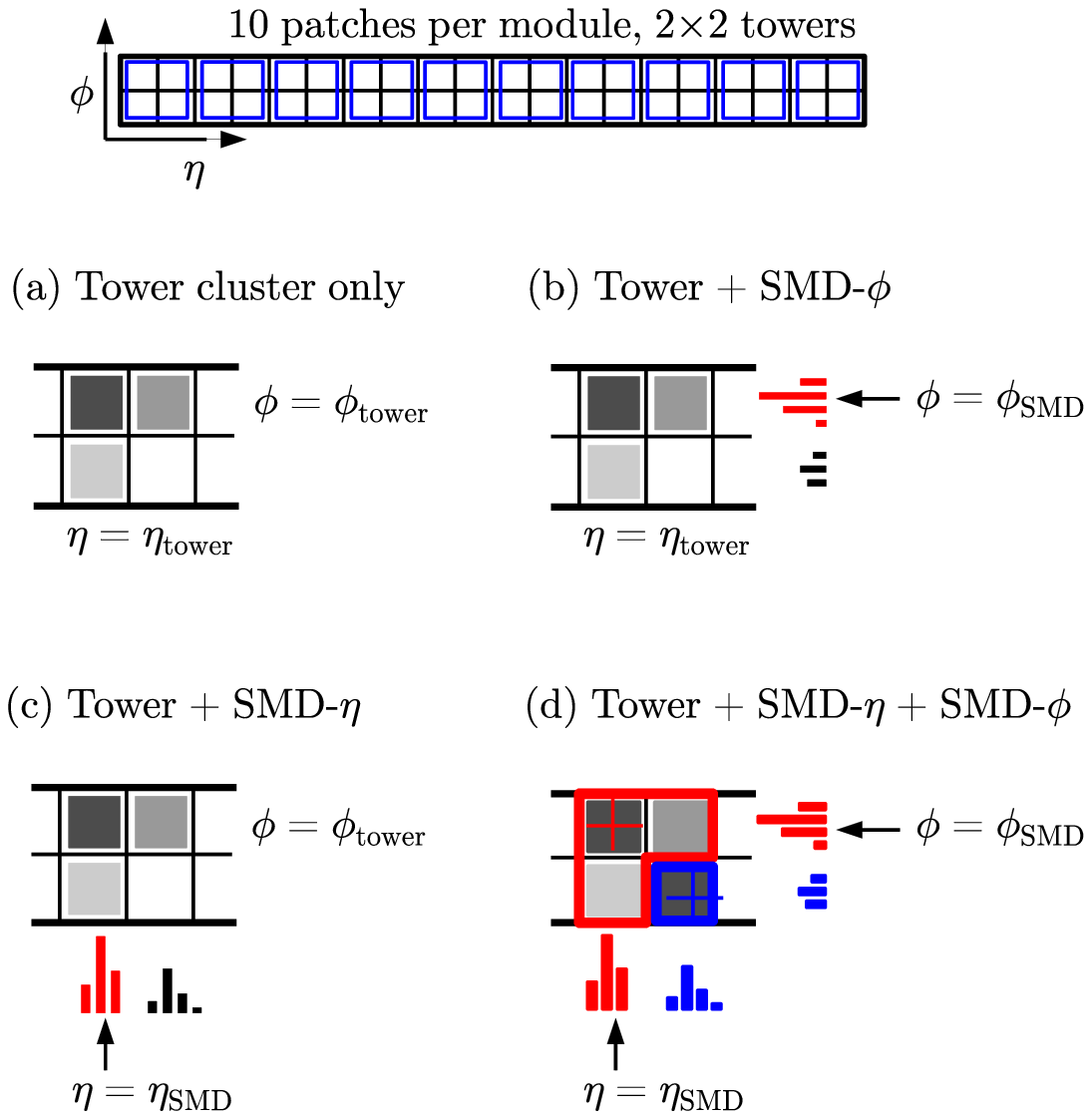}
}}
\caption {Combining tower and \SMD\ clusters into BEMC points.}
\label {fig_point_maker}
\end {figure}

After the tower, preshower, and \SMD\  clusters are found, 
the next step is to combine them into so\tsp{0.3}-called \textit{BEMC points} 
that should correspond as closely as possible to the impact point and energy deposit of a
photon that traversed the calorimeter.
This procedure treats \MM{2 \TIMES{} 2} tower patches corresponding to the \SMDp\ segmentation,
as shown by the top diagram in Figure~\ref {fig_point_maker}\@.

It is required that every reconstructed BEMC point contains a tower cluster, since the energy deposit of the incident particle is measured in the BEMC towers.
Adding information from the \SMD\ leads to a variety of combinations, as shown schematically by the diagrams (a)--(c) in Figure~\ref {fig_point_maker}\@.
In the following paragraphs we describe how each case leads to the reconstruction of a BEMC point.

\paragraph {\boldmath Tower, SMD-$\eta$ and SMD-$\phi$ clusters \unboldmath}
The algorithm calculates for all combinations of \SMDe\ and \SMDp\ clusters in a patch
the energy asymmetry
$$
\DELTA{} = \DIV{\ABS{\SUB{\energy}{\etacoord} - \SUB{\energy}{\phiangle}}}{(\SUB{\energy}{\etacoord} + \SUB{\energy}{\phiangle})}
,
$$
where \SUB{\energy}{\etacoord} and \SUB{\energy}{\phiangle} are, respectively, 
the energy deposits measured in the \SMDe\ and \SMDp\ planes.

The cluster assignment constitutes a well known problem in combinatorics (Assignment problem~\cite {ref_assignment_problem}) 
which we solve by a call to the CERN library routine \texttt {ASSNDX}~\cite {ref_cernlib} 
that combines objects into pairs in a way that minimizes the total cost.
In the present algorithm the cost function is defined as energy asymmetry \DELTA\ between clusters.

Each associated \SMD\ pair is matched to the tower cluster closest in \etacoord\ and \phiangle\@.
The total tower energy in a patch (including unassociated) is shared between points, weighted by their average \SMD\ energy,
that is, each \IT{i}-th pair will produce a point with energy
$$
\SUB{\energy}{\IT{i}} = \SUP{\SUB{\energy}{\RM{t}}}{\RM{total}} 
\TIMES{} 
\FRAC
    {\SUP{\SUB{\energy}{\SMD,\ \IT{i}}}{\RM{assoc}}}
    {\SUB{\ensuremath{\sum}}{\IT{j}} \SUP{\SUB{\energy}{\SMD,\ \IT{j}}}{\RM{assoc}}}
,
$$
where \MM{\SUP{\SUB{\energy}{\SMD,\ \IT{i}}}{\RM{assoc}} = \DIV{(\SUB{\energy}{\etacoord,\ \IT{i}} + \SUB{\energy}{\phiangle,\ \IT{i}})}{2}}\@.
The \etacoord\ and \phiangle\ coordinates are that of the \SMD\ clusters.

This procedure works well, provided that the occupancies of the \MM{2 \TIMES{} 2} tower patches are low. 
Indeed, the number of tower or \SMD\ clusters reconstructed even in the most central \deuterongold\ events is below \MM{30} in 
the complete half-barrel, corresponding to a mean number of \MM{12} clusters per event and an average \linebreak
occupancy of \MM{2\unitns{\%}} per patch.

\paragraph {\boldmath Tower and SMD-$\eta$ clusters \unboldmath}
In this case, the tower and \SMDe\ clusters are associated 
by the same algorithm as used above, except that here the cost function is defined by the energy asymmetry 
$$
\DELTA{} = \DIV{\ABS{\SUB{\energy}{\RM{t}} - \SUB{\energy}{\etacoord}}}{(\SUB{\energy}{\RM{t}} + \SUB{\energy}{\etacoord})}
,
$$
where \SUB{\energy}{\RM{t}} is the energy deposit in a tower 
and \SUB{\energy}{\etacoord} is the energy deposit measured in the \SMDe\ plane.
The total energy of tower clusters in a patch is shared between associated pairs, weighted by their tower energy:
$$
\SUB{\energy}{\IT{i}} = \SUP{\SUB{\energy}{\RM{t}}}{\RM{total}} 
\TIMES{} 
\FRAC
    {\SUP{\SUB{\energy}{\RM{t},\ \IT{i}}}{\RM{assoc}}}
    {\SUB{\ensuremath{\sum}}{\IT{j}} \SUP{\SUB{\energy}{\RM{t},\ \IT{j}}}{\RM{assoc}}}
.
$$

The \etacoord\ coordinate associated to the BEMC point is taken directly from the \SMDe\ cluster, 
while the \phiangle\ coordinate is taken from the tower cluster.

\paragraph {\boldmath Tower and SMD-$\phi$ clusters \unboldmath}
This case is treated as described above. 
The resulting BEMC points will have the \phiangle\ coordinate from the \SMDp\ clusters 
and the \etacoord\ coordinate from the tower clusters.

\paragraph {Tower clusters only}
If there are no \SMD\ clusters in a patch that contains the tower cluster position, 
the energy and coordinates of the BEMC point are taken to be those of the tower cluster.

The relative occurances of these four cases are approximately in proportion of \linebreak
\MM{70:5:5:20\unit{\%}} for clusters with energy above \MM{4\unit{\GeV}}, 
and \MM{25:10:10:55\unit{\%}} at the lower energies.

The information about the shower shape in the \SMD\ is in principle available but not used in the present clustering algorithm.

\section {BEMC cluster cuts}
\label {cluster_cuts}

\begin {figure} [tb]
\centerline {\hbox {
\includegraphics [width=0.9\textwidth] {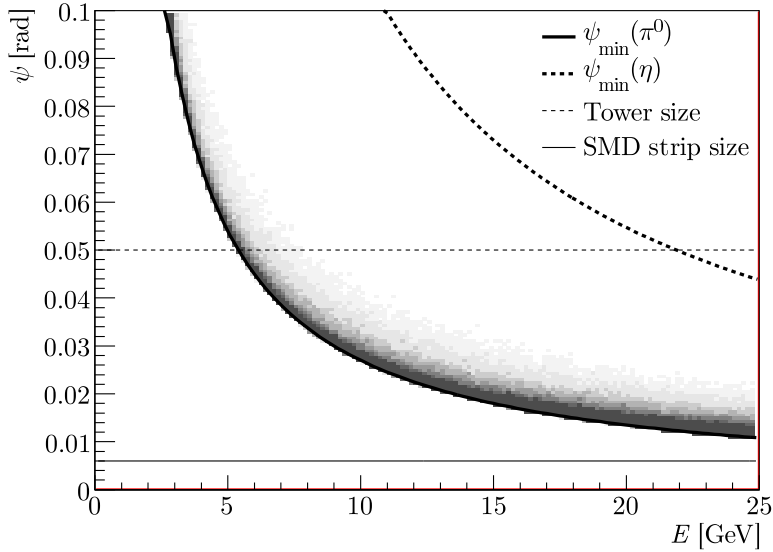}
}}
\caption {The minimal opening angle between \pizero\ and \etameson\ decay photons, compared to the tower and \SMD\ strip size.
Also shown is the actual distribution of the opening angles.
}
\label {fig_pt_reach}
\end {figure}

After clustering, only the BEMC points containing tower and both \SMDe\ and \SMDp\ clusters were kept to be used in the further analysis 
of the HighTower-triggered data.
In the analysis of MinBias data all reconstructed BEMC points were used, even when they do not contain SMD clusters.
From the decay kinematics in the laboratory it follows that the opening angle between the photons 
is smallest when these photons equally share the energy of the parent.
In Figure~\ref {fig_pt_reach}
is shown this minimal opening angle versus the energy of the parent \pizero\ or \etameson\ and compared to the tower and \SMD\ strip size.
It is seen that the spatial resolution of better than a calorimeter tower is needed to resolve the decay photons of 
neutral pions with momenta larger than \MM{5\unit{\GeVc}}\@.
For this reason, the \SMD\ information is essential in this analysis.
\enlargethispage{\baselineskip}

It is seen from beam tests~\cite {ref_emc_beamtest} that the \SMD\ efficiency decreases rapidly 
with energy of the traversing particle and is smaller than \MM{50\unitns{\%}} at \MM{\energy{} \LESS{} 2\unit{\GeVc}}\@.
The energy resolution 
\MM{\DIV{\ensuremath{\delta}\energy}{\energy} = 12\unitns{\%} \ensuremath{\oplus} \DIV{86\unitns{\%}}{\SQRT{\energy}}} 
is also poor at low energy, so that significant fluctuations in the strip readout are expected.
Therefore, an \SMD\ cluster is required to contain signals from at least two strips in order 
to be accepted in the \HighTowerOne\ data.
This cut rejects a large fraction of the distorted and falsely split \SMD\ clusters, 
and reduces a possible effect of poor \SMD\ response simulation at low energies.

It has been found that the tower calibration is less reliable at the edges of the calorimeter acceptance.
For this reason, we only keep the reconstructed clusters in the range \MM{0.1 \LESS{} \etacoord{} \LESS{} 0.9} for the further analysis,
excluding two tower \etacoord-rings at each side of the calorimeter half-barrel.

A charged particle veto (CPV) cut is applied to reject the charged hadrons that are detected in the calorimeter.
These charged hadrons can be recognized as BEMC clusters with a pointing TPC track.
The cluster was rejected if the distance between the BEMC point and the closest TPC track (\IT{D}) 
was smaller than \MM{0.04} in the \etacoord--\phiangle\ coordinates,
$$
\IT{D} = \SQRT{\DELTA\SUP{\etacoord}{2} + \DELTA\SUP{\phiangle}{2}} \LESS{} 0.04
.
$$
The BEMC points remaining after this cut are considered to be photon candidates, which are
combined into pairs, defining the set of \pizero\ candidates.

\begin {figure} [tb]
\centerline {\hbox {
\includegraphics [width=0.9\textwidth] {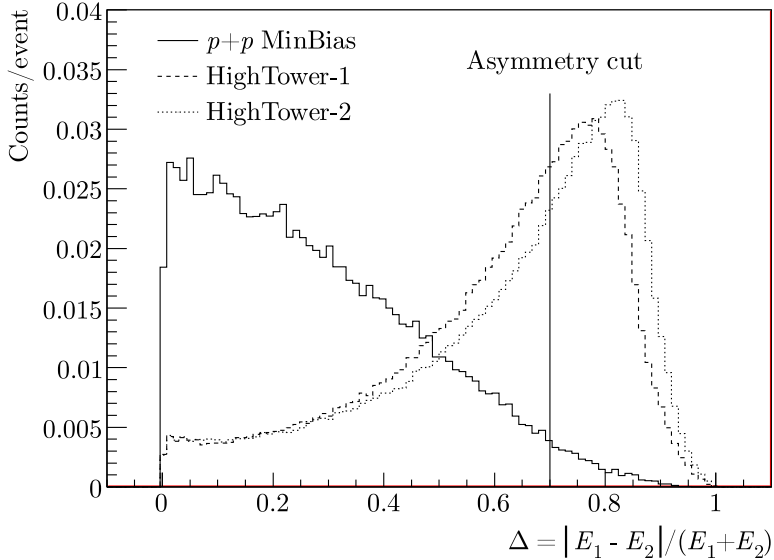}
}}
\caption {The energy asymmetry of the photon pairs reconstructed in \protonproton\ data.}
\label {fig_asymmetry}
\end {figure}

The asymmetry of the two\tsp{0.3}-body decay of neutral mesons is defined as
$$
\DELTA{} = \FRAC {\ABS{\SUB{\energy}{1} - \SUB{\energy}{2}}} {\SUB{\energy}{1} + \SUB{\energy}{2}}
,
$$
where \SUB{\energy}{1} and \SUB{\energy}{2} are the energies of the decay photons.
From the decay \linebreak
kinematics it follows that this energy asymmetry is uniformly distributed \linebreak 
between \MM{0} and \MM{1}~\cite {ref_stolpovsky_thesis}\@.
In Figure~\ref {fig_asymmetry} 
we show the distribution of the asymmetry of photon pairs reconstructed in \protonproton\ data.
In the MinBias data the distribution is not flat because of the acceptance effects --- photons from the asymmetric \linebreak
decay have a large opening angle and there is a large probability that one of them escapes the barrel.
It is also seen that the HighTower energy threshold biases the asymmetry to the higher values, 
because it is easier for an asymmetric decay to pass the trigger.
In this analysis, the \pizero\ candidates were only accepted if the asymmetry was less than \MM{0.7}, 
in order to reject very asymmetric decays, where one of the BEMC points has low energy, 
and to reject a significant part of the low mass background (this background will be described in the following sections)\@.
It turns out that the asymmetry cut improves the signal to background ratio by approximately a factor of \MM{1.5}\@.

Finally, for the HighTower-triggered data the requirement is made that 
at least one of the reconstructed decay photons alone satisfies this trigger.
This requirement is made to guarantee that the trigger efficiency is the same in both real and simulated data,
as was already mentioned in section~\ref {ref_HT_filter}\@.

\section {Invariant mass distribution}

\begin {figure} [tb]
\centerline {\hbox {
\includegraphics [width=0.5\textwidth] {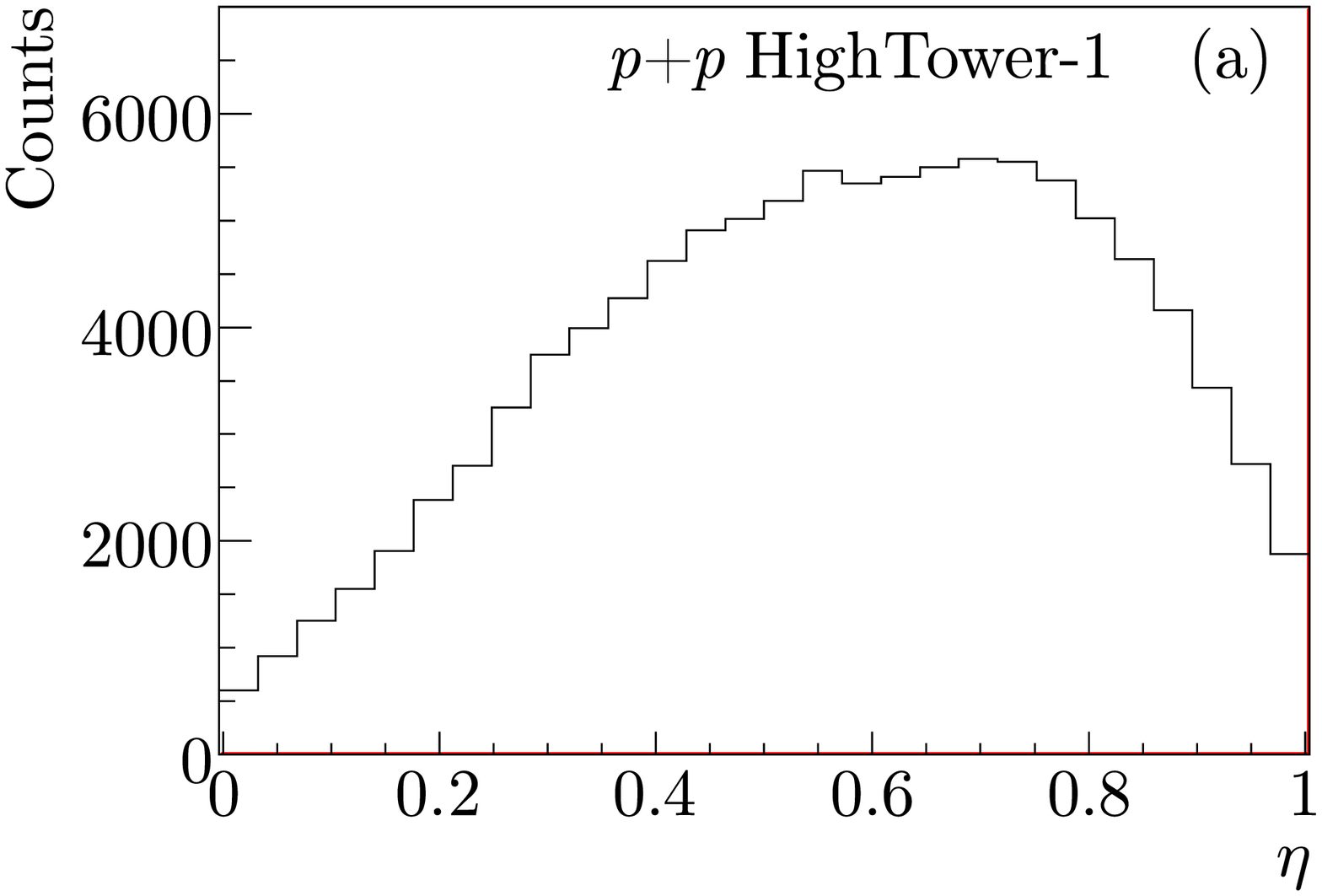}\includegraphics [width=0.5\textwidth] {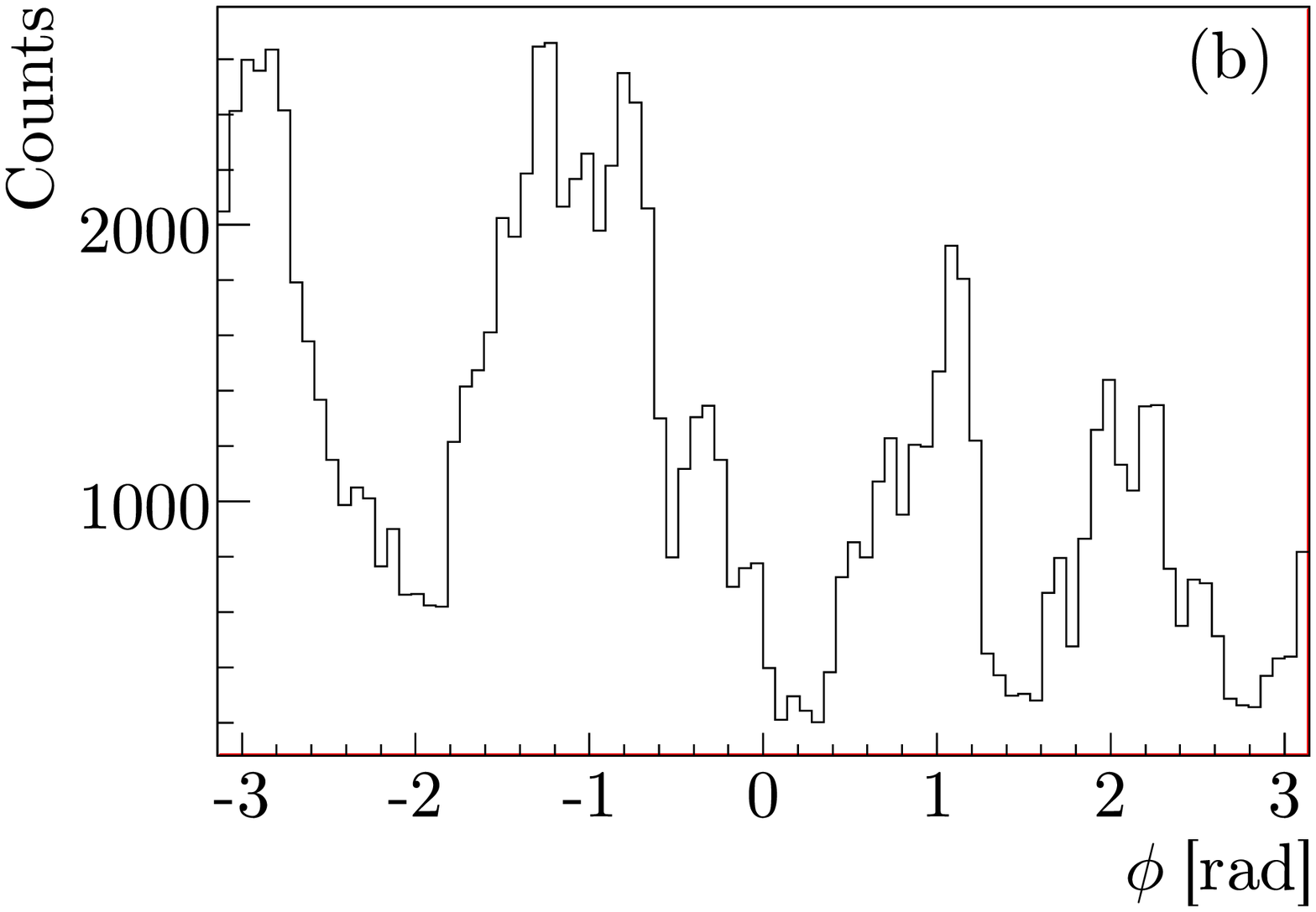}
}}
\centerline {\hbox {
\includegraphics [width=0.5\textwidth] {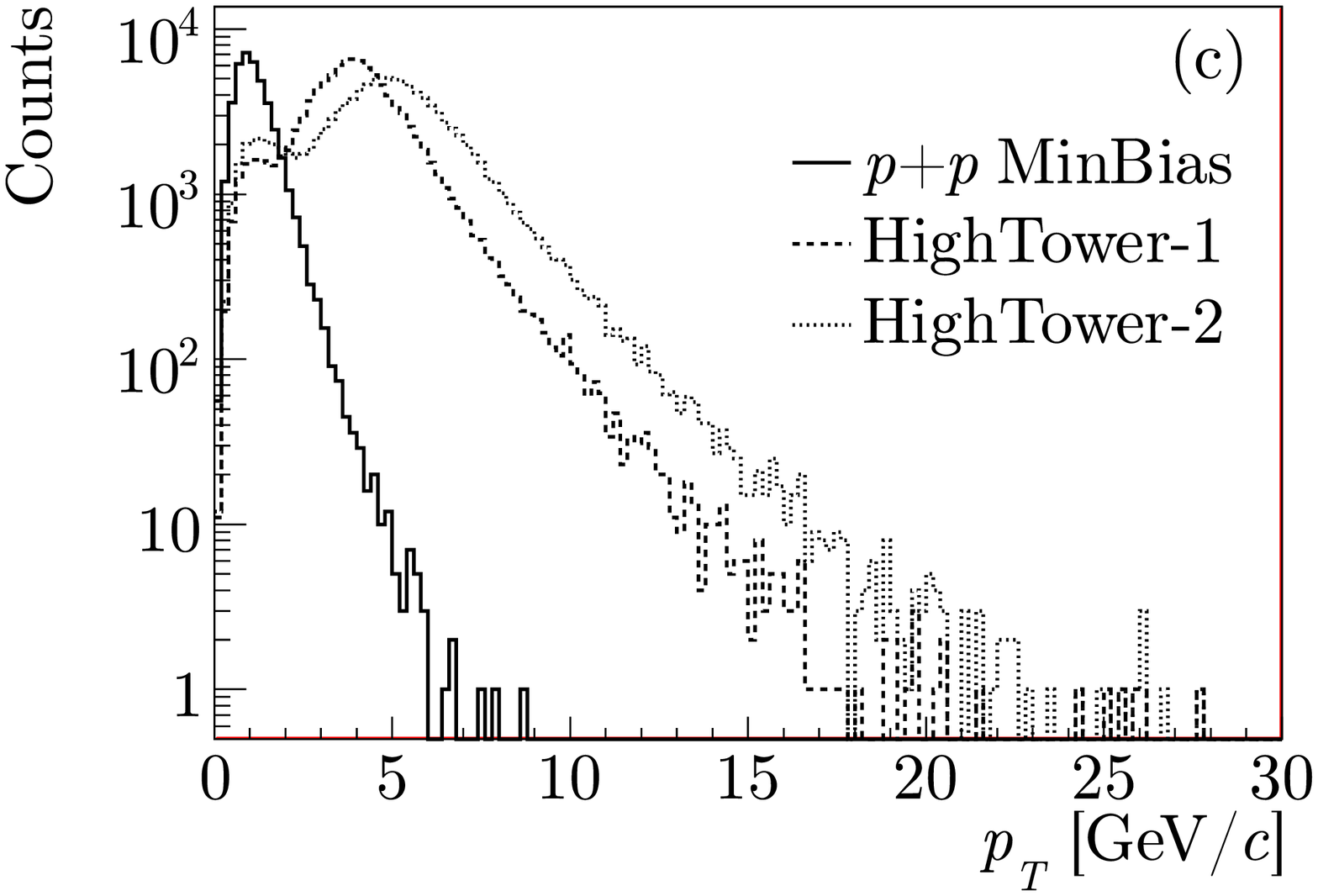}\includegraphics [width=0.5\textwidth] {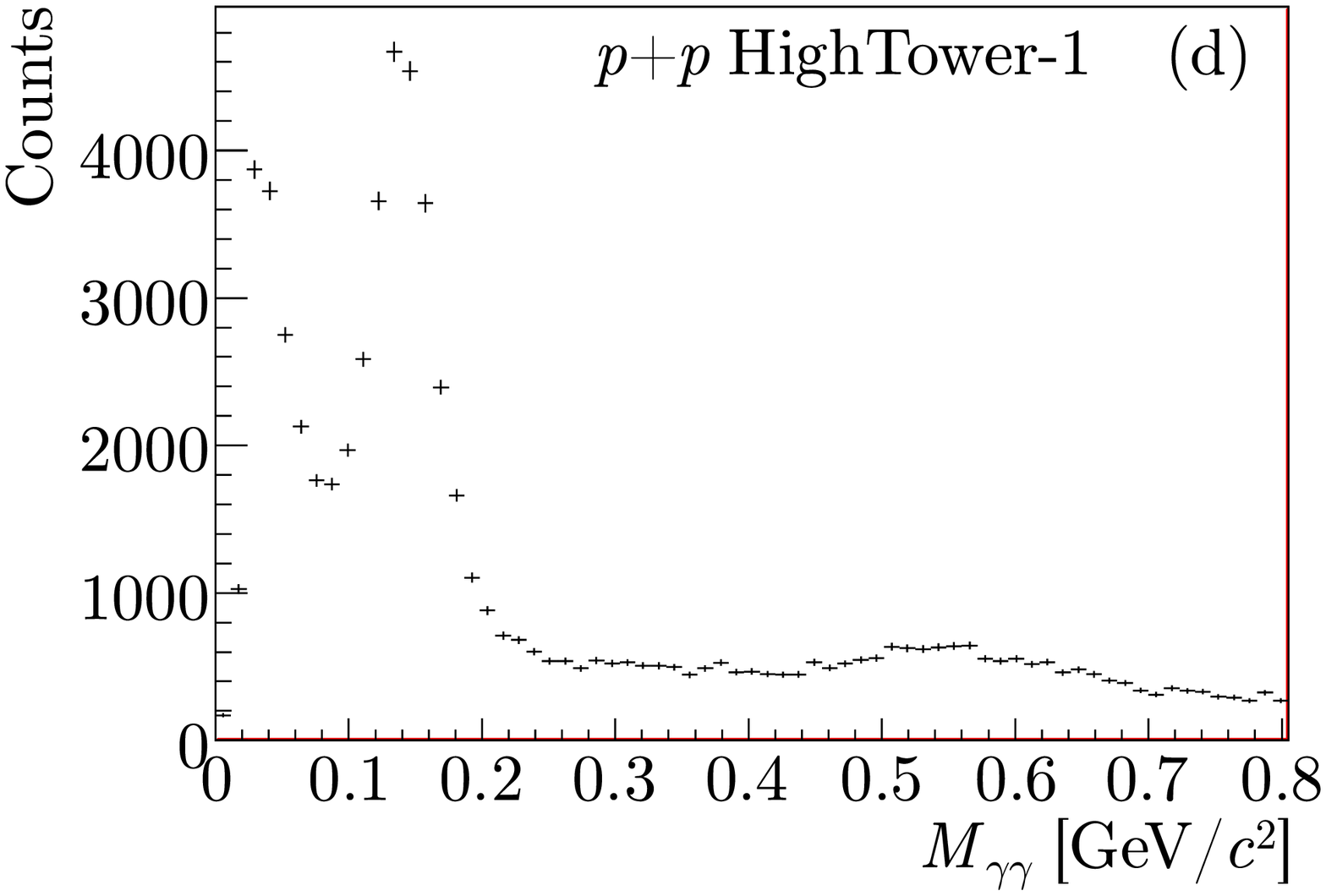}
}}
\caption {Distribution of \pizero\ candidates as a function of \etacoord\ and \phiangle\ (top), 
\pT\ and \Mgg\ (bottom) obtained from the \protonproton\ data.}
\label {fig_data_candidates}
\end {figure}

After cuts, the pairs of BEMC points are turned into 4\tsp{0.3}-vectors by assuming that 
the decay photons originate from the reconstructed main vertex.
For each \pizero\ candidate, the pseudorapidity \etacoord, the azimuth \phiangle, 
the transverse momentum \pT, and the invariant mass \Mgg\ (Eq.~\ref {eq:inv_mass}) are calculated.
In Figure~\ref {fig_data_candidates}
we show the \etacoord, \phiangle, \pT, and \Mgg\ distributions of the \pizero\ candidates in the \protonproton\ dataset.
For the \deuterongold\ data these distributions look similar as those shown for \protonproton\@.

The \etacoord\ distribution shows the decrease of the calorimeter acceptance at the edges, because there it is likely than one 
of the decay photons escapes the calorimeter.
The asymmetry is due to the fact that the calorimeter half-barrel is positioned asymetrically with respect to the interaction point.
The structure seen in the \phiangle\ distribution reflects the azimuthal dependence of the calorimeter acceptance
caused by failing \SMD\ modules.

In Figure~\ref {fig_data_candidates}(c) is shown the \pT\ distribution of the photon pairs 
separately for the MinBias and HighTower datasets. 
It is seen that the HighTower triggers significantly increase the rate of pion candidates at large \pT\@.
The \pT-integrated invariant mass distribution in Figure~\ref {fig_data_candidates}(d) clearly shows 
the \pizero\ and \etameson\ peaks superimposed on a broad background distribution.
This background has a combinatorial and a low mass component.
In the next two sections we will discuss each background component in detail.

\section {Combinatorial background}

\begin {figure} [tb]
\centerline {\hbox {
\includegraphics [width=0.5\textwidth] {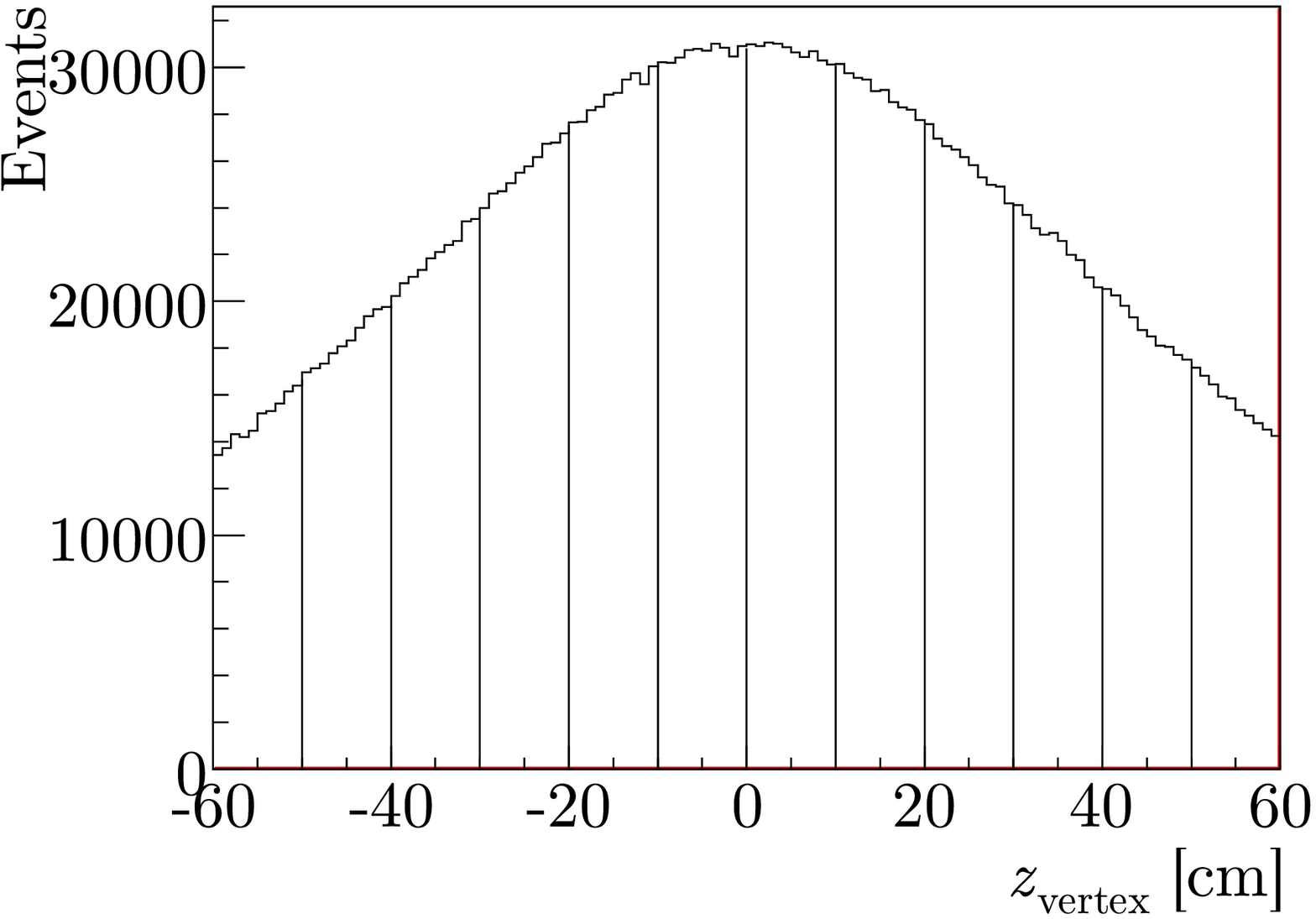}\includegraphics [width=0.5\textwidth] {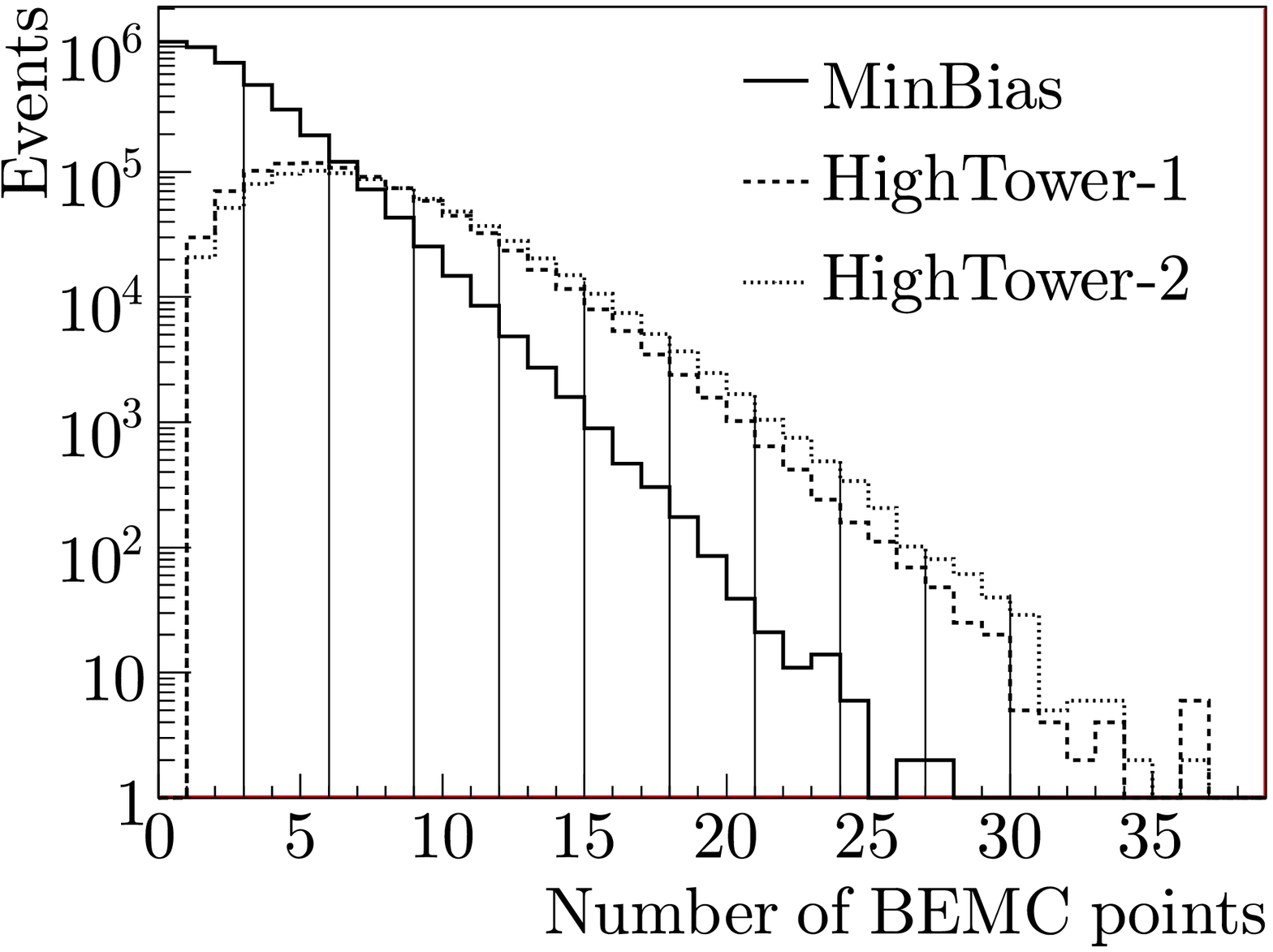}
}}
\caption {Distributions of the number of events, as a function of vertex \zcoord\ (left) and number of BEMC points (right),
obtained from the \protonproton\ dataset.
In both figures are shown the intervals used to define the event mixing classes.}
\label {fig_mixing_classes}
\end {figure}

The combinatorial background in the invariant mass distribution originates from pairs of photon clusters 
that are not produced in a single \pizero\ decay.
To describe the shape of the combinatorial background, we use the event mixing technique,
where photon clusters from two different events are combined.
To mix only similar event topologies, the data were subdivided into the mixing classes 
based on the vertex position, BEMC multiplicity, and trigger type (MinBias, \HighTowerOne, and \HighTowerTwo)\@.
In Figure~\ref {fig_mixing_classes} 
we show the \protonproton\ vertex and multiplicity distributions, and the bins defining the mixing classes.

\begin {figure} [p]
\centerline {\hbox {
\includegraphics [width=0.9\textwidth] {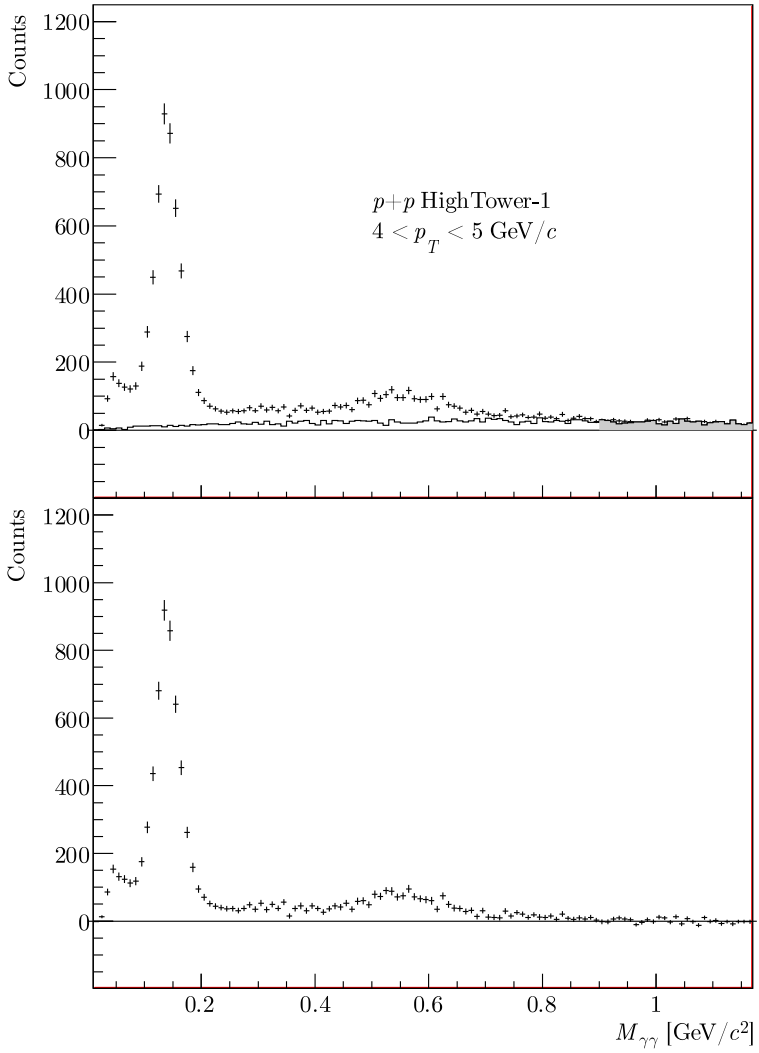}
}}
\caption {The same-event invariant mass distribution and the combinatorial background obtained from the random event mixing (top) 
and the background subtracted distribution (bottom)\@.
The shaded area in the top plot indicates the region where the mixed event background is normalized to the data.}
\label {fig_inv_rndmix}
\end {figure}

Figure~\ref {fig_inv_rndmix} 
shows an example of an invariant mass distribution in the \linebreak
\MM{4 \LESS{} \pT{} \LESS{} 5\unit{\GeVc}} bin, 
obtained from the \HighTowerOne\ \protonproton\ data,
together with the combinatorial background obtained from the event mixing.
The mixed event background distribution is normalized to the same-event distribution 
in the invariant mass region \MM{0.9 \LESS{} \Mgg{} \LESS{} 1.2\unit{\GeVcc}}\@.
In the bottom panel of this figure the background subtracted distribution is shown.

It can be seen that there is still some residual background 
in the interval \MM{0.2 \LESS{} \Mgg{} \LESS{} 0.4\unit{\GeVcc}}, 
which could be caused by the fact that the mixing procedure
does not fully take into account the correlation structure of the event.
For example, an important source of particle correlations is the jet structure, 
which is not present in the sample of mixed events.
In order to preserve jet-induced correlations, the jet axes in both events are aligned before mixing, as described below.

To determine the \MM{(\etacoord,\phiangle)} position of the most energetic jet in every event, 
the standard STAR jet finding algorithm~\cite {ref_jet_asymmetry} was used. 
The mixed pion candidates were constructed by taking two photons from different events, 
where one of the events was displaced in \etacoord\ and \phiangle\ by 
\MM{\DELTA\etacoord{} = \SUB{\etacoord}{2} - \SUB{\etacoord}{1}} 
and \MM{\DELTA\phiangle{} = \SUB{\phiangle}{2} - \SUB{\phiangle}{1}}, respectively.
Here \MM{\SUB{\etacoord}{1,2}} and \MM{\SUB{\phiangle}{1,2}} are the jet orientations in the two events. 

\begin {figure} [tb]
\centerline {\hbox {
\includegraphics [width=0.9\textwidth] {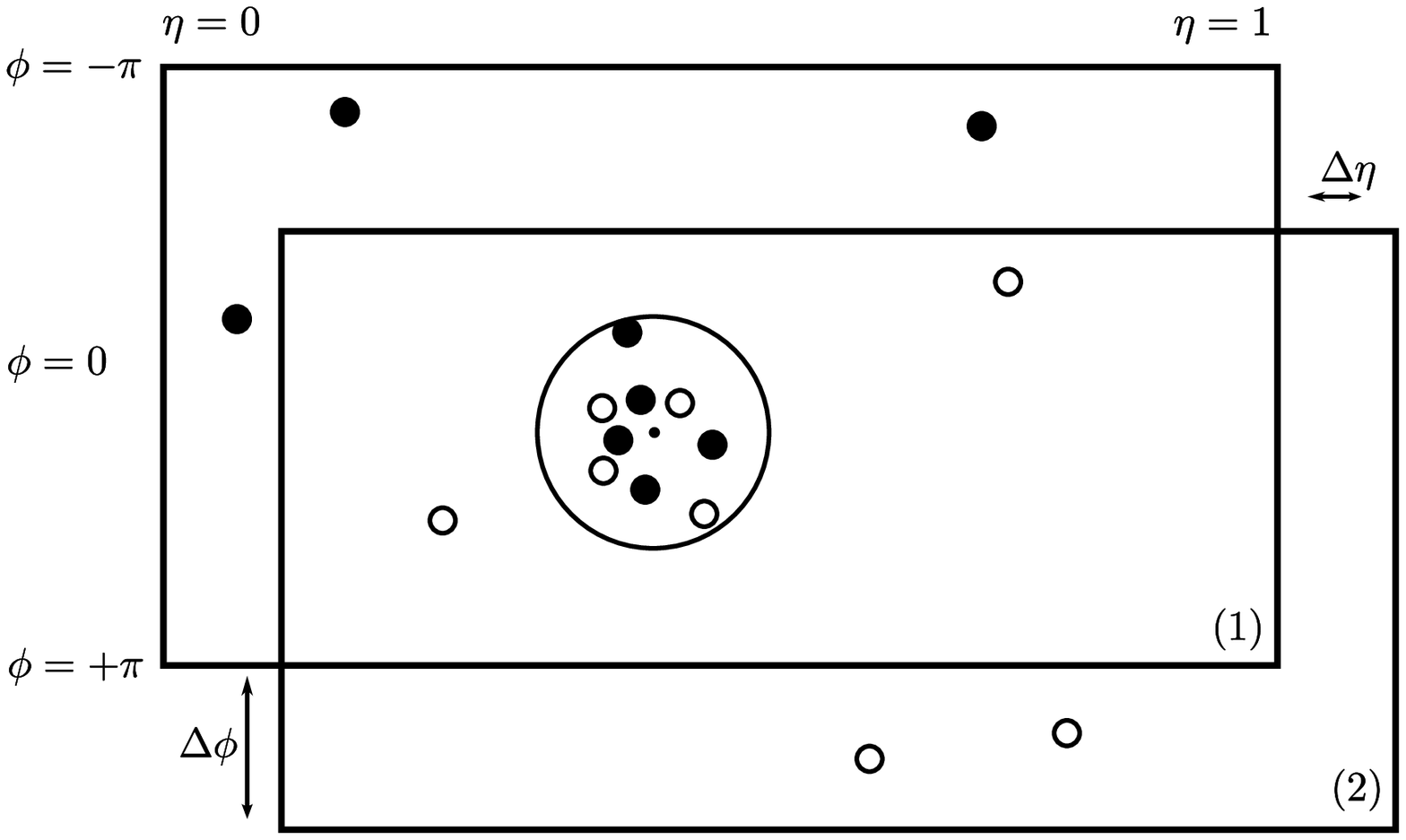}
}}
\caption {A schematic view of two superimposed events, where the jet axes are aligned.}
\label {fig_jetmix}
\end {figure}

In Figure~\ref {fig_jetmix}
we show a schematic view of two superimposed events, where the jet axes are aligned.
In order to minimize acceptance distortions, the events were divided into mixing classes in the jet \etacoord\ coordinate.
By mixing only events in the same class, the shift \MM{\DELTA\etacoord} was kept smaller than \MM{0.1}\@.
Because the calorimeter has a cylindrical shape, the shift in \phiangle\ does not induce any significant acceptance distortion.

However, a side effect of this procedure is that correlations are induced if there is no real jet structure, because the 
jet finding algorithm will then simply pick the most energetic track in the event.
To reduce possible bias introduced by such correlations, we assume that 
a jet structure is associated with large \pT\ pions but not with low \pT\ pions.
The combinatorial background is then taken as a \pT-dependent linear combination of the distributions obtained by
random mixing and jet-aligned mixing,
$$
\IT{B}(\Mass,\pT) = \IT{A}(\pT)\tsp{0.5}\SUB{\IT{B}}{\IT{J}}(\Mass, \pT) + (1 - \IT{A}(\pT))\tsp{0.5}\SUB{\IT{B}}{\IT{R}}(\Mass, \pT)
.
$$
Here \MM{\SUB{\IT{B}}{\IT{J}}(\Mass, \pT)} 
and \MM{\SUB{\IT{B}}{\IT{R}}(\Mass, \pT)} are the background spectra from, respectively, 
the jet-aligned and random event mixing in a given \pT\ bin.
The interpolation coefficient \MM{\IT{A}(\pT)} is given by
\begin {equation} \label {eq:aptdef}
\IT{A}(\pT) = 
\ensuremath{\left\{} 
\begin {array}{ll}
\IT{a} \pT + \IT{b} & \RM{for\ } \pT{} \LESS{} 10\unit{\GeVc} \\
1         & \RM{otherwise,}
\end {array} 
\ensuremath{\right.}
\end {equation}
where the coefficients are \MM{\IT{a} = 0.097\unit{\GeVcinv}} and \MM{\IT{b} = -0.117}\@.
\label {syst_mixed_bg}
We assign a systematic uncertainty of \MM{10\unitns{\%}} to \IT{A}, 
which propagates into a systematic uncertainty of \MM{5\unitns{\%}} on the \pizero\ and \MM{1\unitns{\%}} on the \etameson\ yields.

\begin {figure} [p]
\centerline {\hbox {
\includegraphics [width=0.9\textwidth] {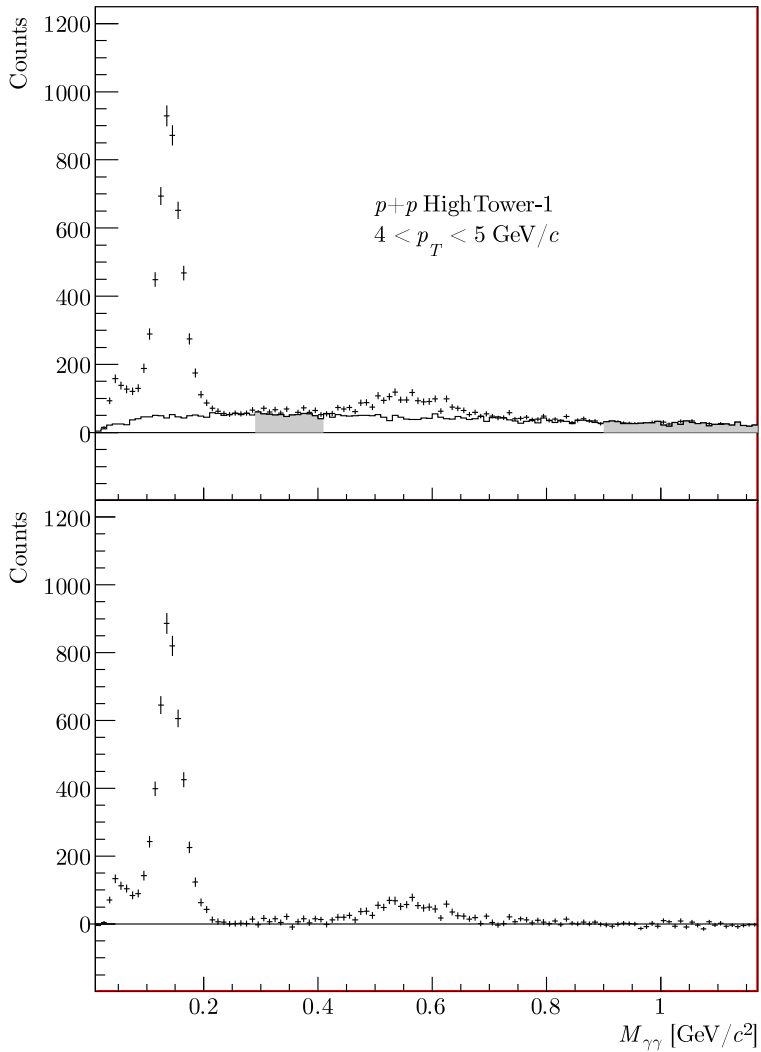}
}}
\caption {The same-event invariant mass distribution and the combinatorial background obtained from the jet-aligned event mixing (top) 
and the background subtracted distribution (bottom)\@.
The shaded regions in the top plot indicate where the mixed event background is normalized to the data.}
\label {fig_inv_jetmix}
\end {figure}

In Figure~\ref {fig_inv_jetmix} 
we plot the same invariant mass spectrum as that shown in Figure~\ref {fig_inv_rndmix},
with the background estimated by the combined random and jet-aligned event mixing.
The mixed event background is normalized to same-event distribution in the ranges
\MM{0.3 \LESS{} \Mgg{} \LESS{} 0.4} and \MM{0.8 \LESS{} \Mgg{} \LESS{} 1.6\unit{\GeVc}}\@.
\label {syst_comb_bg}
By changing the subtracted background within the normalization uncertainty 
we obtained a systematic error on the \pizero\ and \etameson\ yields.
This error was found to increase with \pT\ from \MM{0.5} to \MM{3\unitns{\%}} for the \pizero\ 
and from \MM{10} to \MM{50\unitns{\%}} for the \etameson\ yield.

In the bottom panel of Figure~\ref {fig_inv_jetmix} the background subtracted spectrum is plotted, which still shows 
a residual background component at low invariant mass. 
The origin of this background is described in the next section.

\section {Low-mass background}
\label {section_low_mass}

In Figure~\ref {fig_cluster_finder} we have shown a double peaked hit pattern, 
which will be reconstructed by the clustering algorithm as two separate adjacent clusters.
However, it is possible that random fluctuations will accidentally generate such a two peak structure, 
so that the clustering algorithm will incorrectly split the cluster.
These random fluctuations enhance the yield of pairs with minimal 
angular separation and thus contribute to the lowest di-\tsp{-0.2}photon invariant mass region, as can be seen in Figure~\ref {fig_inv_jetmix}\@.
However, at a given small opening angle the invariant mass increases with increasing energy of the photons,
so that the low mass background spectrum will extend to larger values of \Mgg\ with increasing \pT\ of the parent particle.

The shape of the low mass background was obtained from a simulation as follows.
Single photons were generated with flat distributions in \phiangle, \MM{-0.2 \LESS{} \etacoord{} \LESS{} 1.2} 
and \MM{0 \LESS{} \pT{} \LESS{} 25\unit{\GeVc}}\@.
These photons were tracked through a detailed \linebreak
description of the STAR geometry with the GEANT program~\cite {ref_geant}\@.
A detailed \linebreak
simulation of the electromagnetic shower development in the calorimeter was used to generate 
realistic signals in the towers and the \SMD\@.
The simulated \linebreak
signals were processed by the same reconstruction chain as the real data. \linebreak
Photons with more than one reconstructed cluster were observed, and the \linebreak
invariant mass and \pT\ of such cluster pairs were calculated.
The invariant masses were histogrammed with each entry weighted by the \pT\ spectrum of photons in the real data, corrected for the 
photon detection efficiency.

\begin {figure} [p]
\centerline {\hbox {
\includegraphics [width=0.85\textwidth] {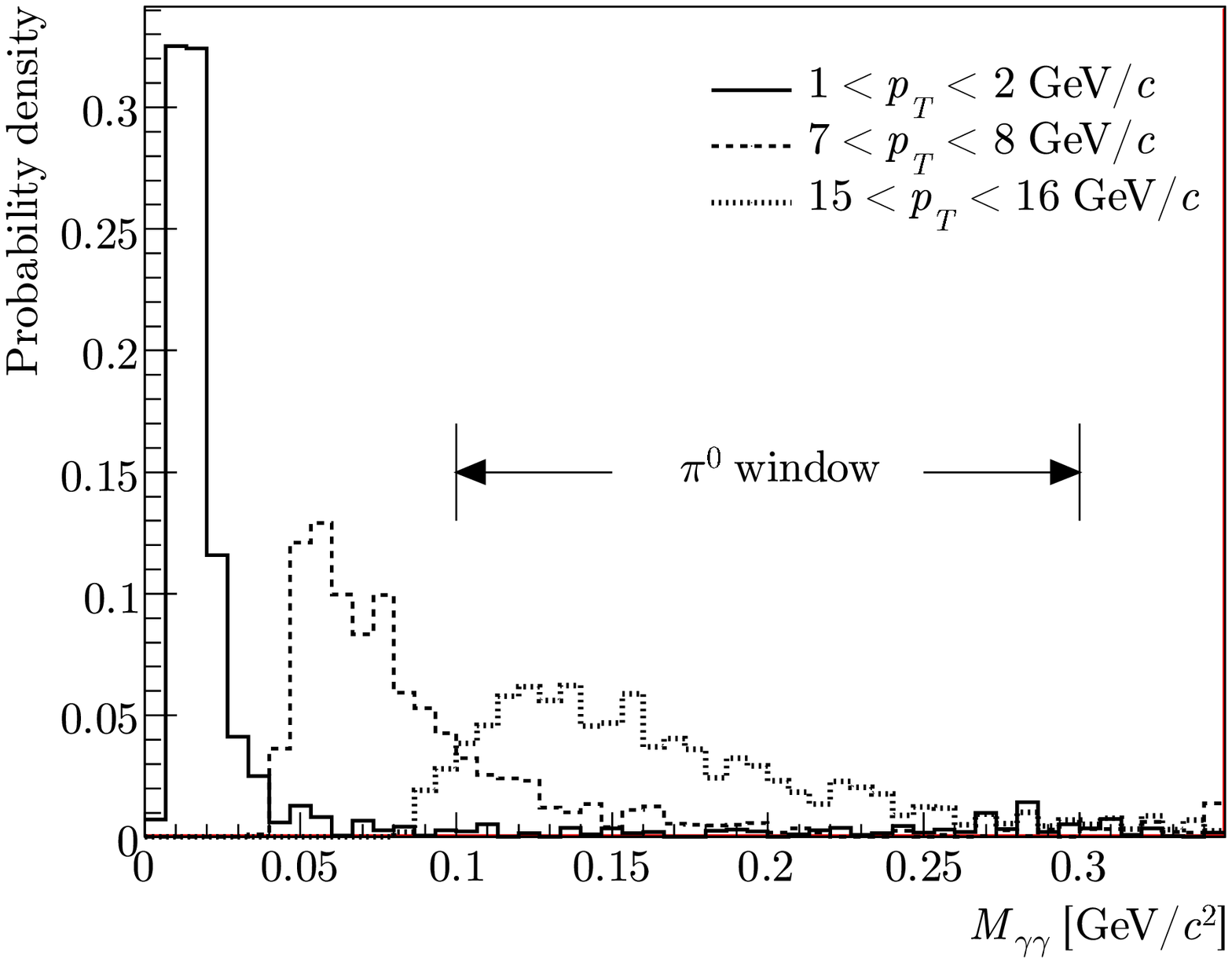}
}}
\centerline {\hbox {
\includegraphics [width=0.85\textwidth] {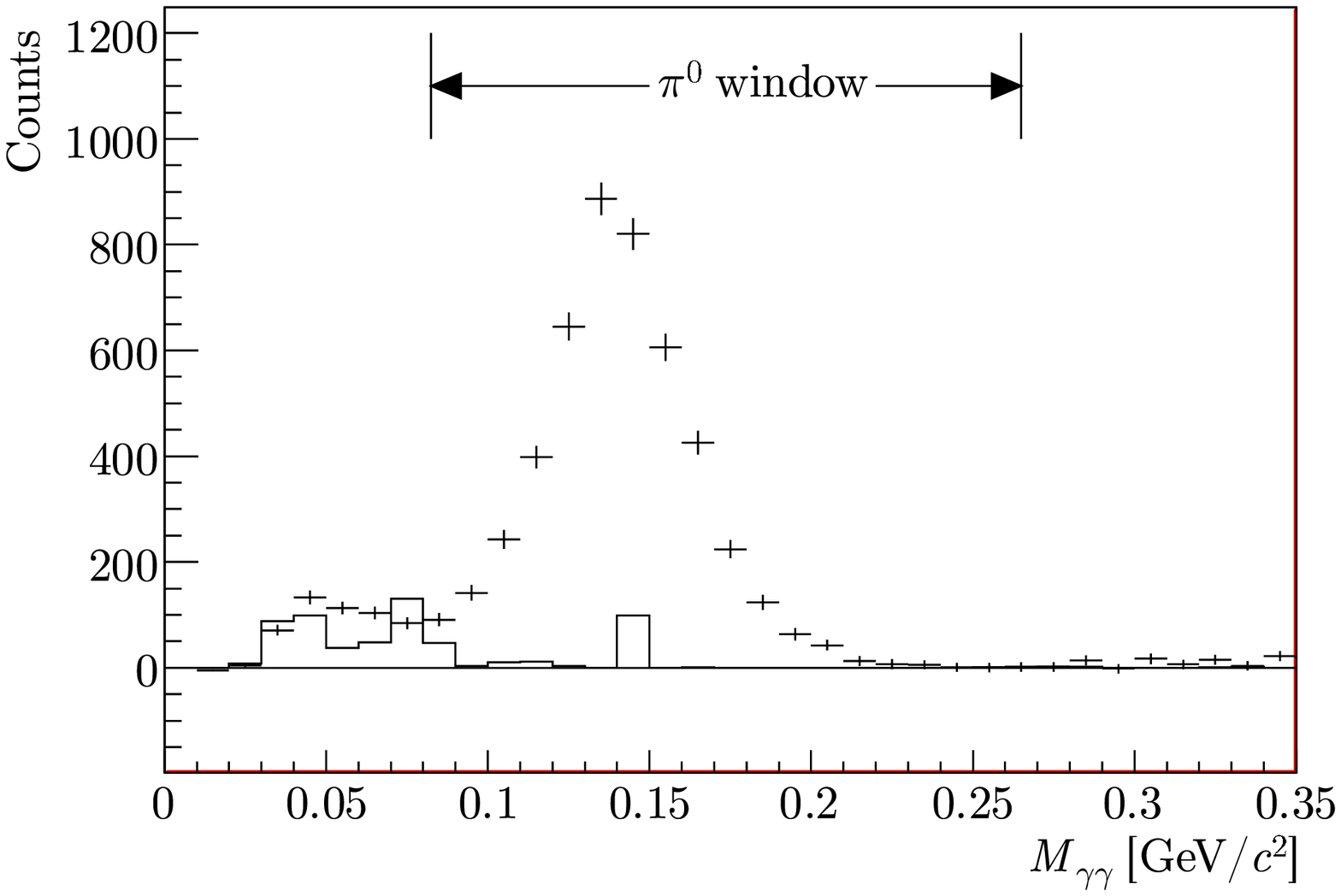}
}}
\caption {The low mass background distributions from erroneous splitting of single photons
in three bins of the reconstructed pair \pT\ (top)\@.
The distributions extend to larger invariant masses with increasing \pT\@.
The low mass background component in the invariant mass distribution obtained from \protonproton\ data (bottom)\@.}
\label {fig_inv_gamma}
\end {figure}

In the top plot of Figure~\ref {fig_inv_gamma}
we show the low mass background distributions in three bins of the reconstructed pair \pT\@.
It is seen that the distributions indeed move to larger invariant masses with increasing \pT\ and extend far into the pion window at large \pT\@.
For this reason, it is not possible to estimate this background from a phenomenological fit to the data, so that
we have to rely on the Monte Carlo simulation to subtract the low mass background.

The second significant source of neutral clusters in the calorimeter are the neutral hadrons 
produced in the collision, mostly antineutrons. As a first \linebreak
attempt to account for the additional low mass background from these hadrons,
simulations of antineutrons were performed in the same way as photons, and the reconstructed invariant mass distribution was 
added according to the realistic proportion \MM{\antineutron/\gama = \antineutron/2\pizero}\@.
The ratio \MM{\antineutron/\pizero} was taken to be equal 
to the average value of \MM{\antiproton/\piminus} from the STAR measurement \cite {ref_star_idhadrons} in 
the \pT\ range covered by each of MinBias and HighTower datasets.
In the bottom plot of Figure~\ref {fig_inv_gamma} we compare the simulated low mass background (histogram) to the data.



\begin {figure} [p]
\centerline {\hbox {
\includegraphics [width=0.9\textwidth] {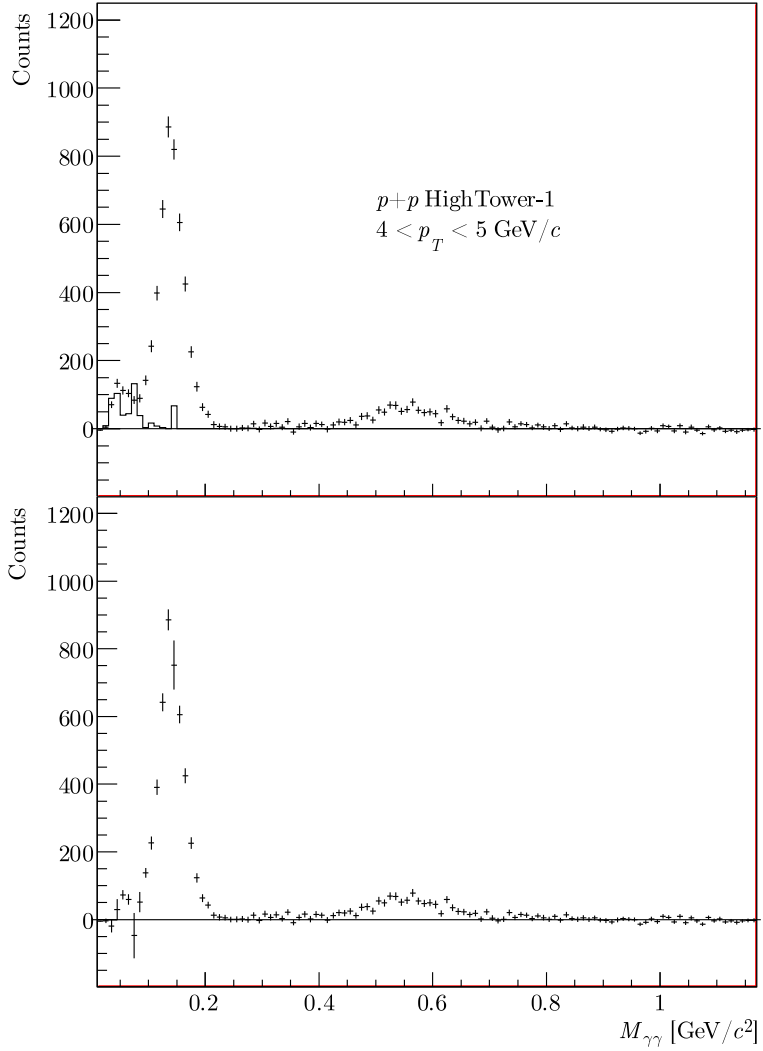}
}}
\caption {The invariant mass distribution before (top) and after the low mass background subtraction (bottom)\@.}
\label {fig_inv_subtracted}
\end {figure}

In Figure~\ref {fig_inv_subtracted} 
we show the invariant mass spectra and the low mass \linebreak
background component (top), together
with the final background subtracted spectrum (bottom)\@.

\section {Yield extraction}

\begin {figure} [p]
\centerline {\hbox {
\includegraphics [angle=90,width=0.9\textwidth] {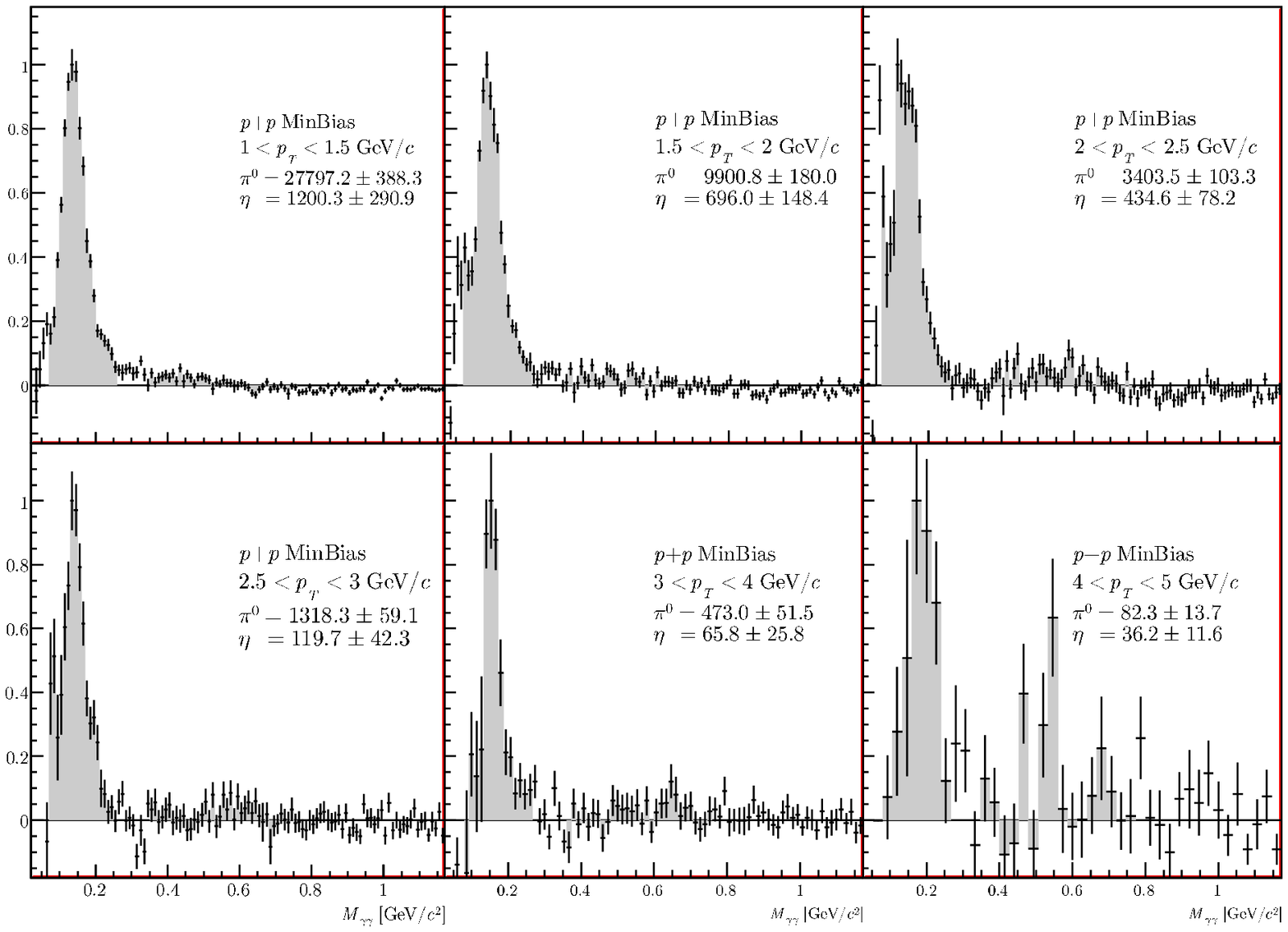}
}}
\caption {Invariant mass distributions in all \pT\ bins, \protonproton\ MinBias data.
The spectra are normalized to the bin content in the \pizero\ peak.}
\label {fig_invmass_ppMB}
\end {figure}

\begin {figure} [p]
\centerline {\hbox {
\includegraphics [angle=90,width=0.9\textwidth] {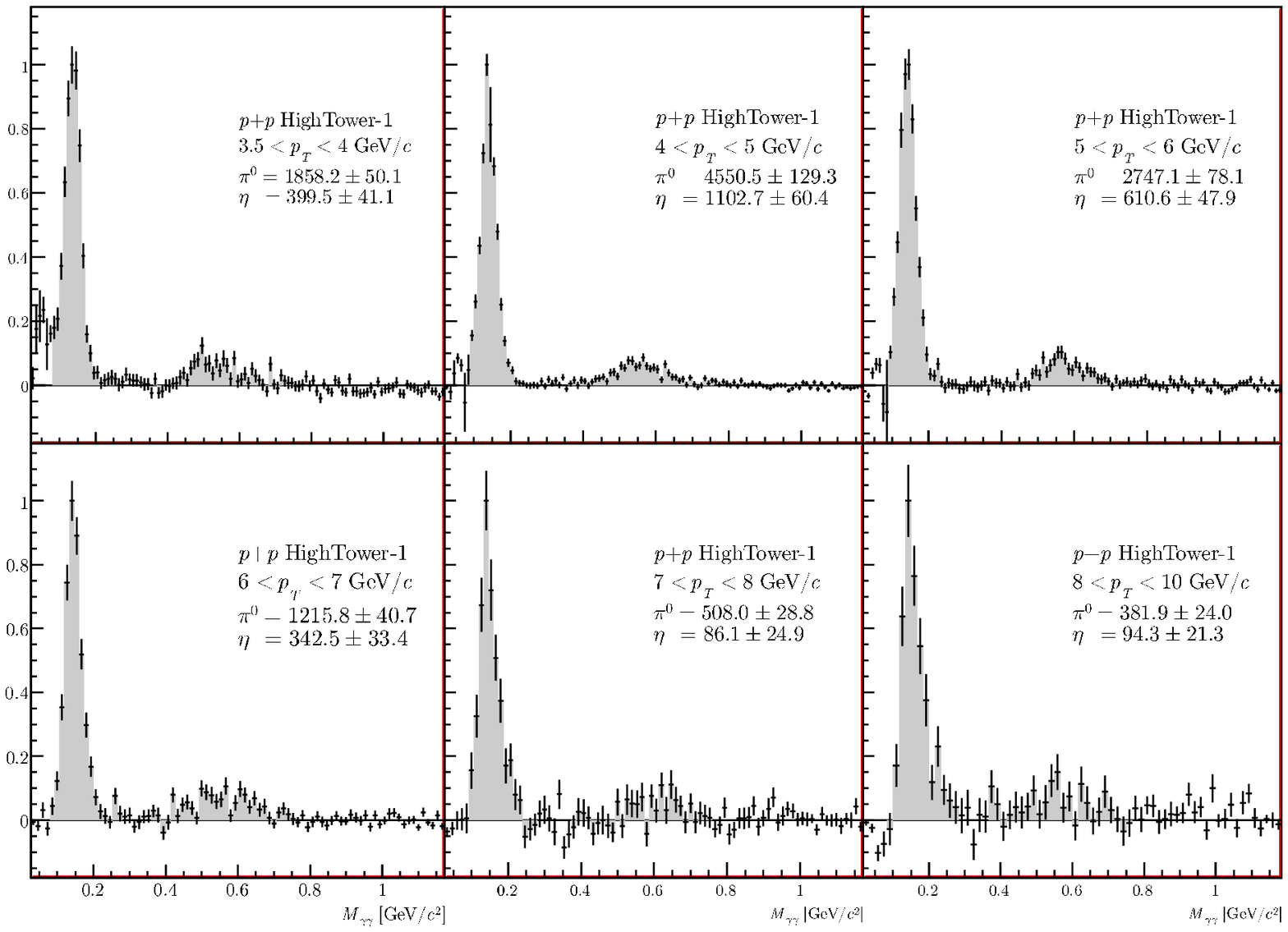}
}}
\caption {Invariant mass distributions in all \pT\ bins, \protonproton\ \HighTowerOne\ data.
The spectra are normalized to the bin content in the \pizero\ peak.}
\label {fig_invmass_ppHT1}
\end {figure}

\begin {figure} [p]
\centerline {\hbox {
\includegraphics [angle=90,width=0.9\textwidth] {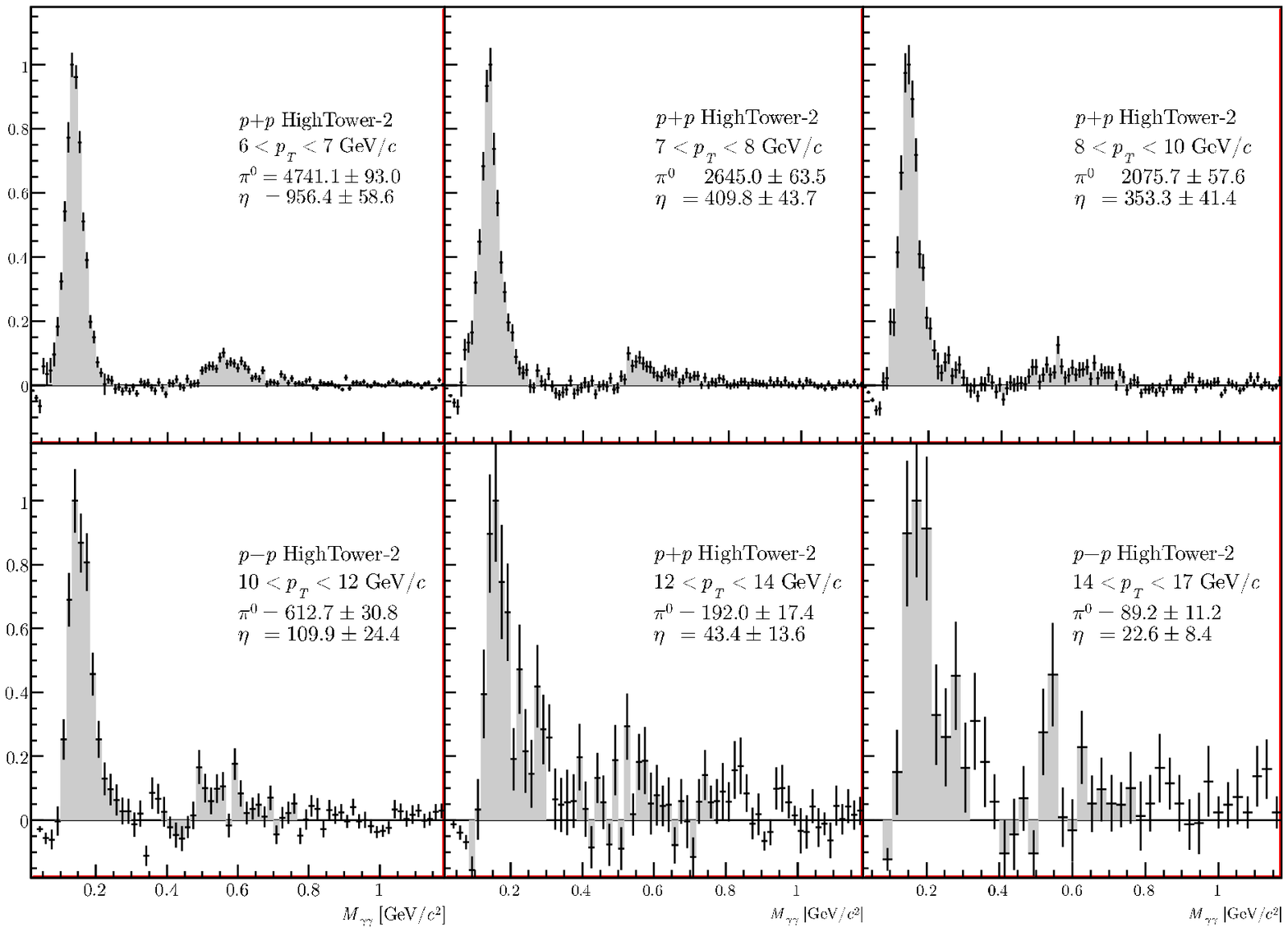}
}}
\caption {Invariant mass distributions in all \pT\ bins, \protonproton\ \HighTowerTwo\ data.
The spectra are normalized to the bin content in the \pizero\ peak.}
\label {fig_invmass_ppHT2}
\end {figure}

\begin {figure} [p]
\centerline {\hbox {
\includegraphics [angle=90,width=0.9\textwidth] {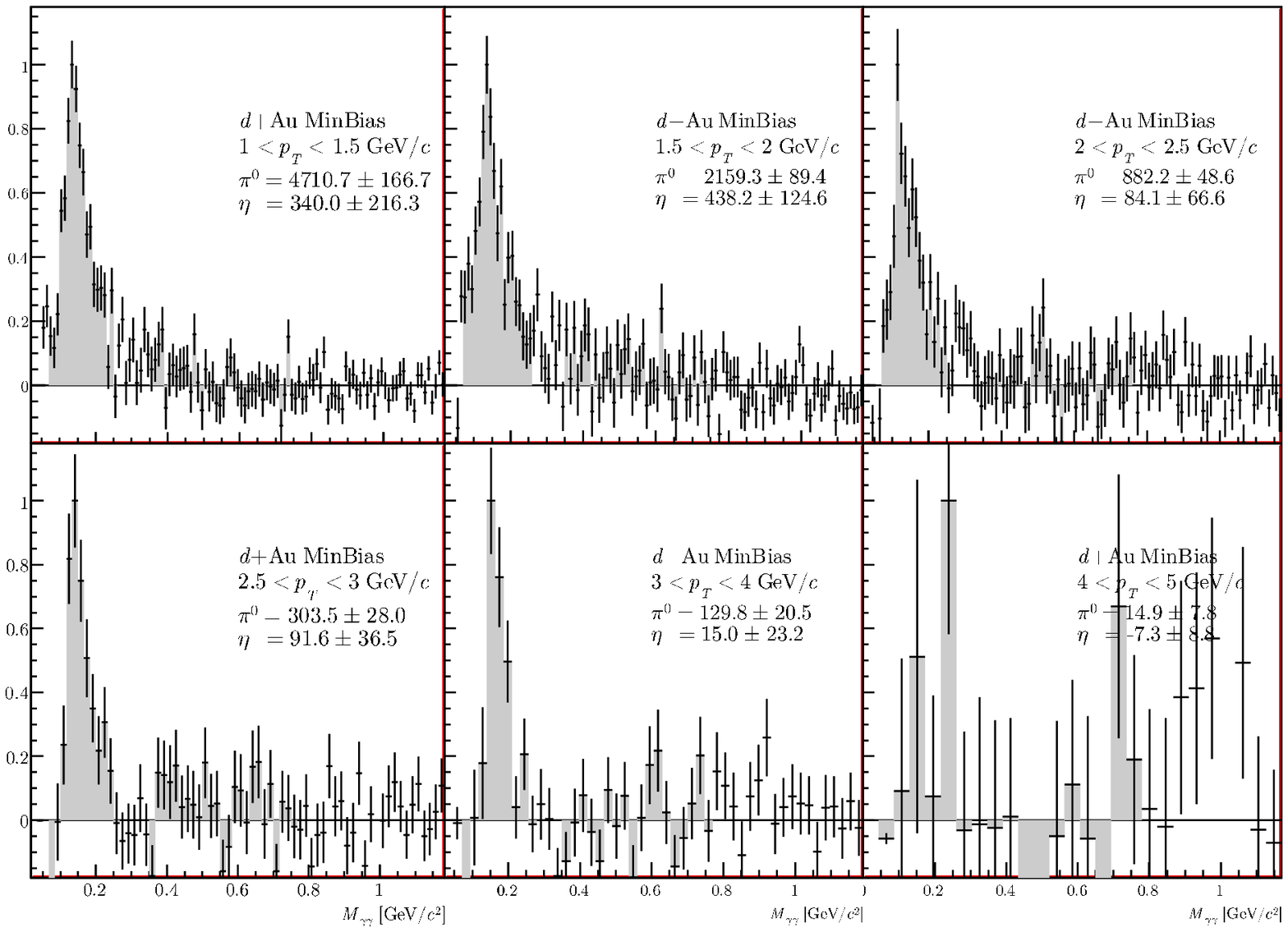}
}}
\caption {Invariant mass distributions in all \pT\ bins, \deuterongold\ MinBias data.
The spectra are normalized to the bin content in the \pizero\ peak.}
\label {fig_invmass_dAuMB}
\end {figure}

\begin {figure} [p]
\centerline {\hbox {
\includegraphics [angle=90,width=0.9\textwidth] {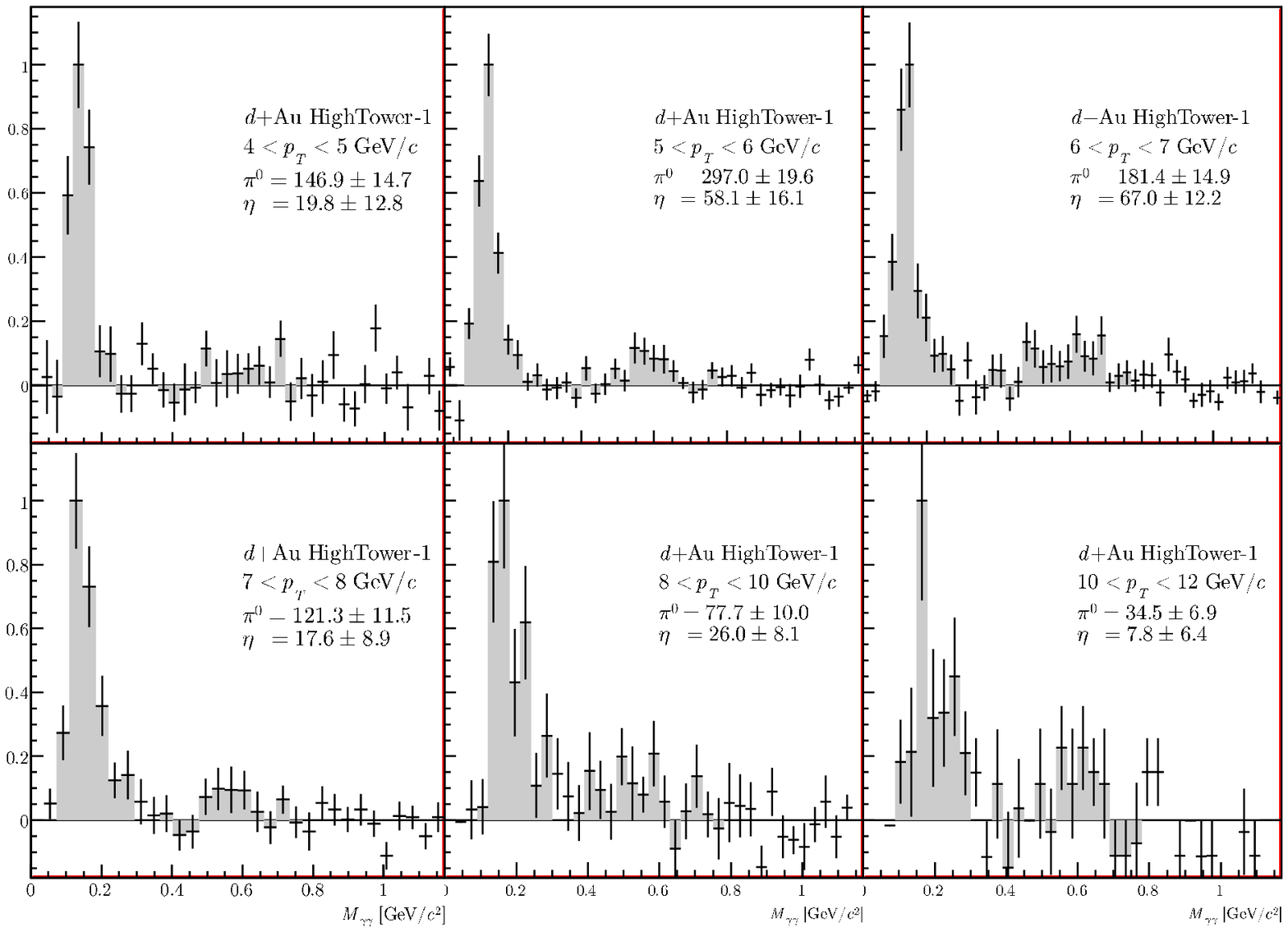}
}}
\caption {Invariant mass distributions in all \pT\ bins, \deuterongold\ \HighTowerOne\ data.
The spectra are normalized to the bin content in the \pizero\ peak.}
\label {fig_invmass_dAuHT1}
\end {figure}

\begin {figure} [p]
\centerline {\hbox {
\includegraphics [angle=90,width=0.9\textwidth] {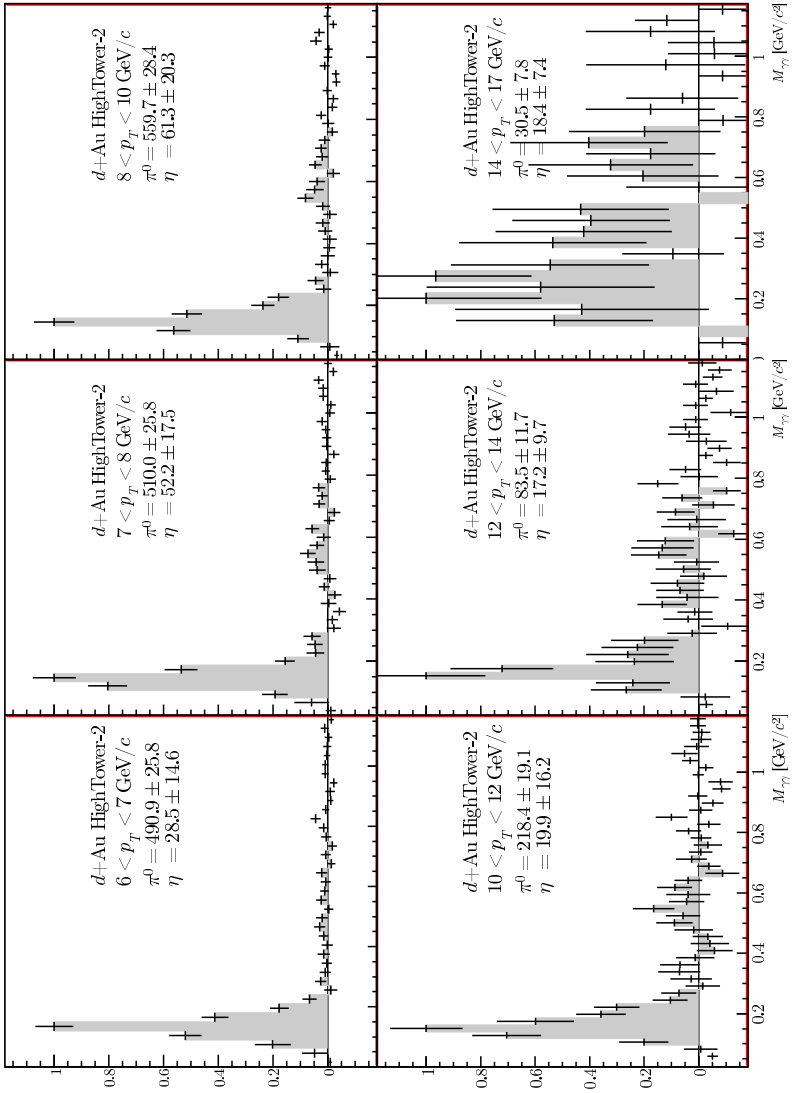}
}}
\caption {Invariant mass distributions in all \pT\ bins, \deuterongold\ \HighTowerTwo\ data.
The spectra are normalized to the bin content in the \pizero\ peak.}
\label {fig_invmass_dAuHT2}
\end {figure}

The complete set of invariant mass spectra for all \pT\ bins, triggers, and datasets are shown 
in Figures~\ref {fig_invmass_ppMB}--\ref {fig_invmass_dAuHT2}\@.
For display purposes, the spectra are normalized to the bin content in the \pizero\ peak. 
The shaded areas in the figures indicate the \pizero\ and \etameson\ peak regions 
where the yields are calculated simply by adding up the bin contents.

The left border of the \pizero\ peak region was taken to be a linear function of \pT, common for all datasets and triggers.
It was adjusted in a way that most of the yield is captured, while 
the low mass background and its associated uncertainty is avoided as much as possible.
The right border also linearly increases with \pT, in order to cover the asymmetric right tail of the peak.
Similarly, the \etameson\ peak region is a \pT-dependent window that captures most of the signal.
For completeness, we give below the parametrization of the \pizero\ and \etameson\ windows:
$$
\begin {array}{r@{\ <\ }l@{\ <\ }l}
75 + 1.7 \pT & \SUB{\Mass}{\pizero} & 250 + 3.3 \pT \quad \unitns{\MeVcc}, \\
350 + 3.3 \pT & \SUB{\Mass}{\etameson} & 750 \quad \unitns{\MeVcc}. \\
\end {array} 
$$
The stability of the yields was determined by varying the vertex position cut, 
the energy asymmetry cut, and the yield integration window.
\label {syst_analysis_cuts}
From the observed variations, a point\tsp{0.2}-to\tsp{0.4}-point systematic error of \MM{5\unitns{\%}} was assigned to the yields.

\chapter {Invariant yield calculation}

The invariant yield of the neutral pions and \etameson\ mesons per one minimum bias collision, 
as a function of the transverse momentum \pT, is given by
\begin {equation}
\energy \FRAC{\SUP{\der}{3}\Number}{\der\SUP{\momentumbold}{3}} = 
\FRAC{\SUP{\der}{3}\Number}{\pT\tsp{0.5}\der\pT\tsp{0.5}\der\rapidity\tsp{0.5}\der\phiangle} = 
\FRAC{\SUP{\der}{2}\Number}{2\PI\pT\tsp{0.5}\der\pT\tsp{0.5}\der\rapidity},
\end {equation}
where in the last equation isotropic production in azimuth is assumed.
Using the experimentally measured quantities, the invariant yield is calculated as
\begin {equation} \label {eq:yielddef}
\energy \FRAC{\SUP{\der}{3}\Number}{\der\SUP{\momentumbold}{3}} = 
\FRAC{1}{2\PI\pT}
\FRAC{\SUB{\EPS}{\RM{vertex}}}{\SUB{\Number}{\RM{trig}} \SUB{\IT{K}}{\RM{trig}} (1 - \SUB{\EPS}{\RM{beam}})} 
\FRAC{\DELTA\IT{Y}}{\DELTA\pT\tsp{0.5}\DELTA\rapidity} 
\FRAC{1}{\SUB{\EPS}{\RM{acc}}} 
\FRAC{1}{\SUB{\EPS}{\RM{cpv}}}
\FRAC{1}{\DIV{\SUB{\ensuremath{\Gamma}}{\gama\gama}}{\ensuremath{\Gamma}}},
\end {equation}
where:
\begin {itemize}
\item {\MM{\DELTA\IT{Y}}} is the raw yield measured in the bin \MM{\DELTA\pT\tsp{0.5}\DELTA\rapidity};
\item {\MM{\SUB{\Number}{\RM{trig}}}} is the number of triggers recorded;
\item {\MM{\SUB{\IT{K}}{\RM{trig}}}} is the trigger prescale factor that is unity for the MinBias events and larger than unity for the HighTower data.
The product \MM{\SUB{\Number}{\RM{trig}}\SUB{\IT{K}}{\RM{trig}}} then gives the equivalent number of minimum bias events that produced the yield \MM{\DELTA\IT{Y}};
\item {\MM{\SUB{\EPS}{\RM{vertex}}}} is the vertex finding efficiency in minimum bias events;
\item {\MM{\SUB{\EPS}{\RM{beam}}}} is the beam background contamination in minimum bias events;
\item {\MM{\DELTA\pT}} is the width of the \pT\ bin for which the yield is calculated;
\item {\MM{\DELTA\rapidity}} is the rapidity range of the measurements, in this analysis \MM{\DELTA\rapidity{} = 1};
\item {\MM{\SUB{\EPS}{\RM{acc}}}} is the BEMC acceptance and efficiency correction factor;
\item {\MM{\SUB{\EPS}{\RM{cpv}}}} is a correction for random vetoes;
\item {\MM{\DIV{\SUB{\ensuremath{\Gamma}}{\gama\gama}}{\ensuremath{\Gamma}}}} is the branching ratio of the di-\tsp{-0.2}photon decay channel, 
equal to \MM{0.988} for \pizero\ and \MM{0.392} for \etameson~\cite {ref_pdg}\@.
\end {itemize}

Each of these corrections is described in detail in one the following sections.

\section {Acceptance and efficiency correction}
\label {eff_correction}

To calculate the acceptance and efficiency correction factor \SUB{\EPS}{\RM{acc}},  
a Monte Carlo simulation of the detector was used, 
where neutral pions and their decay photons were tracked through the STAR detector geometry using GEANT~\cite {ref_geant}\@.
The simulated signals were passed through the same analysis chain as the real data.

The pions were generated in the pseudorapidity region \MM{-\tsp{2.0}0.3 \LESS{} \etacoord{} \LESS{} +1.3}, which is sufficiently large 
to account for edge effects caused by the calorimeter \linebreak
acceptance limits of \MM{0\tsp{-0.5}\LESS\tsp{-0.5}\etacoord\tsp{-0.5}\LESS{}\tsp{-0.5}1}\@.
The~azimuth~was~generated~flat~in~\mbox{\MM{-\PI\tsp{-0.5}\LESS\tsp{-0.5}\phiangle\tsp{-0.5}\LESS{}\tsp{-1.5}+\PI}}\@.\linebreak
The \pT\ distribution was taken to be flat between zero and \MM{25\unit{\GeVc}}, which amply covers the measured 
pion \pT\ range of up to \MM{17\unit{\GeVc}}\@. The vertex \linebreak
distribution of the generated pions was taken to be Gaussian in \zcoord, with a spread 
of \MM{\SIGMA{} = 60\unit{\cm}} and centered at \MM{\zcoord{} = 0}\@.

The generated pions were allowed to decay into \MM{\pizero\TO\gama\gama}\@.
The GEANT simulation accounts for all interaction of the decay photons with the detector, such as
pair conversion into \epluseminus\ and showering in the calorimeter or in the material in front.

\begin {figure} [tb]
\centerline {\hbox {
\includegraphics [width=0.5\textwidth] {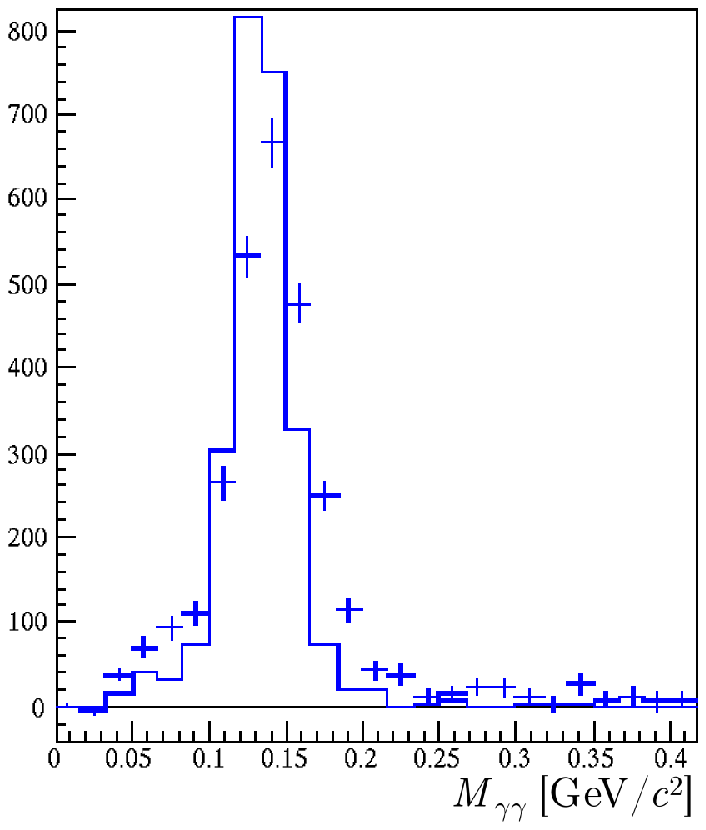}\ \ \includegraphics [width=0.5\textwidth] {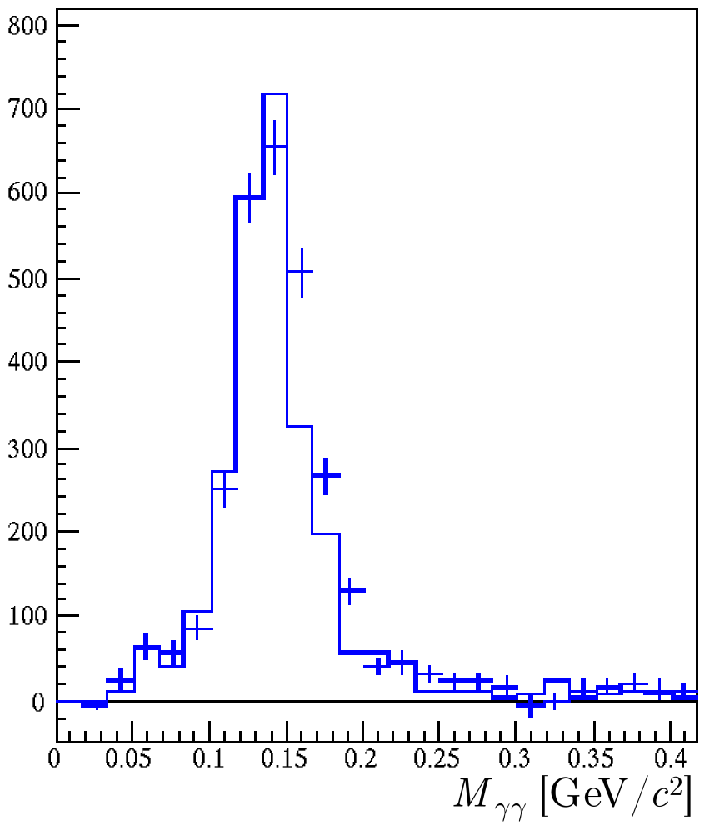}
}}
\caption {The invariant mass distribution in the real \protonproton\ data (crosses) and in the simulation (histogram)\@.
The Monte Carlo produces a narrower \pizero\ peak (left) than is observed in the data, 
so that an additional energy smearing was introduced to reproduce the calorimeter resolution (right)\@.}
\label {fig_calib_spread}
\end {figure}

To reproduce a realistic energy resolution of the calorimeter, 
an additional smearing has to be applied to the energy deposit generated by GEANT in the towers.
The effect of this can be seen in Figure~\ref {fig_calib_spread},
where the simulated \pizero\ invariant mass peak is shown in comparison to the \protonproton\ data with and without smearing.
An additional spread of \MM{5\unitns{\%}} was used to reproduce the \protonproton\ data and \MM{10\unitns{\%}} for the \deuterongold\ data.

To reproduce the \pT\ spectrum of pions in the data, each Monte Carlo event was weighted by a \pT-dependent function.
Such weighting technique allows to sample the whole \pT\ range with good statistical power, while, at the same time, 
the bin migration effect caused by the finite detector energy resolution is reproduced.
A next-to\tsp{0.4}-leading order QCD calculation~\cite {ref_vogelsang_private} provided the initial weight function, 
parametrized as described in Section~\ref {cross_section_calculation}, 
which was subsequently \linebreak
adjusted in an iterative procedure. 

\begin {figure} [tb]
\centerline {\hbox {
\includegraphics [width=0.5\textwidth] {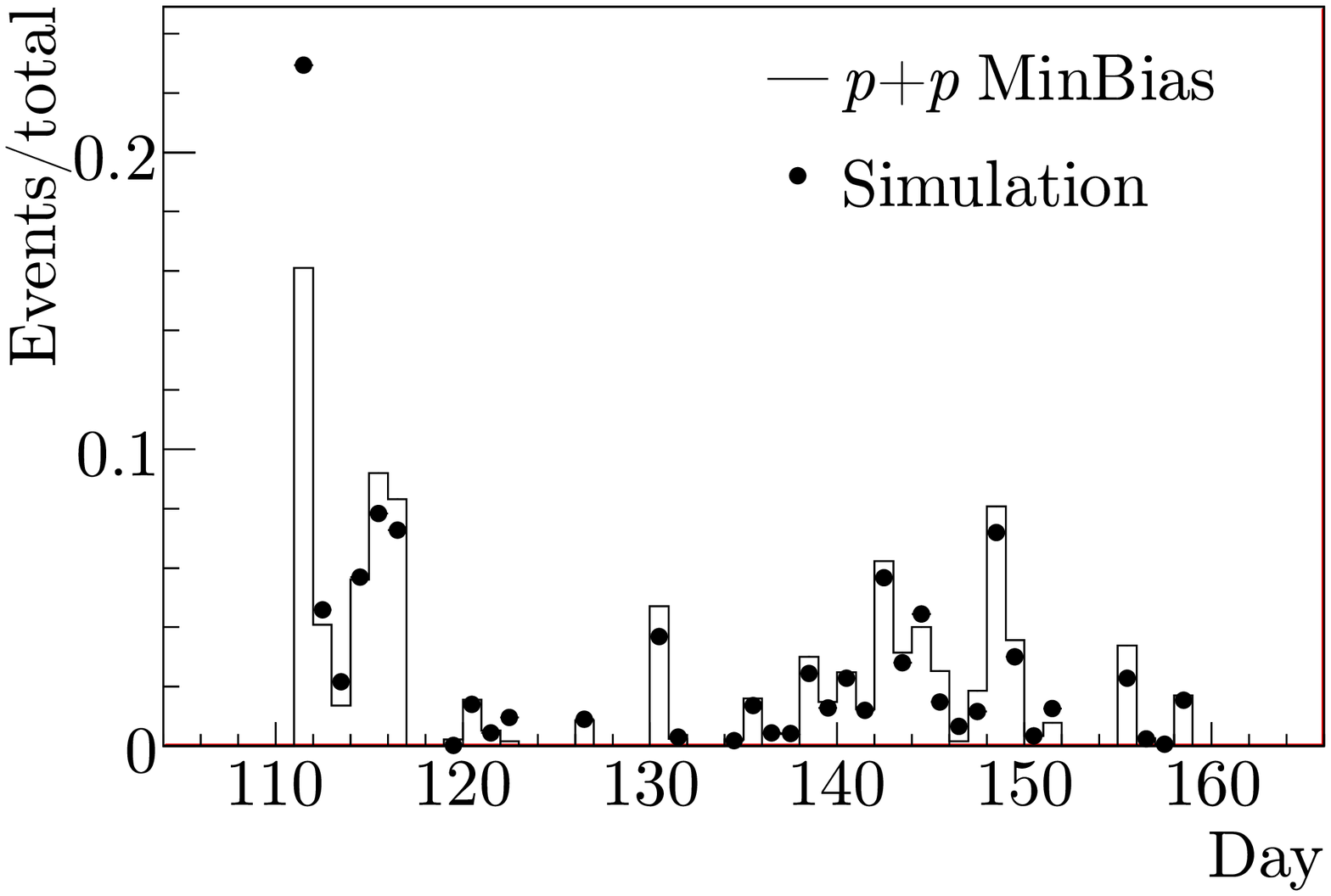}\includegraphics [width=0.5\textwidth] {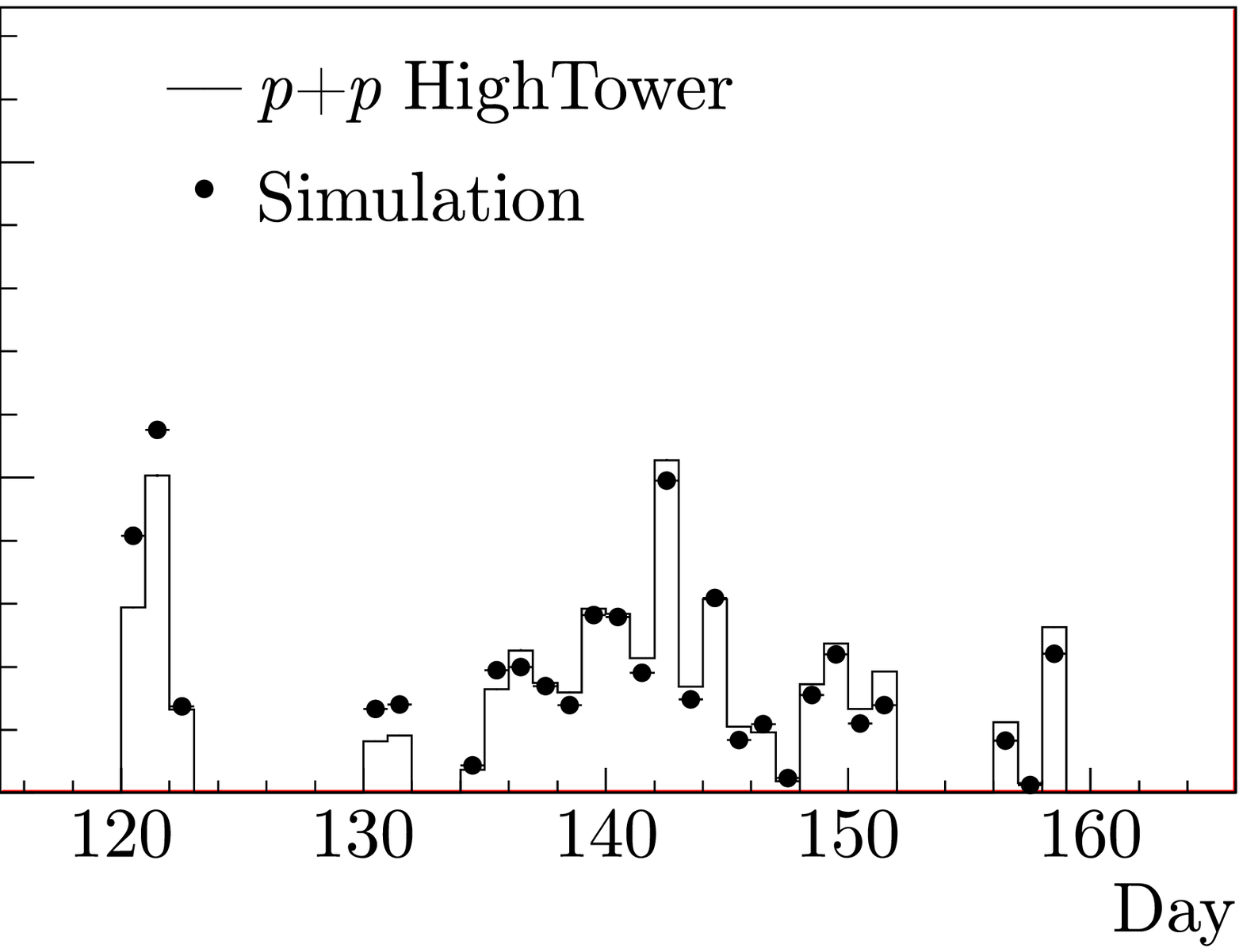}
}}
\caption {Statistics accumulated per day (histogram) and simulated in the Monte Carlo (full circles), 
for \protonproton\ MinBias (left) and HighTower data (right)\@.}
\label {fig_stat_day}
\end {figure}

As mentioned in Section~\ref {BEMC_status_tables_preparation},
the time dependence of the calorimeter \linebreak
acceptance is stored in data tables, which are fed into the analysis.
In order to reproduce this time dependence in the Monte Carlo, the simulated events were 
assigned time stamps that follow the timeline of the real data taking. 
In Figure~\ref {fig_stat_day} 
is shown, separately for MinBias and HighTower data, the accumulated real data statistics per day (histogram),
together with the time distribution of the simulated events (full circles)\@.
In this way, the geometrical calorimeter acceptance (fraction of good towers) was reproduced in the Monte Carlo 
with a precision of better than~\MM{0.5\unitns{\%}}\@.

In the real data analysis, we use vertices reconstructed from the TPC tracks with a sub\tsp{0.4}-millimiter resolution,
as well as vertices derived from the BBC time of flight measurement with a precision of about \MM{40\unit{\cm}}\@.
To account for this poor resolution, a fraction of the simulated pions had their point of origin artificially smeared in the \zcoord\ direction.
This fraction was taken to be \MM{35\unitns{\%}} of the generated pions in case of the \protonproton\ MinBias analysis,
and taken to be zero for all the other datasets since no BBC vertex was used in these sets (see Chapter~\ref {ref_event_reconstruction})\@.

\begin {figure} [p]
\centerline {\hbox {
\includegraphics [width=0.9\textwidth] {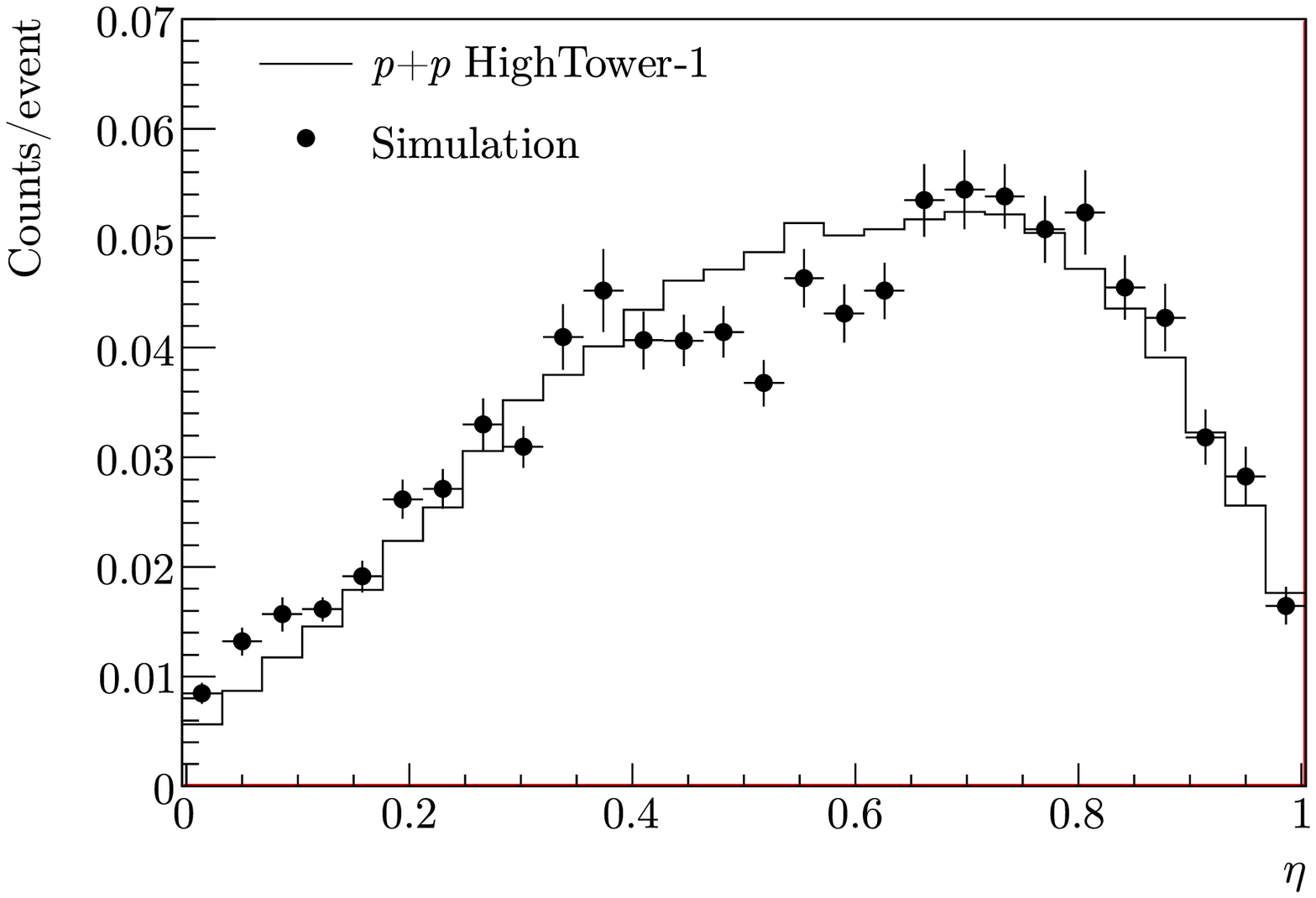}
}}
\centerline {\hbox {
\includegraphics [width=0.9\textwidth] {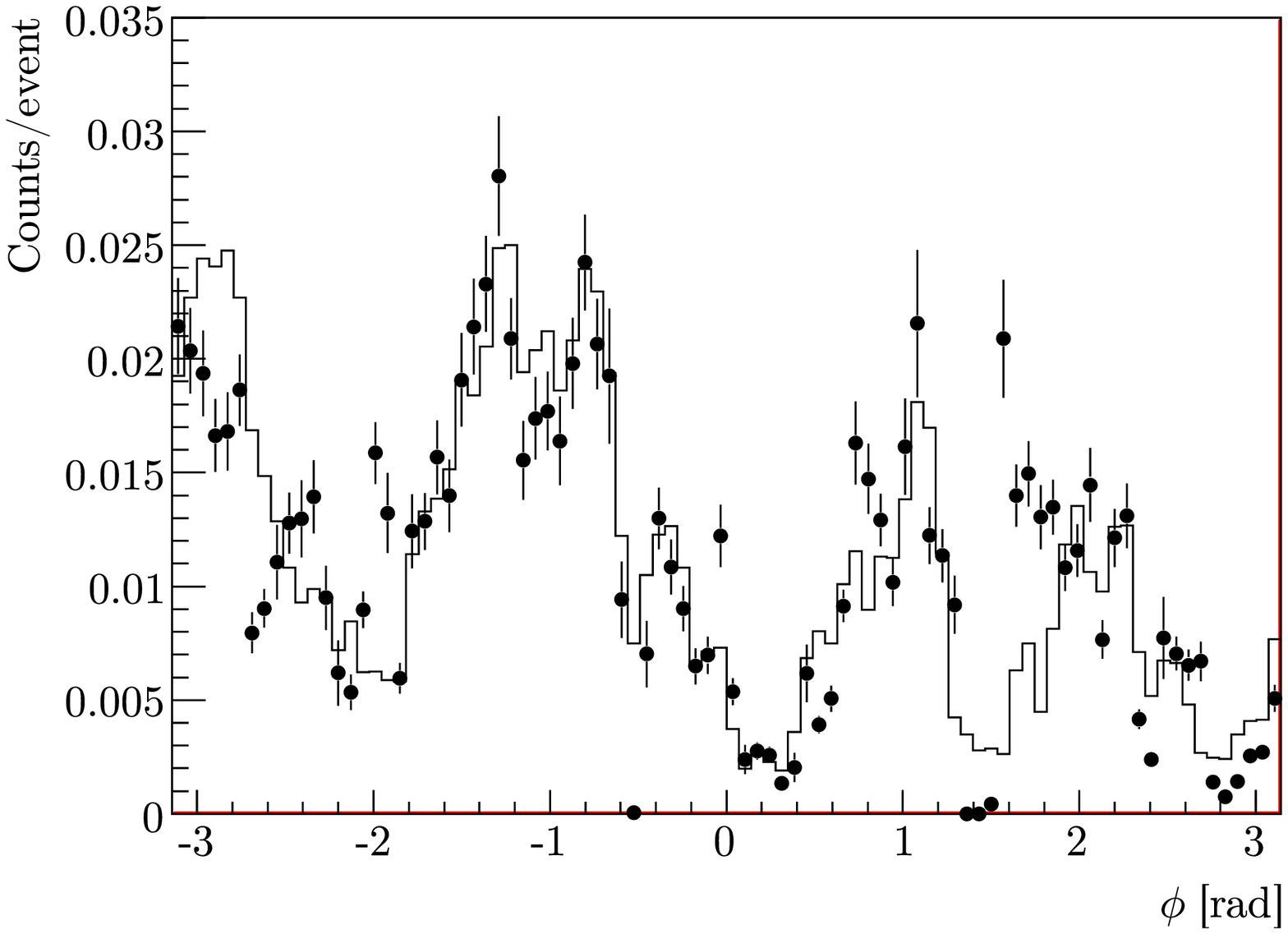}
}}
\caption {Distributions of \etacoord\ (top) and \phiangle\ (bottom) coordinates of the reconstructed Monte Carlo pions, 
compared to the pion candidates in the \protonproton\ \HighTowerOne\ data.
The structure seen in the \phiangle\ distribution reflects the azimuthal dependence of the calorimeter acceptance,
caused by failing \SMD\ modules. 
This structure is well reproduced in the simulations.}
\label {fig_compare_sim_etaphi}
\end {figure}

\begin {figure} [tb]
\centerline {\hbox {
\includegraphics [width=0.9\textwidth] {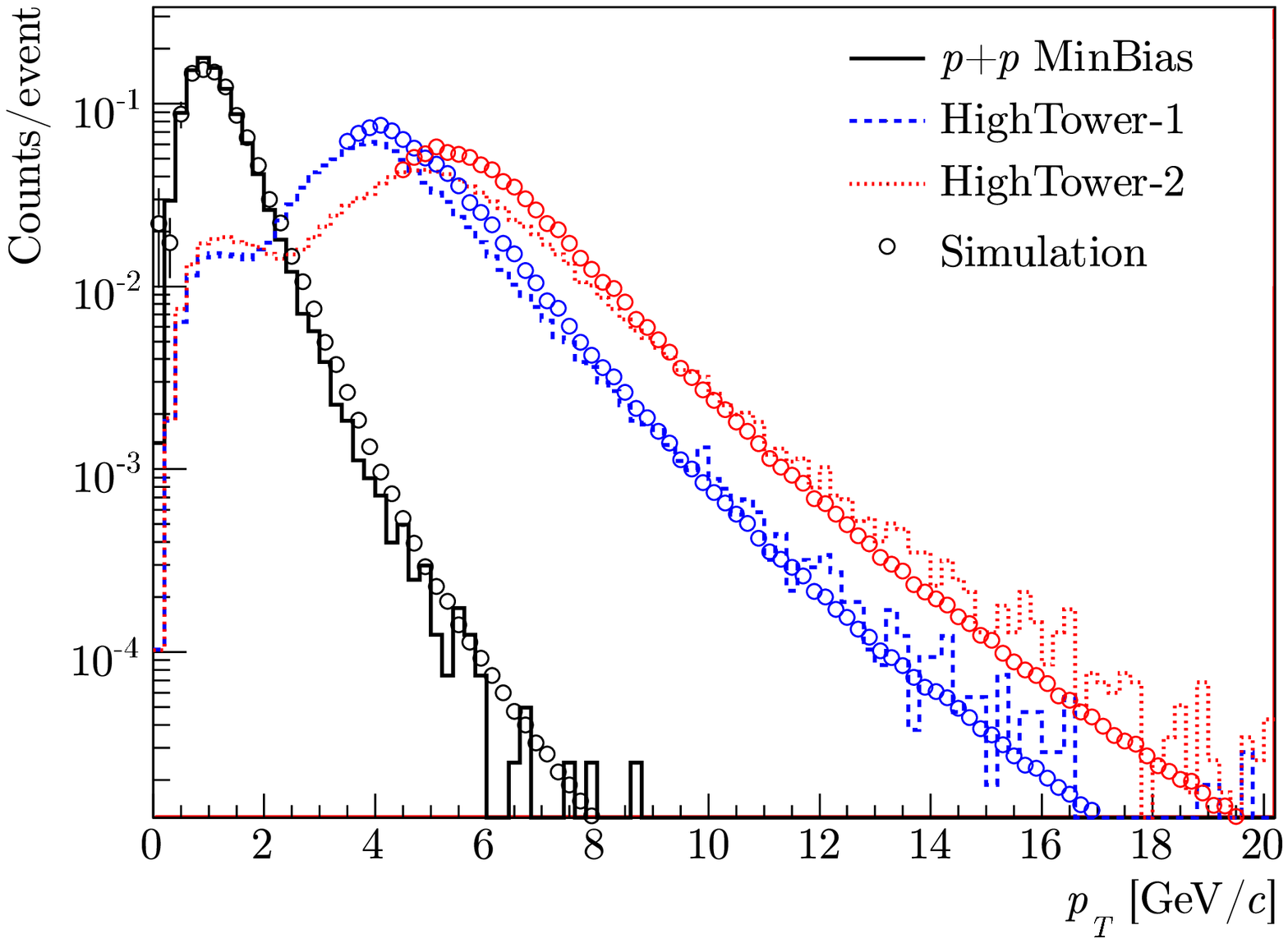}
}}
\caption {Distributions of the reconstructed \pT\ of the Monte Carlo MinBias, \HighTowerOne, and \HighTowerTwo\ pions, compared to the 
\protonproton\ data.}
\label {fig_compare_sim_pt}
\end {figure}

In Figure~\ref {fig_compare_sim_etaphi} 
we show the \etacoord\ and \phiangle\ distributions of the reconstructed Monte Carlo pions in comparison to the 
\protonproton\ data.
The agreement is satisfactory, indicating that the calorimeter acceptance is well reproduced in the simulation.
In Figure~\ref {fig_compare_sim_pt} 
the reconstructed \pT\ of simulated pions is compared to that of pion candidates from the \protonproton\ data.
It is seen that the HighTower trigger threshold effects are reasonably well reproduced.

\begin {figure} [p]
\centerline {\hbox {\includegraphics [width=0.8\textwidth] {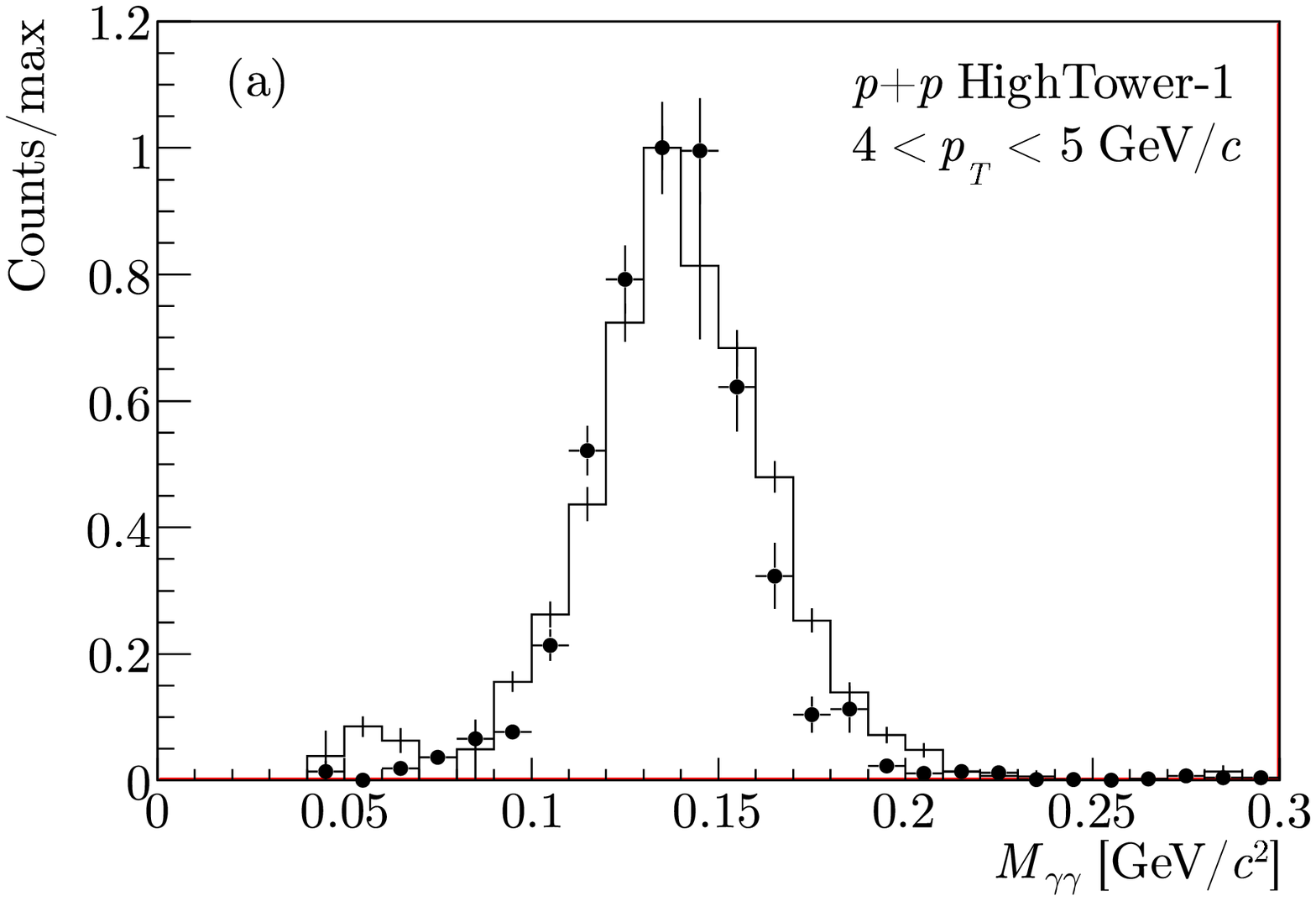}}}\centerline {\hbox {\includegraphics [width=0.8\textwidth] {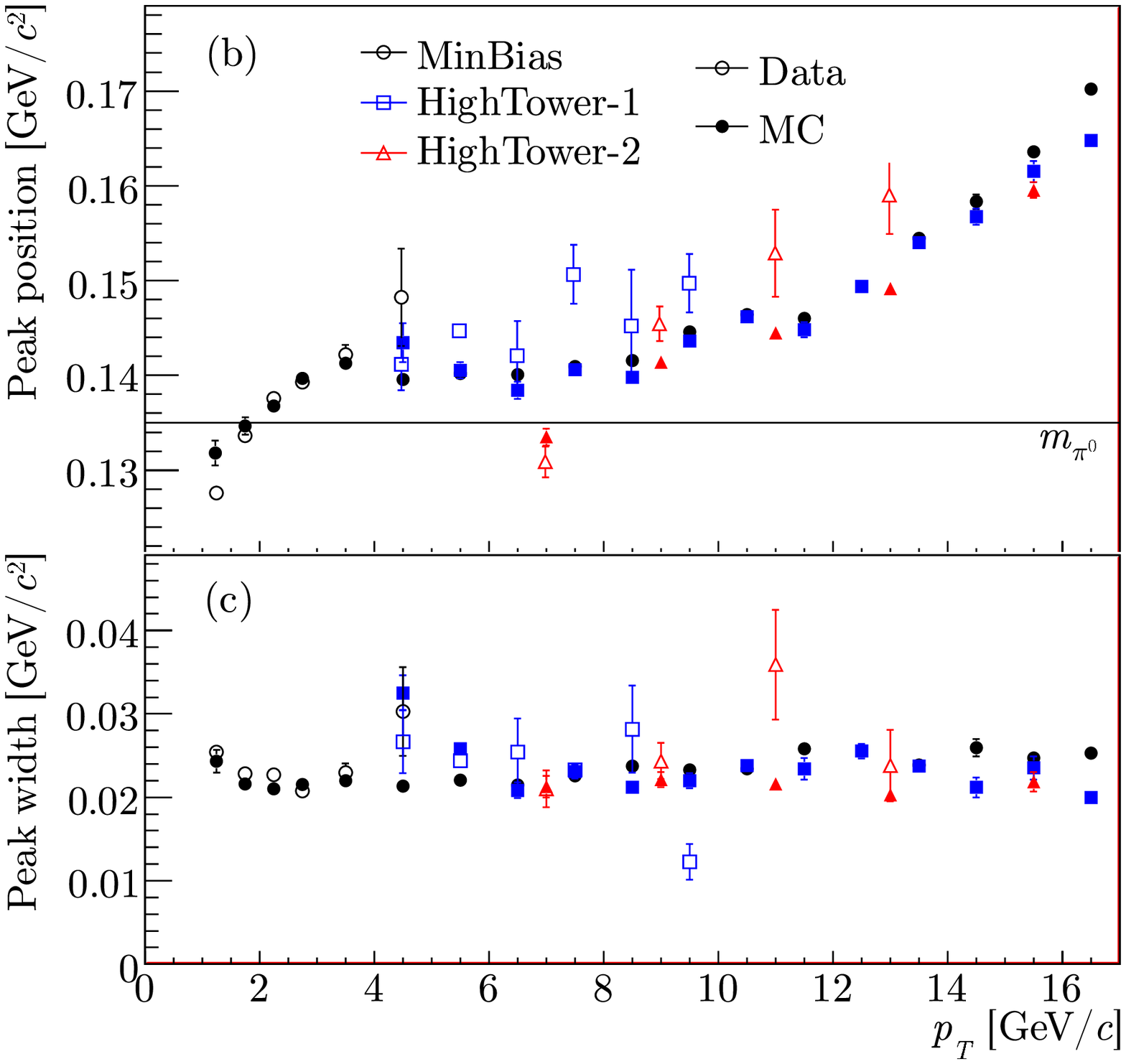}}}
\caption {
Invariant mass spectrum reconstructed in the simulation, in 
comparison to the \protonproton\ \HighTowerOne\ data in \MM{4 \LESS{} \pT{} \LESS{} 5\unit{\GeVc}} bin (top)\@.
Peak position (middle) and width (bottom) in the real data and MC simulation.}
\label {fig_compare_m}
\end {figure}

In Figure~\ref {fig_compare_m}(a)
the background subtracted invariant mass distribution is shown in the region \MM{4 \LESS{} \pT{} \LESS{} 5\unit{\GeVc}} obtained from 
the \protonproton\ \HighTowerOne\ data, together with the corresponding distribution from the Monte Carlo.
In order to compare the real and simulated invariant mass distributions for all bins in \pT\ and for all datasets, 
we have estimated the position and width of the peaks by Gaussian fits in the peak region.
In Figure~\ref {fig_compare_m}(b) are shown the peak positions obtained from the fit to the \protonproton\ data.
It is seen that the peak position shifts towards higher masses with increasing \pT\@.
This shift is a manifestation of bin migration effects that originate from statistical fluctuations in the calorimeter response.
Due to the steeply falling \pT\ spectrum the energy resolution will cause a net migration towards larger \pT\@.
Since larger values of \pT\ imply larger values of \Mgg, the migration effect will bias the invariant mass peak towards larger values.
The good agreement between the data and Monte Carlo indicates that such resolution and migration effects are well reproduced.

\enlargethispage{\baselineskip}
In Figure~\ref {fig_compare_m}(c) is shown the comparison of the \pizero\ peak width in data and simulation.
The peak width is well reproduced in simulation, 
which is not surprising since additional smearing was introduced 
to improve the comparison between data and Monte Carlo, see Figure~\ref {fig_calib_spread}\@.

The acceptance and efficiency correction factor was calculated from the Monte Carlo simulation 
as the ratio of the raw yield of neutral pions reconstructed in a \pT\ bin, to the 
number of simulated pions with the true \pT\ in that bin. 
This was done separately for each trigger, using the same pion reconstruction cuts 
as was done in the real data analysis.
In particular, the reconstructed value of pseudorapidity was required to fall in the range \MM{0 \LESS{} \etacoord{} \LESS{} 1} in both 
the data and the Monte Carlo, while in the latter the generated value of \etacoord\ was also required to fall in this range.

\begin {figure} [p]
\centerline {\hbox {
\includegraphics [width=0.5\textwidth] {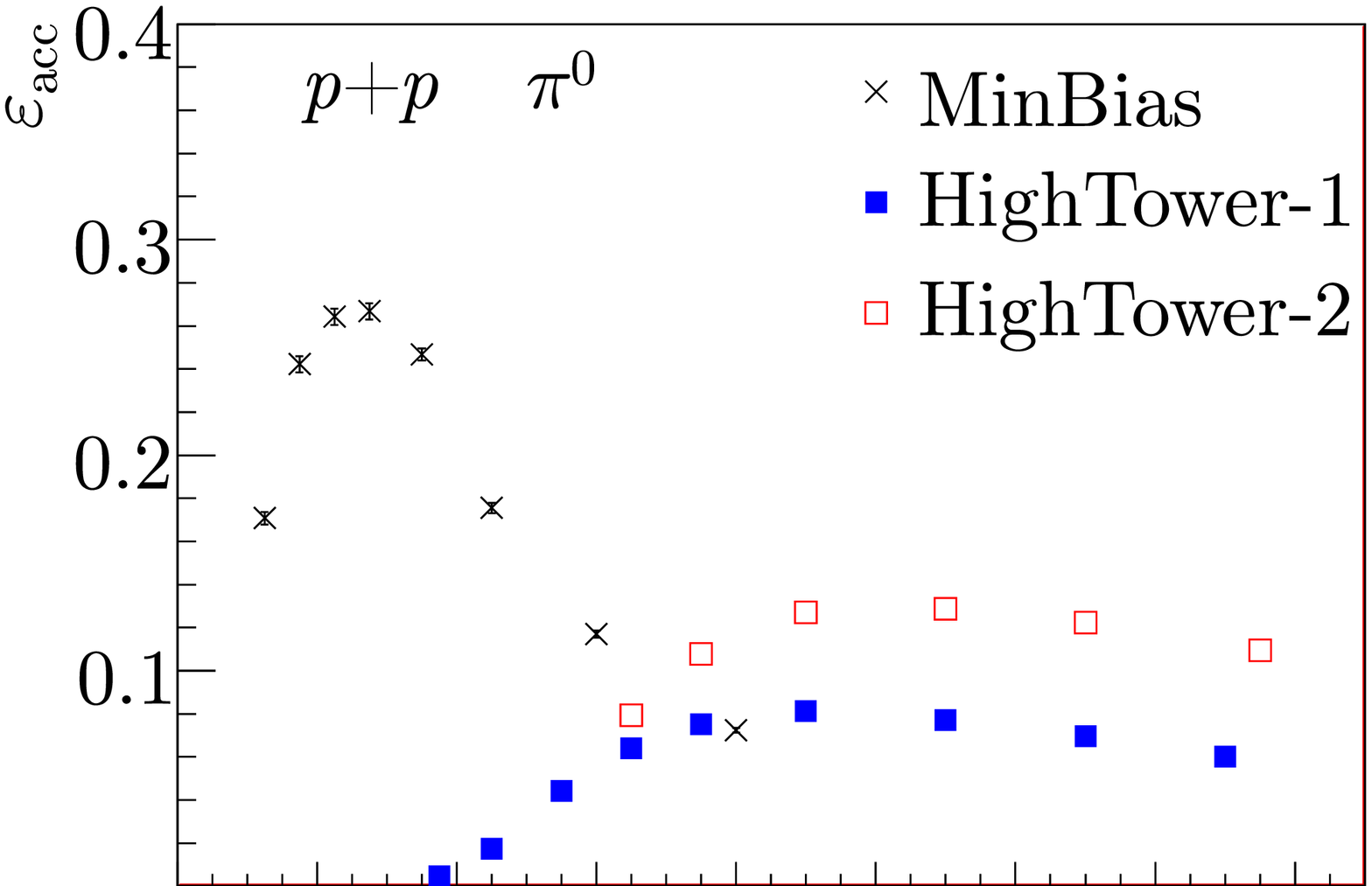}\includegraphics [width=0.5\textwidth] {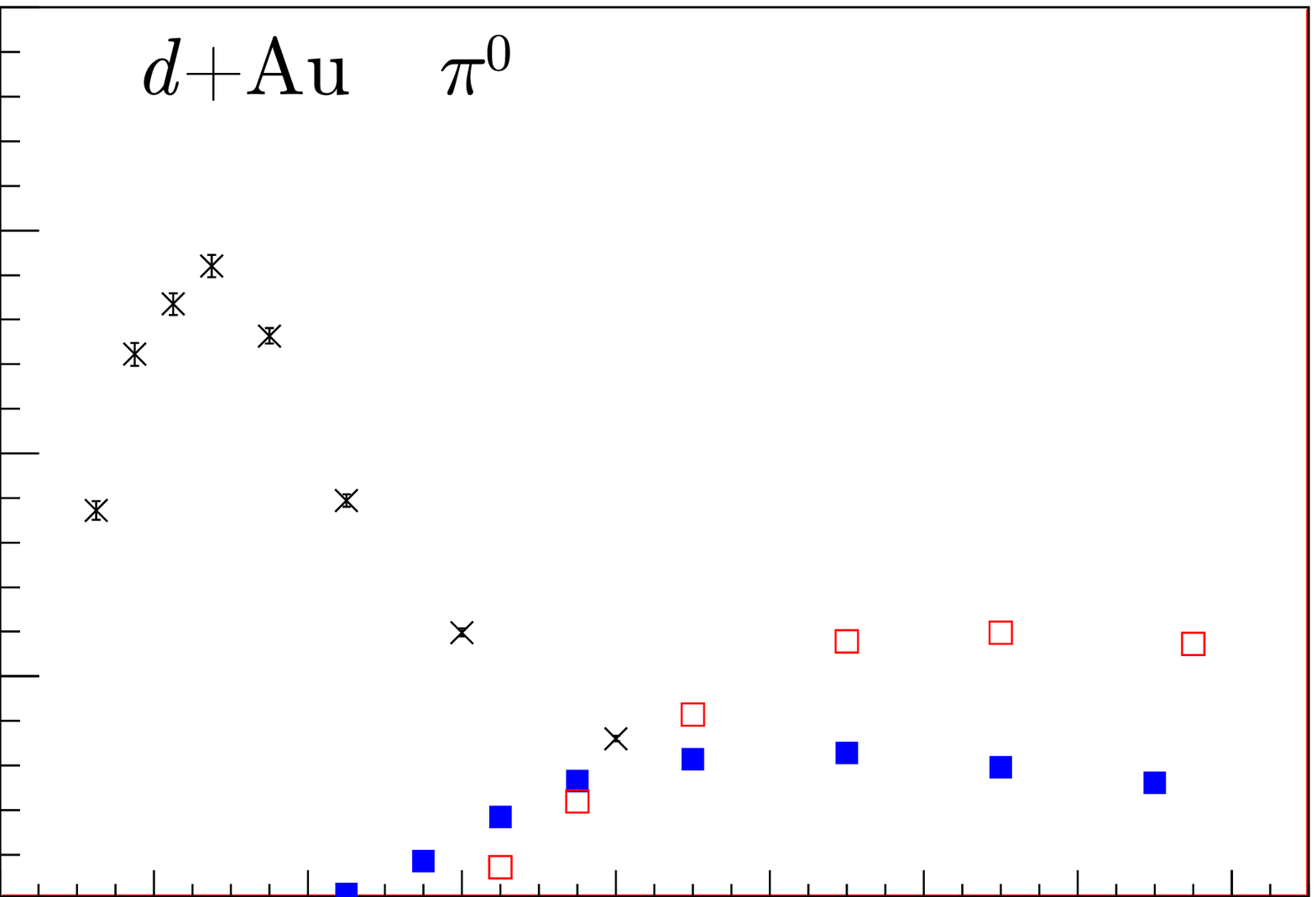}
}}
\centerline {\hbox {
\includegraphics [width=0.5\textwidth,clip] {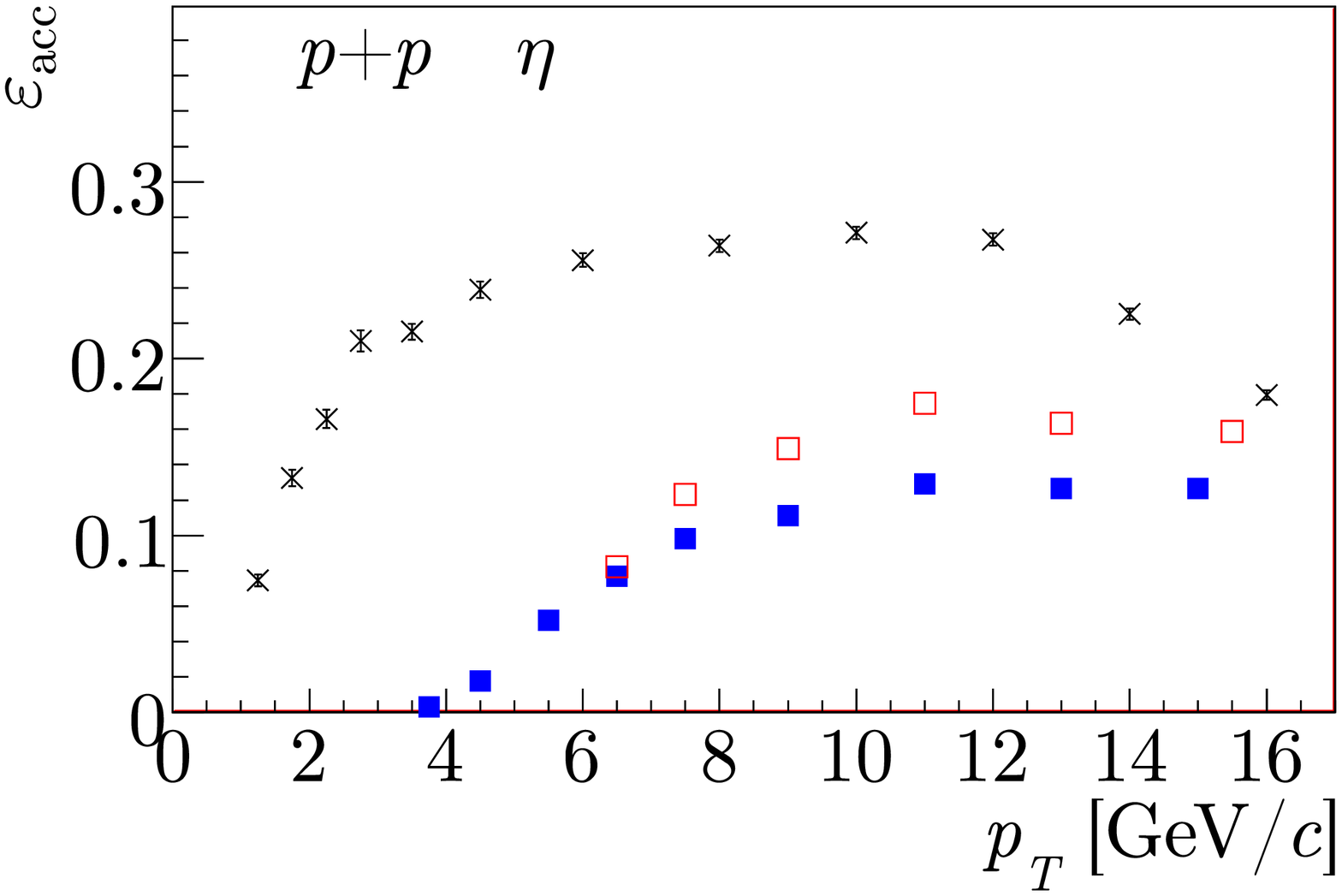}\includegraphics [width=0.5\textwidth,clip] {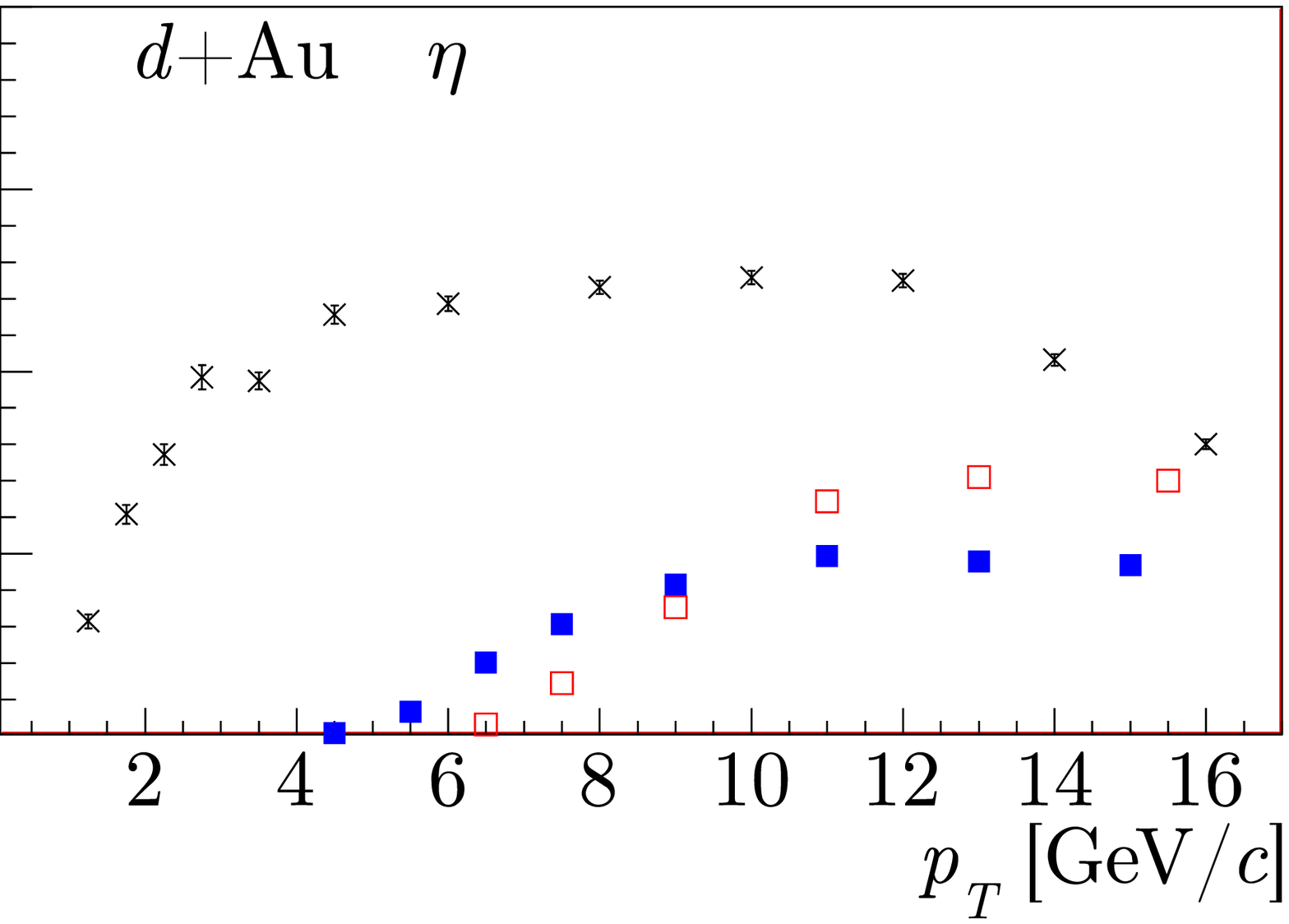}
}}
\caption {Acceptance and efficiency factor \SUB{\EPS}{\RM{acc}} calculated from the Monte Carlo simulation 
for the \protonproton\ (left\tsp{0.2}-hand plots) and \deuterongold\ datasets (right\tsp{0.2}-hand plots)\@.
The \pizero\ and \etameson\ efficiencies are shown separately in the top and bottom plots, respectively.}
\label {fig_eff}
\end {figure}

In Figure~\ref {fig_eff} 
are shown the \pizero\ and \etameson\ correction factors for all datasets and triggers used in this analysis.

\enlargethispage{2\baselineskip}
The large difference between the MinBias and HighTower correction factors is caused by the SMD requirement in the HighTower data, 
while in the MinBias data we accept all reconstructed BEMC points. 
The absense of the SMD \linebreak
information also reduces the \pizero\ reconstruction efficiency at \MM{\pT{} \GREATER{} 3\unit{\GeVc}}, 
where the decay photons are separated by less than two towers.
\enlargethispage{\baselineskip}
The \etameson\ reconstruction starts being affected at larger values of \pT\@.
\clearpage
\begin {figure} [p]
\centerline {\hbox {
\includegraphics [width=0.9\textwidth] {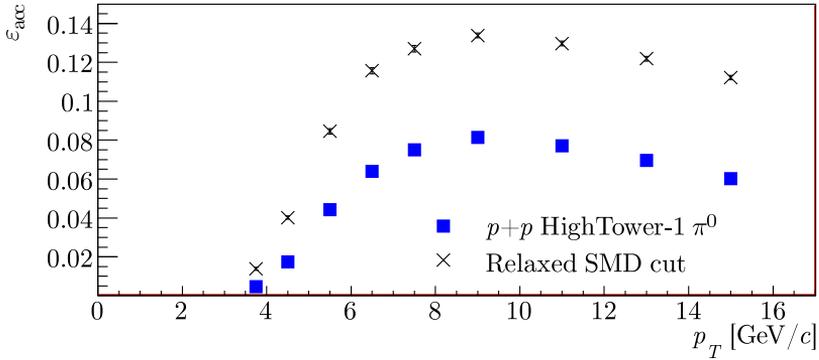}
}}
\caption {Acceptance and efficiency correction for the \protonproton\ \HighTowerOne\ data, with standard set of cuts 
and with \SMD\ quality cut removed.}
\label {fig_eff_diff}
\end {figure}
\clearpage
We have checked the effect of the \SMD\ quality requirement (at least two adjacent strips in a cluster) 
on the correction factor for HighTower triggered data.
In Figure~\ref {fig_eff_diff}
is shown the correction factor calculated for the \protonproton\ \HighTowerOne\ dataset 
with (full squares) and without (crosses) the \SMD\ quality requirement.
It is seen that this requirement reduces the number of accepted \pizero\ candidates by about \MM{45\unitns{\%}}\@.
This explains the difference between the \HighTowerOne\ and \HighTowerTwo\ (no \SMD\ quality cut) 
correction factors at large \pT\ seen in Figure~\ref {fig_eff}\@.

To verify a possible dependence of the acceptance correction on the track multiplicity and thus on the 
centrality, we have analyzed a sample of generated neutral pions embedded in real \deuterongold\ data.
These embedded data are centrally produced by the STAR offline group and are used by several analyses in STAR~\cite {ref_simon_thesis}\@.
No significant centrality dependence was found, so that same correction factors were applied to the 
various centrality classes in the \deuterongold~data.

\section {Corrections for random vetoes}

\begin {figure} [tb]
\centerline {\hbox {
\includegraphics [width=0.9\textwidth] {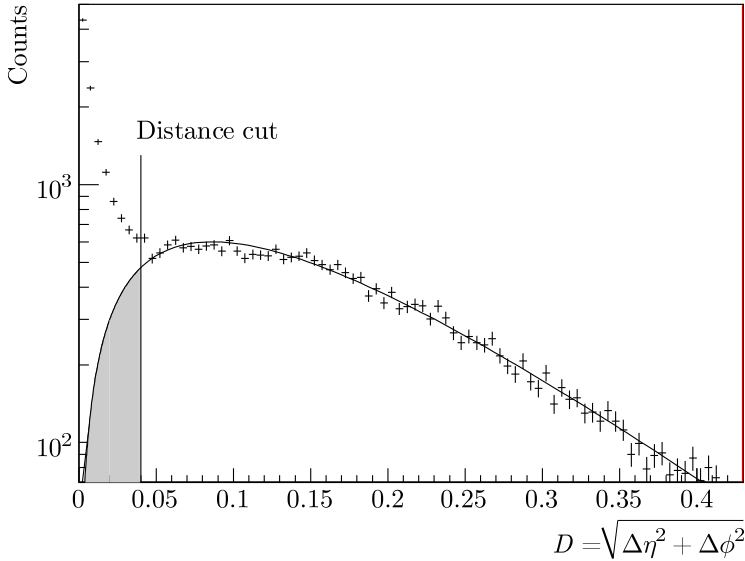}
}}
\caption {Distribution of the distance between BEMC points and the closest track, 
obtained from \protonproton\ \HighTowerOne\ data in the bin \MM{4 \LESS{} \pT{} \LESS{} 5\unit{\GeVc}}\@. 
The curve shows a fit to Eq.~(\ref {eq:cpv_distr}) and the vertical line indicates the CPV cut.}
\label {fig_cpv_cut}
\end {figure}

\begin {figure} [tb]
\centerline {\hbox {
\includegraphics [width=0.5\textwidth] {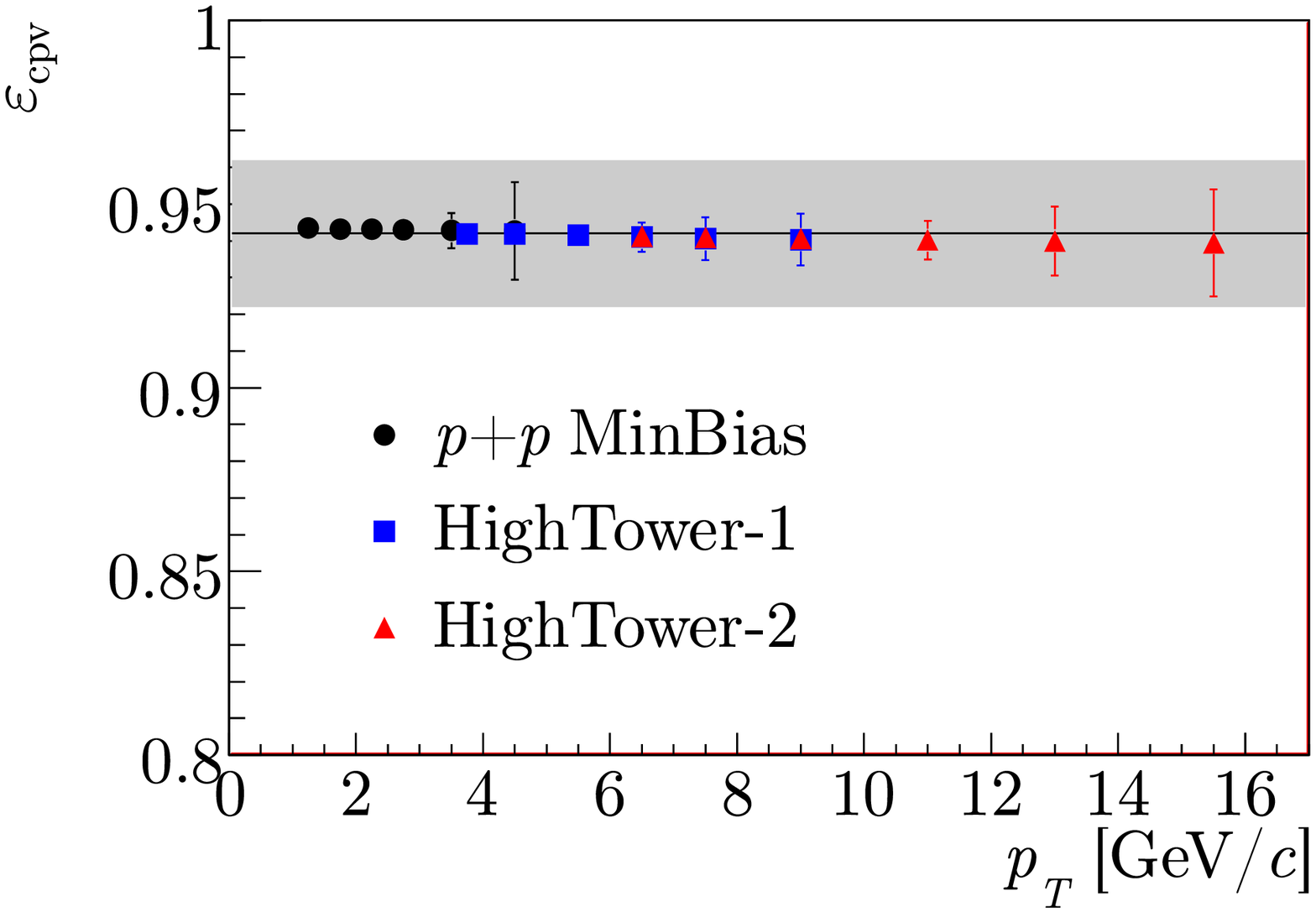}\includegraphics [width=0.5\textwidth] {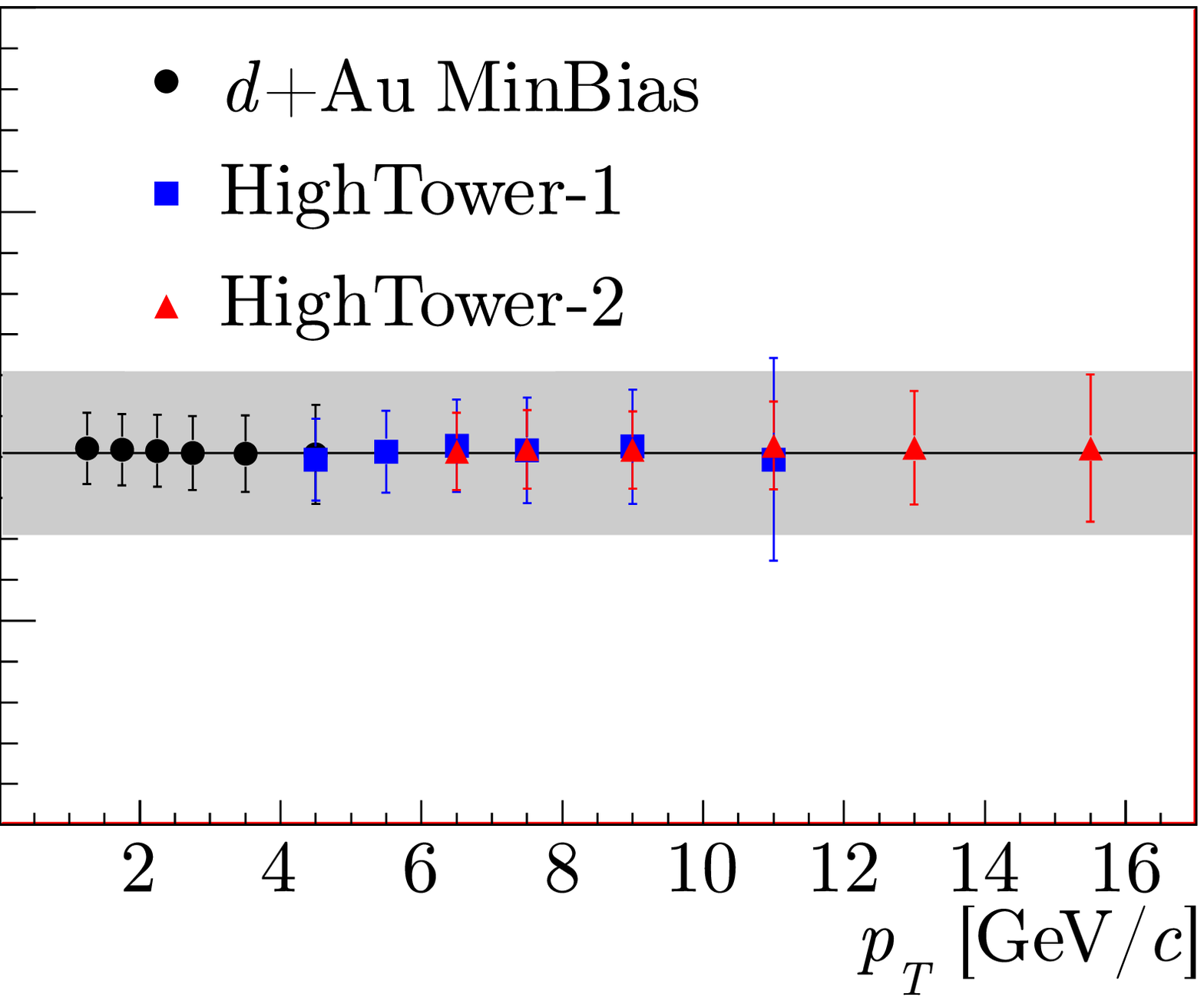}
}}
\caption {Charged particle veto correction in \protonproton\ (left) and \deuterongold\ (right) data.
The horizontal line indicates the correction factor \SUB{\EPS}{\RM{cpv}} applied to the data,
while the shaded band corresponds to the systematic uncertainty assigned to the correction factors.}
\label {fig_cpv_corr}
\end {figure}

This analysis uses the TPC as a veto detector to reject charged particles, 
which introduces false rejection of photon clusters 
if an unrelated charged particle happens to hit the calorimeter nearby the cluster. 
In Figure~\ref {fig_cpv_cut}
we plot the distribution of distances between the BEMC point and the closest charged track in the event.
In this plot one easily distinguishes the peak of real charged particles at small distances,
superimposed on a random component, which shows up as a shoulder at larger distances.
Assuming that the charged tracks are uniformly distributed in \etacoord\ and \phiangle\ around the BEMC point,
it follows that the radial distribution is given by
\begin {equation} \label {eq:cpv_distr}
\IT{f}(\IT{D}) = 
\IT{D}
\tsp{1.0}
\SUP{\IT{e}}{-\IT{D} \ensuremath{\tau}},
\end {equation}
where the parameter \ensuremath{\tau} has the meaning of the local track density in the region where the photon probes it.
This parameter is obtained from a simultaneous fit to the data in all bins of the event multiplicity \IT{M},
assuming its linear dependence on the multiplicity \MM{\ensuremath{\tau} = \IT{a} + \IT{b}\IT{M}}\@.
The parametrization (\ref {eq:cpv_distr}) well describes the random component, as shown by the full curve in Figure~\ref {fig_cpv_cut}\@.
The relative amount of random coincidences is then obtained by integrating the fitted curve up to the distance cut 
and weighting with the multiplicity distribution observed in each \pT\ bin.
Separate sets of correction factors were calculated for the various triggers in the \protonproton\ and \deuterongold\ data.
The results are shown in Figure~\ref {fig_cpv_corr}
as a function of \pT\@.
We have applied a correction factor of 
\MM{\SUB{\EPS}{\RM{cpv}} = 0.94 \PLMN{} 0.02} 
to the \protonproton\ datasets 
and of 
\MM{\SUB{\EPS}{\RM{cpv}} = 0.89 \PLMN{} 0.02} 
to the \deuterongold\ datasets.
\label {syst_cpv}
The errors assigned to these corrections contribute to a \pT\ independent 
systematic error on the corrected \pizero\ and \etameson\ yields.

\section {HighTower trigger scale factors}

We have shown in Figure~\ref {fig_data_candidates}(c) the \pT\ distribution of \pizero\ candidates for the \protonproton\ 
MinBias, \HighTowerOne, and \HighTowerTwo\ data.
To match the HighTower spectra to those of the MinBias, a \pT-independent scale factor (\SUB{\IT{K}}{\RM{trig}}) was applied.
These scale factors were estimated as the ratio of observed MinBias to HighTower event rates
\begin {equation} \label {eq:HTprescale}
\SUB{\IT{K}}{\RM{trig}} = \FRAC {\ensuremath{\sum} \SUB{\Number}{\RM{MB}} \SUB{\IT{S}}{\RM{MB}}} {\ensuremath{\sum} \SUB{\Number}{\RM{HT}} \SUB{\IT{S}}{\RM{HT}}}.
\end {equation}
Here \SUB{\Number}{\RM{MB}} and \SUB{\Number}{\RM{HT}} are the numbers of MinBias and HighTower triggers
that pass the event selection cuts described in Chapter~\ref {ref_event_reconstruction}\@.
The factors \SUB{\IT{S}}{\RM{MB}} and \SUB{\IT{S}}{\RM{HT}} are the hardware prescale factors 
adjusted on a run-by-run basis to accomodate the DAQ bandwidth.
In Eq.~(\ref {eq:HTprescale}), the sums are taken over all runs in which both the MinBias and HighTower triggers were active.

To check the results, the scale factors were also estimated using another method.
Here the HighTower software filter (see section~\ref {ref_HT_filter}) was applied to the minimum bias data.
The scale factors were then obtained as the ratio of the total number of MinBias events 
to the number of those that passed the filter.
To obtain a more precise \HighTowerOne/\tsp{0.5}\HighTowerTwo\ relative scale factor,
the software filter was applied to the \HighTowerOne\ dataset.

The results from the two methods agree within \MM{3\unitns{\%}} for \HighTowerOne\ data and within \MM{5\unitns{\%}} for \HighTowerTwo\ data.
\label {syst_ht_prescale}
This is taken as the systematic uncertainties on the trigger scale factors.

\section {Vertex finding efficiency}

In the \pizero\ reconstruction it is assumed that the decay photons originate from the vertex.
It is therefore required that each event entering the analysis has a reconstructed vertex.
In the \protonproton\ dataset this requirement is always fulfilled, because we use the BBC 
timing information in case the TPC vertex reconstruction fails (this happens in about \MM{35\unitns{\%}} of the minimum bias events)\@.

In the \deuterongold\ HighTower data, the charged track multiplicities are large enough to always have a reconstructed TPC vertex.
However, a TPC vertex is missing in about \MM{7\unitns{\%}} of the minimum bias events and cannot be recovered from BBC information 
because the BBC is not included in the \deuterongold\ minimum bias trigger.
Minimum bias events without vertex have low charged track multiplicity, and the contribution from these very soft 
events to the \pizero\ yield above \MM{1\unit{\GeV}} is assumed to be negligible~\cite {ref_vertex_efficiency}\@.
\label {syst_vertex_eff}
Therefore, the correction for vertex inefficiency is applied as a constant normalization factor to the yield 
and its uncertainty contributes to the total normalization uncertainty of the measured cross sections.

The vertex efficiencies were determined to be \MM{0.93 \PLMN{} 0.01} from a full simulation of \deuterongold\ minimum bias events 
as described in~\cite {ref_vertex_efficiency}\@.
However, this efficiency depends on the centrality, and we assume that central events are \MM{100\unitns{\%}} efficient.
Scaling the above efficiency by the ratio of peripheral to total number of \deuterongold\ events, we obtain 
an efficiency correction factor of \MM{0.88 \PLMN{} 0.02} for the sample of peripheral events.

Note, that the difference between vertex finding efficiencies in MinBias and HighTower data is effectively absorbed
in the scale factor \SUB{\IT{K}}{\RM{trig}} defined in the previous section.
The vertex finding efficiency correction is, therefore, applied to the minimum bias data, as well as to the scaled HighTower trigger data.

\section {Residual beam background contamination}

The beam background contamination in the \deuterongold\ minimum bias trigger has been estimated from an 
analysis of the RHIC empty bunches to be \MM{5 \PLMN{} 1\unit{\%}}~\cite {ref_dAu_1}\@.
In our analysis, the beam background in \deuterongold\ events is rejected when the energy deposit in the calorimeter is
much larger than the total energy of all charged tracks reconstructed in the TPC, see Section~\ref {event_selection}\@.
To estimate the residual beam background in our data, we have analysed a sample of \MM{3\e{5}} minimum bias triggers from unpaired RHIC bunches.
These events were passed through the same analysis cuts and reconstruction procedure as the real data.
We observed that about \MM{10\unitns{\%}} of the fake triggers passed all cuts, and that none of these contained a reconstructed \pizero\@.
The residual beam background contamination is thus estimated to be \MM{0.1 \TIMES{} 5 = 0.5\unitns{\%}}, which is considered to be negligible.

In the \protonproton\ data the beam background contamination to the minimum bias trigger rate
is also estimated to be negligible due to the BBC coincidence requirement in the trigger and the cut on the BBC vertex position.

\section {Bin centering scale factors}

To assign a value of \pT\ to the yield measured in a \pT\ bin, the procedure from~\cite {ref_where_to_stick_your_data_points} was applied.
Here the measured yield, initially plotted at the bin centers, is approximated by a power law function of the form
\begin {equation} \label {eq:power_law}
\IT{f}(\pT) = \FRAC {\IT{A}} { \SUP{\ensuremath{\left(} 1 + \DIV{\pT}{\momentum_{0}} \ensuremath{\right)}}{\IT{n}}}.
\end {equation}
To each bin a momentum \SUP{\pT}{\tsp{0.7}*} was assigned as calculated from the equation
\begin {equation}
\IT{f}(\SUP{\pT}{\tsp{0.7}*}) = \FRAC{1}{\DELTA\pT} \SUB{\ensuremath{\int}}{\DELTA\pT} \tsp{-5.0} \IT{f}(\xcoord)\tsp{0.5}\der\xcoord.
\end {equation}
The function~(\ref {eq:power_law}) is then re\tsp{0.1}-fitted taking \SUP{\pT}{\tsp{0.7}*} as the abscissa.
This procedure was re\tsp{0.3}-iterated until the values of \SUP{\pT}{\tsp{0.7}*} were stable (typically after three iterations)\@.
Final fitted curves are shown in Figures~\ref {fig_pi0_invyield}~and~\ref {fig_eta_invyield}\@.

For convenience of comparing results from the various datasets,
the yields were scaled to the \pT\ bin centers by the ratio
\begin {equation}
\IT{K} = \FRAC {\IT{f}(\SUP{\pT}{\tsp{0.7}*})} {\IT{f}(\pT)},
\end {equation}
where \pT\ is the center of the bin.
The statistical and systematic errors were also scaled by the same factor.

\section {Jacobian correction}

All calculations in this analysis were performed in the defined pesudorapidity region \MM{0 \LESS{} \pseudorapidity{} \LESS{} 1} 
that corresponds to the rapidity region \MM{0 \LESS{} \rapidity{} \LESS{} \SUB{\rapidity}{0}}, 
where the rapidity limit \SUB{\rapidity}{0} is well approximated by pseudorapidity for a particle 
with momentum much larger than its mass.

\begin {figure} [tb]
\centerline {\hbox {
\includegraphics [width=\textwidth] {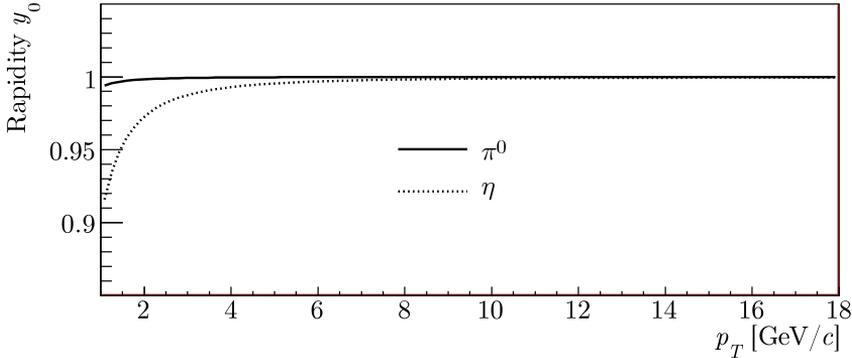}
}}
\caption {Jacobian correction that accounts for the rapidity limit \SUB{\rapidity}{0} being 
not equal to pseudorapidity \MM{\pseudorapidity = 1}\@.}
\label {fig_jacobian_corr}
\end {figure}

The correction was applied to account for the rapidity limit \SUB{\rapidity}{0} being not equal to pseudorapidity \MM{\pseudorapidity{} = 1}, 
as shown in Figure \ref {fig_jacobian_corr}\@.
This correction is smaller than \MM{10\unitns{\%}} for the \etameson\ data points at \MM{\pT{} \LESS{} 3\unit{\GeVc}}, 
and is negligible for the other data points.

\section {Fully corrected yields}
\label {corrected_yields}

\begin {figure} [p]
\centerline {\hbox {\includegraphics [width=0.9\textwidth] {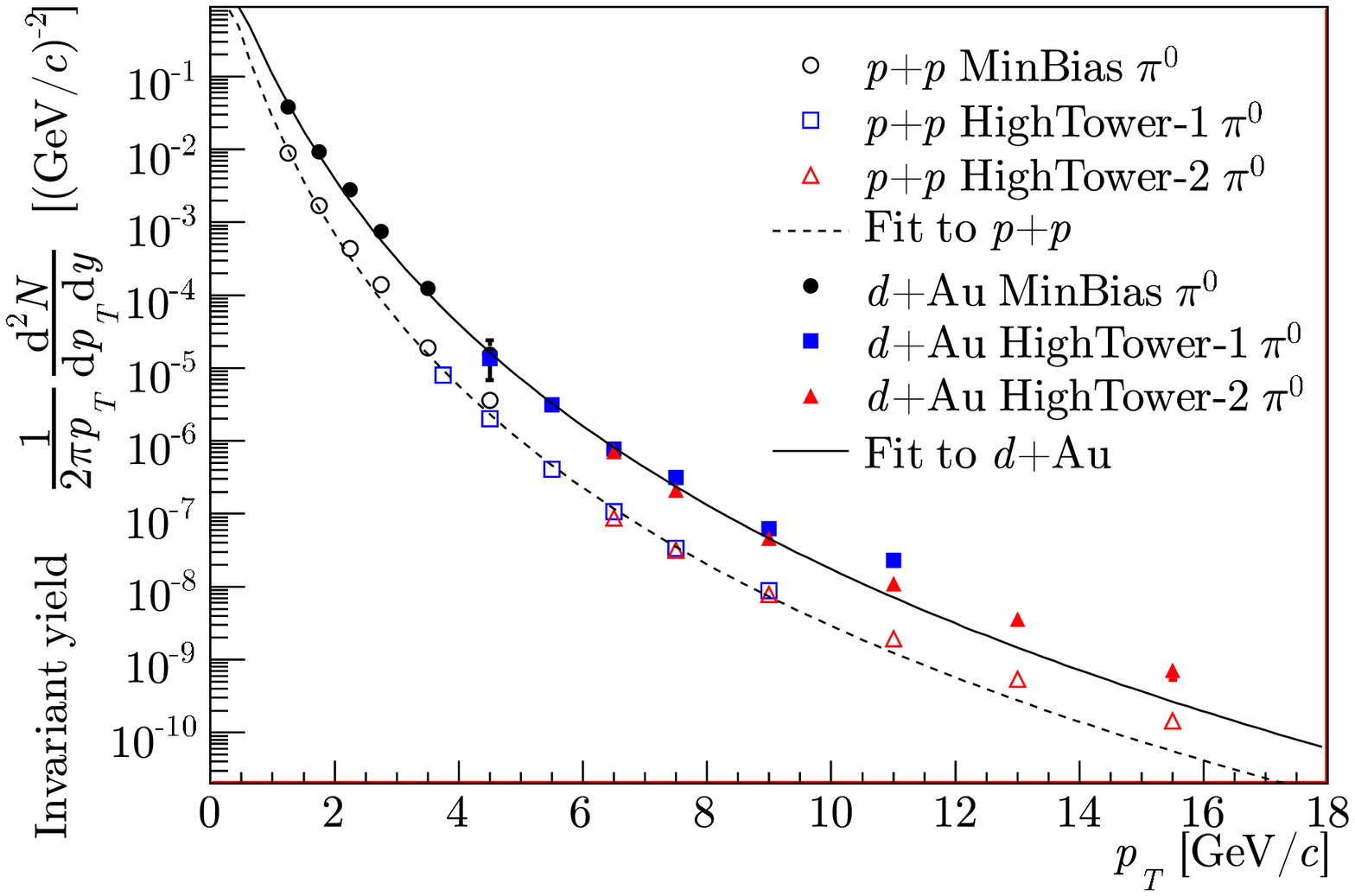}}}\centerline {\hbox {\includegraphics [width=0.9\textwidth] {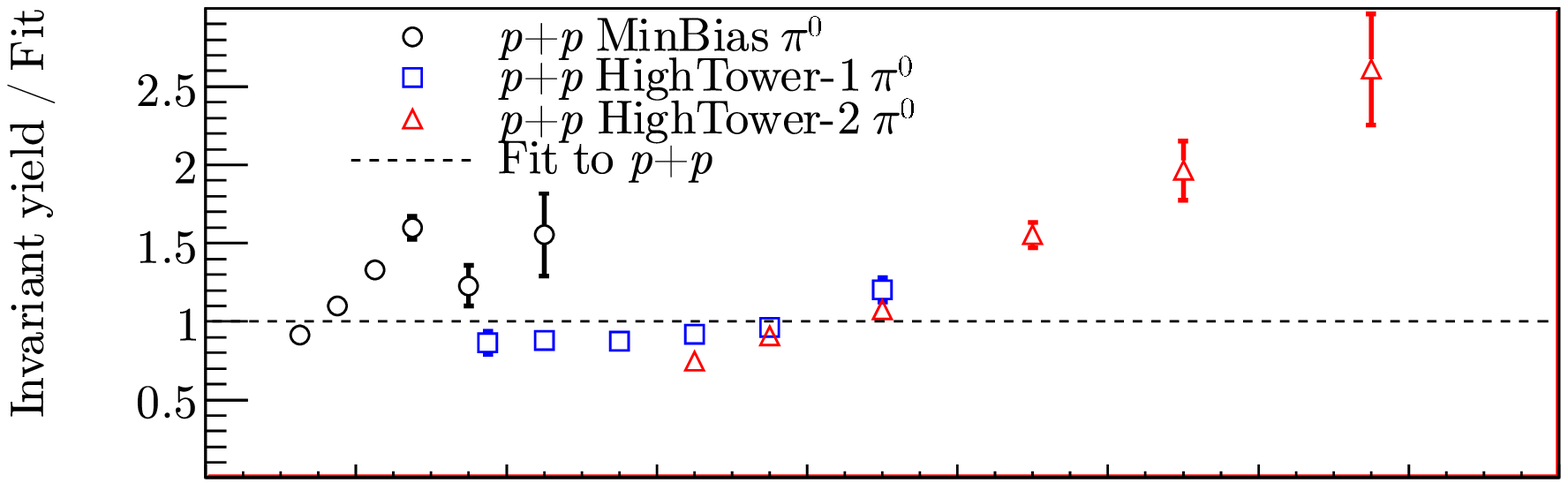}}}\centerline {\hbox {\includegraphics [width=0.9\textwidth,clip] {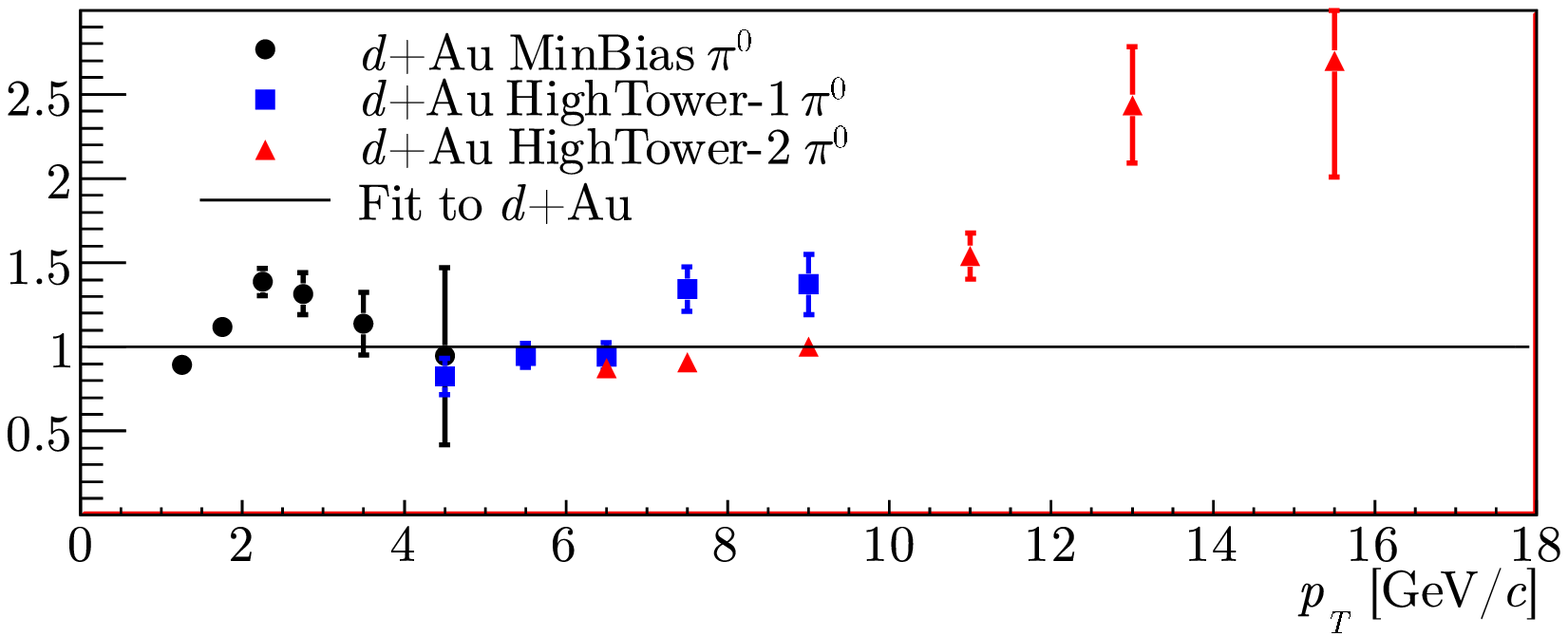}}}
\caption {Invariant yield of \pizero\ per minimum bias event in \protonproton\ and \deuterongold\ collisions (top)\@. 
Curves are the power law fits given in the text.
Invariant yield divided by the fit to the \protonproton\ (middle) and \deuterongold\ (bottom) data.
The errors shown are statistical only.
}
\label {fig_pi0_invyield}
\end {figure}

\begin {figure} [p]
\centerline {\hbox {\includegraphics [width=0.9\textwidth] {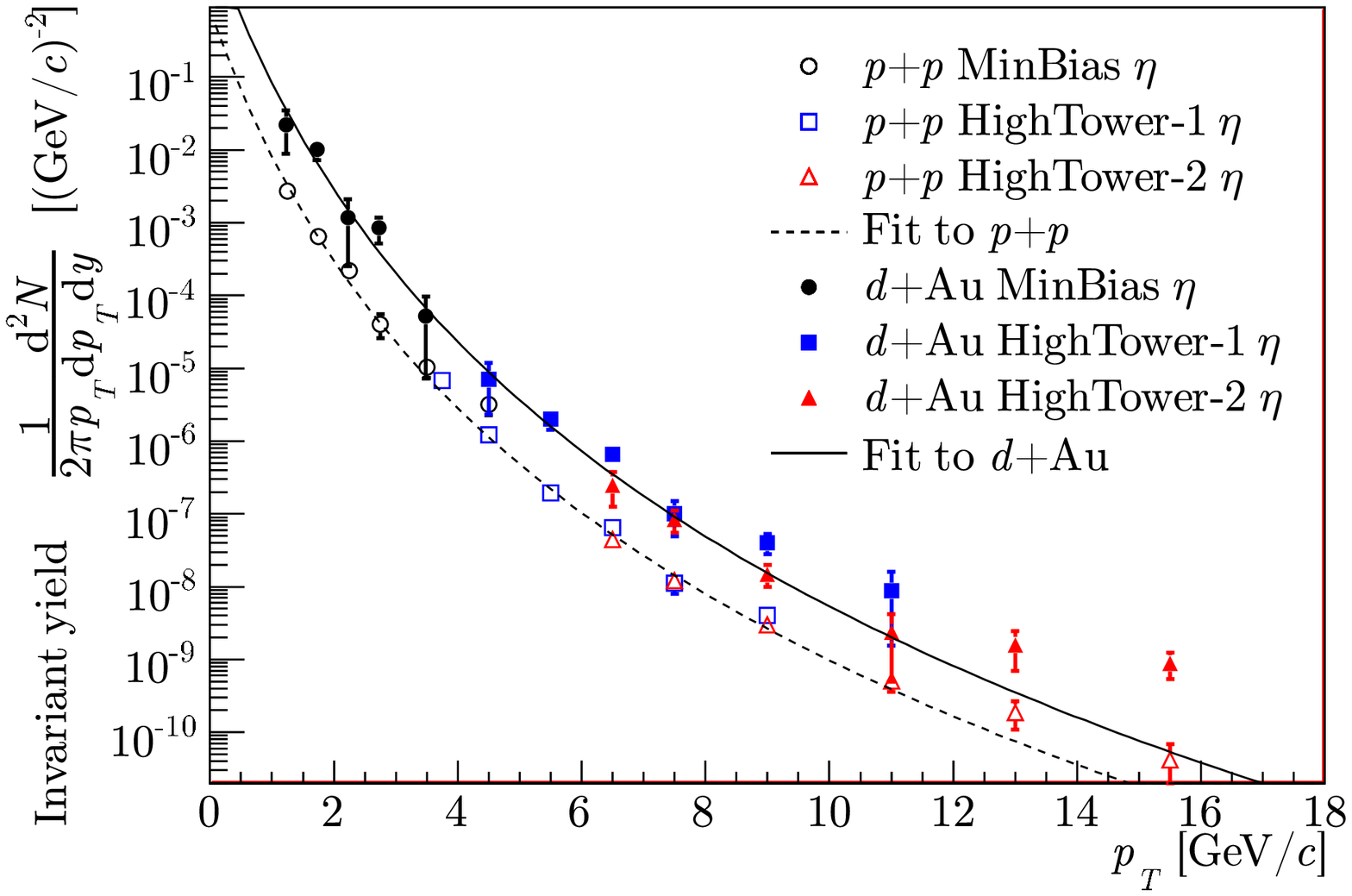}}}\centerline {\hbox {\includegraphics [width=0.9\textwidth] {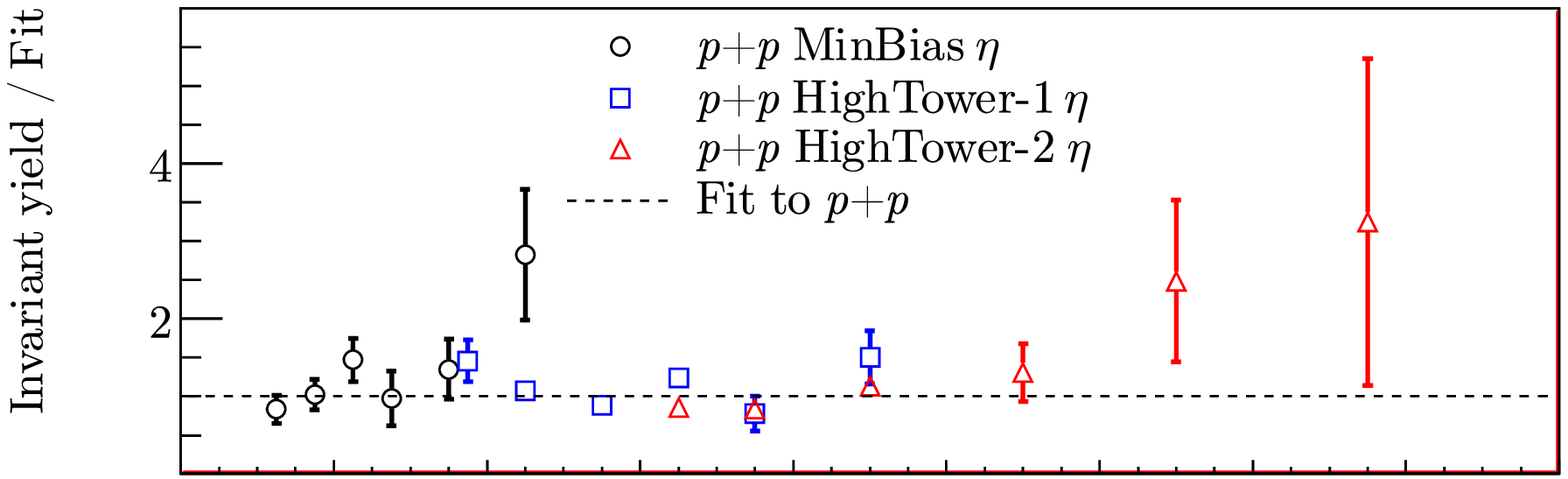}}}\centerline {\hbox {\includegraphics [width=0.9\textwidth,clip] {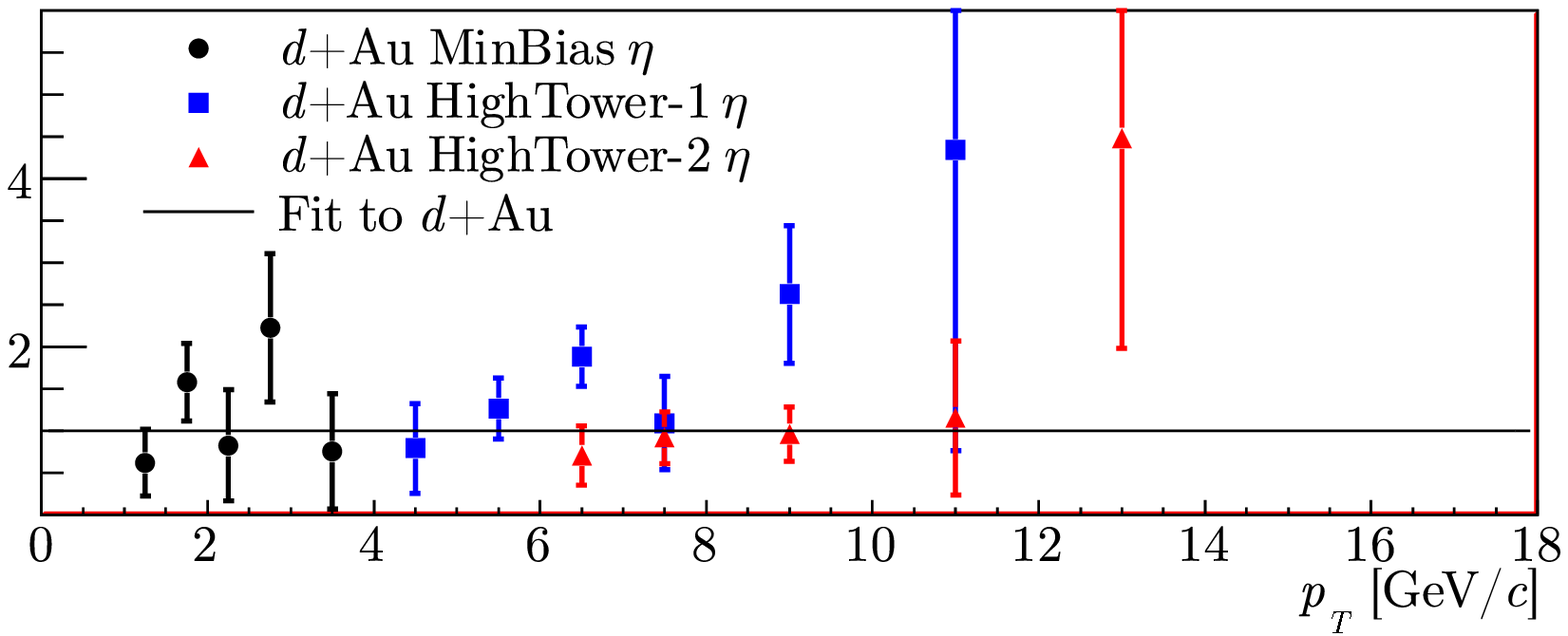}}}
\caption {Invariant yield of \etameson\ meson per minimum bias event in \protonproton\ and \deuterongold\ collisions (top)\@.
Invariant yield divided by the fit to the \protonproton\ (middle) and \deuterongold\ (bottom) data.
The errors shown are statistical only.}
\label {fig_eta_invyield}
\end {figure}

The fully corrected \pizero\ invariant yields per minimum bias event in \protonproton\ and \deuterongold\ collisions 
were calculated from Eq.~(\ref {eq:yielddef}) and
are shown in the top plots of Figure~\ref {fig_pi0_invyield} and Figure~\ref {fig_eta_invyield}\@.
The curves in these figures represent a fit of Eq.~(\ref {eq:power_law}) to the data.
In the bottom plots are shown the ratios between the data and the fit.
From these plots it is seen that the agreement between the datasets taken with the three triggers is satisfactory.


For the calculation of the final cross section results and cross section ratios, 
the data from three triggers were merged together and 
only one data point was chosen in each overlapping \pT\ bin. 
The \HighTowerOne\ points were preferred over MinBias, and \HighTowerTwo\ over \HighTowerOne, because at high \pT\ data samples 
are highly correlated, while HighTower datasets typically have smaller statistical error.

\label {syst_energy_scale}
The systematic uncertainty due to the calorimeter calibration was estimated from 
$$
\ensuremath{\delta} \IT{f}(\pT) = \ensuremath{\left|} \FRAC{\der\IT{f}}{\der\pT} \ensuremath{\right|} \ensuremath{\delta}\pT
,
$$
where \MM{\ensuremath{\delta}\pT} was taken to be \MM{5\unitns{\%}} in the \deuterongold\ and 
\protonproton\ data (see Section~\ref {energy_calibration}),
and where the derivative was calculated from the fitted function, Eq.~(\ref {eq:power_law})\@.
This \pT-dependent systematic uncertainty is, on average, \MM{38\unitns{\%}} in the \protonproton\ data 
and \MM{44\unitns{\%}} in the \deuterongold\ data.

\clearpage
All systematic error contributions mentioned in this and in the previous
sections are summarized in Table~\ref {table_systematic_uncertainty},
\begin {table} [b!]
\begin {center}
\caption {\normalsize Systematic error contributions. 
The classifications A, B, C and N are defined in the text. 
The error contributions to the cross section, the \etatopi\ ratio, \Rcp\ and \RdA\ 
are indicated in the respective columns.
The last column refers to the section where each source of systematic error is described.
}
\label {table_systematic_uncertainty}
\begin {tabular*} {\textwidth} {@{\extracolsep{\fill}}l@{\extracolsep{\fill}}c@{\extracolsep{\fill}}c@{\extracolsep{\fill}}c@{\extracolsep{\fill}}c@{\extracolsep{\fill}}c@{\extracolsep{\fill}}c@{\extracolsep{\fill}}}
Source                     & Type & \MM{\energy\ \DIV{\SUP{\der}{3}\SIGMA}{\der\SUP{\momentumbold}{3}}} & \etatopi & \Rcp & \RdA & Section \\
\hline
Combinatorial background   & A & + & + & + & + & \ref {syst_comb_bg} \\
Mixed-event background     & C & + & + &   &   & \ref {syst_mixed_bg} \\
Random vetoes              & N & + &   & + & + & \ref {syst_cpv} \\
HighTower scale factors    & B & + &   & + & + & \ref {syst_ht_prescale} \\
Analysis cuts              & A & + & + & + & + & \ref {syst_analysis_cuts} \\
Energy scale               & B & + &   &   & + & \ref {syst_energy_scale} \\
Vertex finding efficiency  & N & + &   & + & + & \ref {syst_vertex_eff} \\
Min.\ bias cross section    & N & + &   &   & + & \ref {syst_minbias_xsec} \\
Glauber model \Ncoll       & N &   &   & + & + & \ref {syst_glauber_model} \\
\end {tabular*}
\end {center}
\end {table}
classified into the following categories:
\label {syst_classification}
\begin {itemize}
\item {A} point-by-point systematic uncertainty;
\item {B} point-by-point \pT-correlated systematic uncertainty, but uncorrelated \linebreak between datasets;
\item {C} point-by-point \pT-correlated systematic uncertainty, also correlated \linebreak between datasets;
\item {N} normalization uncertainty, uncorrelated between datasets.
\end {itemize}

\chapter {Results and discussion}

\section {Cross section}

\label {cross_section_calculation}
The invariant differential cross section for \pizero\ and \etameson\ production 
in inelastic \protonproton\ interactions is given by
\begin {equation} \label {eq:crosssectiondef}
\energy \FRAC{\SUP{\der}{3}\SUP{\SUB{\SIGMA}{\RM{inel}}}{\protonproton}}{\der\SUP{\momentumbold}{3}} = 
\energy \FRAC{\SUP{\der}{3}\SUP{\SUB{\SIGMA}{\RM{NSD}}}{\protonproton}}{\der\SUP{\momentumbold}{3}} = 
\SUP{\SUB{\SIGMA}{\RM{NSD}}}{\protonproton} 
\tsp{1.0} 
\FRAC{\SUP{\der}{2}\Number}{2\PI\pT\tsp{0.5}\der\pT\tsp{0.5}\der\rapidity}.
\end {equation}
It has been shown that the singly diffractive (SD) contribution to the inelastic cross section 
is negligible at \MM{\pT{} \GREATER{} 1\unit{\GeVc}}~\cite {ref_star_idhadrons1}, so that
we can assume that the differential inelastic cross section is equal to the 
differential NSD cross section in our \pT\ range.
The total NSD cross section in \protonproton\ collisions was taken to be
\MM{\SUP{\SUB{\SIGMA}{\RM{NSD}}}{\protonproton} = 30.0 \PLMN{} 3.5\unit{\mb}}, 
as described in Section~\ref {section_triggers}\@.
The total hadronic cross section in \deuterongold\ collisions was taken to be
\MM{\SUP{\SUB{\SIGMA}{\RM{hadr}}}{\deuterongold} = 2.21 \PLMN{} 0.09\unit{\barn}}~\cite {ref_star_dAu_evidence}\@.

In tables~\ref {table_crossection_pi0_pp}
and~\ref {table_crossection_pi0_dAu}
we list the cross sections calculated from Eq.~(\ref {eq:crosssectiondef}) for the \protonproton\ and \deuterongold\ datasets.
In the third column of these tables are given the statistical errors, 
while in the remaining columns the quadratic sum of the systematic errors are given separately 
for each group defined in Table~\ref {table_systematic_uncertainty}\@.
In addition to these \pT-dependent systematic errors, the quadratic sum of the normalization uncertainties 
is found to be \MM{12.2\unitns{\%}} for the \protonproton\ and \MM{5.6\unitns{\%}} for the \deuterongold\ data.

\pagebreak

\begin {table} [p!]
\begin {center}
\caption {\normalsize Invariant cross section of \pizero\ production measured in \protonproton\ collisions.
Systematic errors classification given in Section~\ref {syst_classification}\@.
Normalization uncertainty of \MM{12.2\unitns{\%}} is not included.
}
\label {table_crossection_pi0_pp}
\small
\begin {tabular} {cr@{\TIMES}lr@{\TIMES}lr@{\TIMES}lr@{\TIMES}lr@{\TIMES}l}
\pT & \multicolumn{2}{c}{\MM{\energy\ \DIV{\SUP{\der}{3}\SIGMA}{\der\SUP{\momentumbold}{3}}}} & \multicolumn{2}{c}{Statistical} & \multicolumn{6}{c} {Systematic errors} \\
\MM{[\unitns{\GeVc}]} & \multicolumn{2}{c}{\MM{[\unitns{\mb}\SUP{\unit{\GeV}}{-2}\SUP{\unit{\cspeed}}{3}]}} & \multicolumn{2}{c}{error} & \multicolumn{2}{c}{A} & \multicolumn{2}{c}{B} & \multicolumn{2}{c}{C} \\
\hline
1.25\rule[9pt]{0pt}{0.6pt} & 2.646&\ee{-1}    & 5.973&\ee{-3}     & 1.336&\ee{-2}     & 9.835&\ee{-2}     & 1.323&\ee{-2}     \\ 
1.75    & 5.095&\ee{-2}    & 1.217&\ee{-3}     & 2.579&\ee{-3}     & 1.898&\ee{-2}     & 2.548&\ee{-3}     \\ 
2.25    & 1.314&\ee{-2}    & 4.349&\ee{-4}     & 6.669&\ee{-4}     & 4.902&\ee{-3}     & 6.569&\ee{-4}     \\ 
2.75    & 4.154&\ee{-3}    & 1.948&\ee{-4}     & 2.115&\ee{-4}     & 1.553&\ee{-3}     & 2.077&\ee{-4}     \\ 
3.50    & 5.724&\ee{-4}    & 6.062&\ee{-5}     & 2.928&\ee{-5}     & 2.146&\ee{-4}     & 2.862&\ee{-5}     \\ 
4.50    & 6.076&\ee{-5}    & 2.717&\ee{-6}     & 3.131&\ee{-6}     & 2.293&\ee{-5}     & 3.038&\ee{-6}     \\ 
5.50    & 1.223&\ee{-5}    & 4.403&\ee{-7}     & 6.353&\ee{-7}     & 4.632&\ee{-6}     & 6.113&\ee{-7}     \\ 
6.50    & 3.246&\ee{-6}    & 1.248&\ee{-7}     & 1.702&\ee{-7}     & 1.234&\ee{-6}     & 1.623&\ee{-7}     \\ 
7.50    & 9.592&\ee{-7}    & 2.638&\ee{-8}     & 5.081&\ee{-8}     & 3.681&\ee{-7}     & 4.796&\ee{-8}     \\ 
9.00    & 2.362&\ee{-7}    & 6.919&\ee{-9}     & 1.272&\ee{-8}     & 9.114&\ee{-8}     & 1.181&\ee{-8}     \\ 
11.00\ \   & 5.797&\ee{-8}    & 3.029&\ee{-9}     & 3.198&\ee{-9}     & 2.253&\ee{-8}     & 2.898&\ee{-9}     \\ 
13.00\ \   & 1.632&\ee{-8}    & 1.564&\ee{-9}     & 9.250&\ee{-10}     & 6.389&\ee{-9}     & 8.162&\ee{-10}     \\ 
15.50\ \   & 4.357&\ee{-9}    & 5.918&\ee{-10}     & 2.559&\ee{-10}     & 1.720&\ee{-9}     & 2.178&\ee{-10}     \\ 

\end {tabular}
\end {center}
\end {table}
\begin {table} [p!]
\begin {center}
\caption {\normalsize Invariant cross section of \pizero\ production measured in \deuterongold\ collisions.
Systematic errors classification given in Section~\ref {syst_classification}\@.
Normalization uncertainty of \MM{5.6\unitns{\%}} is not included.
}
\label {table_crossection_pi0_dAu}
\small
\begin {tabular} {cr@{\TIMES}lr@{\TIMES}lr@{\TIMES}lr@{\TIMES}lr@{\TIMES}l}
\pT & \multicolumn{2}{c}{\MM{\energy\ \DIV{\SUP{\der}{3}\SIGMA}{\der\SUP{\momentumbold}{3}}}} & \multicolumn{2}{c}{Statistical} & \multicolumn{6}{c}{Systematic errors} \\
\MM{[\unitns{\GeVc}]} & \multicolumn{2}{c}{\MM{[\unitns{\mb}\SUP{\unit{\GeV}}{-2}\SUP{\unit{\cspeed}}{3}]}} & \multicolumn{2}{c}{error} & \multicolumn{2}{c}{A} & \multicolumn{2}{c}{B} & \multicolumn{2}{c}{C} \\
\hline
1.25\rule[9pt]{0pt}{0.6pt} & 8.487&\ee{1}    & 3.716&\ee{0}     & 4.286&\ee{0}     & 3.133&\ee{1}     & 4.244&\ee{0}     \\ 
1.75    & 2.052&\ee{1}    & 9.744&\ee{-1}     & 1.039&\ee{0}     & 7.648&\ee{0}     & 1.026&\ee{0}     \\ 
2.25    & 6.116&\ee{0}    & 3.602&\ee{-1}     & 3.104&\ee{-1}     & 2.302&\ee{0}     & 3.058&\ee{-1}     \\ 
2.75    & 1.643&\ee{0}    & 1.556&\ee{-1}     & 8.365&\ee{-2}     & 6.245&\ee{-1}     & 8.215&\ee{-2}     \\ 
3.50    & 2.709&\ee{-1}    & 4.373&\ee{-2}     & 1.386&\ee{-2}     & 1.044&\ee{-1}     & 1.354&\ee{-2}     \\ 
4.50    & 3.000&\ee{-2}    & 3.898&\ee{-3}     & 1.546&\ee{-3}     & 1.182&\ee{-2}     & 1.500&\ee{-3}     \\ 
5.50    & 6.924&\ee{-3}    & 5.282&\ee{-4}     & 3.598&\ee{-4}     & 2.778&\ee{-3}     & 3.462&\ee{-4}     \\ 
6.50    & 1.573&\ee{-3}    & 1.002&\ee{-4}     & 8.249&\ee{-5}     & 6.457&\ee{-4}     & 7.864&\ee{-5}     \\ 
7.50    & 4.717&\ee{-4}    & 2.663&\ee{-5}     & 2.499&\ee{-5}     & 1.971&\ee{-4}     & 2.359&\ee{-5}     \\ 
9.00    & 1.014&\ee{-4}    & 5.453&\ee{-6}     & 5.462&\ee{-6}     & 4.347&\ee{-5}     & 5.071&\ee{-6}     \\ 
11.00\ \    & 2.439&\ee{-5}    & 2.172&\ee{-6}     & 1.346&\ee{-6}     & 1.081&\ee{-5}     & 1.220&\ee{-6}     \\ 
13.00\ \    & 7.896&\ee{-6}    & 1.118&\ee{-6}     & 4.475&\ee{-7}     & 3.614&\ee{-6}     & 3.948&\ee{-7}     \\ 
15.50\ \    & 1.585&\ee{-6}    & 4.056&\ee{-7}     & 9.311&\ee{-8}     & 7.542&\ee{-7}     & 7.926&\ee{-8}     \\ 

\end {tabular}
\end {center}
\end {table}

\clearpage

To parametrize the \pT\ dependence, the measured \pizero\ cross sections were fitted to
the power law function from Eq.~(\ref {eq:power_law}), resulting in the following \linebreak
parameters:
$$
\begin {array} {cccc}
\RM{Dataset}& \IT{A} & \SUB{\momentum}{0} & \IT{n} \\
              & [\unitns{\mb}\SUP{\unit{\GeV}}{-2}\SUP{\unit{\cspeed}}{3}] & [\unitns{\GeVc}] & \\
\hline
\protonproton\rule[10pt]{0pt}{0.6pt} & 7.53\e{2} & 0.95 & 9.31 \\
\ \:\:\deuterongold & 4.10\e{4} & 1.60 & 10.43\mbox{\ \,} \\
\end {array}
$$

\begin {figure} [p]
\centerline {\hbox {
\includegraphics [width=0.9\textwidth] {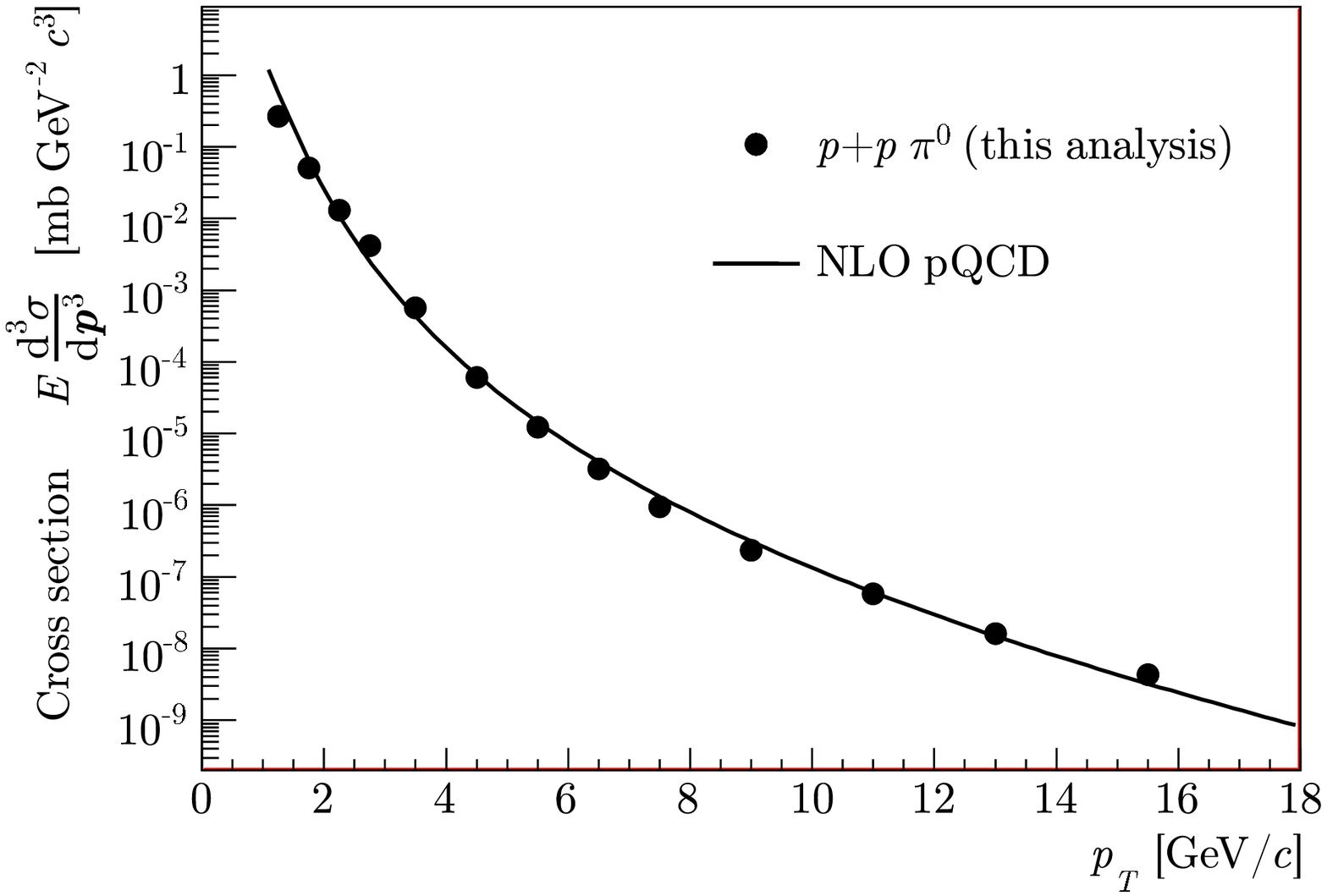}
}}
\centerline {\hbox {
\includegraphics [width=0.9\textwidth] {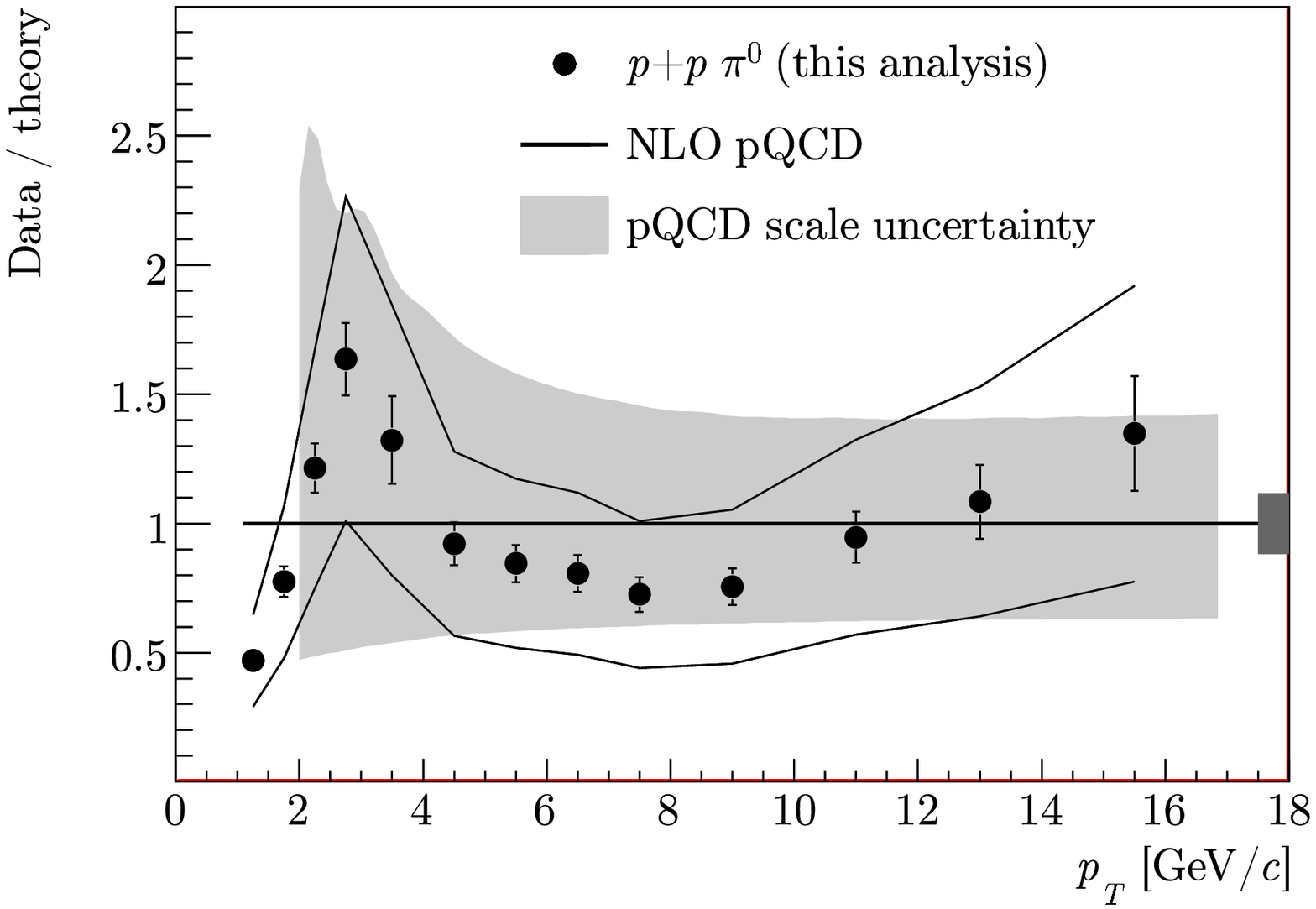}
}}
\caption {Cross section of the neutral pion production in \protonproton\ collisions (top), divided by the pQCD calculation (bottom)\@.
The errors are statistical and point\tsp{0.2}-to\tsp{0.4}-point systematic, 
excluding the energy calibration uncertainty shown as the outer lines (bottom)\@.
Normalization uncertainty is indicated by a shaded band around unity (bottom)\@.}
\label {fig_crossection_pp_theory}
\end {figure}

The measured cross sections for \pizero\ production in \protonproton\ collisions are shown in Figure~\ref {fig_crossection_pp_theory}, 
compared to the NLO pQCD calculation from Ref.~\cite {ref_phenix_pi0_pp}\@.
Input to this calculation are the CTEQ6M parton densities~\cite {ref_cteq6} and the KKP fragmentation functions~\cite {ref_kkp_ff}\@.
The factorization scale \ensuremath{\mu} was set equal to \pT\ and was varied by a factor of two to estimate the scale uncertainty, 
as indicated by the shaded band in the bottom plot of Figure~\ref {fig_crossection_pp_theory} 
that shows the ratio of the measured cross sections to the QCD prediction. 

The errors shown in the plot are the statistical and point\tsp{0.2}-to\tsp{0.4}-point systematic uncertainties added in quadrature,
excluding the uncertainty due to the energy calibration of the calorimeter. 
This additional uncertainty is shown by the outer lines around the data points on the lower plot.
The normalization uncertainty is indicated by shaded band around unity on the right hand side of the plot.

\begin {figure} [p]
\centerline {\hbox {
\includegraphics [width=0.9\textwidth] {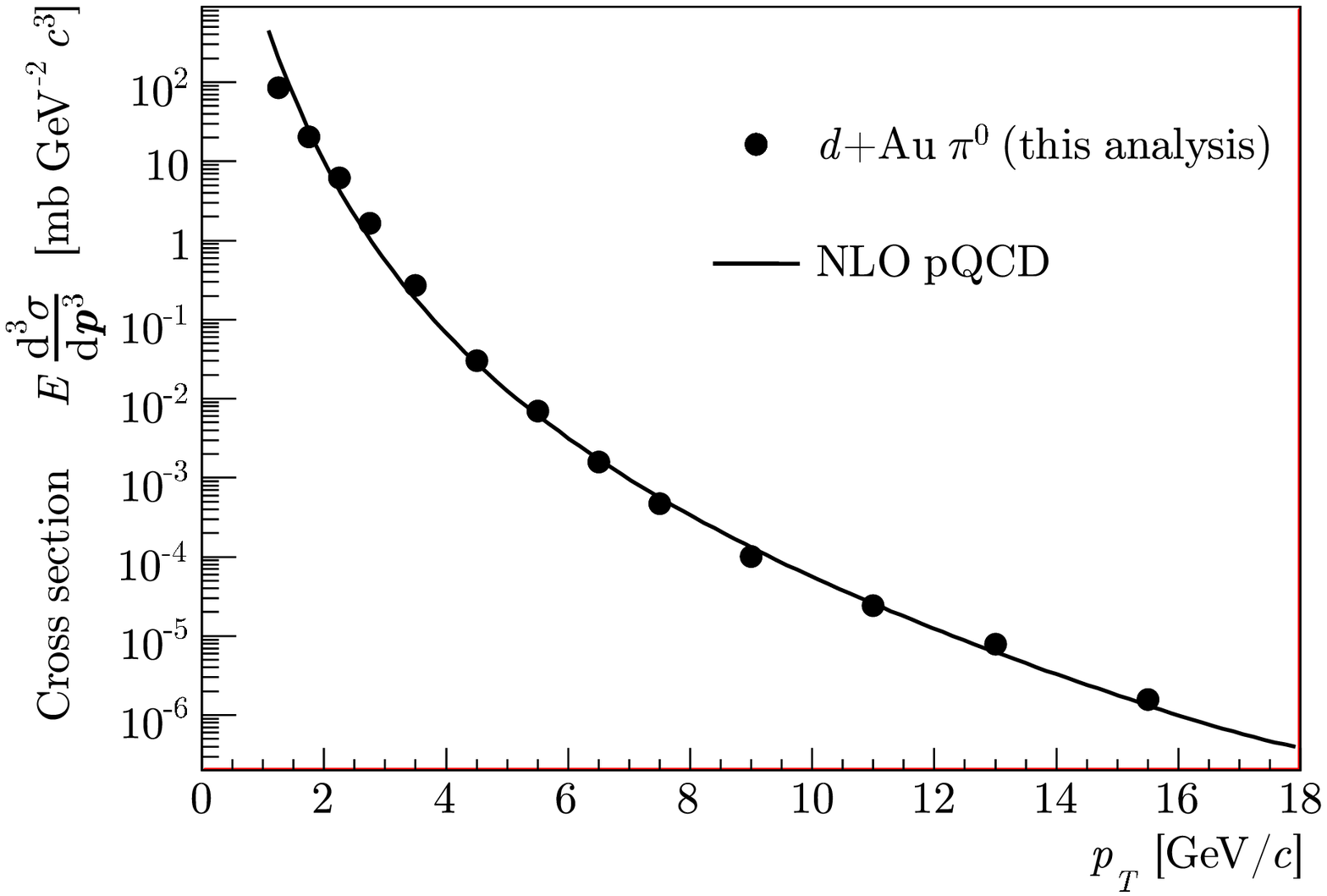}
}}
\centerline {\hbox {
\includegraphics [width=0.9\textwidth] {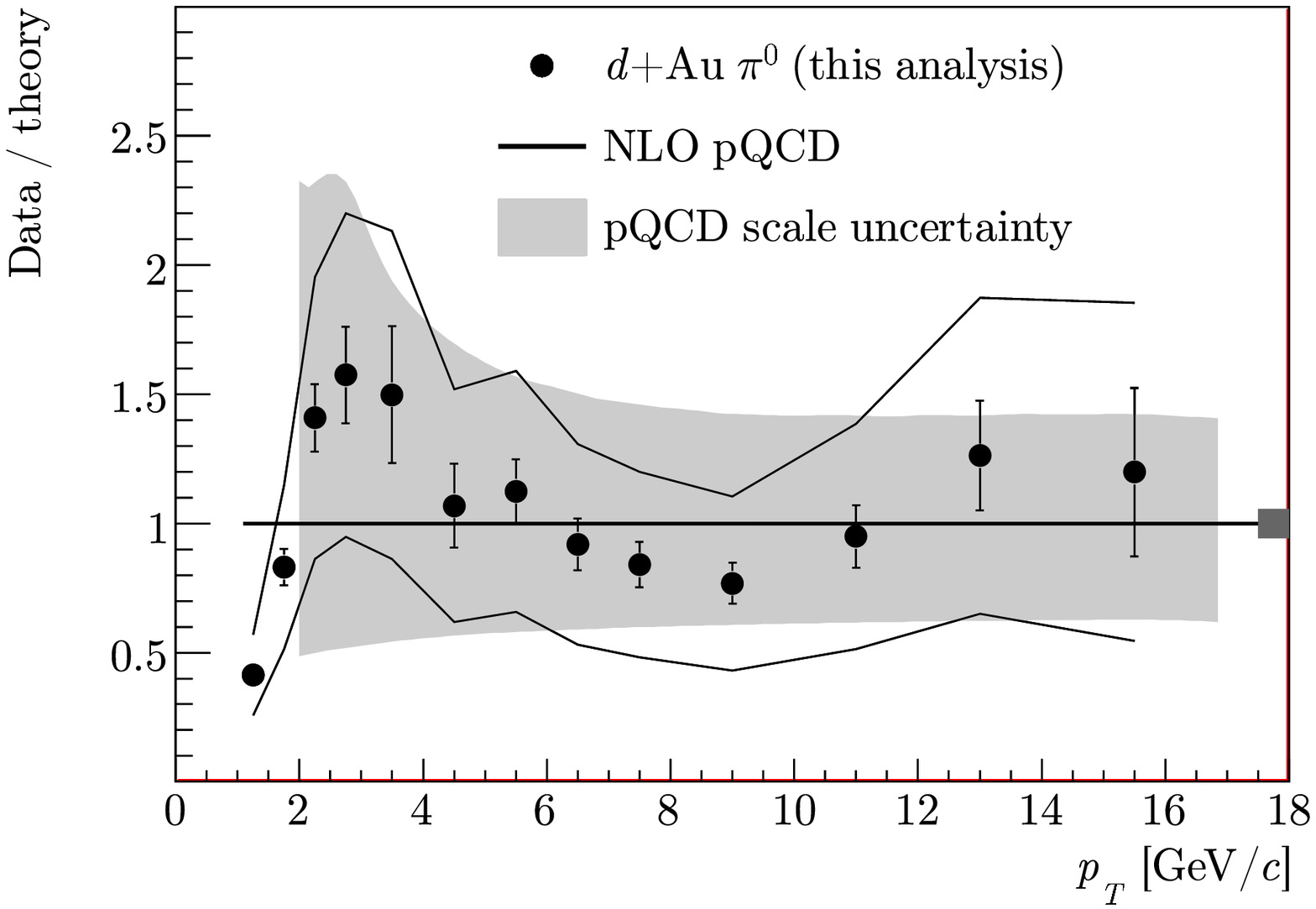}
}}
\caption {Cross section of the neutral pion production in \deuterongold\ collisions (top), divided by the pQCD calculation (bottom)\@.
The errors are statistical and point\tsp{0.2}-to\tsp{0.4}-point systematic, 
excluding the energy calibration uncertainty shown as the outer lines (bottom)\@.
Normalization uncertainty is indicated by a shaded band around unity (bottom)\@.}
\label {fig_crossection_dAu_theory}
\end {figure}

The \pizero\ cross section measured in \deuterongold\ collisions is shown in Figure~\ref {fig_crossection_dAu_theory} 
and compared to the NLO pQCD calculations of~\cite {ref_vogelsang_private}\@.
Here were used the KKP fragmentation functions, the CTEQ6M parton distributions for deuterium 
and the nuclear parton distributions for \gold~\cite {ref_au_pdf_1,ref_au_pdf_2,ref_au_pdf_3}\@. 
The errors shown in the plot are defined in the same way as in Figure~\ref {fig_crossection_pp_theory} for \protonproton\@.

It is seen that the measured \pizero\ cross section in both the \protonproton\ and \deuterongold\ collisions
is well described by the pQCD calculations in the full \pT\ range.
The possible excess relative to the theory seen in the \deuterongold\ data at low \pT\ may be 
an indication of the Cronin effect, which was not included in the pQCD calculations.

\begin {figure} [p]
\centerline {\hbox {
\includegraphics [width=0.9\textwidth] {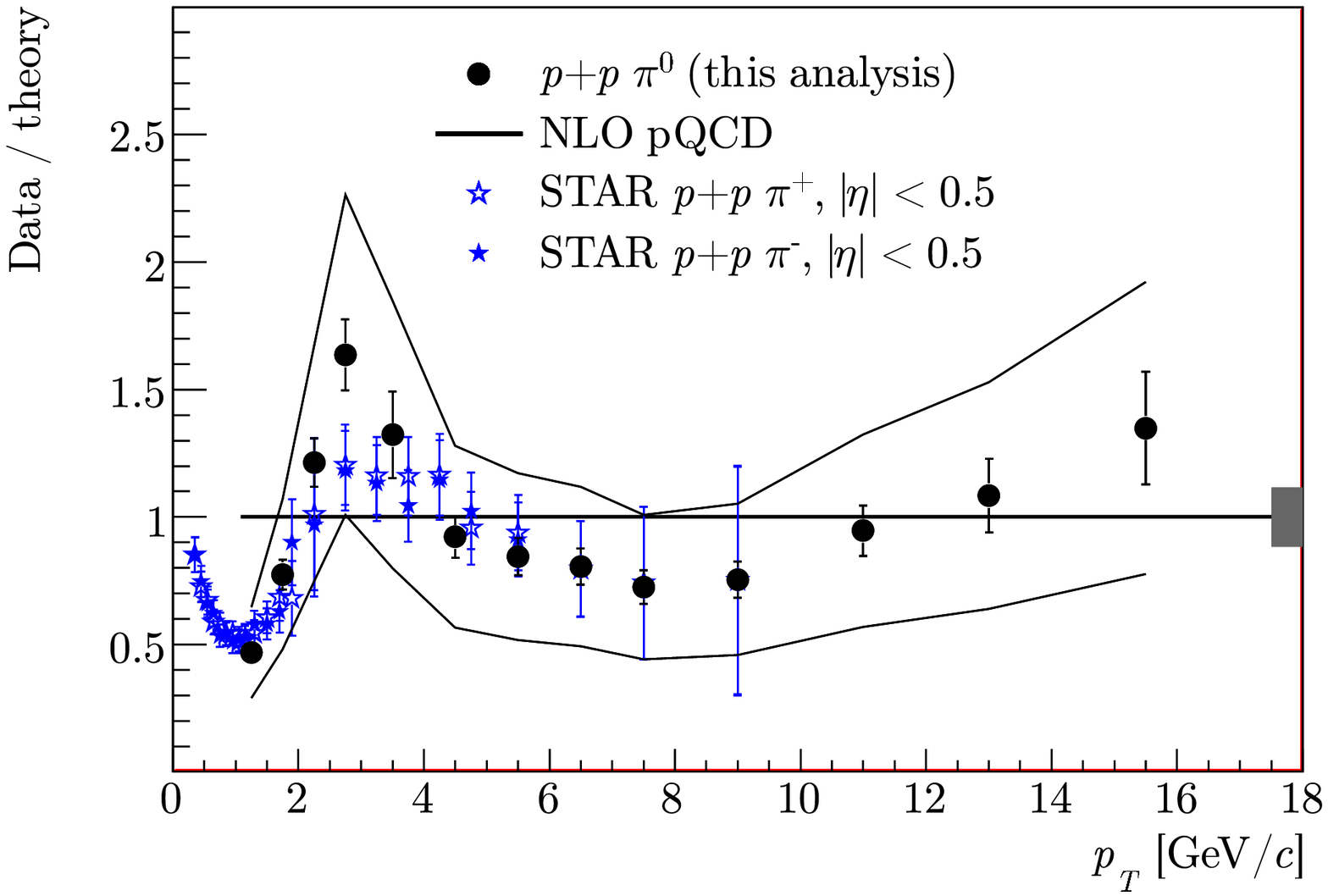}
}}
\centerline {\hbox {
\includegraphics [width=0.9\textwidth] {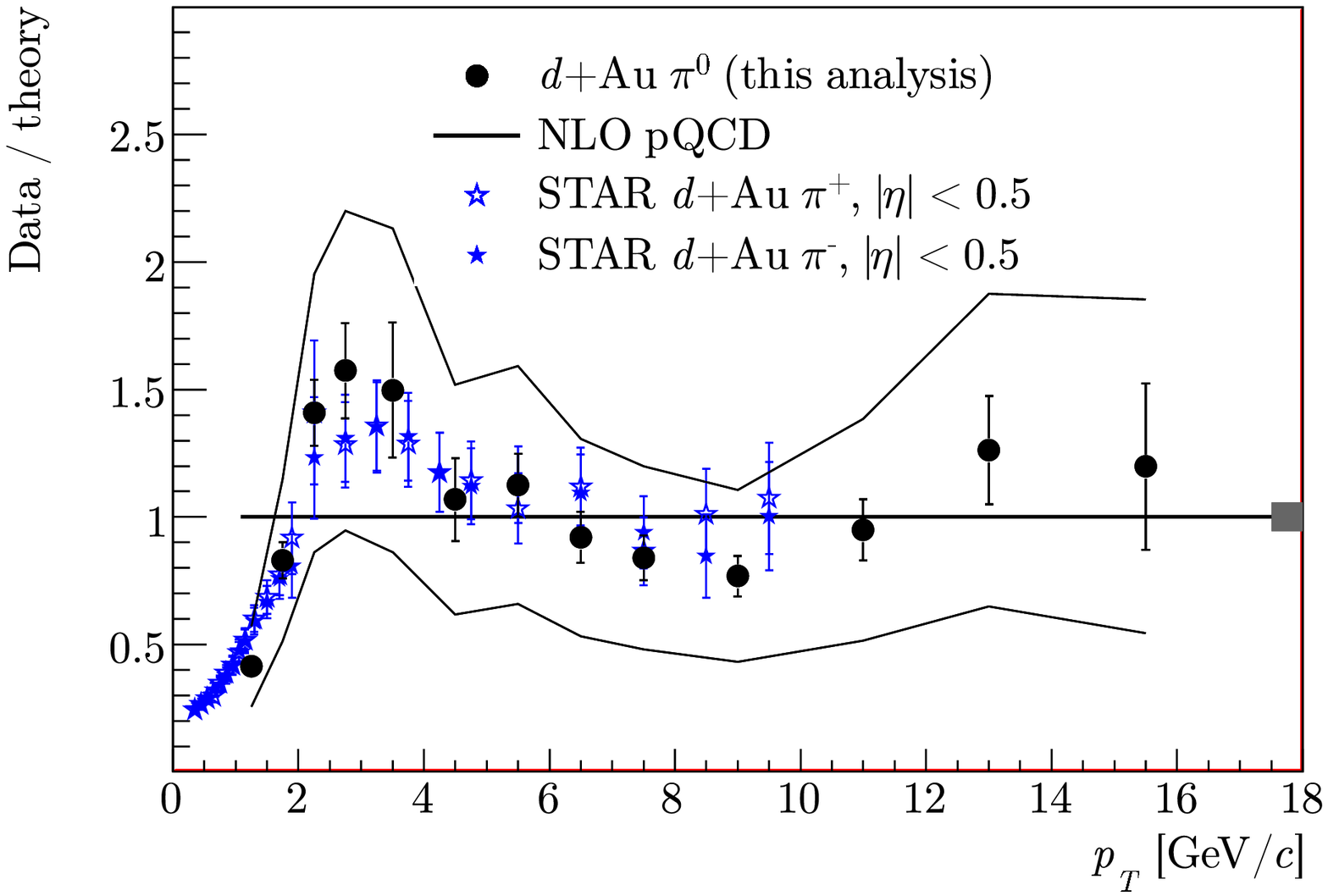}
}}
\caption {Cross section of the neutral pion production in \protonproton\ (top) and \deuterongold\ (bottom) collisions divided by the pQCD calculation,
compared to the STAR \piplusminus~\protect\cite {ref_star_idhadrons}\@.
The errors are statistical and point\tsp{0.2}-to\tsp{0.4}-point systematic, 
excluding the calorimeter energy calibration uncertainty shown as the outer lines.
Common normalization uncertainty is indicated by a shaded band around unity.}
\label {fig_crossection_star}
\end {figure}

In Figure~\ref {fig_crossection_star} 
we compare the \pizero\ measurement in \protonproton\ and \deuterongold\ with 
previous measurements of charged pions by STAR~\cite {ref_star_idhadrons}\@.
For ease of comparison, the \piplusminus\ data points are divided by the \pizero\ pQCD curves.
Note, that the normalization uncertainty shown by the grey bands in the figure 
are largely correlated between the \pizero\ and the \piplusminus\ data points.
It is seen that the neutral and charged pion spectra agree with each other very well 
in both \protonproton\ and \deuterongold\ datasets.

\begin {figure} [p]
\centerline {\hbox {
\includegraphics [width=0.9\textwidth] {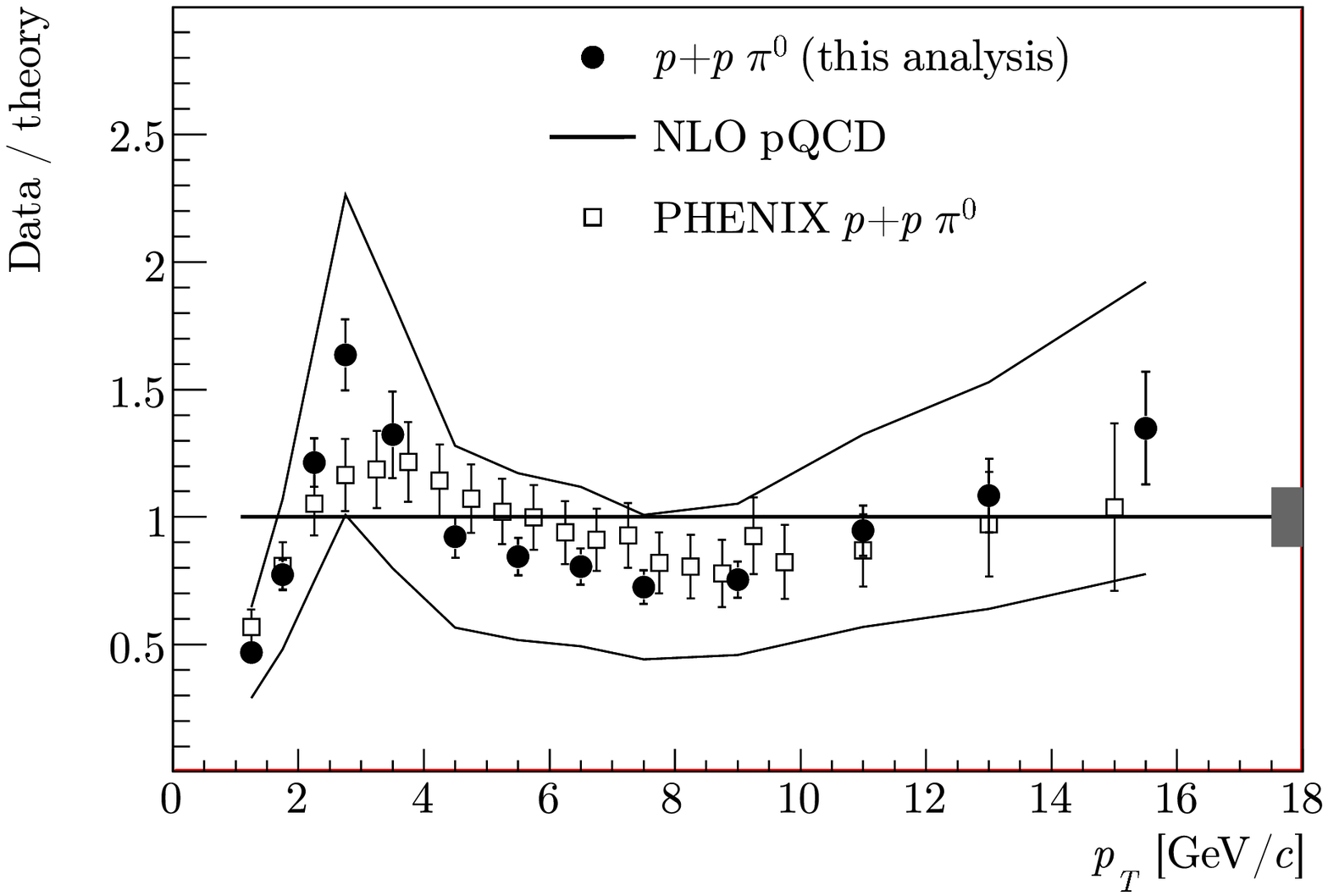}
}}
\centerline {\hbox {
\includegraphics [width=0.9\textwidth] {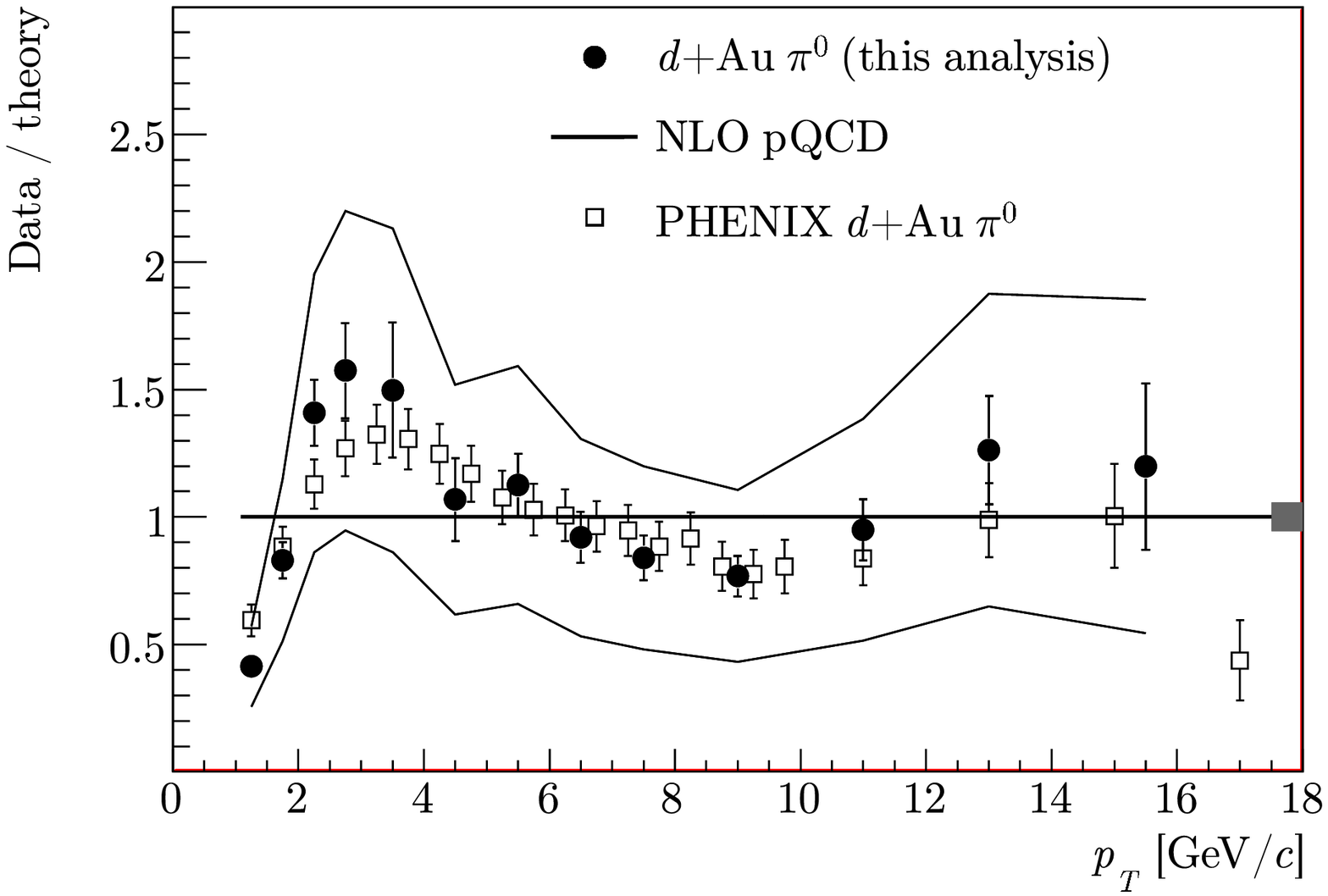}
}}
\caption {Cross section of the neutral pion production in \protonproton\ (top) and \deuterongold\ (bottom) collisions divided by the pQCD calculation,
compared to the PHENIX \pizero\ measurements in \protonproton~\protect\cite {ref_phenix_pi0_pp} and \deuterongold~\protect\cite {ref_phenix_pi0_dAu}\@.
The errors are statistical and point\tsp{0.2}-to\tsp{0.4}-point systematic, 
excluding the energy calibration uncertainty shown as the outer lines.
Normalization uncertainty is indicated by a shaded band around unity.}
\label {fig_crossection_phenix}
\end {figure}

In Figure~\ref {fig_crossection_phenix} 
we compare the present \pizero\ measurements with the neutral pion results from PHENIX~\cite {ref_phenix_pi0_pp,ref_phenix_pi0_dAu}\@.
This comparison indicates a good agreement, within errors, between the results of the two experiments.

\section {Eta to pion ratio}

\begin {figure} [p]
\centerline {\hbox {
\includegraphics [width=\textwidth] {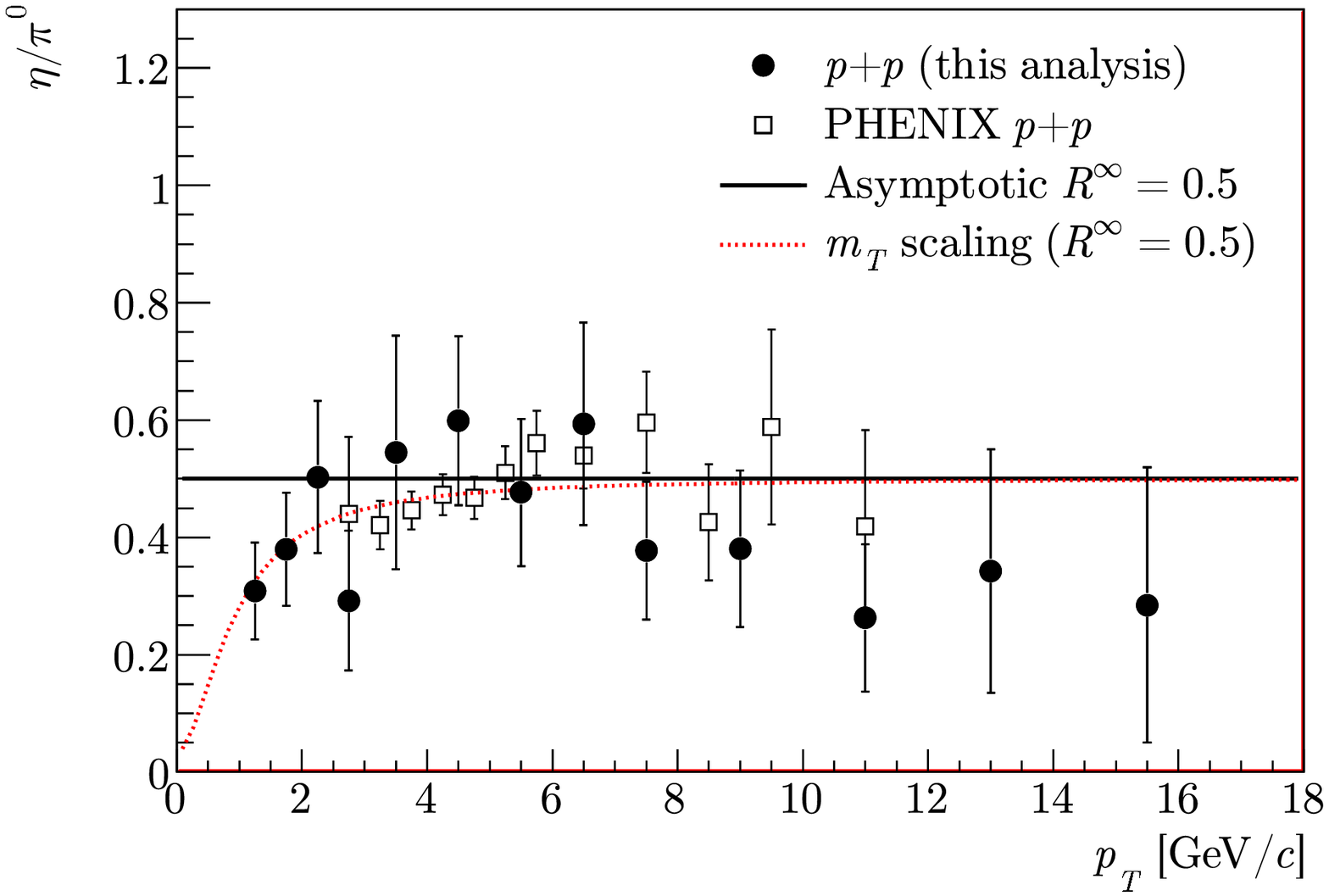}
}}
\centerline {\hbox {
\includegraphics [width=\textwidth] {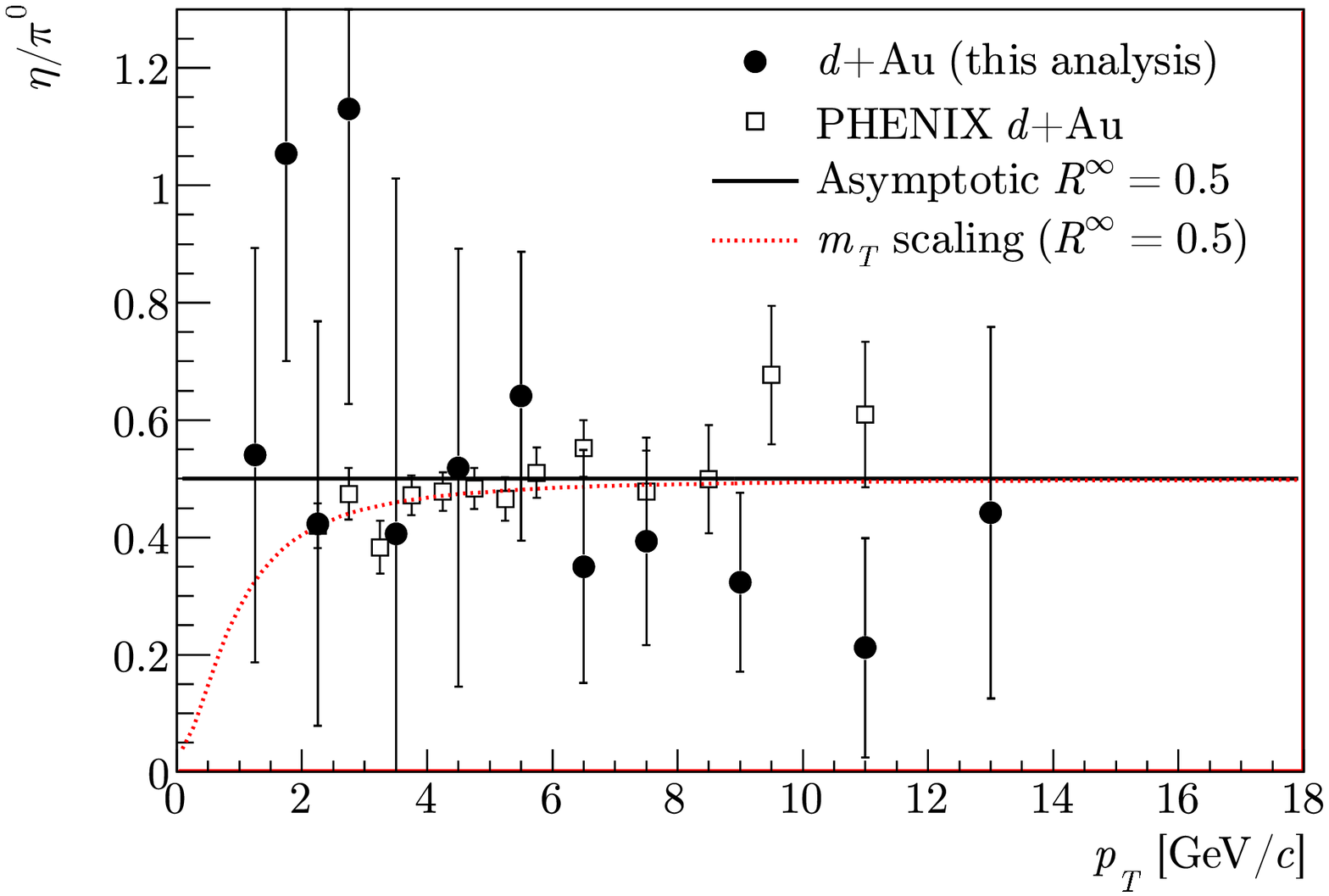}
}}
\caption {\etatopi\ ratio measured in \protonproton\ (top) and \deuterongold\ (bottom) collisions,
compared to the PHENIX measurements~\protect\cite {ref_phenix_eta}\@.
Errors are statistical and systematic combined.}
\label {fig_etatopi}
\end {figure}

The \etameson\ measurement is presented here as the ratio of \etameson\ to \pizero\ invariant yields,
which allows many systematic uncertainties to cancel, see Table~\ref {table_systematic_uncertainty} in Section~\ref {corrected_yields}\@.
The \etatopi\ ratios measured in \protonproton\ and \deuterongold\ collisions 
are listed in Tables~\ref {table_etatopi0_pp} 
\begin {table} [tbp]
\begin {center}
\caption {\normalsize \etatopi\ ratio measured in \protonproton\ collisions.
Systematic errors classification given in Section~\ref {syst_classification}\@.
}
\label {table_etatopi0_pp}
\begin {tabular} {ccccc}
\pT & \etatopi & Statistical & \multicolumn{2}{c}{Systematic errors} \\
\MM{[\unitns{\GeVc}]} & & error & A & C \\
\hline
1.25    & 0.308    & 0.069     & 0.045     & 0.012     \\ 
1.75    & 0.380    & 0.075     & 0.059     & 0.015     \\ 
2.25    & 0.503    & 0.097     & 0.084     & 0.020     \\ 
2.75    & 0.292    & 0.107     & 0.052     & 0.012     \\ 
3.50    & 0.545    & 0.166     & 0.107     & 0.022     \\ 
4.50    & 0.599    & 0.054     & 0.131     & 0.024     \\ 
5.50    & 0.477    & 0.045     & 0.116     & 0.019     \\ 
6.50    & 0.593    & 0.064     & 0.158     & 0.024     \\ 
7.50    & 0.378    & 0.038     & 0.110     & 0.015     \\ 
9.00    & 0.381    & 0.044     & 0.125     & 0.015     \\ 
11.00\ \   & 0.263    & 0.076     & 0.099     & 0.011     \\ 
13.00\ \   & 0.343    & 0.147     & 0.146     & 0.014     \\ 
15.50\ \   & 0.285    & 0.188     & 0.139     & 0.011     \\ 

\end {tabular}
\end {center}
\end {table}
and~\ref {table_etatopi0_dAu} 
\begin {table} [tbp]
\begin {center}
\caption {\normalsize \etatopi~\mbox{ratio}~\mbox{measured}~in~\mbox{\deuterongold}~\mbox{collisions}.~\mbox{Systematic}~\mbox{errors}~\mbox{classification} \protect\linebreak
given in Section~\ref {syst_classification}\@.
}
\label {table_etatopi0_dAu}
\begin {tabular} {ccccc}
\pT & \etatopi & Statistical & \multicolumn{2}{c}{Systematic errors} \\
\MM{[\unitns{\GeVc}]} & & error & A & C \\
\hline
1.25    & 0.540    & 0.344     & 0.078     & 0.022     \\ 
1.75    & 1.054    & 0.310     & 0.164     & 0.042     \\ 
2.25    & 0.423    & 0.337     & 0.071     & 0.017     \\ 
2.75    & 1.130    & 0.459     & 0.202     & 0.045     \\ 
3.50    & 0.406    & 0.601     & 0.079     & 0.016     \\ 
4.50    & 0.519    & 0.355     & 0.114     & 0.021     \\ 
5.50    & 0.641    & 0.189     & 0.156     & 0.026     \\ 
6.50    & 0.350    & 0.175     & 0.093     & 0.014     \\ 
7.50    & 0.393    & 0.134     & 0.114     & 0.016     \\ 
9.00    & 0.323    & 0.109     & 0.106     & 0.013     \\ 
11.00\ \   & 0.212    & 0.168     & 0.080     & 0.008     \\ 
13.00\ \   & 0.442    & 0.254     & 0.188     & 0.018     \\ 

\end {tabular}
\end {center}
\end {table}
and are shown in Figure~\ref {fig_etatopi}\@.
The error definitions in the tables and in the plot are the same as described above for the differential cross sections.
The present measurement agrees very well with previous PHENIX \linebreak
results~\cite {ref_phenix_eta}, as shown by the open squares in the plot.
The full curves in Figure~\ref {fig_etatopi} show the asymptotic ratio \MM{\SUB{\SUP{\IT{R}}{\ensuremath{\infty}}}{\etatopi} = 0.5} 
consistent with the world \etatopi\ measurements.
The constant fit to our data at \MM{\pT{} \GREATER{} 4\unit{\GeVc}} gives 
\MM{\SUB{\SUP{\IT{R}}{\protonproton}}{\etatopi} = 0.42 \PLMN{} 0.05} 
and 
\MM{\SUB{\SUP{\IT{R}}{\deuterongold}}{\etatopi} = 0.37 \PLMN{} 0.08}\@. 
The dotted curves in Figure~\ref {fig_etatopi} show the prediction based on empirical \mT-scaling observation~\cite {ref_mT_scaling} 
that the hadron production cross sections have the same shape as a function 
of the transverse mass of the produced particle \MM{\mT{} = \SQRT{\SUP{\SUB{\mass}{\mbox{\rule{0pt}{4pt}}}}{2} + \SUP{\pT}{2}}}\@. 
It is seen that the data are consistent with such scaling behaviour.

\section {Nuclear modification factor}

We calculate the \RdA\ ratio, defined by Eq.~(\ref {eq:rabdef}) and (\ref {eq:tabdef}), as
\begin {equation} \label {eq:rdadef}
\RdA{} = 
\FRAC 
    {\SUB{\SUP{\SIGMA}{\nucleon\nucleon}}{\RM{inel}} \  \DIV{\SUP{\der}{2}\SUB{\Number}{dA}}{\der\pT\tsp{0.5}\der\rapidity}} 
    {\Ncollmean \  \DIV{\SUP{\der}{2}\SUP{\SIGMA}{\protonproton}\ensuremath{\!}}{\der\pT\tsp{0.5}\der\rapidity}}
,
\end {equation}
where the nucleon-nucleon inelastic cross section \MM{\SUB{\SUP{\SIGMA}{\nucleon\nucleon}}{\RM{inel}}} 
is taken to be \MM{42\unit{\mb}} 
and \MM{\Ncollmean{} = 7.5 \PLMN{} 0.4} is calculated from the Glauber model, as described in Section~\ref {section_centrality_selection}\@.

\begin {figure} [p]
\centerline {\hbox {
\includegraphics [width=0.9\textwidth] {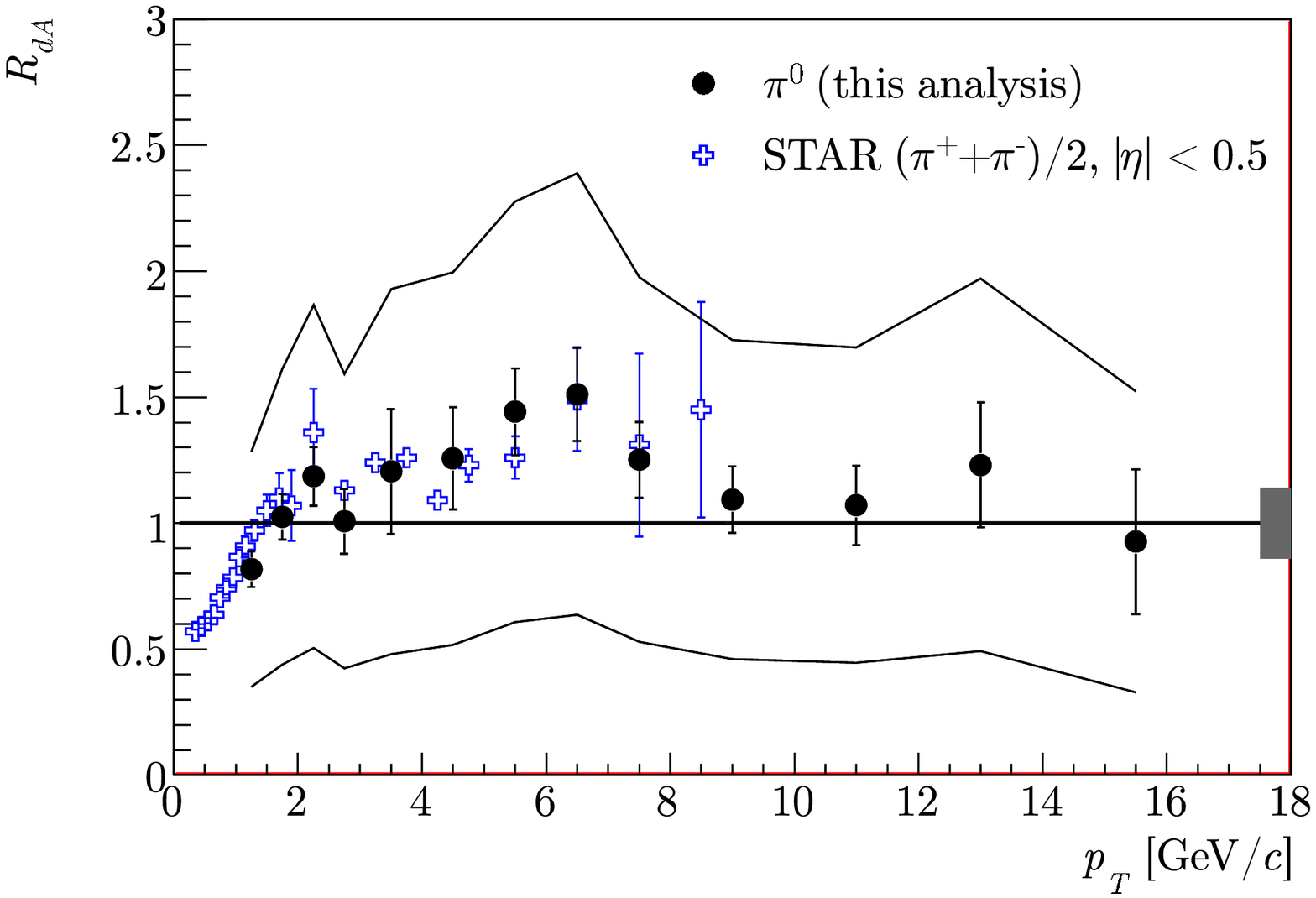}
}}
\centerline {\hbox {
\includegraphics [width=0.9\textwidth] {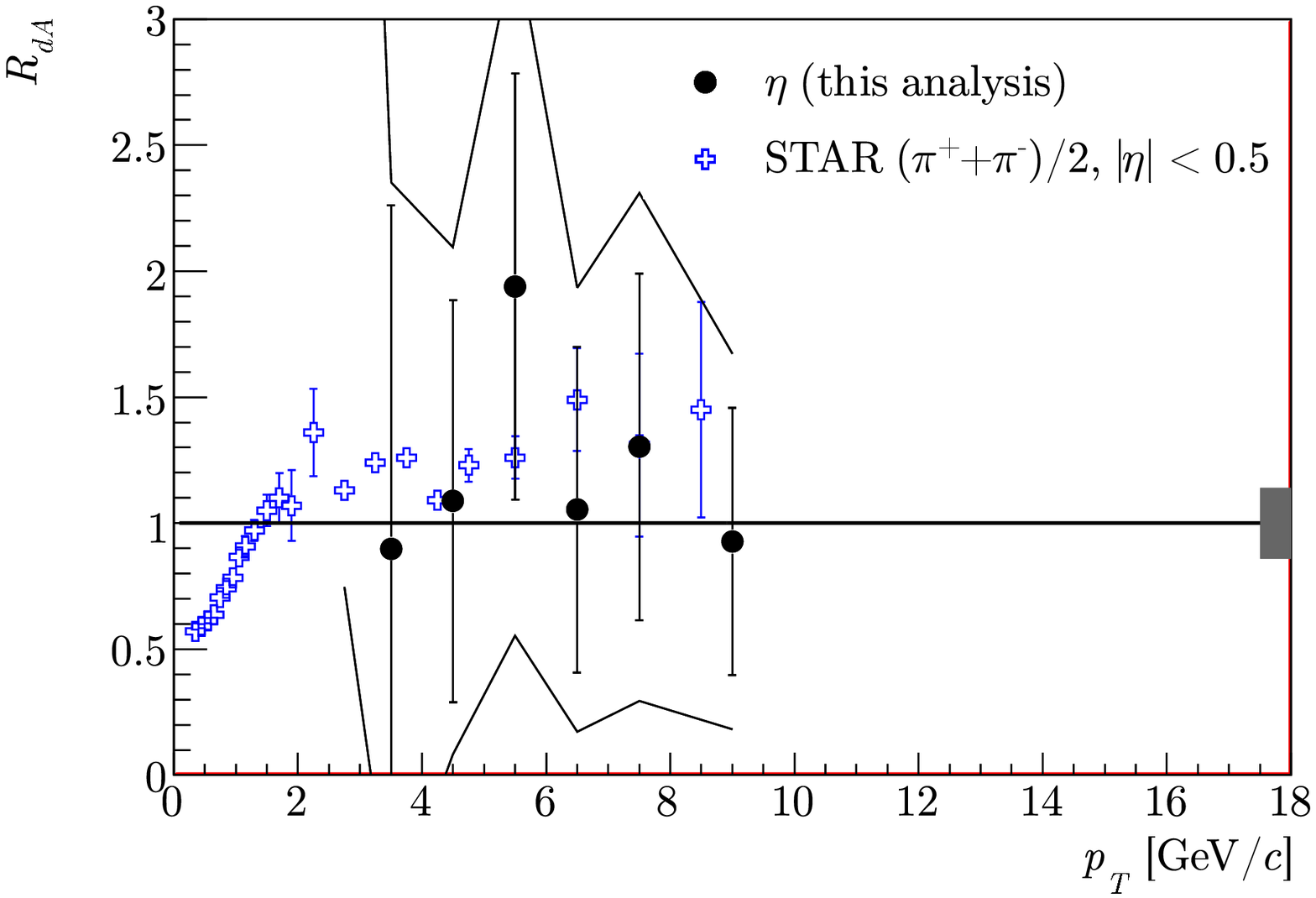}
}}
\caption {\RdA\ ratio for \pizero\ (top) and \etameson\ meson (bottom),
compared to the STAR \piplusminus~\protect\cite {ref_star_idhadrons}\@.
Errors are statistical and point\tsp{0.2}-to\tsp{0.4}-point systematic, 
excluding the calorimeter energy calibration uncertainty shown as the outer lines.
Common normalization uncertainty is indicated by a shaded band around unity.}
\label {fig_rda_star}
\end {figure}

The nuclear modification factors for \pizero\ and \etameson\ are listed in Tables~\ref {table_pi0_RdA} 
\begin {table} [tbp]
\begin {center}
\caption {\normalsize \mbox{Nuclear}~\mbox{modification}~\mbox{factor}~\RdA~for~\pizero.\ \protect\nolinebreak
\mbox{Systematic}~\mbox{errors}~\mbox{classification} \protect\linebreak
given in Section~\ref {syst_classification}\@.
Normalization uncertainty of \MM{14.5\unitns{\%}} is not included.
}
\label {table_pi0_RdA}
\begin {tabular} {ccccc}
\pT & \RdA & Statistical & \multicolumn{2}{c}{Systematic errors} \\
\MM{[\unitns{\GeVc}]} & & error & A & B \\
\hline
1.25    & 0.817    & 0.040     & 0.058     & 0.462     \\ 
1.75    & 1.025    & 0.055     & 0.073     & 0.580     \\ 
2.25    & 1.185    & 0.080     & 0.085     & 0.670     \\ 
2.75    & 1.007    & 0.106     & 0.072     & 0.570     \\ 
3.50    & 1.205    & 0.233     & 0.087     & 0.681     \\ 
4.50    & 1.257    & 0.173     & 0.091     & 0.713     \\ 
5.50    & 1.442    & 0.122     & 0.105     & 0.818     \\ 
6.50    & 1.512    & 0.104     & 0.111     & 0.862     \\ 
7.50    & 1.252    & 0.079     & 0.093     & 0.714     \\ 
9.00    & 1.093    & 0.067     & 0.082     & 0.623     \\ 
11.00\ \   & 1.071    & 0.111     & 0.083     & 0.611     \\ 
13.00\ \   & 1.232    & 0.211     & 0.098     & 0.702     \\ 
15.50\ \   & 0.926    & 0.268     & 0.076     & 0.528     \\ 

\end {tabular}
\end {center}
\end {table}
and~\ref {table_eta_RdA} 
\begin {table} [tbp]
\begin {center}
\caption {\normalsize Nuclear modification factor \RdA\ for \etameson\@.
Systematic errors classification given in Section~\ref {syst_classification}\@.
Normalization uncertainty of \MM{14.5\unitns{\%}} is not included.
}
\label {table_eta_RdA}
\begin {tabular} {ccccc}
\pT & \RdA & Statistical & \multicolumn{2}{c}{Systematic errors} \\
\MM{[\unitns{\GeVc}]} & & error & A & B \\
\hline
2.75    & 3.896    & 2.088     & 0.840     & 2.204     \\ 
3.50    & 0.898    & 1.345     & 0.218     & 0.508     \\ 
4.50    & 1.088    & 0.736     & 0.303     & 0.617     \\ 
5.50    & 1.939    & 0.578     & 0.612     & 1.100     \\ 
6.50    & 1.053    & 0.525     & 0.371     & 0.600     \\ 
7.50    & 1.302    & 0.456     & 0.507     & 0.742     \\ 
9.00    & 0.927    & 0.325     & 0.413     & 0.529     \\ 

\end {tabular}
\end {center}
\end {table}
and shown in Figure~\ref {fig_rda_star}\@.
Again, the definition of the errors is as given for the differential cross sections in Section~\ref {corrected_yields}\@.
Also shown in Figure~\ref {fig_rda_star} are the results of \RdA\ for charged pions measured by STAR~\cite {ref_star_idhadrons}\@.
A good agreement between STAR neutral and charged pions is observed.

\begin {figure} [p]
\centerline {\hbox {
\includegraphics [width=0.9\textwidth] {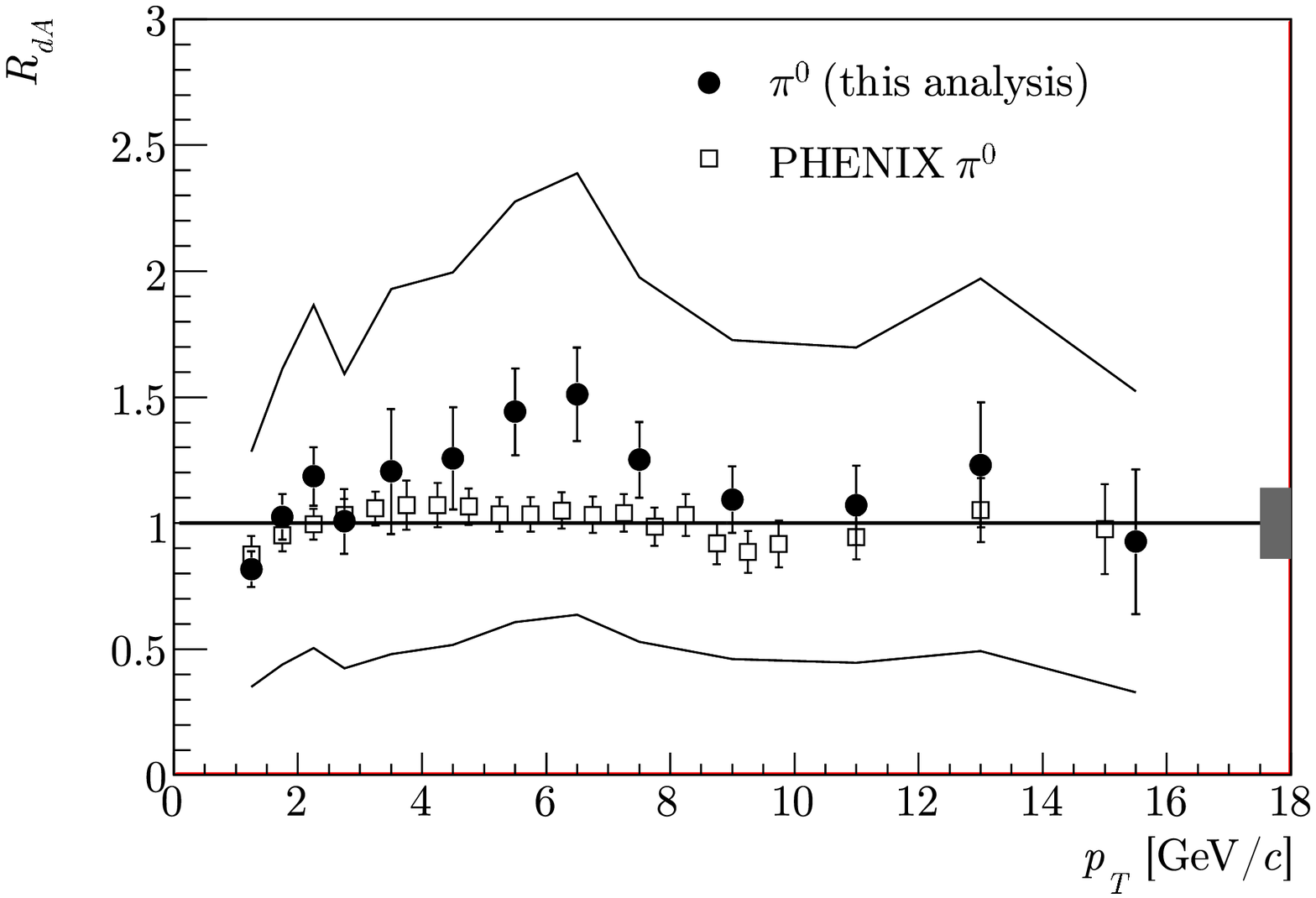}
}}
\centerline {\hbox {
\includegraphics [width=0.9\textwidth] {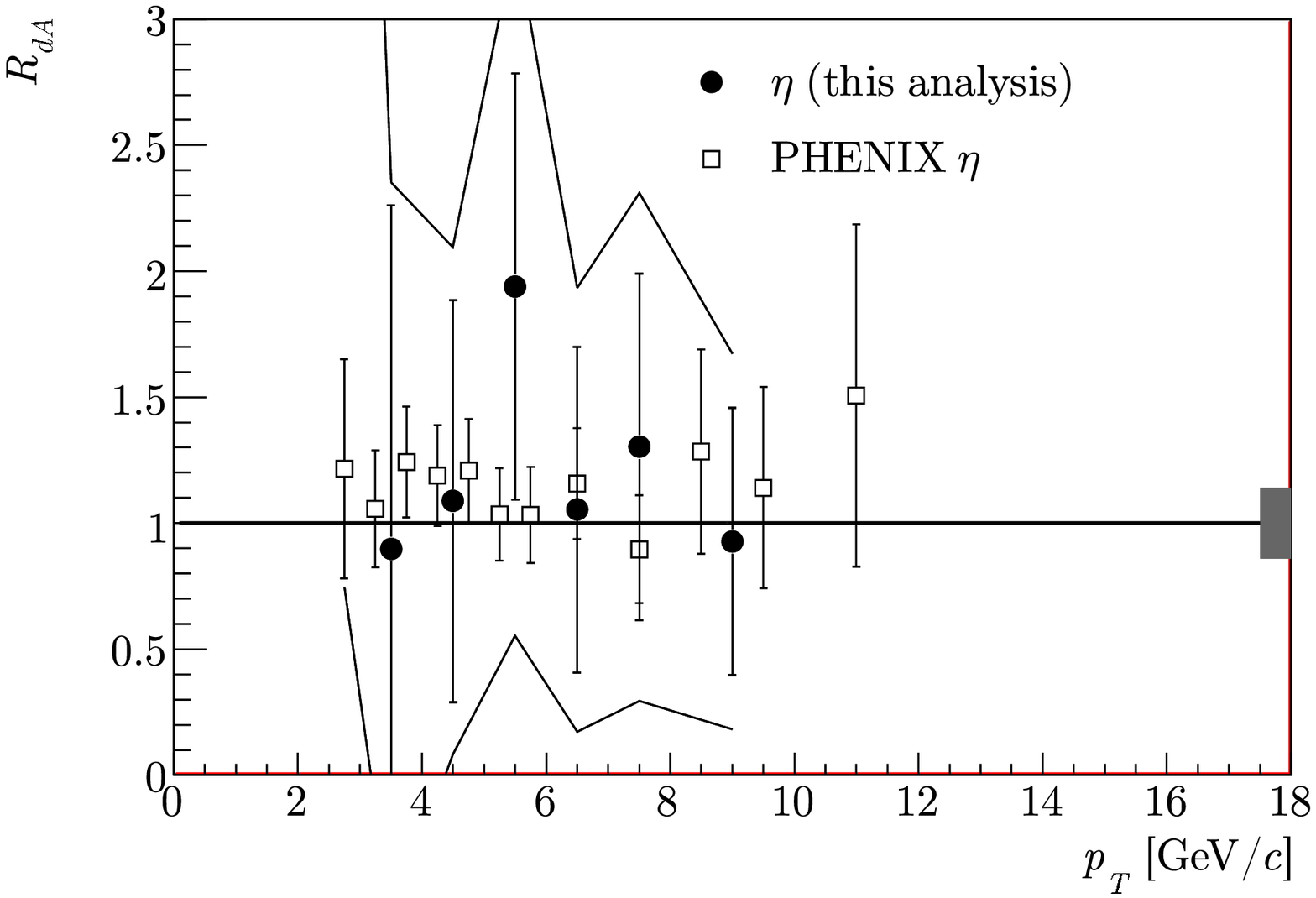}
}}
\caption {\RdA\ ratio for \pizero\ (top) and \etameson\ meson (bottom),
compared to the PHENIX measurements~\protect\cite {ref_phenix_eta,ref_phenix_eta_pi0_dAu}\@.
Errors are statistical and point\tsp{0.2}-to\tsp{0.4}-point systematic, 
excluding the energy calibration uncertainty shown as the outer lines.
Normalization uncertainty is indicated by a shaded band around unity.}
\label {fig_rda_phenix}
\end {figure}

In Figure~\ref {fig_rda_phenix}
we compare the \RdA\ ratio for \pizero\ (top panel) and \etameson\ (bottom panel)
to the corresponding PHENIX measurements~\cite {ref_phenix_eta,ref_phenix_eta_pi0_dAu}\@.
Our data agree reasonably well with PHENIX, 
except at \MM{\pT{} \LESS{} 8\unit{\GeVc}}, where the present results seem to be systematically higher by about \MM{30\unitns{\%}}\@.

\begin {table} [t!]
\begin {center}
\caption {\normalsize \mbox{Nuclear}~\mbox{modification}~\mbox{factor}~\Rcp~for~\pizero.~\mbox{Systematic}~\mbox{errors}~\mbox{classification} \protect\linebreak
given in Section~\ref {syst_classification}\@.
Normalization uncertainty of \MM{11.1\unitns{\%}} is not included.
}
\label {table_pi0_Rcp}
\begin {tabular} {ccccc}
\pT & \Rcp & Statistical & \multicolumn{2}{c}{Systematic errors} \\
\MM{[\unitns{\GeVc}]} & & error & A & B \\
\hline
1.25    & 1.032    & 0.094     & 0.073     & 0.000     \\ 
1.75    & 1.177    & 0.111     & 0.084     & 0.000     \\ 
2.25    & 1.265    & 0.164     & 0.090     & 0.000     \\ 
2.75    & 1.059    & 0.238     & 0.076     & 0.000     \\ 
3.50    & 1.211    & 0.610     & 0.087     & 0.000     \\ 
4.50    & 1.428    & 0.236     & 0.103     & 0.061     \\ 
5.50    & 1.153    & 0.151     & 0.084     & 0.049     \\ 
6.50    & 0.859    & 0.149     & 0.063     & 0.036     \\ 
7.50    & 1.119    & 0.106     & 0.083     & 0.079     \\ 
9.00    & 0.913    & 0.085     & 0.069     & 0.065     \\ 
11.00\ \   & 1.095    & 0.169     & 0.085     & 0.077     \\ 
13.00\ \   & 0.840    & 0.211     & 0.067     & 0.059     \\ 
15.50\ \   & 1.021    & 0.335     & 0.084     & 0.072     \\ 

\end {tabular}
\end {center}
\end {table}
\begin {figure} [t!]
\centerline {\hbox {
\includegraphics [width=0.9\textwidth,clip] {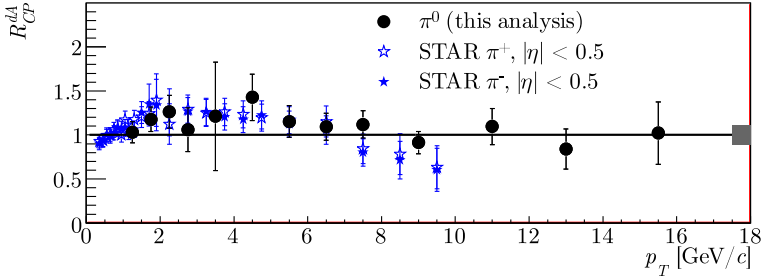}
}}
\caption {\Rcp\ ratio measured in \deuterongold\ collisions,
compared to STAR charged pions~\protect\cite {ref_star_idhadrons}\@.
Errors are statistical and point\tsp{0.2}-to\tsp{0.4}-point systematic.
Common normalization uncertainty is indicated by a shaded band around unity.}
\label {fig_rcp}
\end {figure}
The \Rcp\ ratio for \pizero\ is listed in Table~\ref {table_pi0_Rcp} 
and shown in Figure~\ref {fig_rcp} 
compared to the STAR charged pions~\cite {ref_star_idhadrons}\@.
It is seen that the agreement between the neutral and charged pion measurements in STAR is very good.
The ratio stays constant at a value consistent with unity beyond \MM{\pT{} = 8\unit{\GeVc}}\@.
The indication of a decrease from the charged pion data is not supported by this measurement.
\clearpage

\section {Conclusions and outlook}

There is a good agreement between the \pizero\ cross sections in \protonproton\ and \deuterongold\ collisions 
and nuclear modification factors measured in the present analysis,
and the charged pions previousely measured in STAR at \MM{\pT{} \LESS{} 10\unit{\GeVc}}\@.
This demonstrates a consistency between the charged and neutral pion results,
in spite of very different analysis methods and detectors (BEMC versus TPC) used for the measurements.
This analysis extends the \pT\ range of identified hadron measurements in STAR up to \MM{17\unit{\GeVc}}\@.
There is also a good agreement with the corresponding \pizero\ cross sections measured by the PHENIX experiment 
and with those calculated in NLO pQCD\@.

From the measurement of the nuclear modification factor~\RdA,~no~\mbox{suppression} \linebreak
of the \pizero\ production is seen in the 
\deuterongold\ collisions compared to \protonproton\ collisions.
This is in line with the observation made elsewhere~\cite {ref_star_dAu_evidence} that the large suppression seen in the 
central \goldgold\ collisions is due to the final state effects.

This analysis presents the first \etameson\ meson measurement in STAR\@.
The cross section, presented as an \etatopi\ ratio, is in agreement with the PHENIX measurement 
and with the \mT-scaling assumption.

There are several important and unique features in the present analysis.
First, the technique of estimating the low invariant mass background using the single photon simulation allows 
to remove the \pizero\ contamination at high \pT, where this type of background is indistinguishable from the signal.
A possible further improvement would require a better handle on the \SMD\ response simulation.
Second, the jet-aligned event mixing method very well reproduces the combinatorial background 
in the \pizero\ and \etameson\ peak region, so that no residual background subtraction is necessary.
This eliminates the systematic uncertainty usually related to a residual background parametrization.

It is seen that the experimental uncertainties can be significantly reduced by improving the calorimeter energy calibration.
Better measurement in the low \pT\ region would also require improvements in the simulation of the \SMD\ response.
Furthermore, calorimeter-based measurements using data taken since \yr{2006}\ will 
benefit from the full \MM{\etacoordabs{} \LESS{} 1} BEMC acceptance coverage.

In summary, these \protonproton\ and \deuterongold\ results provide a baseline measurement for the future \goldgold\ measurements.
These measurements are interesting to shed light on quark number scaling in particle production at intermediate \pT\ and 
to study the origin of suppression phenomena at large \pT\@.

\appendix

\chapter {BEMC electronics operation}
\label {BEMC_electronics_operation}

The tower phototubes are powered by Cockroft-$\!$Walton (CW) bases that are able to keep the high voltage up to a high precision.
The bases are programmed through the serial line from a dedicated computer in the Control Room.
The analog signals from the phototubes are routed to the tower digitizer crates mounted on the outer side of the magnet.

The tower digitizer crate contains five boards that take \MM{32} analog PMT inputs each and digitize it to \MM{12} bit on each RHIC bunch crossing, 
storing in the digital pipeline until a \levelZero\ trigger arrives.
The crate controller board then sends the data packets to the Tower Data Collector on the platform that feeds it to the DAQ.
The crate controller is also responsible for the slow control communication.

The STAR \levelZero\ trigger uses the BEMC data in the form of trigger primitives calculated by the tower digitizer boards, instead of the full tower data.
Two trigger primitives are calculated for each tower patch of \MM{0.2 \TIMES{} 0.2} in \MM{\etacoord{} \TIMES{} \phiangle} 
(\MM{4 \TIMES{} 4} towers) using pedestal subtracted tower \ADC:
\begin {itemize}
\item \emph {High Tower} - single largest tower signal in a patch
\item \emph {Patch Sum} - sum of all \MM{16} towers signal in a patch
\end {itemize}

\begin {figure} [tbp]
\centerline {\hbox {
\includegraphics [width=\textwidth] {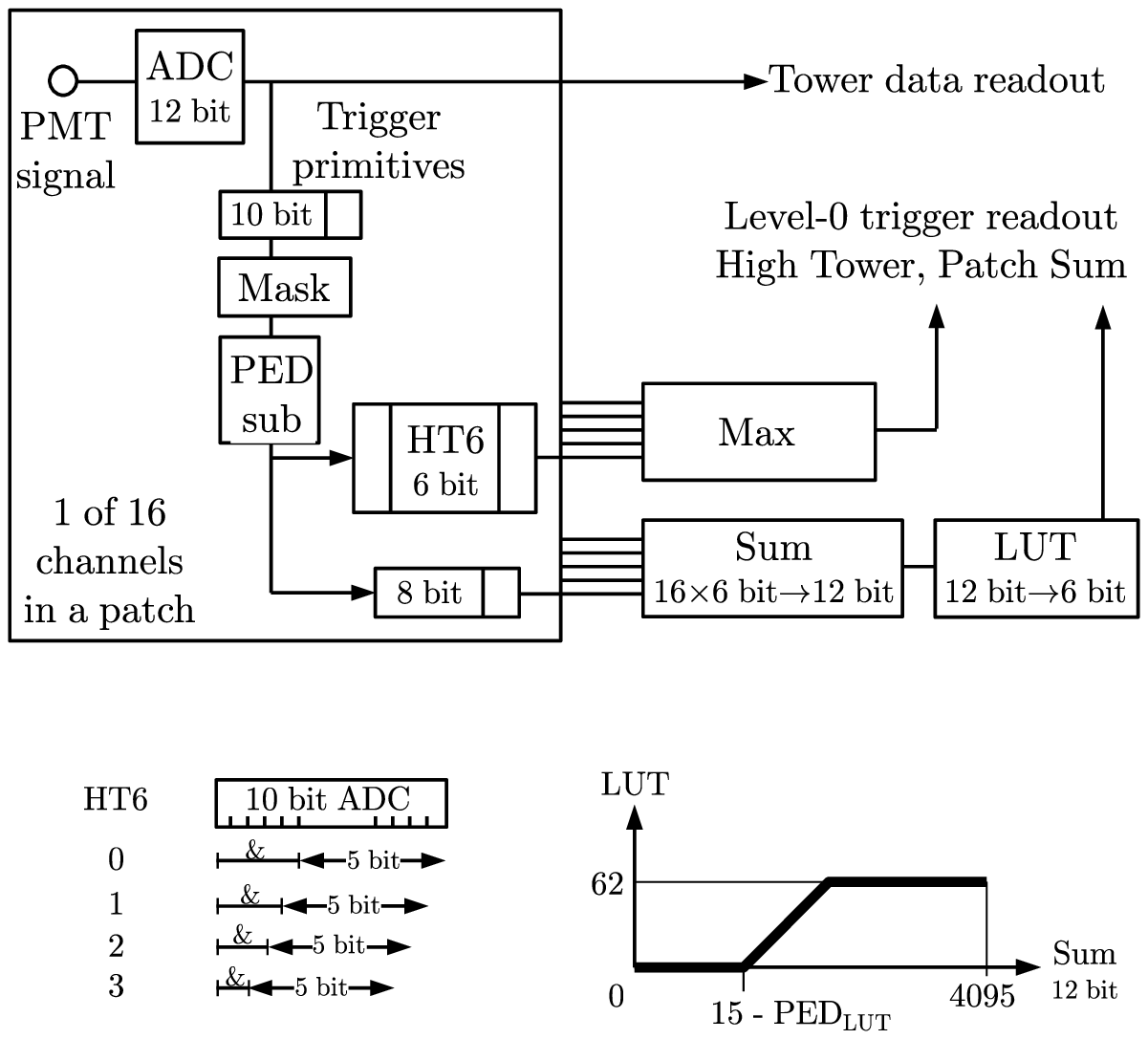}
}}
\caption {The digital processing in the tower digitizer boards.}
\label {fig_tower_adc}
\end {figure}

In the process of calculating those primitives, the on-board FPGA algorithm performs the following operations, as illustrated on Figure~\ref {fig_tower_adc}:
\begin {enumerate}
\item Drop the last \MM{2} bits of the tower \ADC, which becomes a \MM{10} bit signal.
\item Subtract the stored pedestal \PED\ from the \ADC, mask the channel out if necessary. The pedestals are calculated in a special way as described below.
\item For the High Tower: convert \MM{10} bits into \MM{6}, using one of four methods (\MUX\ selector), then select the largest value of all \MM{16} towers as output.
\item For the Patch Sum: drop the last \MM{2} bits to make it an \MM{8} bit value, then sum those from all \MM{16} channels into a \MM{12} bit value and 
transform it into the \MM{6} bit output value.
The transformation function has a special shape which is described in details later. Internally, it uses a lookup table (\LUT); 
the \MM{6} bit number stored in the \LUT\ is the output.
The PatchSum trigger sensitivity is, therefore, fixed:
$$
\ADCPatchSum{} = 16 \TIMES{} \ADCtower
.
$$
\end {enumerate}
Finally, two \MM{6} bit numbers are sent to the trigger Data Storage and Manipulation (DSM) boards upon recceiving a trigger signal.

The following \MUX\ methods are available to select \MM{6} of \MM{10} bits for the High Tower output with varying degree of sensitivity, 
almost equivalent to selecting a constant attenuation factor:
\begin {enumerate}
\item [(\MM{0})] Select \MM{6} lowest bits, combine five highest bits by logical ``and'' into the highest bit of the result. 
This is the most sensitive trigger setting, one HighTower trigger \ADC\ count is equal to \MM{4} raw tower \ADC\ counts.
\item [(\MM{1})] Select \MM{6} lower bits starting from \MM{1}, combine four highest bits by logical ``and'' into the highest bit of the result:
\MM{\ADCHighTower{} = 8 \TIMES{} \ADCtower}
\item [(\MM{2})] Select \MM{6} lower bits starting from \MM{2}, combine three highest bits by logical ``and'' into the highest bit of the result:
\MM{\ADCHighTower{} = 16 \TIMES{} \ADCtower}
\item [(\MM{3})] Select \MM{6} lower bits starting from \MM{3}, combine two highest bits by logical ``and'' into the highest bit of the result.
This is the least sensitive trigger setting, one HighTower trigger \ADC\ count is equal to \MM{32} raw tower \ADC\ counts.
\end {enumerate}

The tower pedestals and masks, the \MUX\ selectors, and the \LUT\ arrays are prepared and uploaded into the on-board registers via the slow control program.

The tower pedestals are being specially prepared in a way that puts the \MM{6} bit High Tower and Patch Sum pedestals at \MM{1} (not zero) to be observed during the run.
For each tower, the calculation starts from the exact value of the pedestal, which is measured by issuing the software triggers to FEE via slow control
in a periods between data taking when there is no beam in the machine.
The global ``pedestal shift'' variable \PedestalShift\ gets subtracted from the tower pedestal, in order to center the pedestal-subtracted tower signal at \PedestalShift\@.
Finally, the pedestal gets rounded to the nearest multiple of four and two last bits are removed.
The last four bits of the result are used, together with the sign, to fill the \MM{5} bit pedestal register \PED\ in the FEE.

During \yr{2003}\ data taking run, the pedestal subtraction scheme was not yet \linebreak
implemented in FEEs, and the HighTower sensitivity was chosen to be \MM{\MUX = 3}\@. \linebreak
For the \yr{2005}\ data, the settings were \MM{\PedestalShift{} = 24} and \MM{\MUX{} = 2}, which defined \MM{\ADCHighTower{} = \ADCPatchSum{} = 16 \TIMES{} \ADCtower}
and thus aligned the HighTower and PatchSum readings at the center of bin \MM{1} in the absense of tower physics signal.

The \LUT\ arrays are prepared in a way that gives a linear response to the patch sum in the range from \MM{0} to \MM{62}, to allow diagnosing the broken
cables by observing ``all ones'' bit pattern \MM{63} during the run.
With the nominal setting of \MM{\PedestalShift{} = 24}, each tower contributes a pedestal value \MM{\DIV{24}{16} = 1} to the sum, so the \LUT\ is constructed from the following three pieces:

\begin {itemize}

\item \emph {Zero}, if the sum of \MM{16} towers is below \MM{16}

\MM{\LUT(\IT{s}) = 0, \quad\quad\quad\  0 \LESSEQ{} \IT{s} \LESS{} 16}

\item \emph {Linear rise} in response to the sum in the \MM{6} bit range, excluding \MM{63}

\MM{\LUT(\IT{s}) = \IT{s} - 15, \quad 16 \LESSEQ{} \IT{s} \LESS{} 16 + 63}

\item \emph {Saturation} at \MM{62}

\MM{\LUT(\IT{s}) = 62, \quad\quad\quad\ensuremath{\!} \IT{s} \GREATEREQ{} 16 + 63}

\end {itemize}

If a tower is masked out of the PatchSum trigger, the \LUT\ is modified to accomodate the loss of its pedestal, in this case it starts rising one count earlier
than the nominal \MM{16}. Therefore, the so\tsp{0.3}-called \LUT\ pedestal (\LUTPED) is equal to the number of masked towers in a patch.

The \SMD\ electronics (FEE) board is mounted on the \MM{\etacoordabs{} = 1} side of each module.
At the FEE board, the amplified cathode strip signals are buffered in a switched capacitor array (SCA)
before being delivered to external digitizer boards outside of the STAR magnet.

The signals from the pads of the \SMD\ are amplified and stored in an analog pipeline, composed of switched capacitor arrays, to await the \levelZero\ trigger. 
Upon \levelZero\ trigger, the \SMD\ analog signals are queued with multiplexing ratio of \MM{80:1} to the \MM{10}-bit \SMD\ digitizers. 
\SMD\ digitized signals are first available in STAR \levelTwo\ trigger processors in \MM{200\unit{\mus}}, still well ahead of digital information from the TPC.

The digitizing electronic boards and crates for preshower detector are identical to the ones used in the \SMD.

\begin {figure} [p]
\centerline {\hbox {
\includegraphics [width=1.1\textwidth] {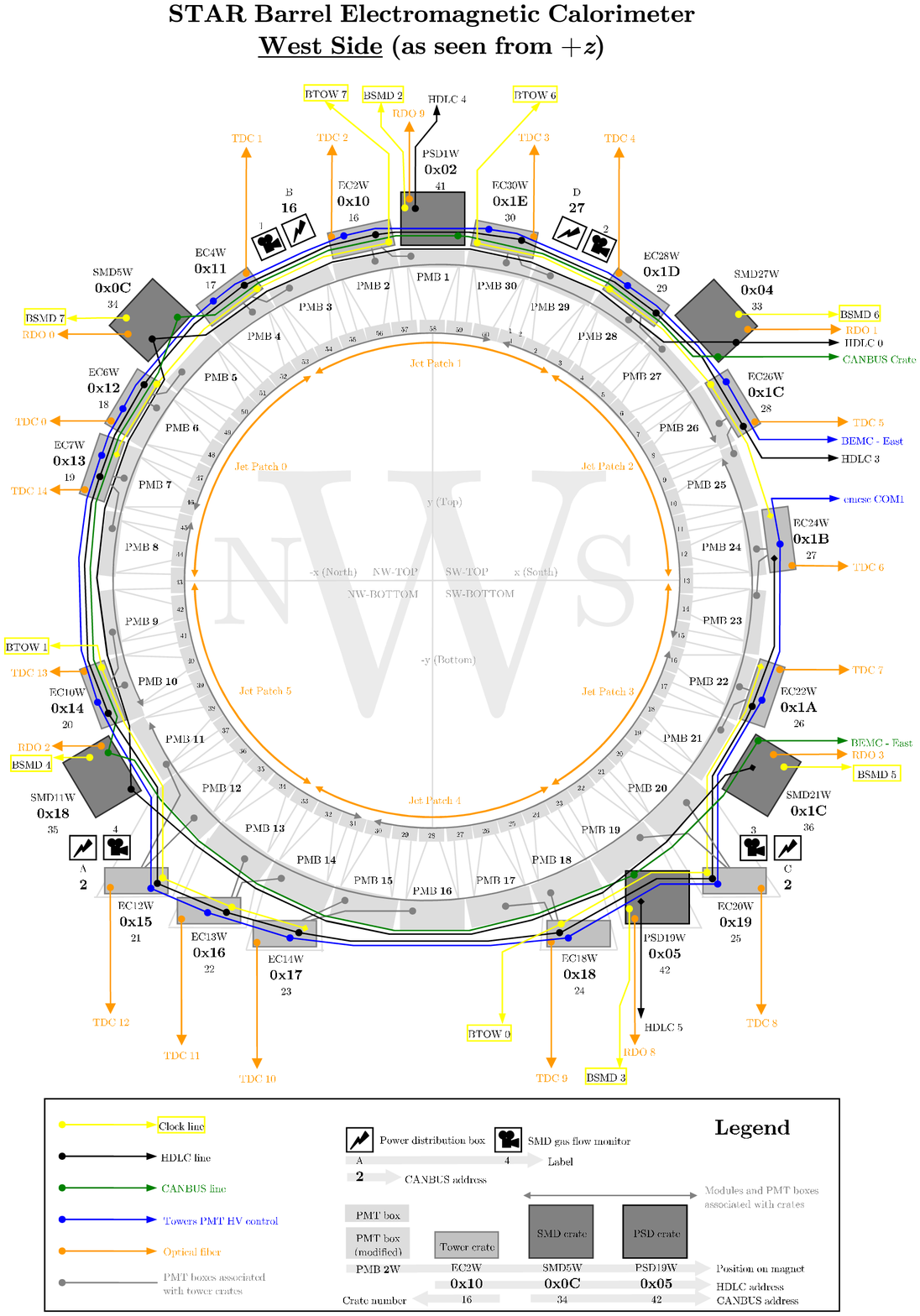}
}}
\label {fig_bemc_view6}
\end {figure}

\phantomsection
\addcontentsline{toc}{chapter} {Bibliography}
\bibliography {bibliography}
\markboth {} {\textsl{BIBLIOGRAPHY}\hfill}

\chapter* {Summary}
\addcontentsline{toc}{chapter} {Summary}
\markboth {} {}
This thesis presents a measurement of neutral pion and eta meson production in \protonproton\ and
\deuterongold\ collisions at a center-of-mass energy of \MM{\sNN{} = 200\unit{\GeV}}, 
measured with STAR detector at Relativistic Heavy Ion Collider (RHIC) in Brookhaven National Laboratory (BNL, USA).

The present neutral pion spectrum complements that of the charged pions measured in STAR
in the transverse momentum range \MM{0.35 \LESS{} \pT{} \LESS{} 10\unit{\GeVc}} and extends up to \MM{\pT{} = 17\unit{\GeVc}}.
There is a good agreement between the \linebreak
neutral and charged pion cross sections in STAR, 
in spite of very different methods and detectors used in the analysis.
The neutral pion cross section also agrees well with the measurements of PHENIX, another large detector at RHIC, 
and with the theoretical NLO pQCD calculations.

This thesis also presents the first measurements by STAR of eta meson production, 
which are in agreement with the PHENIX measurements and with the \mT-scaling assumption.

Possible medium-induced modifications of particle production in a nucleus-nucleus collision, 
compared with an incoherent superposition of nucleon-nucleon collisions, 
can be observed by measuring the so-called nuclear modification \linebreak
factor \IT{R}.
This thesis presents the measurements of \RdA, where the neutral pion production in \deuterongold\ collisions is 
compared to that in the \protonproton, and \Rcp, the comparison between central and peripheral \deuterongold\ collisions.
Both results are in a good agreement with the charged pion measurements previousely done by STAR.

The \protonproton\ and \deuterongold\ results presented here 
provide a baseline measurement for the future \goldgold\ measurements.
These measurements are interesting to shed light on 
quark number scaling in particle production at intermediate \pT\ and
to study the origin of suppression phenomena at large \pT.

\chapter* {Samenvatting}
\addcontentsline{toc}{chapter} {Samenvatting}
\markboth {} {}
Dit proefschrift beschrijft metingen van neutrale pion en eta meson productie in 
\protonproton\ en \deuterongold\ botsingen bij een
zwaartepuntsenergie van \MM{\sNN{} = 200\unit{\GeV}}, gemeten met de STAR detector bij de
Relativistic Heavy Ion Collider (RHIC) in Brookhaven National Laboratory (BNL, USA).

Het neutrale pion spectrum is complementair aan dat van de geladen pionen gemeten
met STAR in het transversale impuls gebied \MM{0.35 \LESS{} \pT{} \LESS{} 10\unit{\GeVc}} 
en loopt door tot \MM{\pT{} = 17\unit{\GeVc}}. 
Er is een goede overeenstemmingtussen de neutrale en geladen werkzame doorsnede in
STAR, ondanks de zeer verschillende methoden en detectoren die gebruikt zijn in de analyse. De
gemeten werkzame doorsnede van de \pizero\ stemt ook goed overeen met de metingen van PHENIX, een andere detector
bij RHIC, en met NLO pQCD berekeningen.

Dit proefschrift beschrijft ook de eerste metingen met STAR van eta meson productie,
die in overeenstemming zijn met de PHENIX metingen en met de \mT-schaling aanname.

De door het medium geinduceerde verandering van deeltjes productie in een kern-kern
botsing, vergeleken met incoherente superpositie van individuele \linebreak
nucleon-nucleon botsingen, kan waargenomen
worden door de zogenaamde \linebreak
``nuclear modification factor'' \IT{R} te meten. Dit proefschrift beschrijft de metingen
van \RdA, waar \pizero\ productie in \deuterongold\ botsingen vergeleken worden met die 
in \protonproton\ botsingen, en \Rcp, de vergelijking tussen centrale 
en perifere \deuterongold\ botsingen. Beide resultaten zijn in goede overeenstemming met
eerdere metingen aan geladen pionen met STAR.

De \protonproton\ en \deuterongold\ resultaten die hier gepresenteerd worden zijn een referentie voor
toekomstige \goldgold\ metingen. Deze metingen zijn interessant omdat ze licht werpen op de schaling
met het aantal quarks in deeltjes productie bij intermediaire \pT\ en voor de studie
naar de oorsprong van onderdrukkingsfenomenen bij hoge \pT.

\chapter* {Acknowledgments}
\addcontentsline{toc}{chapter} {Acknowledgments}
\markboth {} {}

This thesis would not be finished without saying thanks to people who helped me.
Our research group at NIKHEF and Utrecht University created a friendly but demanding atmosphere,
it was a real pleasure to be part of it for the last four years.

Help from my promotor Thomas Peitzmann and co\tsp{0.3}-promotor Michiel Botje 
and a good advise from Raimond Snellings certainly made my work more \linebreak
enjoyable and the research fruitful.
I am especially thankful to Michiel for the guidance in the preparation of this thesis, 
it would be almost unreadable without his extensive comments.

Andre Mischke provided the strong support at the beginning of this work and stayed interested in it ever after.
The fellow student Martijn Russcher, doing the research on a related subject, 
was very helpful in discussing the details of the analysis.

Participating in a large experiment, such as STAR, is a very special experience.
Here I would like to name people with whom I have spent more than a year at Brookhaven Lab in \yr{2004}--\yr{2005}\@.
Stephen Trentalange, Oleg Tsai and Alexander Stolpovsky provided a very good and productive working environment,
which was also an excellent learning place.

On a separate note, I want to mention here my alma mater, Kiev National University in Ukraine, 
where I received the master's degree five years ago.
In these early days Igor Kadenko and Gennady Zinovjev helped me to choose the path of science, 
for which I will always be grateful.

\end {document}